%% file: paper-sls-3-rev.tex
\documentclass[twoside,12pt]{article}
\usepackage{graphicx}
\usepackage{lineno,hyperref}
\modulolinenumbers[1]
\usepackage{amsmath}
\usepackage{relsize} 
\usepackage{afterpage}
\usepackage{rotating}
\usepackage{slashed}
\usepackage{amssymb}
\usepackage{color}
\definecolor{revcolor}{rgb}{0,0,0}
\usepackage{float}

\usepackage{ifthen} 
\newboolean{uprightparticles}
\setboolean{uprightparticles}{false} 
\usepackage{xspace}
\input{lhcb-symbols-def} 

\def\PDF{\mathcal{P}}
\def\Mat{\mathcal{M}}
\def\PDFt{\mathcal{F}_t}
\def\H{{\cal H}}
\def\ZP{P_c}
\def\twolnL {\ensuremath{-2\ln{\cal L}}\xspace}
\newcommand{\Nstar}{\ensuremath{N^{*}}\xspace}

\usepackage{cite} 
\usepackage{mciteplus}
\def\Q{{\mathcal{Q}}}
\def\P{{\mathcal{P}}}
\def\bm{\boldmath}
\def\tbl{\caption}
\def\Hline{%
  \noalign{\ifnum0=`}\fi\hrule \@height 2\arrayrulewidth \futurelet
   \@tempa\@xhline}
   
\def\Luescher{L\"{u}scher}
\def\Meissner{Mei{\ss}ner}
\def\Tornqvist{T\"{o}rnqvist}

\newcommand{\BaBar}{\mbox{\slshape B\kern -0.1em{\smaller A}\kern-0.1em B\kern-0.1em{\smaller A\kern -0.2em R }}}
\begin{document}


\title{{\hspace{5.2in}\normalsize DESY~17-071}\\ \vspace{1cm}Exotics: Heavy Pentaquarks and Tetraquarks}
\author{{Ahmed Ali,$^1$ Jens S\"oren Lange,$^2$ and Sheldon Stone$^3$}\\
\small{$^1${DESY Theory Group, Notkestrasse 85, D-22607 Hamburg, Germany}}\\
\small{$^2${Justus-Liebig-Universit\"at Giessen, II.\ Physikalisches Institut, Heinrich-Buff-Ring 16, }}\\
 \small{D-35392 Giessen, Germany}\\
\small{$^3${Physics Department, 201 Physics Building, Syracuse University, Syracuse, NY 13244-1130, USA}}}
\maketitle

\begin{abstract}
For many decades after the invention of the quark model in 1964 there was no evidence that hadrons are formed from anything other than the simplest pairings of quarks and antiquarks, mesons being formed of a quark-antiquark pair and baryons from three quarks. In the last decade, however, in an explosion of data from both $e^+e^-$ and hadron colliders, there are many recently observed states that do not fit into this picture. These new particles are called generically ``exotics". They can be either mesons or baryons. Remarkably, they all decay into at least one meson formed of either a $c\overline{c}$ or $b\overline{b}$ pair.  In this review, after the introduction,  we explore each of these new discoveries in detail first from an experimental point of view, then subsequently give a theoretical discussion. These exotics can be explained if the new mesons contain two-quarks and two-antiquarks (tetraquarks), while the baryons contain four-quarks plus an antiquark (pentaquarks). The theoretical explanations for these states take three divergent tracks: tightly bound objects, just as in the case of normal hadrons, but with more constituents, or loosely bound ``molecules" similar to the deuteron, but formed from two mesons, or a meson or baryon, or more wistfully, they are not multiquark states but appear due to kinematic effects caused by different rescatterings of virtual particles; most of these models have all been post-dictions. Both the tightly and loosely bound models predict the masses and related quantum numbers of new, as yet undiscovered states. Thus, future experimental discoveries are needed along with theoretical advances to elucidate the structure of these new exotic states.

\end{abstract}
\setcounter{page}{1}
\tableofcontents
\clearpage


\input{Sec-Introduction-rev}

\input{Sec-Conventional-spectroscopy-rev}

\input{Sec-Experimental-evidence-for-pentaquarks}

\input{Sec-Experimental-evidence-for-tetraquarks-rev}

\input{Sub-sec-4140}

\input{Sec-Theoretical-models-for-tetraquarks-rev}

\input{Sec-Theoretical-models-for-pentaquarks-rev}

\input{Sec-Acknowledgments}

\clearpage



\input{paper-sls-3-rev.bbl}
\end{document}

%% file: lhcb-symbols-def.tex
\def\MagUp {\mbox{\em Mag\kern -0.05em Up}\xspace}

\ifthenelse{\boolean{uprightparticles}}%
{
 
 \def\Pgamma      {\ensuremath{\upgamma}\xspace}

 \def\Ppi         {\ensuremath{\uppi}\xspace}

 \def\Ppsi        {\ensuremath{\uppsi}\xspace}

 \def\PDelta      {\ensuremath{\Delta}\xspace}                 
 \def\PXi      {\ensuremath{\Xi}\xspace}                 
 \def\PLambda      {\ensuremath{\Lambda}\xspace}                 
 \def\PSigma      {\ensuremath{\Sigma}\xspace}                 
 \def\POmega      {\ensuremath{\Omega}\xspace}                 
 \def\PUpsilon      {\ensuremath{\Upsilon}\xspace}                 
 
 \def\PB      {\ensuremath{\mathrm{B}}\xspace}                 
                  
 \def\PD      {\ensuremath{\mathrm{D}}\xspace}

 \def\PJ      {\ensuremath{\mathrm{J}}\xspace}                 
 \def\PK      {\ensuremath{\mathrm{K}}\xspace}

 \def\PZ      {\ensuremath{\mathrm{Z}}\xspace}                 
                  
 \def\Pb      {\ensuremath{\mathrm{b}}\xspace}

 \def\Ps      {\ensuremath{\mathrm{s}}\xspace}

}
{
 
 \def\Pgamma      {\ensuremath{\gamma}\xspace}

 \def\Ppi         {\ensuremath{\pi}\xspace}

 \def\Ppsi        {\ensuremath{\psi}\xspace}                 
                  
 \mathchardef\PDelta="7101
 \mathchardef\PXi="7104
 \mathchardef\PLambda="7103
 \mathchardef\PSigma="7106
 \mathchardef\POmega="710A
 \mathchardef\PUpsilon="7107
                  
 \def\PB      {\ensuremath{B}\xspace}                 
                  
 \def\PD      {\ensuremath{D}\xspace}

 \def\PJ      {\ensuremath{J}\xspace}                 
 \def\PK      {\ensuremath{K}\xspace}

 \def\PZ      {\ensuremath{Z}\xspace}                 
                  
 \def\Pb      {\ensuremath{b}\xspace}

 \def\Ps      {\ensuremath{s}\xspace}

}

\makeatletter
\ifcase \@ptsize \relax
  \newcommand{\miniscule}{\@setfontsize\miniscule{4}{5}}
\or
  \newcommand{\miniscule}{\@setfontsize\miniscule{5}{6}}
\or
  \newcommand{\miniscule}{\@setfontsize\miniscule{5}{6}}
\fi
\makeatother

\DeclareRobustCommand{\optbar}[1]{\shortstack{{\miniscule (\rule[.5ex]{1.25em}{.18mm})}
  \\ [-.7ex] $#1$}}

\def\g      {{\ensuremath{\Pgamma}}\xspace}
\def\H      {{\ensuremath{\PH^0}}\xspace}

\def\Z      {{\ensuremath{\PZ}}\xspace}

\def\squark    {{\ensuremath{\Ps}}\xspace}

\def\bquark    {{\ensuremath{\Pb}}\xspace}

\def\pion   {{\ensuremath{\Ppi}}\xspace}

\def\pip    {{\ensuremath{\pion^+}}\xspace}

\def\kaon    {{\ensuremath{\PK}}\xspace}
  \def\Kbar    {{\kern 0.2em\overline{\kern -0.2em \PK}{}}\xspace}

\def\KorKbar    {\kern 0.18em\optbar{\kern -0.18em K}{}\xspace}

\def\Km      {{\ensuremath{\kaon^-}}\xspace}

\def\KS      {{\ensuremath{\kaon^0_{\rm\scriptscriptstyle S}}}\xspace}

  \def\Dbar    {{\kern 0.2em\overline{\kern -0.2em \PD}{}}\xspace}
\def\D       {{\ensuremath{\PD}}\xspace}

\def\DorDbar    {\kern 0.18em\optbar{\kern -0.18em D}{}\xspace}

\def\Dstarb  {{\ensuremath{\Dbar{}^*}}\xspace}

\def\Bbar    {{\ensuremath{\kern 0.18em\overline{\kern -0.18em \PB}{}}}\xspace}

\def\BorBbar    {\kern 0.18em\optbar{\kern -0.18em B}{}\xspace}

\def\Bzb     {{\ensuremath{\Bbar{}^0}}\xspace}

\def\Bsb     {{\ensuremath{\Bbar{}^0_\squark}}\xspace}
\def\Bdb     {{\ensuremath{\Bbar{}^0}}\xspace}

\def\jpsi     {{\ensuremath{{\PJ\mskip -3mu/\mskip -2mu\Ppsi\mskip 2mu}}}\xspace}

  \def\Y#1S{\ensuremath{\PUpsilon{(#1S)}}\xspace}

\def\Lz          {{\ensuremath{\PLambda}}\xspace}
\def\Lbar        {{\ensuremath{\kern 0.1em\overline{\kern -0.1em\PLambda}}}\xspace}
\def\LorLbar    {\kern 0.18em\optbar{\kern -0.18em \PLambda}{}\xspace}

\def\Lb      {{\ensuremath{\Lz^0_\bquark}}\xspace}
\def\Lbbar   {{\ensuremath{\Lbar{}^0_\bquark}}\xspace}

\def\to                 {\ensuremath{\rightarrow}\xspace}

\def\CP                {{\ensuremath{C\!P}}\xspace}

\def\AT#1     {\ensuremath{A_{\mathrm{T}}^{#1}}\xspace}           

\def\C#1      {\ensuremath{\mathcal{C}_{#1}}\xspace}                       
\def\Cp#1     {\ensuremath{\mathcal{C}_{#1}^{'}}\xspace}                    
\def\Ceff#1   {\ensuremath{\mathcal{C}_{#1}^{\mathrm{(eff)}}}\xspace}        
\def\Cpeff#1  {\ensuremath{\mathcal{C}_{#1}^{'\mathrm{(eff)}}}\xspace}       
\def\Ope#1    {\ensuremath{\mathcal{O}_{#1}}\xspace}                       
\def\Opep#1   {\ensuremath{\mathcal{O}_{#1}^{'}}\xspace}                    

\newcommand{\ket}[1]{\ensuremath{|#1\rangle}}              

\newcommand{\tev}{\ensuremath{\mathrm{\,Te\kern -0.1em V}}\xspace}
\newcommand{\gev}{\ensuremath{\mathrm{\,Ge\kern -0.1em V}}\xspace}
\newcommand{\mev}{\ensuremath{\mathrm{\,Me\kern -0.1em V}}\xspace}
\newcommand{\kev}{\ensuremath{\mathrm{\,ke\kern -0.1em V}}\xspace}
\newcommand{\ev}{\ensuremath{\mathrm{\,e\kern -0.1em V}}\xspace}
\newcommand{\gevc}{\ensuremath{{\mathrm{\,Ge\kern -0.1em V\!/}c}}\xspace}
\newcommand{\mevc}{\ensuremath{{\mathrm{\,Me\kern -0.1em V\!/}c}}\xspace}
\newcommand{\gevcc}{\ensuremath{{\mathrm{\,Ge\kern -0.1em V\!/}c^2}}\xspace}
\newcommand{\gevgevcccc}{\ensuremath{{\mathrm{\,Ge\kern -0.1em V^2\!/}c^4}}\xspace}
\newcommand{\mevcc}{\ensuremath{{\mathrm{\,Me\kern -0.1em V\!/}c^2}}\xspace}

\def\gsim{{~\raise.15em\hbox{$>$}\kern-.85em
          \lower.35em\hbox{$\sim$}~}\xspace}
\def\lsim{{~\raise.15em\hbox{$<$}\kern-.85em
          \lower.35em\hbox{$\sim$}~}\xspace}

\def\PDF {PDF\xspace}

\def\sPlot{\mbox{\em sPlot}\xspace}

\def\pt         {\mbox{$p_{\rm T}$}\xspace}

\def\tell1  {TELL1\xspace}
\def\ukl1   {UKL1\xspace}

\newcommand{\eg}{\mbox{\itshape e.g.}\xspace}
\newcommand{\ie}{\mbox{\itshape i.e.}\xspace}

%% file: Sec-Introduction-rev.tex
\section{Introduction}
\label{sec:Introduction}

When the two $B$ factory experiments \BaBar and Belle were proposed, well over a quarter century ago, their main
goal was to measure the matter-antimatter asymmetry in the $B$-meson sector.\footnote{They followed the exploratory work done by the ARGUS and CLEO experiments operating in the $\Upsilon$ resonance region.}
This goal they indeed accomplished very well, together with other related key  measurements in flavor physics,  which 
greatly helped in testing the charged current weak interactions in the standard model \cite{Bevan:2014iga}. 
While in depth studies of hadronic structure was not on the agenda of these experiments, they also succeeded in uncovering a new facet of QCD.
The discovery of the exotic hadron $X(3872)$ in 2003 in the decays $B^0 \to \jpsi\pi^+\pi^-  K$ by Belle \cite{Choi:2007wga} came as a complete surprise.\footnote{Mention of a specific decay mode implies use of the charge-conjugate mode as well.}  Confirmation soon thereafter by  \BaBar \cite{Aubert:2004ns,Aubert:2008gu}, 
CDF II \cite{Acosta:2003zx,Abulencia:2005zc,Aaltonen:2009vj}, D0 \cite{Abazov:2004kp}, and later by the 
LHCb \cite{x3872lhcb} and CMS \cite{x3872cms}, established  the $X(3872)$ as a genuine resonance, as opposed to a threshold effect (called a ``cusp"). Its discovery turned out to be 
 the harbinger of a new direction in hadron physics. Thus, out of the debris of $B$-meson decays emerged a second layer of 
strongly interacting particles called tetraquarks (four-quark states), all containing a $c\overline{c}$ quark pair. Likewise, in the decays of the
$\Lambda_b$-baryon, LHCb later established the existence of two pentaquark states $P_c^+(4380)$ and
$P_c^+(4450)$ in the $J/\psi p$ decay mode \cite{Aaij:2015tga}. As the minimum valence quark content of such a state is
$c\bar{c} uud$, the newly discovered baryons have five quarks.
Meanwhile, well over two dozen exotic hadrons have found entries in the Particle Data Group  \cite{pdg}. Tentatively called  $X$, $Y$, $Z$ and $P_c$, they have various $J^{PC}$ quantum numbers, and come both as charged and neutrals, as discussed in detail in this review
and elsewhere \cite{Lebed:2016hpi,Esposito:2016noz,Guo:2017jvc,Chen:2016qju}. At least two $b\bar{b}$ counterparts
of the $c\bar{c}$ states, $Z_b^+(10610)$ and $Z_b^+(10650)$, have also been reported by Belle  \cite{Adachi:2011ji}.  
Their valence quark content is $b\bar{b} u\bar{d}$. A closely related question is whether
tetraquarks and pentaquarks also come with a single charm or bottom valence quark, was reviewed recently  \cite{Chen:2016spr}.

Quarkonium physics is a well-studied system theoretically, having its roots in the non-relativistic quarkonium potential \cite{Eichten:1978tg,Eichten:1979ms}.  Its present formulation is in the context of effective field theories based on QCD \cite{Brambilla:2004wf,Brambilla:2010cs} and lattice-QCD 
\cite{Gray:2005ur,Dowdall:2011wh,Bali:2011rd,Liu:2012ze}. The progress in lattice-based techniques
is impressive and allows us to connect the observed hadronic properties  with the fundamental parameters in QCD.
In particular, hadron spectra below the strong decay thresholds are calculated accurately, and extensive results
for multiplets are available. However, so far there are no reliable lattice results for flavor exotic states, though possible $ c\bar{c}\bar{d}u$ candidates with $ J^{PC}=1^{+-} $ have been searched for in the vicinity of the $D\bar{D}^*$
 threshold \cite{Ikeda:2016zwx},
where the $Z_c^+(3900)$ is found experimentally. No $c\bar{c}s\bar{s}$ resonance has been found in the $J/\psi \phi$
scattering on the lattice, though their experimental evidence has mounted, 
and no $cc \bar{d}\bar{u}$ bound state is found in the $DD^*$ scattering either  \cite{Prelovsek:2014zga}. We briefly review the current lattice simulations in 
section~\ref{sec:Spectroscopy}, but we shall be mostly using the available phenomenological
approaches for the theoretical discussion of tetraquarks and pentaquarks in this review. 

For the spectra of the $c\bar{c}$ and $b\bar{b}$ bound states below their corresponding open flavor thresholds, $D\bar{D}$ and $B\bar{B}$, potential models are reliable, as they successfully reproduce the observed 1S, 2S, and 1P states, in the $c\bar{c}$ -, and many more higher
states in the $b\bar{b}$-sector. The most popular of these, the Cornell potential 
\cite{Eichten:1978tg,Eichten:1979ms}, incorporates a color Coulomb term at short distances and a linear confining term at large distances, with the spin-dependent interquark
 interaction responsible for the splitting of the states in these multiplets governed by a one-gluon exchange Breit-Fermi
 Hamiltonian. The quarkonium-potential-based studies were substantially extended to cover the $c\bar{c}$ and $b\bar{b}$ states above the open
 flavor thresholds (see, for example, Ref. \cite{BarnesGodfreySwanson}). The concordance between experiments and such theoretical estimates is remarkable in both the charm and bottom sectors,
 predicting correctly a large number of observed states. While empirical, 
 these potential-model based studies provide useful benchmarks for the quarkonia states, as they provide the background upon which exotic particles are to be searched for and  characterized. 
 
 What are the criteria for an observed hadron to be termed as exotic?
  The clues to novel features in QCD may come in a  number of ways which we discuss here briefly.
  For example, the exotic hadrons  may have
 unconventional $J^{PC}$ quantum numbers, such as having $J^{PC}=0^{--},\; 0^{+-}, \;1^{+-}$,
 which are not allowed for non-relativistic $q\bar{q}$ states. Also, more hadrons  with the same $J^{PC}$
 quantum numbers could be  discovered  than are allowed by the quark model counting. The spectrum of the
 charmonium states around 4~GeV is particularly interesting in this context, as it contains, in addition to the anticipated
 conventional $c\bar{c}$ states, such as $\psi(3770)$, $\psi(3823)$, the states $X(3872)$ ($J^{PC}=1^{++}$),
 $X(3940)$   ($J^{PC}=0^{++}$), the $Z_c^0(3900)$ ($J^{PC}=1^{+-}$), and the $X(3940)$, whose $J^{PC}$
quantum numbers have yet to be determined. At least some of them are good candidates for non-$q\bar{q}$ states.
They may also show systematic different patterns in their intrinsic properties, such as the spin-spin 
and spin-orbit interactions, compared to what is known for the quarkonium multiplets.  
The charged states, such as  $Z_c^\pm(3900)$, $Z_c^\pm(4020)$, $Z_b^\pm(10610)$ and $Z_b^\pm(10650)$,
are manifestly non-$q\bar{q}$ due to the minimum valence quark composition required for their discovery modes.
  They may also exhibit
   stark differences in their decay characteristics, such as the total and partial widths as well as their final
   state profiles,  compared to the other similar and well-understood systems. The dipion invariant mass
    distributions in $\Upsilon(10860) \to \Upsilon(nS) \pi^+ \pi^-$, with $nS=1S, 2S, 3S$,  as well as the $\Upsilon(nS) \pi^\pm$  invariant mass distribution \cite{Belle:2011aa}, are two cases in point. Dalitz plot studies in these decays reveal unusual structures in both these channels. The decay rate and dipion mass spectrum in the
  transition of the lower $\Upsilon$ states, such as $\Upsilon(4S) \to \Upsilon(1S) \pi^+ \pi^-$, on the other hand, are well described by the QCD multipole
   expansion \cite{Brown:1975dz, Voloshin:1975yb}, with no structure seen in the Dalitz distributions, as expected
   from Zweig-forbidden transitions. In line with this,
   the partial  decay width $\Gamma(\Upsilon(4S) \to \Upsilon(1S) \pi^+ \pi^-)$ as well as 
   $\Gamma(\Upsilon(4S) \to \Upsilon(2S) \pi^+ \pi^-)$ and  
   $\Gamma(\Upsilon(nS) \to \Upsilon(mS) \pi^+ \pi^-)$, involving the pairs $(3S,1S)$, $(3S,2S)$, $(2S,1S)$, are all
   of O(1) keV. They are typically two orders of magnitude smaller than the $\Upsilon(10860) \to \Upsilon(1S,2S,3S) \pi^+ \pi^-$ partial decay widths  \cite{pdg}.  Likewise, decay rates for the heavy quark spin-flip transitions 
  $\Upsilon(10860) \to (h_b(1P), h_b(2P)) \pi^+ \pi^-$ are found experimentally comparable to the heavy quark spin
  non-flip transitions  $\Upsilon(10860) \to \Upsilon(nS) \pi^+ \pi^-$ \cite{Abdesselam:2015zza} -  in apparent
  violation of the heavy quark symmetry. These are  all hints of anomalous phenomena, not predicted by the QCD-based phenomenology of the conventional quarkonium systems, and in all likelihood they point to
 hitherto unexplored and novel facets of QCD.

 The emergence of four-quark (more precisely two quarks and two antiquarks) and five-quark (
 four quarks and an antiquark) hadrons has provided new challenges for QCD. As discussed in section  \ref{sec:Spectroscopy}, it is
 too early for a definite statement on their properties from lattice QCD, and we review several competing 
 phenomenological models
 put forward in the literature  to accommodate them. They range from the mundane (kinematic artifacts called
 cusps \cite{Tornqvist:1995kr,Bugg:2011jr}) to compact tetraquarks \cite{Vijande:2007rf}, which consist of 
 a $Q\bar{Q}$ pair and a light quark and an antiquark, bound in a compact color-singlet tetraquark
  $(Q\bar{Q} q \bar{q})_1$. The phenomenolgy of compact tetraquarks is developed in the framework where the structure consists of a color-antitriplet diquark $(Qq)_{\bar{3}}$ and a color-triplet
  antidiquark $(\bar{Q}\bar{q})_{3}$, bound by a gluon. They are called in the literature
  diquark-onium or simply diquark models. The main idea, going back to earlier papers by Jaffe \cite{Jaffe:1976ig}
   and Jaffe and Wilczek \cite{Jaffe:2003sg}, is that a tightly bound colored diquark plays a fundamental role in hadron spectroscopy \cite{Wilczek:2004im}, and possibly other areas, such as QCD in high baryon density
   and color superconductivity \cite{Alford:1997zt}.
  In the context of the $X,Y,Z$ tetraquarks, and the $P_c$ pentaquarks, 
   we will follow here mainly the works presented in ~\cite{Maiani:2004uc,Maiani:2004vq,Brodsky:2014xia,Maiani:2015vwa}.
  They  will be called diquark models in this review. There are other constructs, motivated by different theoretical scenarios. The hadroquarkonium models 
  \cite{Voloshin:1976ap,Dubynskiy:2008mq} are  motivated by the analogy with the hydrogen atom, in that the exotic hadrons have a heavy quarkonium core ($J/\psi, \psi^\prime, \eta_c$ ...), with light $q\bar{q}$ pair around it. 
Other models are motivated by analogy with the Deuteron, a well known hadron molecule, with
 a binding energy of about 2.2 MeV. Their possible existence in the charmonium sector was already suggested 
 very early on \cite{DeRujula:1976qd}. They were reinvented subsequently by 
  T\"{o}rnqvist \cite{Tornqvist:1991ks,Tornqvist:1993ng}, who proposed 
 the existence of possible Deuteron-like states of two mesons, called Deusons (see also \cite{Ericson:1993wy}).
  Bound together essentially by pion exchange, they provide a plausible template for some of the exotica in the charmonium and bottomonium sectors.
These ideas, borrowed essentially from nuclear bindings, have been  further developed
 by a number of groups, in which the constituent heavy mesons rescatter to give rise to the final states in 
 which the exotic hadrons are discovered. They are generically called 
  hadronic molecules, reviewed comprehensively in a recent paper \cite{Guo:2017jvc}. The last theoretical construct that we  briefly discuss are QCD hybrids \cite{Close:2005iz}, consisting of a $Q\bar{Q}g$, where $g$ is a constituent gluon, typically having a mass of about 1 GeV. The $J^{PC}=1^{--}$ $Y(4260)$ is the usual suspect here. 
  Thus, some of the exotic hadrons may eventually turn out to be hybrids.  In the theoretical part of this review, we discuss the salient features of these theoretical proposals, but elaborate the compact diquark model in somewhat more detail,
  as this represents the most radical and far reaching departure from the $q\bar{q}$ orthodoxy.
 
 The main  physics interest in this field is driven by the following questions:
  Are we at the threshold of a new frontier of QCD - with a vast and unexplored  hadronic landscape -
 in which the compact diquarks and/or the hybrids play a central role? Or, are we witnessing the effects of a residual chromodynamic van der Waals force, surfacing in the form of hadronic molecules? Given enough data, the diquark
 models can be distinguished from the molecular interpretation, as in the former we expect complete $SU(3)_F$ multiplets,
 unlike the hadronic molecules. This becomes evident later in this review, as we work out the spectroscopy 
 of the multiquark states in the diquark model. There are other differences in the production mechanisms and decays,
 such as the cross sections, $p_T$ distributions, and final state configurations. 
 The cusp-based interpretation can be confirmed, or debunked, based
 on the phase motion of the amplitudes in question. This difference between a genuine resonance (such as a Breit-Wigner)
 and a cusp is well known and will be discussed later as well.
 In the coming years, the current experiments at the Large Hadron Collider (LHC), in particular LHCb, and BESIII, an experiment at the Beijing electron positron collider, will be joined by Belle II, at KEK in Japan, JLAB in the US,
 and ${\rm \overline{P}ANDA}$ at the planned $p\bar{p}$ facility at GSI, Darmstadt. The high luminosity LHC phase with an upgraded LHCb (and other detectors) will surely contribute greatly to this field. They will subject the current measurements and competing theoretical models to stringent tests. That this is a vibrant field is underscored by the observation of five new narrow $\Omega_c^0$ states in one go decaying to $\Xi_c K^-$  \cite{Aaij:2017nav}, which are in all 
 likelihood quark-model states \cite{Wang:2017vnc,Padmanath:2017lng,Aliev:2017led,Karliner:2017kfm,Agaev:2017lip}, 
 and provide valuable information on the diquark binding in charm baryons. Their dynamics, interpreted in terms of a heavy
 (charm) quark and a light diquark (two strange quarks), can be worked out in the heavy-quark symmetry
 approach, and this may  help elucidate the role of diquarks  within tetraquarks and pentaquarks
 involving heavy quarks.\footnote{Luciano Maiani (private communication).}

This paper is organized as follows: In section \ref{sec:Spectroscopy}, we briefly recall the salient features of the potential models in use in the quarkonium spectroscopy and review some of the ongoing lattice simulations in the excited
and exotic charmonia sector.
In section~\ref{sec:Experimental-evidence-for-pentaquarks}, we discuss in detail the experimental evidence for the pentaquark states $P_c^+(4380)$ and $P_c^+(4450)$ from LHCb in the decays $\Lambda_b \to p J/\psi K^- $
and $\Lambda_b \to p J/\psi \pi^- $. We also discuss briefly the impending studies in the photoproduction process.
In section~\ref{sec:Experimental-evidence-for-tetraquarks} we summarize the experimental evidence for tetraquarks, starting with the $X(3872)$, and review  various theoretical interpretations. Next in line is the state $X(3940)$, observed by Belle in double charmonium production $e^+ e^- \to c\bar{c}\; c \bar{c}$. We then discuss the $Y$ states, in particular the $Y(4260)$, measured in the initial state radiation process $e^+ e^- \to \gamma_{ISR} \gamma_V $, with  $V \to Y(4260)$, observed by Belle in the $m(J/\psi \pi^+ \pi^-)$
invariant mass . The $Z_c$ states are taken up next, first observed by Belle in the decays
$ B^0 \to \psi^\prime K^\pm  \pi^\pm $, and review  $Z_c(4430)$ and $Z_c(4200)$ in detail. The state $Z_c(3900)$,
observed by BESIII in the decay $Y(4260) \to Z_c(3900)^+ \pi^- $ is discussed next (calling it the $Z_c$ state of type II).  
The $Z_b$ states in the bottomonium sector, $Z_b(10610)$ and $Z_b(10650)$, discovered by Belle in the decays
$\Upsilon(10860) \to Z_b(10610)^+ \pi^-$ and $\Upsilon(5S) \to Z_b(10650)^+ \pi^-$ are the last of the
$Z$ states we review. We close this section by discussing the resonant $J/\psi \phi$ states, discovered in a number of experiments (CDF, Belle, D0, CMS, LHCb). Theoretical models for tetraquarks are reviewed in section
\ref{sec:Theoretical-models-for-tetraquarks}, which include the
models discussed earlier, and the corresponding models for the pentaquarks are reviewed in section 
\ref{sec:Theoretical-models-for-pentaquarks}. We conclude
with a brief summary and outlook in section 7.

%% file: Sec-Conventional-spectroscopy-rev.tex
\section{Theoretical techniques for quarkonia and exotic hadrons spectroscopy}
\label{sec:Spectroscopy}
The current interest in the Quarkonium spectroscopy has shifted to excited states, in which the gluonic and light mesonic
degrees of freedom are present in the Fock space description of a hadron. This includes the hybrid and tetraquark states, with the latter taking the form of  two heavy-light mesons states, bound by mesonic exchange (hadronic molecule), or  genuine 
compact four-quark (tetraquark), states bound by gluons. Likewise, with the discovery of the two pentaquarks
$P_c^\pm (4380)$ and $P_c^\pm (4450)$ by LHCb, we have experimental evidence for the $c\bar{c} uud$ states.
The observed pentaquarks are found to have masses very close to a number of baryon-meson thresholds, as discussed
quantitatively later. They could be the manifestation of these nearby thresholds in the scattering matrix, but they
may also be genuine compact five-quark states. In this case,  one anticipates also the  $b\bar{b} uud$ states, and, in general, complete  $SU(3)_F$ multiplets. So, in line with the experimental developments, the next
frontier is to map out the spectrum of the $Q\bar{Q} g$,   $Q\bar{Q}q \bar{q}$, and $Q\bar{Q} q q q$ states in
QCD, and figure out the underlying dynamics.

In this section, we review briefly the various theoretical techniques which have been used to calculate the spectroscopy of the excited
quarkonia and exotic states. These include the potential models, which reflect the general features of QCD at short
and long-distances and include spin-spin interactions among the quarks, and lattice QCD, in terms of the Born-Oppenheimer approximation, and simulations using  current correlators with mesonic and diquark degrees of freedom. Also, very popular are the QCD sum rules,
which have played an important role in understanding the properties of the light and heavy mesons, and which
have also been applied for the studies of the exotic hadrons. However,  as 
they have been covered extensively in reviews \cite{Chen:2016qju,Nielsen:2009uh}, we will not discuss them here.
We also recall theoretical attempts in studying the spectrum of the nucleons and the strange baryons in the context
of chiral models based on Goldstone Boson Exchange (GBE) \cite{Glozman:1995fu}. Some versions of the GBE chiral model
have been used to study the stability of tetraquarks $QQ \bar{q}\bar{q}$ ($Q=c,b$)\cite{Pepin:1996id}, albeit with color blind
forces, and in studying the positive-parity pentaquarks $uudd \bar{Q}$ \cite{Stancu:1998sm,Helminen:2000jb}. More
recently, this model has been applied to hidden charm pentaquarks $qqqc\bar{c}$ \cite{Yuan:2012wz}, preceding the
LHCb discovery. Along the same lines, pentaquarks with the charm quantum number $C=\pm 1$ have been studied in the
Skyrme model \cite{Riska:1992qd,Oh:1994np,Wu:2004wg}.
      
We start by briefly reviewing the Cornell non-relativistic
potential \cite{Eichten:1978tg,Eichten:1979ms} and the Godfrey-Isgur model \cite{Godfrey:1985xj}, which is its relativistic version, as both of them had a major impact on understanding the spectrum of the observed states and the transitions connecting the various quarkonium states. 

\subsection{Quarkonium potentials}
\label{subsec:Quarkonium-potentials}
The central Cornell potential is the standard color Coloumb plus a scalar linear form, and includes a hyperfine interaction.
For the sake of definiteness we concentrate on the charmonium sector:
\begin{equation}
V_0^{c\bar{c}} (r)= -\frac{4}{3}\frac{\alpha_s}{r} + b r + \frac{32 \pi \alpha_s}{9 m_c^2}\delta_\sigma(r) 
\vec{S}_c  \cdot \vec{S}_{\bar{c}},
\label{eq:Cornell-central}
\end{equation}
where $\delta_\sigma(r) =(\sigma/\pi)^3  e^{-\sigma^2 r^2}$ is a Gaussian-smeared spin-spin contact interaction,
a form taken from a later work \cite{BarnesGodfreySwanson}. There are four parameters, the effective strong coupling
$\alpha_s$, string tension, $b$, charm quark mass, $m_c$, and the hyperfine coupling strength, $\sigma$.
The remaining spin-independent terms of the potential are included in the leading order perturbation theory, and
include the one-gluon-exchange spin-orbit and tensor interactions:
\begin{equation}
V^{c\bar{c}}_{\alpha_s}(r)= \frac{1}{m_c^2}\left[ \left(\frac{2\alpha_s}{r^3} - \frac{b}{2r}\right)\vec{L} \cdot \vec{S} + \frac{4 \alpha_s}{r^3}T        \right].
\label{eq:Cornell-pert}
\end{equation}  
There are just three independent spin-dependent operators, shown in Eqs. \ref{eq:Cornell-central} and
\ref{eq:Cornell-pert}.  Of these,
 the tensor operator is defined in the configuration space as
\begin{equation}
\vec{T} \equiv  (\vec{S}_c \cdot \hat{r})(\vec{S}_{\bar{c}}  \cdot \hat{r}) -\frac{1}{3} \vec{S}_c  \cdot \vec{S}_{\bar{c}},
\label{eq:Cornell-tensor}
\end{equation}
and the spin-orbit operator is $\vec{L} \cdot \vec{S}$, where $\vec{S}=  \vec{S}_c + \vec{S}_{\bar{c}}$.
They transform as $S=2$ and $S=1$, respectively. Hence, for states with $S=0$, their matrix elements vanish.
The matrix elements also vanish for $L=0$ states. For $L >0$ and $S=1$, the total angular quantum number 
can assume the values $J=L-1, L$ and $L+1$.
 
The quarkonium potential-model studies  were updated in  2005~\cite{BarnesGodfreySwanson} to incorporate the data from CLEO and the two $B$-factories. The parameters of the potential
 $V^{c\bar{c}}_{\rm NR}(r)= V_0^{c\bar{c}}(r) + V^{c\bar{c}}_{\alpha_s}(r)$ are fitted from the well-established $c\bar{c}$ states $1S$ - $4S$, $1P$, and the two $D$-states $\psi(3770)$ and $\psi(4159)$.
With only four parameters, $(\alpha_s, b, m_c, \sigma)$, the NR-Cornell potential model is very predictive, and the entire normal charmonium spectrum was worked out. In particular,
this allowed to predict the charmonium spectrum above the $D\bar{D}$ threshold, which includes $2P$, $3P$,
$1D$, $2D$, $1F$, $2F$, and $1G$ states. 

The Godfrey-Isgur model \cite{Godfrey:1985xj} assumes a relativistic dispersion relation for the quark kinetic energy, and an effective
potential of the $c\bar{c}$ system, containing a Lorentz vector one-gluon exchange short-distance and a scalar
linear confining term for the long distance part, 
\begin{equation}
{\cal H}^{c\bar{c}}_{GI}(p,r) = 2 \sqrt{m_c^2 + \mathbf{p}^2} + V^{c\bar{c}}_{\rm eff}(\mathbf{p},r).
\label{eq:GI-pot}
\end{equation}
The effective potential $V^{c\bar{c}}_{\rm eff}(\mathbf{p},r) $ is derived from the on-shell $c\bar{c}$ amplitude, and given in Ref.~ \cite{Godfrey:1985xj}.
In the non-relativistic limit, it  reduces to
the potential  $V^{c\bar{c}}_{\rm NR}(r)$ given above, with $\alpha_s$ replaced by the running coupling constant. Hence, the fits
yield different values for the parameters $b$ (string tension) and $m_c$. The spectra predicted by 
 $V^{c\bar{c}}_{\rm NR}(r)$
and ${\cal H}^{c\bar{c}}_{GI}(p,r)$ are very similar for the $S$- and $P$-wave states \cite{BarnesGodfreySwanson}. Both models predicted the  charmonium spectroscopy quite accurately, including some of the states which lie above the corresponding open heavy meson thresholds.

 However, not all the predicted charmonium states have been found, and several $2P$ states in the charmonium sector, such as the $0^{++}$ $\chi_{c0}(2P)$, the $1^{++}$ $\chi_{c1}(2P)$, and the $1^{+-}$ $h_{c}(2P)$, are still missing or not identified unambiguously. The potential model-based estimates may receive mass shifts due to other
effects, such as the couplings to open flavor channels \cite{Eichten:2004uh,Barnes:2007xu}, which are important if some of
the states are found to have masses near thresholds. By construction, the potential models  are not made to cover the excited charmonia (and bottomonia) states, and exotica,  which have extra light degrees of freedoms in their Fock space, such as the hybrids ($Q\bar{Q}g)$
and the tetraquark states $Q\bar{Q}q \bar{q}$, where $q \bar{q} $ is a light quark ($q=u,d,s$) pair. Their role is that 
they provide benchmarks to map out the normal quarkonia, and hence are very valuable  in the  search for the excited quarkonium states and exotica.

\subsection{Lattice simulations}
\label{subsec:Lattice-simulations}
The spectroscopy of the mesons containing hidden and open-charm  quarks calculated with the lattice techniques has 
made great strides lately. In particular, the calculations of the lowest-lying states well below the strong decay threshold have attained impressive precision with
various systematic effects under control \cite{Dowdall:2011wh,Dowdall:2012ab,Daldrop:2011aa,Donald:2012ga}. 
In lattice simulations, the mass of a single hadron is extracted from $m=E_{\vec {p}=0}$, 
extracted from the energies obtained from the $q\bar{q}$ or $qqq$ interpolating fields, 
with the lattice spacing $a \to 0$, and the quark mass $m_q^{\rm lattice} \to m_q^{\rm phys}$.
For hadrons with $b$ quarks, lattice simulations use non-relativistic QCD (NRQCD), and expansion in the  
heavy quark velocity, and include four flavors $(u,d,s,c)$ of dynamical quarks.
The current status of the
spectroscopy for the heavy mesonic systems for the charm and bottom quarks is reviewed in 
the Particle Data Group (PDG) and compared with
the lattice results (see Fig.~15.7 of the review by Amsler, DeGrand and Krusche in \cite{pdg}).
However, there are a number of states near the $D\bar{D}$ threshold, like $X(3872)$, $Z_c^\pm(3900)$ and
$X(4140)$, whose quark composition is obscure.

\begin{figure}[t]
\centerline{
\centerline{\includegraphics[width=11cm]{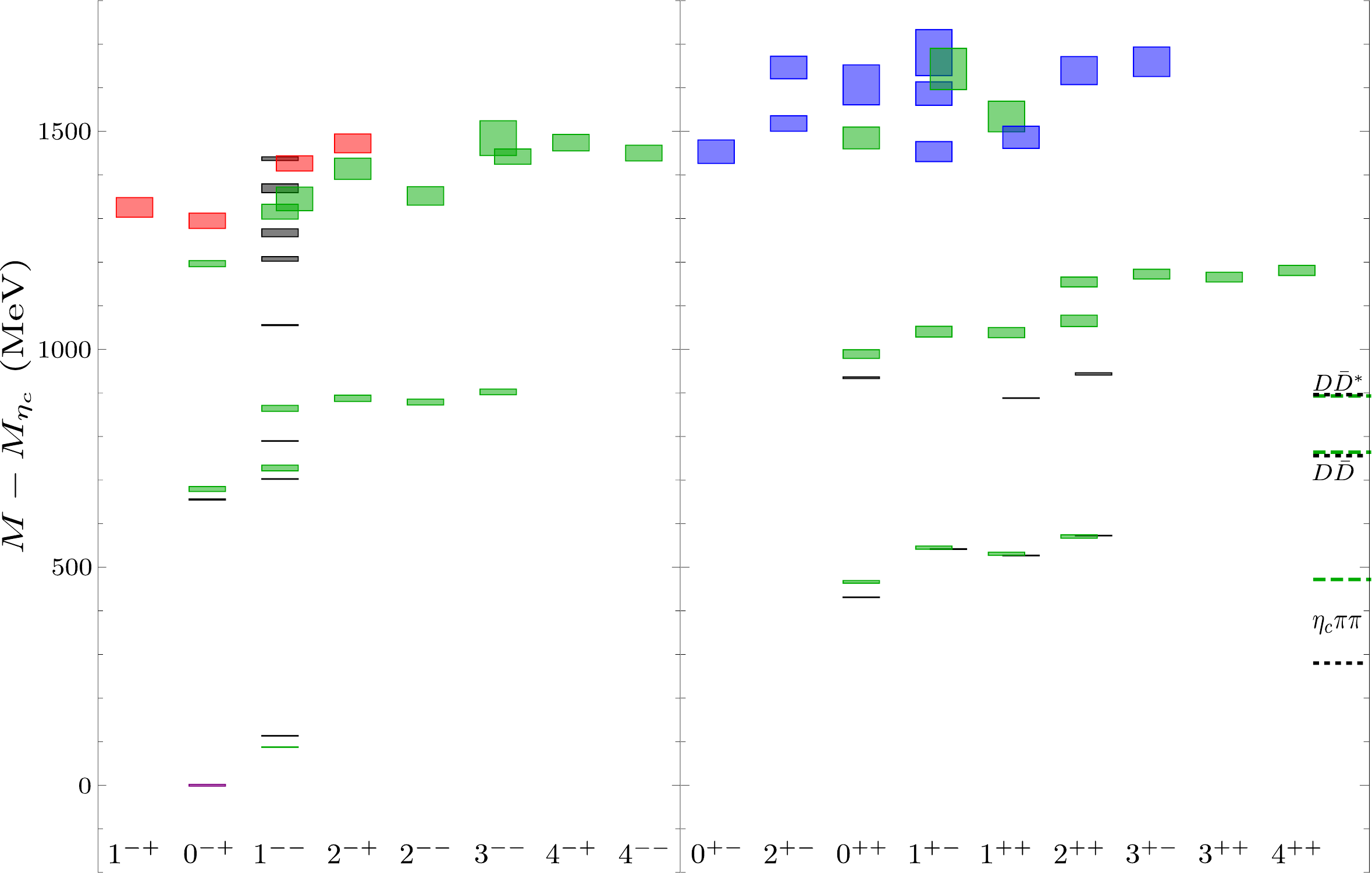}}
}
\caption{Charmonium spectrum up to 4.5 GeV, with the indicated $J^{PC}$ quantum numbers for the negative (positive)
parity states shown in the left (right) panel. The calculated and experimental masses are shown in terms of
$M - M_{\eta_c}$. The vertical size of the boxes represents the $1~\sigma$ statistical error. Red and blue boxes correspond 
 to states identified as  hybrid mesons, grouped into respectively, the lightest and first excited supermultiplet, and
 black boxes are experimental values from PDG. Location of some of the thresholds for strong decays are also indicated
 (from lattice QCD studies \cite{Cheung:2016bym}).  
\label{fig:hsc}}
\end{figure}

A recent lattice work on the excited and exotic charmonium states is from the Hadron Spectrum Collaboration 
\cite{Cheung:2016bym} where the excited spectrum is studied by varying the light quark mass.
Using the $\eta_c$ mass  to estimate the systematic error from tuning the charm quark mass yields an error
of order 1\% from this source. Adopting
 an anisotropic lattice formulation with the spatial lattice spacing $a_s=0.12$ fm, a temporal lattice spacing $a_t=3.5 a_s$, with $N_f=2 +1$ flavor of dynamical quarks, and a pion mass of $m_{\pi} = 240~MeV$, leads to the results for the charmonium spectrum presented in Fig.~\ref{fig:hsc}. The calculated and the experimental masses are compared with each other, with the $\eta_c$ mass subtracted to further reduce the uncertainty from
 the charm quark mass. Many of the states with non-exotic $J^{PC}$ quantum numbers follow the $n^{2S+1}L_J$
pattern, predicted by the charmonium potential models, discussed earlier. All states up to $J=4$ are found.
Figure~\ref{fig:hsc} also shows the states which do not fit the $n^{2S+1}L_J$ pattern. Some of these have exotic 
$J^{PC}$ quantum numbers, $0^{+-}$, $1^{-+}$, $2^{+-}$. These and some extra non-exotic states are interpreted as
hybrid mesons. The lightest hybrids appear about 1.2 - 1.3 GeV above the lightest $S$-wave meson multiplets.
This shows that a rich charmonium hybrid spectrum is waiting to be explored. The lattice mass estimates are in need of
better systematic control to be quantitative, in particular, the current value of the pion mass used in these simulations
is still quite high. 

In the energy region near or above the strong decay thresholds, the masses of the bound-states and resonances
are inferred from the finite-volume scattering matrix of elastic or inelastic scattering. They represent
one-channel and multiple-channel cases, respectively, with the inelastic case obviously technically more challenging.
Various approaches with varying degrees of mathematical rigour are used in the simulation.  The most widely used approach is based on  L\"{u}scher's formulation \cite{Luscher:1990ux,Luscher:1991cf} and its generalisation 
\cite{Doring:2011vk,Gockeler:2012yj, Briceno:2017tce}. Using non-relativistic effective Lagrangians, a framework for the calculation of resonant matrix elements in
lattice QCD is also developed within  L\"{u}scher's finite-volume framework \cite{Bernard:2012bi}.
Other approaches based on finite volume Hamiltonian Effective Field Theory (EFT) are also in use, in which the parameters are extracted by fitting the analytic expressions for the eigenvalues of the finite volume EFT using the lattice spectrum
$E_n$ \cite{Hall:2013qba}. Likewise, the HALQCD approach \cite{HALQCD:2012aa}, which is based on the lattice determination of the potential $V(r)$ between hadrons, employs the Nambu-Bethe-Salpeter equation to extract the masses of the bound states.
Their technical discussion
will take us too far from the phenomenological focus of this review; they are reviewed  in a number of papers
 \cite{Prelovsek:2014zga,Yamazaki:2015nka,Briceno:2014pka}. 

Several lattice results on the
near-threshold bound states, scattering and resonances in the charm sector have been reported. Likewise, 
tetraquark spectroscopy has been investigated in a number of simulations, and there is at least one calculation on the lattice for pentaquarks \cite{Beane:2014sda}. A recent review on these lattice-based developments is Ref.~\cite{Prelovsek:2016dmi}.
In general, investigations of the excited charmonia and excited
open charm mesons have systematic uncertainties, which are not quite accounted for. 
We discuss a sampling of these studies to illustrate the current state of this field.

\vspace*{3mm}
\underline {$\mathbf{X(3872)}$}: The $X(3872)$ lies very close to the $D^0\bar{D}^{*0}$ threshold and it is essential to take into account the effect of
this threshold on the lattice. This represents a shallow bound state from one-channel scattering. 
The first simulation of $D\bar{D}^*$ scattering was done in Ref.~\cite{Prelovsek:2013cra},
where a pole in the scattering matrix was found just below the threshold in the $I=0$, $J^{PC}=1^{++}$ channel. The pole
was associated with the $X(3872)$, and was confirmed subsequently in a simulation by the Fermilab-lattice/MILC
collaboration \cite{Lee:2014uta}. More recently, this topic was reinvestigated \cite{Padmanath:2015era} by including
diquark-antidiquark interpolating fields to determine which Fock components are essential for the $X(3872)$, suggesting
that both the $c\bar{c}$ and the $D^0\bar{D}^{*0}$ threshold are required for the $X(3872)$, 
rather than the diquark-antidiquark correlator. This remains to be  confirmed by theoretically improved calculations with controlled systematic errors.

\vspace*{3mm}
\underline {$\mathbf{Z_c^+(3900)}$}: Lattice simulation of the $Z_c^+(3900)$,  which has the quark flavor
content $c\bar{c}u\bar{d}$ and has the quantum numbers $I^G(J^{PC})=1^+(1^{+-})$, is reported in
 Ref.~\cite{Ikeda:2016zwx}. This is studied by the method of coupled-channel scattering
involving the final states $D\bar{D}^*$, $\pi J/\psi$ and $\rho \eta_c$, for which the HALQCD approach
\cite{HALQCD:2012aa} is used. It involves a calculation of the potential for the $3 \times 3$ scattering matrix. 
First the potential $V_{\pi J/\psi \to \pi J/\psi}(r)$ related to the Nambu-Bethe-Salpeter
equation is determined between the $J/\psi$ and $\pi$ as a function of their separation $r$. The potential for the
other two channels $D\bar{D}^*$ and $\rho \eta_c$ and the off-diagonal elements involving these channels are likewise
calculated, and used to determine the three-body decay $Y(4260) \to J/\psi \pi \pi$ and
$Y(4260) \to D\bar{D}^*$ in a phenomenological way. Indeed, a peak around the $Z_c^+(3900)$ is found.
If the coupling between the $D\bar{D}^*$ and $\pi J/\psi$  is switched off, the peak disappears, which reflects that
the potential for the off-diagonal element  $D\bar{D}^*$ and $\pi J/\psi$ is larger than the
other potentials. Hence, this study suggests that also the $Z_c^+(3900)$ is possibly a rescattering effect. However, the simulation in  Ref.~\cite{Ikeda:2016zwx}
is not done using the rigorous L\"{u}scher's formulation, and hence the conclusions are tentative \cite{Prelovsek:2016dmi}.

\vspace*{3mm}
\underline {\bf Pentaquarks}: Attempts to simulate the LHCb-type pentaquarks on the lattice are also under way. The NPLQCD collaboration
\cite{Beane:2014sda} has
presented first evidence for a $\eta_c N$ bound state approximately 20 MeV below the $\eta_c N$ threshold - again a case
of a shallow one-channel scattering, similar to the $X(3872)$, but now in the baryonic sector. The simulation is done
for $m_\pi=800$ MeV. This is far afield from the physical mass of the
pion, and it is not clear if the evidence for the bound state will persist for more realistic pion mass. The observed pentaquarks $P_c^+(4380)$
and $P_c^*(4450)$ are, on the other hand, about 400 MeV above the $J/\psi p$ threshold. The lattice simulation of these
pentaquarks is much more challenging as it is a multi-channel problem, with several open thresholds nearby. It would be
exciting if experimentally a bound state is found in the $\eta_c N$ channel hinted by the NLPQCD simulation.

\subsection{Born-Oppenheimer tetraquarks}
\label{subsec:B-O-Tetraquarks}
The Born-Oppenheimer (B-O)approximation, which was introduced in 1927 to study the binding of atoms into molecules
\cite{Born:1927bo}, makes use of the large ratio of the masses of the atomic nucleus  and the electron. The nuclei
can be approximated by static sources for the electric field, and the electrons respond almost instantaneously to the
motion of the nuclei. The implicit adiabatic approximation reduces the rather intricate dynamics to the tractable
problem of calculating the B-O potentials, which are  defined by the Coulomb energy of the nuclei and the energy of
the electrons. The QCD analog of this is that the nucleus is replaced by a heavy $Q\bar{Q}$ (or  $QQ$),
pair, and the electron cloud is replaced by the light degrees of freedom, a gluon for a $Q\bar{Q}g$ hybrid, or a light
$q\bar{q}$ (or $\bar{q}\bar{q}$) pair for a tetraquark. If the masses of the two heavy quarks are much larger than the 
QCD scale, $\Lambda_{\rm QCD}$, which is the case for the bottom and charm quarks, then the dynamics can be
described by a quantum mechanical Hamiltonian with an appropriate QCD potential, which can be studied  on the
lattice.

The B-O approximation for $Q\bar{Q}$ mesons in QCD was studied by Juge, Kuti, and Morningstar \cite{Juge:1999ie} 
who investigated the hybrid $b\bar{b}g$ molecules, a bound system of $b\bar{b}$ and excited gluon field, and also
carried out detailed studies in the lattice quenched approximation. The consistency of the result from the two
approaches made a compelling case for the heavy hybrid states. The interest in the B-O approximation 
was revived in Ref.~\cite{Braaten:2014qka} to cover also the flavor-nonsinglet $Q\bar{Q}$ mesons, in particular, 
the $XYZ$ tetraquarks. The B-O potential involves  a single-channel approximation that simplifies 
the Schr\"{o}dinger equation, which can be solved for just one radial wavefunction. In that case, the B-O approximation
offers a reliable template for a coherent description of these hadrons in QCD.  
On the other hand, if the masses of the $XYZ$ tetraquarks are
close to the thresholds for a pair of heavy mesons, which is often the case, then one has to account for the coupling to the
meson pair scattering, and the  Schr\"{o}dinger equation  becomes a multi-channel problem.
 This requires detailed lattice calculations of the B-O potentials to estimate the effects of the couplings between the channels, which has still to be carried out for the $XYZ$ hadrons.

\subsection{Doubly heavy tetraquarks on the lattice}
\label{subsec:doubly-heavy}
The case of two heavy antiquarks ($\bar{Q}\bar{Q}$) and a light $qq$ pair, bound in a hadron, has also
 received a lot of theoretical interest, though so far there is no trace of hadrons, such as $ud \bar{b} \bar{b}$  or $us\bar{b} \bar{b}$ experimentally.  Such exotics are difficult to produce in high energy experiments, with presumably LHC the only
 collider where they may show up albeit with a very small cross section. If found, they would be truly exotic.
 As we focus mainly on the observed pentaquarks and tetraquarks in the rest of this review, 
 we briefly discuss  two recent theoretical estimates of the doubly heavy tetraquarks masses and decay
  widths \cite{Bicudo:2017szl,Francis:2016hui}.
 
In  Ref. \cite{Bicudo:2017szl}, the potential of two static antiquarks $\bar{Q}\bar{Q}$ in the presence of two light quarks $qq$ is parametrized by a screened Coulomb potential 
\begin{equation}
V(r) = -\frac{\alpha}{r} e^{-r^2/d^2},
\label{eq:Cardoso}
\end{equation}
which is inspired by one-gluon exchange at small $\bar{Q}\bar{Q}$ separation $r$ and a screening of the Coulomb potential
due to the formation of two $\overline{B}$-mesons at large $r$. The parameters $\alpha$ and $d$ depend on the isospin $I$ and
the total angular momentum $j$ of the light $qq$ pair. They are determined on the lattice by fitting the ground state potentials in
the attractive channels, and it is found that the $( I=0, j=0 )$ potential is more attractive than
 $( I=1, j=1)$ \cite{Bicudo:2015kna}, yielding $ \alpha=0.34 \pm 0.03$ and   $ d=(0.45^{+0.12}_{-0.10})$ fm.
 With this potential the Schr\"{o}dinger equation is solved:
$(H_0 + V(r) -E)X= -V(r) \psi_0$, where $X$ is defined by the wavefunction splitting $\psi=\psi_0 +X$, with $\psi_0$
the incident and  $X$ the emergent wave. From the asymptotic behavior of $X$, the phase shift, $\delta_\ell$ is determined 
as a function of  the energy $E$, which is continued to the complex $E$-plane. 
 Analyticity is used to determine the pole position of the resonance and its width,
yielding $m =10576 \pm 4$ MeV and $\Gamma=112 ^{+90}_{-103}$ MeV for the 
 $ud \bar{b}\bar{b}$ charged exotic having the quantum numbers $I(J^P)=0(1^-)$ \cite{Bicudo:2017szl}.

The possibility of a doubly heavy tetraquark $qq^\prime \bar{b}\bar{b}$ bound states has also recently been studied
on the lattice, using NRQCD to simulate the bottom quarks  \cite{Francis:2016hui}. A two-point 
lattice QCD correlation function is defined in the Euclidean time
 \begin{equation}
C_{{\cal O}_1 {\cal O}_2} (p,t) = \sum_{n} \langle  0 \mid {\cal O}_1 \mid n \rangle \langle  n \mid {\cal O}_2 \mid 0 \rangle
e^{-E_n(p)t},
\label{eq:Corr}
\end{equation}
where the operators $({\cal O}_i )$ have the appropriate quantum numbers. Thus, for the 
$B(5279)$ meson, with $I(J^P) =\frac{1}{2} (0^-)$, and $B^*(5325)$, with  $I(J^P) =\frac{1}{2} (1^-)$,  the interpolating
fields are the $q\bar{q}$ bilinears $P \sim \bar{b} \gamma_5 q$ and  $V\sim \bar{b} \gamma_i q$, respectively.
They are used to define a two-meson operator $M(x)$ having the quantum number $J^P=1^+$:
\begin{equation}
M(x)= \bar{b}_a^\alpha(x) \gamma_5^{\alpha\beta} u_a^\beta(x) \bar{b}_b^\kappa(x) \gamma_i^{\kappa\rho} d_b^\rho(x)
- \bar{b}_a^\alpha(x) \gamma_5^{\alpha\beta} d_a^\beta(x) \bar{b}_b^\kappa(x) \gamma_i^{\kappa\rho} u_b^\rho(x),
\label{eq:meson-meson}
\end{equation}
and an analogous operator with the $B_sB^*$ structure. Here, $a,b$ are color indices, and $\alpha, \beta,...$ are
Dirac indices .

 The second 
operator $D(x)$ has the diquark-antidiquark structure with $\bar{b}\bar{b}$ a color triplet $3_c$, spin 1, and the light diquark 
having the flavor, spin, color quantum numbers $(\bar{3}_F, 0, \bar{3}_c )$ ,%
\begin{equation}
D(x)= ([u_a^\alpha(x)]^T (C\gamma_5)^{\alpha\beta} q_b^\beta(x)) (\bar{b}_a^\kappa(x) (C\gamma_i)^{\kappa\rho}
[ \bar{b}_b^\rho(x)]^T),
\label{eq:Diquark}
\end{equation}
where $q=d,s$. This yields a $J^P=1^+$ state. With this the binding correlator 
\begin{equation}
G_{{\cal O}_1 {\cal O}_2} (p,t) = \frac{C_{{\cal O}_1 {\cal O}_2} (p,t)}{ C_{PP (t)} C_{VV(t)}},
\label{eq:binding-corr}
\end{equation}
is studied.  For a channel with a tetraquark ground state with (negative) binding energy $\Delta E$, with respect to the
two-meson $PV$ threshold, this correlator grows as $e^{-\Delta Et }$.  With this, the  $(2 \times 2)$ matrix 
with the matrix elements $G_{DD}(t), G_{DM}(t), G_{MD}(t), G_{MM}(t)$ are studied and the two eigenvalues extracted, 
yielding the states  $ud \bar{b}\bar{b}$ and 
$\ell s \bar{b}\bar{b} $, with $\ell=u,d$. They lie, respectively,  $189(10)$ and $98(7)$ MeV below the corresponding
thresholds, with $J^P=1^+$. These double $b$-quark tetraquarks are stable both with respect to the strong and electromagnetic decays, and will decay weakly. The discovery modes are listed as $B^+ \bar{D}^0$ and $J/\psi B^+ K^0$ for the first, and $J/\psi B_s K^+$ and $J/\psi B^+ \phi$ for the second.

In concluding this section on quarkonium potentials and lattice simulations, we note that the two approaches agree
remarkably with each other for the case of single hadrons well below the strong decay thresholds. However, for
the cases where the elastic or the inelastic scatterings are involved, reliable results from the lattice are not yet
quantitative. This problem becomes much more complicated if several channels lie near each other. For
tetraquarks with  doubly heavy quarks $ud \bar{b}\bar{b}$ and 
$\ell s \bar{b}\bar{b} $, with $\ell=u,d$, lattice results have been obtained, and remain to be tested against experiments.
The B-O approximation can yield reliable results for the single-channel case, but requires further study for the
states where scattering thresholds have to be taken into account.

%% file: Sec-Experimental-evidence-for-pentaquarks.tex
\section{Experimental evidence for pentaquarks}
\label{sec:Experimental-evidence-for-pentaquarks}
\def\Like{\mathcal{L}}
\subsection{History}
 The prospect of multi-quark hadrons was raised first by Jaffe  in 1976 \cite{Jaffe:1976ig}, and expanded upon by Strottman \cite{Strottman:1979qu} to include  baryons composed of four-quarks plus one antiquark. The name pentaquark was coined by Lipkin who predicted states with charmed quarks \cite{Lipkin:1987sk}. The concept of stable pentaquarks was mentioned around the same time by Gignoux et al. \cite{Gignoux:1987cn}. 

Pentaquark states composed only of $u$, $d$ and $s$ quarks have been  previously reported to great fanfare, circa 2004. However, they have been debunked by updated analyses.  
The Hicks article \cite{Hicks:2012zz} gives a good review of previous claims and their refutations. A brief summary will be given here. The story starts with observations of a bump in the invariant kaon nucleon mass spectra was observed at about 1.54~GeV in various reactions listed in Table~\ref{tab:thetap}. The widths were rather narrow, around 10~MeV, quite surprising for pentaquarks that could easily decay, and thus should have a larger widths.
\begin{table}[htp]
\centering
\caption{The first four experiments with positive evidence for the $\Theta^+$.}
\vspace{0.2cm}
\begin{tabular}{lccc}
\hline
Experiment & Reaction & Mass (GeV) & Significance\\
\hline
LEPS \cite{Nakano:2003qx} & $\gamma C\to K^+K^-X$ & 1.54$\pm$0.01& 4.6$\sigma$\\
DIANA \cite{Barmin:2003vv} &$K^+ Xe\to \KS p X$ & 1.539$\pm$0.002& 4.4$\sigma$\\
CLAS \cite{Stepanyan:2003qr}& $\gamma d\to K^+K^- p n$ & 1.542$\pm$0.005& (5.2$\pm$0.6)$\sigma$\\
SAPHIR \cite{Barth:2003es}& $\gamma p\to \KS K^+ n$ & 1.540$\pm$0.004& 4.8$\sigma$\\
\hline 
\end{tabular}
\label{tab:thetap}
\end{table}

One notes these were all nuclear physics experiments whose significances did not exceed 5 standard deviation and whose statistics was rather low, so full amplitude analyses were not undertaken. These results were followed by the announcement of $\Xi^{--}$ pentaquark state by NA49 \cite{Alt:2003vb}. This state, however, was not seen by the HERA-B collaboration, that had a larger data sample \cite{Abt:2004tz}.
Soon thereafter the pentaquark fervor was stoked by a paper from the H1 collaboration that showed evidence for a pentaquark containing $udud\bar{c}$ quarks, a charmed pentaquark \cite{Aktas:2004qf}, seen in the $D^{*-}p$ decay mode that was quickly contradicted by the Zeus experiment \cite{Chekanov:2004qm}.

All of the positive results were controversial. There was an attempt to explain the data from kinematical reflections  \cite{Dzierba:2003cm}. There then followed a period with several ``confirmations" of the $\Theta^+$ but also many other experiments that did not see it. CLAS, for example,  with a 20 times larger data sample than their original result, showed that the state was not present. Very high statistics experiments such as BaBar did not see any of these states. (References to these articles can be found in \cite{Hicks:2012zz}). Currently there is no obvious explanation of the positive results, although this may be an additional example of  ``pathological science" \cite{Stone:2000an}.

\subsection{The LHCb observation of pentaquark charmonium states}
\label{sec:LcKp}

\subsubsection{The \boldmath{$\Lb\to \jpsi K^-p$} decay}
\label{sec:jpsiKp}
\begin{figure}[htb]
\begin{center}
    \includegraphics[width=0.9\textwidth]{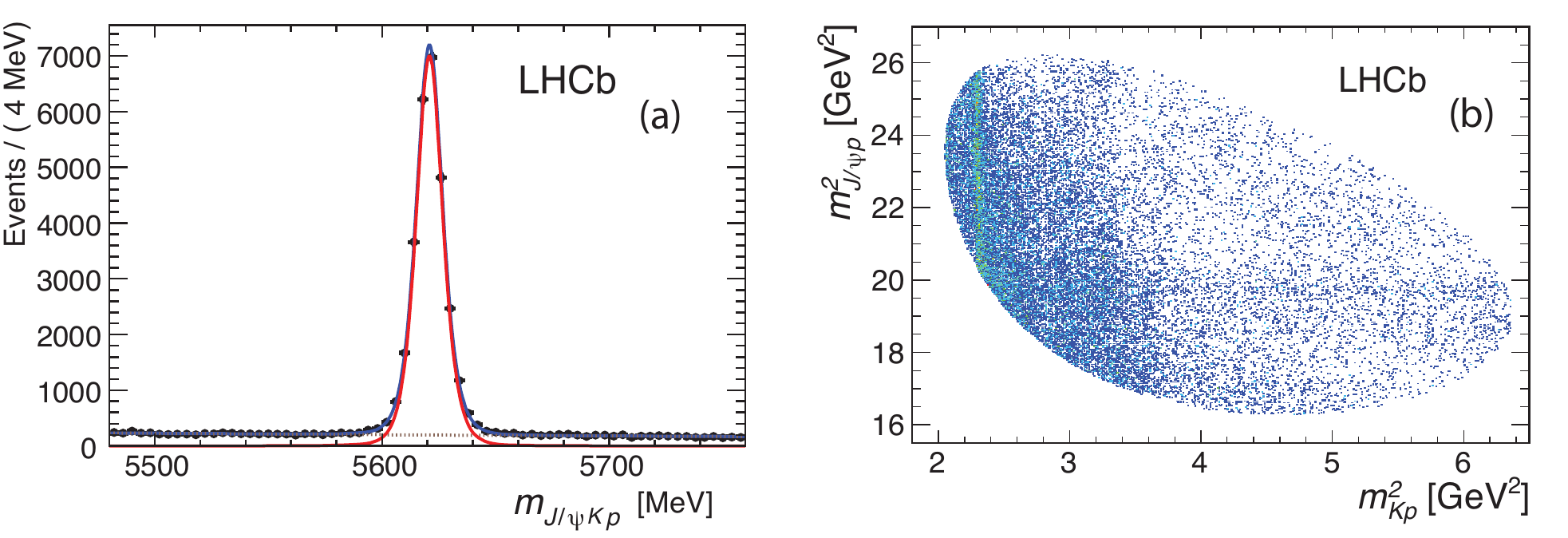}
        \end{center}
\vskip -0.5cm
\caption{(a) Invariant mass spectrum of $\jpsi K^-p$ combinations, with the total fit, signal and background components shown as solid (blue), solid (red) and dashed lines, respectively. (b) Invariant mass  squared of $K^-p$  versus $\jpsi p$ for candidates within $\pm15$~MeV of the \Lb mass (from \cite{Aaij:2015tga}).}
\label{fitmass-newbdt}
\end{figure}

The principal aim of the LHCb experiment is to find physics beyond the Standard Model using rare and \CP-violating decays of $b$-flavored hadrons \cite{LHCb:2011dta}. During one of these studies involving $B\to\jpsi K^+ K^-$ decays a question of a possible background was raised from an as yet to be observed decay  $\Lb\to \jpsi K^-p$, where the $p$ was misidentified as a $K^+$ meson. Note,
at the LHC \Lb baryons are prolifically produced. In the acceptance of the LHCb experiment $\sim\!\!20\%$ of all $b$-flavored hadrons are \Lb's \cite{Aaij:2011jp}. It was quickly realized that this decay mode offered an excellent way to measure the 
\Lb lifetime as it had a large event yield, and had four charged tracks that were used to define the coordinates of the \Lb decay point. The reconstructed $\jpsi K^-p$ invariant mass distribution is shown in Fig.~\ref{fitmass-newbdt}(left) showing about 26,000 signal candidates along with 1,400 background events within $\pm15$~MeV of the mass peak.

A consequence of this discovery was that the \Lb lifetime was measured precisely, which settled an important issue as many previous measurements showed a large difference with other $b$-flavored hadron species  that was not predicted and not confirmed by LHCb \cite{Aaij:2014zyy}. However, examination of the decay products showed an anomalous feature \cite{Aaij:2015tga}. 
The Dalitz like plot \cite{Dalitz:1953cp} shown in Fig.~\ref{fitmass-newbdt}(right) uses the $K^-p$  and $\jpsi p$ invariant masses-squared as independent variables.\footnote{The Dalitz plot was conceived for decays into three scalar mesons, where the phase space is uniform over the plot area, so that the effects of the matrix element governing the decay are directly visible.} There are vertical bands corresponding to $\PLambda^*\to K^-p$ resonant structures, and an unexpected horizontal band near 19.5~GeV$^2$.

\begin{figure}[t!]
\vskip -0.2cm
\begin{center}
\includegraphics[width=0.35\textwidth]{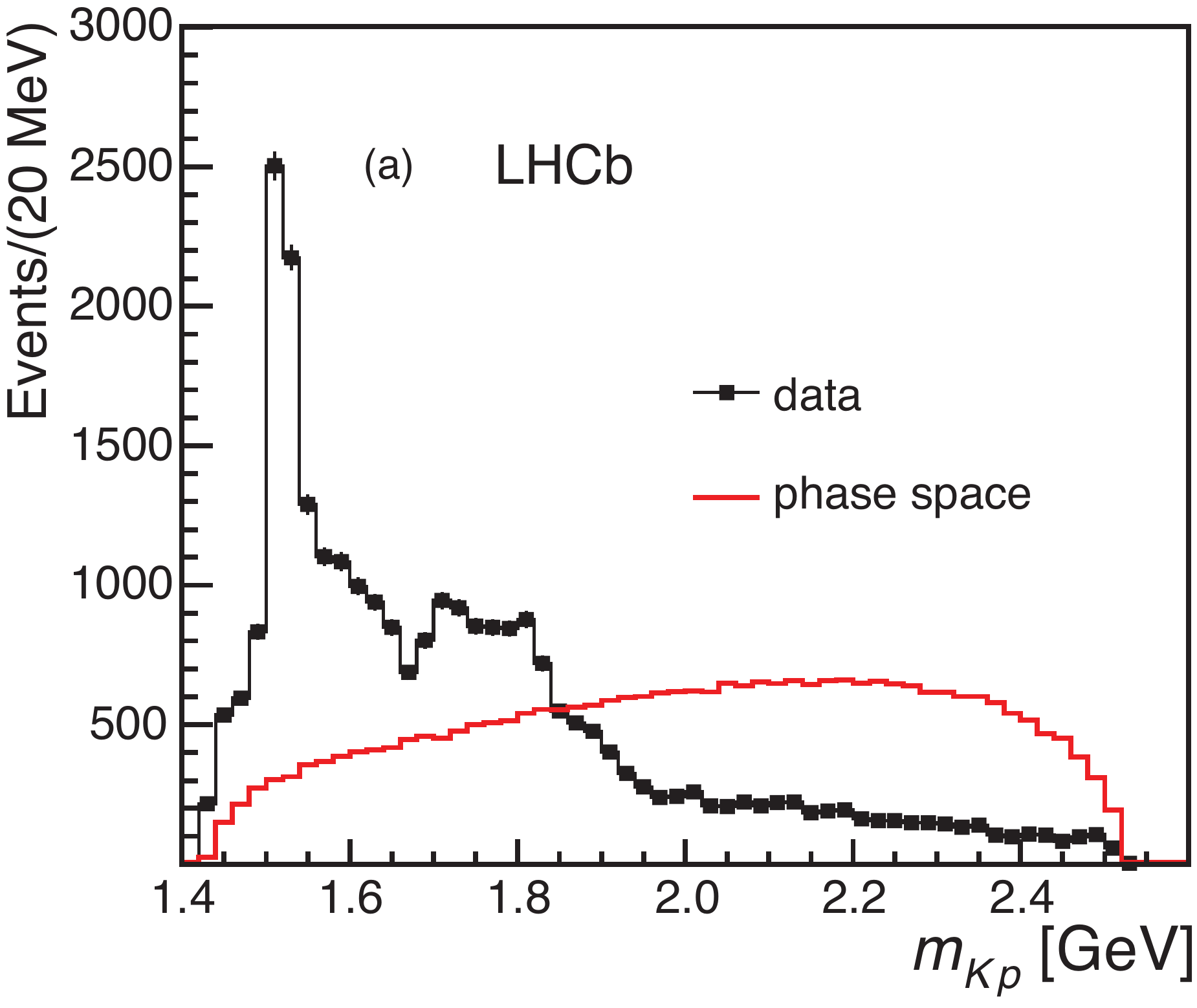}\hspace*{0.99cm}\includegraphics[width=0.35\textwidth]{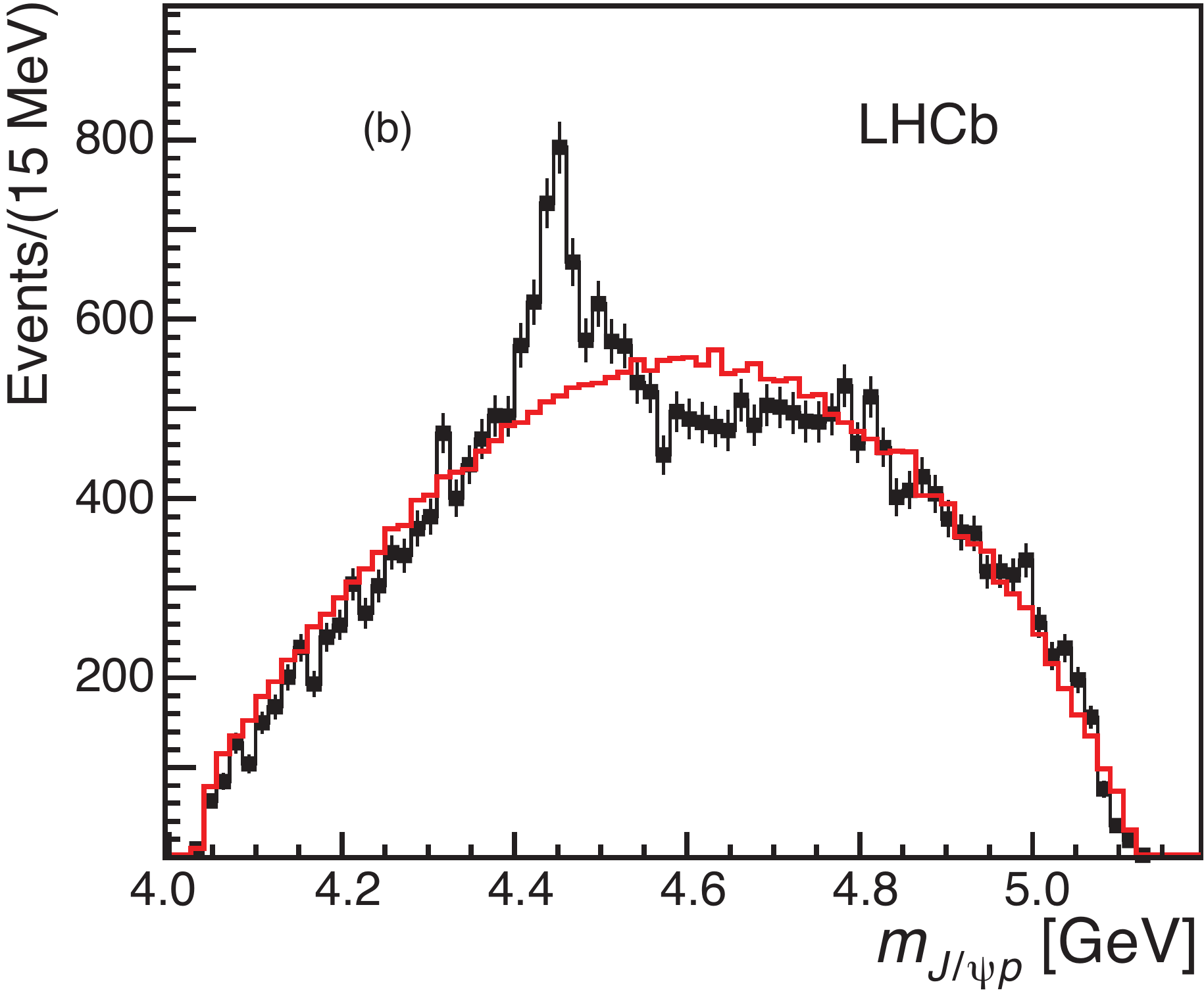}
\end{center}
\vskip -0.5cm
\caption{ Invariant mass of (a) $K^-p$  and (b) $\jpsi p$ combinations from $\Lb\to\jpsi K^-p$ decays. The solid (red) curve is the expectation from phase space. The background has been subtracted (from  \cite{Aaij:2015tga}).}
\label{mpk-mjpsi}
\end{figure}

\begin{figure}[b]
\begin{center}
\includegraphics[width=0.7\textwidth]{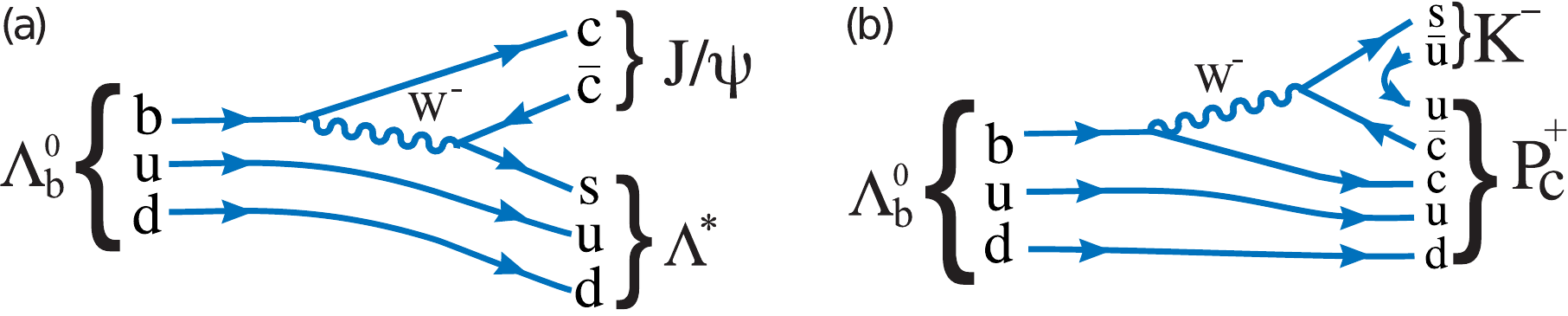}
\end{center}
\vskip -0.3cm
\caption{Feynman diagrams for (a) $\Lb\to \jpsi \Lz^*$ and (b) $\Lb\to P_c^+ K^-$ decay (from \cite{Aaij:2015tga}).}
\label{Feynman-Pc-both-v5}
\end{figure}

The Dalitz plot projections are shown in Fig.~\ref{mpk-mjpsi}. Indeed there are significant structures in the $K^-p$ mass spectrum that differ from phase space expectations, and there is also a peak in the $\jpsi p$ mass spectrum. The leading order Feynman diagrams for $\Lb\to\jpsi \PLambda^*$, and for $\Lb\to K^-P_c^+$, where $P_c^+$ is a possible state that decays into $\jpsi p$, are shown in Fig.~\ref{Feynman-Pc-both-v5}.

This decay can proceed by the diagram shown in Fig.~\ref{Feynman-Pc-both-v5}(a), and is expected to be dominated by $\Lz^*\to K^-p$ resonances, as are evident in the data shown in Fig.~\ref{mpk-mjpsi}(a). It could also have exotic contributions, as indicated by the diagram in Fig.~\ref{Feynman-Pc-both-v5}(b), that could result in resonant structures in the $\jpsi p$ mass spectrum shown in Fig.~\ref{mpk-mjpsi}(b).

In order to establish the resonant content of this \Lb decay it is necessary to consider both $\Lb\to\jpsi\PLambda^*$ and $\Lb\to K^-P_c^+$ decay sequences simultaneously. LHCb, however, had to first consider the possibility that interferences among only the $\PLambda^*$ decay chain caused a peak in the $\jpsi p$ mass spectrum.  So at first only the amplitudes described by the diagram in Fig.~\ref{Feynman-Pc-both-v5}(a) were considered. 

Even with this restriction there are many final states. The decays $\PLambda^*\to K^-p$ and $\PSigma^*\to K^- p$ are both, in principle, possible. Note however, that the isospin of the initial \Lb baryon is zero, as is the isospin of the \jpsi meson, so if the resonant state  were to be a isospin one $\PSigma^*$ rather than a $\PLambda^*$ we would have a change of isospin of one unit in the weak decay, which even though possible is thought to be highly suppressed, similar to the situation in the decay of the kaon into two pions \cite{Donoghue:1979mu} where the isospin changing 3/2 amplitude is highly suppressed with respect to the 1/2 amplitude.  The  $\PLambda^*$ states are listed as the ``Extended" model in Table~\ref{tab:Lstar}. Along with each state the number of possible decay amplitudes with different orbital angular momentum $L$, and spin $S$ quantum numbers are also given.  Of course it is not apriori known which of these states and amplitudes is present, or even if there are additional states that have yet to be discovered which also contribute.

\begin{table}[htb]
\centering
\caption{The $\Lz^*$ resonances used in the different fits. Parameters are taken from the PDG \cite{pdg}. We take $5/2^-$ for the $J^P$ of the $\Lz(2585)$.
The number of $LS$ couplings is also listed for both the ``reduced'' and ``extended'' models. In the reduced model fewer states and amplitudes are considered.
To fix overall phase and magnitude conventions, which otherwise are arbitrary, we set the helicity coupling of the spin-1/2 $\Lz(1520)$ to (1,0).  
A zero entry means the state is excluded from the fit. }
\vspace{0.2cm}
\begin{tabular}{lccccc}
\hline\\[-2.5ex] 
State & $J^P$ & $M_0$ (MeV) & $\Gamma_0$ (MeV)& \# Reduced  & \# Extended \\
\hline \\[-2.5ex] 
$\Lz(1405)$ &1/2$^-$ & $1405.1^{+1.3}_{-1.0}$ & $50.5\pm 2.0$ & 3 & 4 \\
$\Lz(1520)$ &3/2$^-$ &$1519.5\pm 1.0$ & $15.6\pm 1.0$& 5 & 6 \\
$\Lz(1600)$ &1/2$^+$ &1600 & 150 &3 & 4 \\
$\Lz(1670)$ &1/2$^-$ & 1670 & 35 & 3 & 4\\
$\Lz(1690)$ &3/2$^-$ & 1690 & 60 & 5 & 6\\
$\Lz(1800)$ &1/2$^-$ & 1800 & 300 &4 &4 \\
$\Lz(1810)$ &1/2$^+$ &1810& 150&3&4\\
$\Lz(1820)$ &5/2$^+$ & 1820 & 80 &1&6\\
$\Lz(1830)$ &5/2$^-$ & 1830 & 95& 1&6 \\
$\Lz(1890)$ &3/2$^+$ & 1890 & 100 &3&6 \\
$\Lz(2100)$ &7/2$^-$ &2100 & 200&1 & 6\\
$\Lz(2110)$ &5/2$^+$ & 2110 & 200  &1 &6\\
$\Lz(2350)$ &9/2$^+$ & 2350 & 150 &0 &6\\
$\Lz(2585)$ &?& $\approx$2585 & 200 &0 & 6\\\hline
\end{tabular}
\label{tab:Lstar}
\end{table}

\subsubsection{\boldmath{The $\Lb\to \jpsi \PLambda^*$  decay amplitude}}
\label{sec:Lstardecayamp}

\begin{figure}[t]
\begin{center}
\includegraphics[width=.7\textwidth]{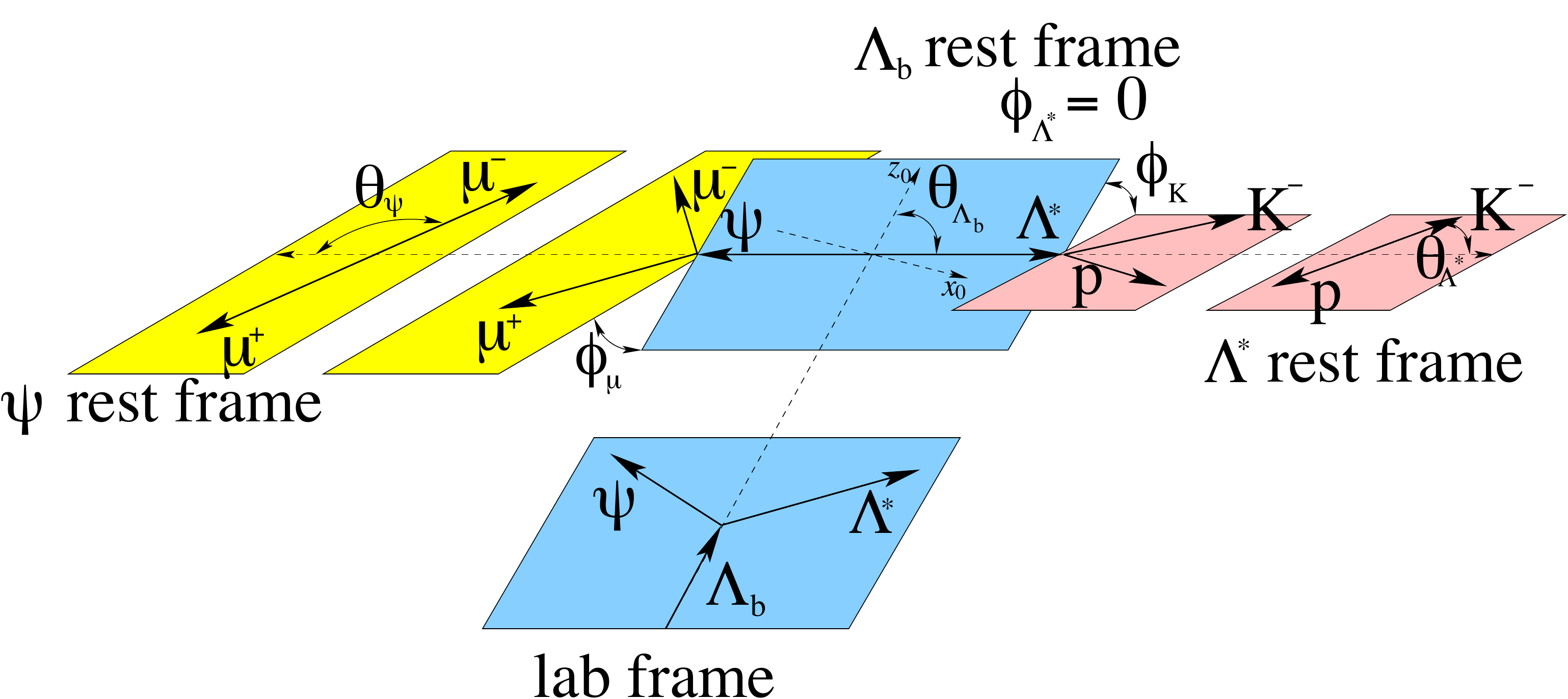}
\end{center}
\vskip -0.5cm
\caption{
Definition of the decay angles in the $\Lz^*$ decay chain (from \cite{Aaij:2015tga}).
}
\label{fig:helicitylambdastar}
\end{figure}

To identify the $\PLambda^*$ states present one could try to fit the $m_{Kp}$ mass distribution with some or all of the resonances listed in Table~\ref{tab:Lstar}. This procedure, however,  would only use part of the information available in the data. The correlations among the decay angular distributions of the final state particles contains information on the spin-parities of intermediate $\PLambda^*$ resonances. In fact, 
the decay of the \Lb into $\jpsi\PLambda$ with $\jpsi\to\mu^+\mu^-$ and $\PLambda^*\to K^- p$ is fully characterized by  $K^-p$ invariant mass and the five angular variables shown in Fig.~\ref{fig:helicitylambdastar}.  The \Lb decay plane is defined by the cross product of the $\psi$ and $\PLambda^*$ three vectors, while the $\psi$ decay plane is defined by the cross product of the $\mu^+$ and $\mu^-$ three vectors, and the $\PLambda^*$ decay plane by the cross product of the $K^-$ and $p$ three vectors. The angle between the \Lb rest frame and the $\psi$ rest frame is denoted as $\phi_{\mu}$, while the angle of the $\mu^-$ in the $\psi$ rest frame with respect to the $\psi$ direction is called $\theta_{\psi}$. Similarly, the angle between the \Lb rest frame and the $\PLambda^*$ rest frame is denoted as $\phi_{K}$, while the angle of the $K^-$ in the $\PLambda^*$ rest frame with respect to the $\PLambda^*$ direction is called $\theta_{\PLambda^*}$.

In order to fit the data to determine its resonant content, it is necessary to express the total decay amplitude  in terms of these variables and the allowed individual decay amplitudes for each allowed $\PLambda^*$ resonance decays listed in Table~\ref{tab:Lstar}.  The mathematical expression for this amplitude is derived in the supplementary material of Ref.~\cite{Aaij:2015tga}. Besides finding the correct angular expressions, the amplitude uses Breit-Wigner functions for the $m_{Kp}$ distributions except for the $\PLambda(1405)$ where a Flatt\'e parameterization \cite{Flatte:1976xu} is taken. These functions are labeled as $R_A(m_{Kp})$ in the expression below and $R_{\Lz^*n}(m_{Kp})$ in Eq.~[\ref{eq:matrix1}]

It is customary to use the helicity formalism \cite{Chung:1971ri,Richman:1984gh,Jacob:1959at} where each sub-decay $A\to B\,C$ contributes 
a term to the amplitude:
\begin{align}
\label{eq:hgeneric}
\H^{A\to B\,C}_{\lambda_{B},\,\lambda_{C}} \,\,
D^{\,\,J_A}_{\lambda_{A},\,\lambda_B-\lambda_C}(\phi_B,\theta_A,0)^*R_A(m_{BC})
= &
\H^{A\to B\,C}_{\lambda_{B},\,\lambda_{C}} \,\,
e^{i\,\lambda_A\,\phi_B}\,\,d^{\,\,J_A}_{\lambda_{A},\,\lambda_B-\lambda_C}(\theta_A)\nonumber\\
&\times R_A(m_{BC}).
\end{align}
The  $\lambda$'s  are the helicity quantum numbers given by the projection of the particles spin in the direction of its momentum vector, and $\H^{A\to B\,C}_{\lambda_{B},\,\lambda_{C}}$ are 
complex helicity amplitudes describing the decay dynamics. Here
$\theta_A$ and $\phi_B$ are the polar and azimuthal angles of $B$ in the rest frame of $A$ ($\theta_A$ is often called the  ``helicity angle" of $A$).
$D$ is the Wigner matrix whose three arguments are Euler angles describing the rotation of the initial coordinate system 
with the $z$-axis along the helicity axis of $A$ 
to the coordinate system with the $z$-axis along the helicity axis of $B$ \cite{pdg}. 
The convention is chosen which sets  the third Euler angle to zero. In Eq.~(\ref{eq:hgeneric}), $d^{J_A}_{\lambda_A,\lambda_B-\lambda_C}(\theta_A)$ is the Wigner reduced rotation matrix.

The helicity couplings  can be written in terms of partial wave amplitudes ($B_{L,S}$), 
where $L$ is the orbital angular momentum
in the decay, and $S$ is the total spin of $A$ plus $B$:
\begin{align}
\H_{\lambda_B,\lambda_C}^{A\to B\,C}=\sum_{L} \sum_{S} 
\sqrt{ \tfrac{2L+1}{2J_A+1} } B_{L,S} 
&\left( 
\begin{array}{cc|c}
 J_{B} & J_{C} & S \\
 \lambda_{B} & -\lambda_{C} & \lambda_{B}-\lambda_{C} 
\end{array}
\right)
\times  \nonumber\\
&\left( 
\begin{array}{cc|c}
 L  & S & J_A \\
 0 & \lambda_{B}-\lambda_{C} & \lambda_{B}-\lambda_{C}   
\end{array}
\right), 
\label{eq:LS}
\end{align}
where the expressions in parentheses are Clebsch-Gordon coefficients \cite{Jacob:1959at}.
For strong decays, possible $L$ values are constrained 
by the conservation of parity: $P_A=P_B\,P_C\,(-1)^L$.

The matrix element for the $\Lb\to\jpsi\Lz^*$ decay sequence is 
\begin{align}
\label{eq:matrix1}
\Mat_{\lambda_{\Lb},\,\lambda_p,\,\Delta\lambda_\mu}^{\Lz^*} \equiv &
\sum\limits_{n}
\sum\limits_{\lambda_{\Lz^*}}
\sum\limits_{\lambda_{\psi}} 
\,\,
\H^{\Lb\to \Lz^*n \psi}_{\lambda_{\Lz^*},\,\lambda_{\psi}} 
D^{\,\,\frac{1}{2}}_{\lambda_{\Lb},\,\lambda_\Lz^*-\lambda_\psi}(0,\theta_{\Lb},0)^*\H^{\Lz^*_n\to K p}_{\lambda_p,\,0} 
\\\nonumber
&
D^{\,\,J_{\Lz^*n}}_{\lambda_{\Lz^*},\,\lambda_p}(
\phi_{K},\theta_{\Lz^*},0)^*
R_{\Lz^*n}(m_{Kp})
D^{\,\,1}_{\lambda_{\psi},\,\Delta\lambda_\mu}(\phi_{\mu},\theta_{\psi},0)^*,
\label{eq:lbtolpsi}
\end{align}
where the $x$-axis, in the coordinates describing the $\Lb$ decay, is chosen 
to fix $\phi_{\Lz^*}=0$.
The sum over $n$ allows several different $\Lz^*_n$ resonances to
contribute to the amplitude. Since the $\jpsi$ decay is electromagnetic, the values of 
$\Delta\lambda_\mu \equiv \lambda_{\mu^+}-\lambda_{\mu^-}$ are restricted to $\pm1$.

\subsubsection{Fits of the reaction \boldmath{$\Lb\to\jpsi \Lz^*$} to the data}

The next step is to fit the square of the matrix element given in Eq.~\ref{eq:lbtolpsi} to the  background subtracted and efficiency corrected data. One method of subtracting the background uses the sidebands on either side of the \Lb mass peak to provide a sample of events. The efficiency correction is done by simulating many events according to a model where the decay $\Lb\to\jpsi K^- p$ has a unit matrix element and so the decay distribution merely represents the available phase space. Then, after the fully simulated events are reconstructed, maps of the efficiencies versus all of the variables are formed.  An unbinned maximum likelihood ($\Like$) fit is then performed. The differences between values of $-2\ln\Like$ are used to discriminate among the fits with various resonant or non-resonant components included. The likelihood $\Like$ is itself a function of $m_{Kp}$ and the five angular variables.

\begin{figure}[b]
\begin{center}
\vskip -4mm
\includegraphics[width=0.35\textwidth]{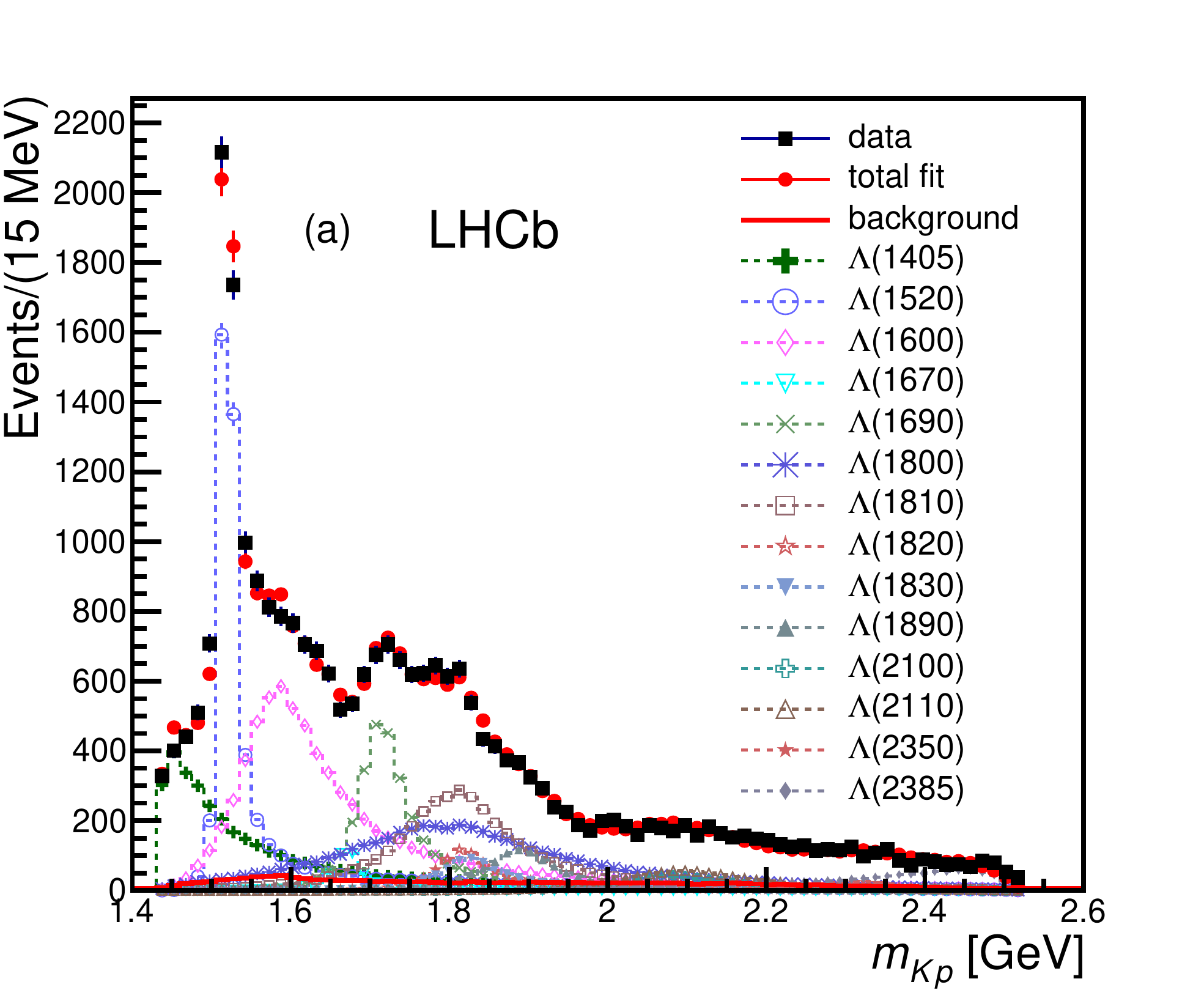}\includegraphics[width=0.35\textwidth]{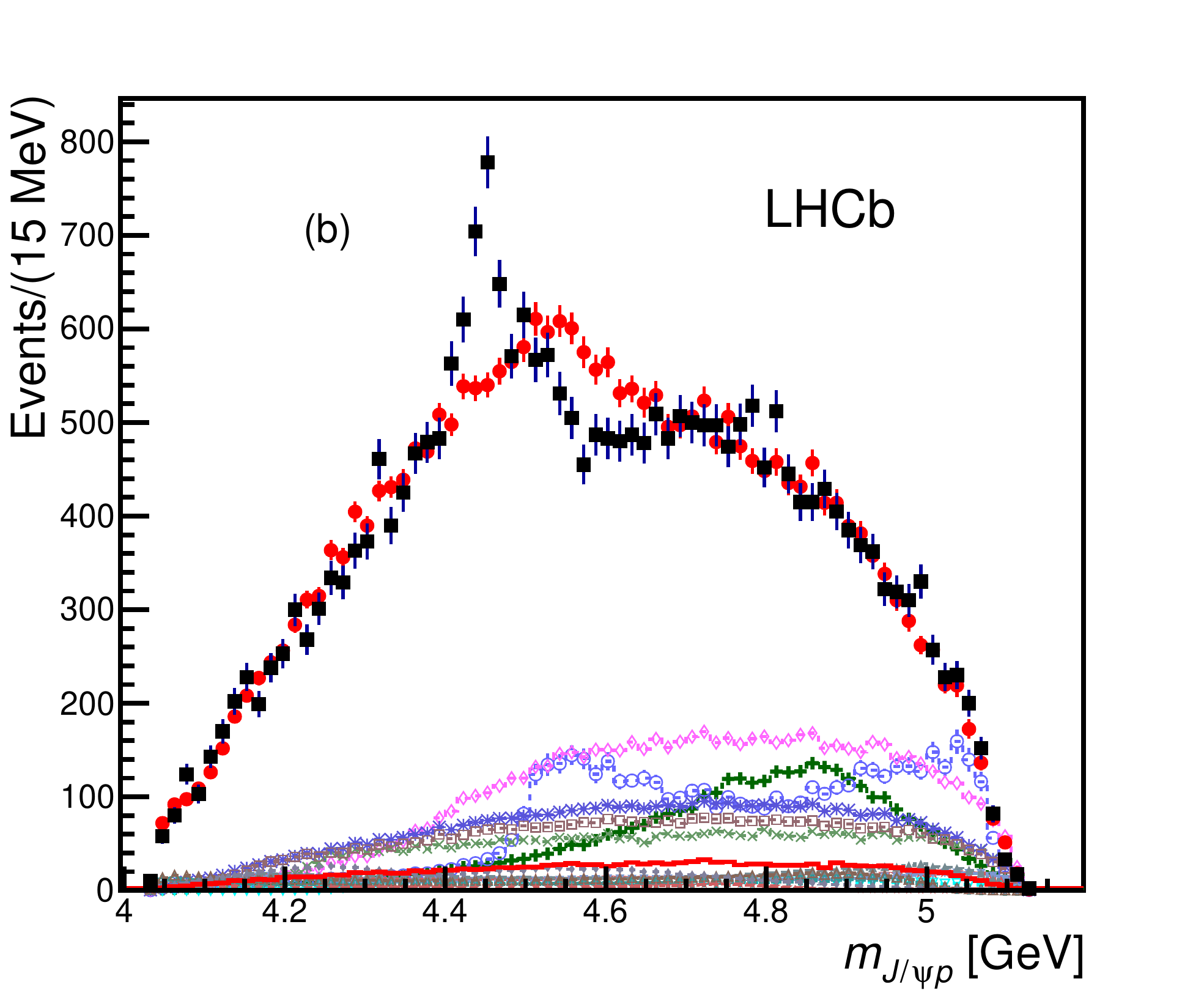}
\end{center}
\vskip -0.5cm
\caption{Results for (a) $m_{Kp}$  and (b) $m_{\jpsi p}$ for the extended $\Lz^*$  model fit without $P_c^+$ states. The data are shown as (black) squares with error bars, while the (red) circles show the results of the fit. The error bars on the points showing the fit results are due to simulation statistics (from \cite{Aaij:2015tga}).}
\label{Mall}
\end{figure}

The results with only $\Lz^*$ without any $P_c^+$ component are shown in Fig.~\ref{Mall}.  The $m_{KP}$ variable is one of the fit variables, while $m_{\jpsi p}$ is calculated from the other variables. The $m_{Kp}$ distribution is reasonably well described by $\Lz^*$ resonances and their interferences, however the peaking structure in $m_{\jpsi p}$ is not reproduced. 

Not satisfied with using all the known $\Lz^*$ states LHCb tried several other different configurations: (i) added all the possible $\PSigma^*$ states, (ii) added two additional $\Lz^*$ allowing their masses and widths to float in the fit and allowed spins up to $5/2$ with both parities, and (iii) added four non-resonant components with $J^P=1/2^+,~1/2^-,3/2^+,$ and $3/2^-$. None of these fits explains the data, indeed the improvements were small. 

\subsubsection{The matrix element for \boldmath{$\Lb\to\jpsi\ZP^+ K^-$} decay}
\label{sec:matrixPc}

\begin{figure}[t]
\begin{center}
\includegraphics[width=.55\textwidth]{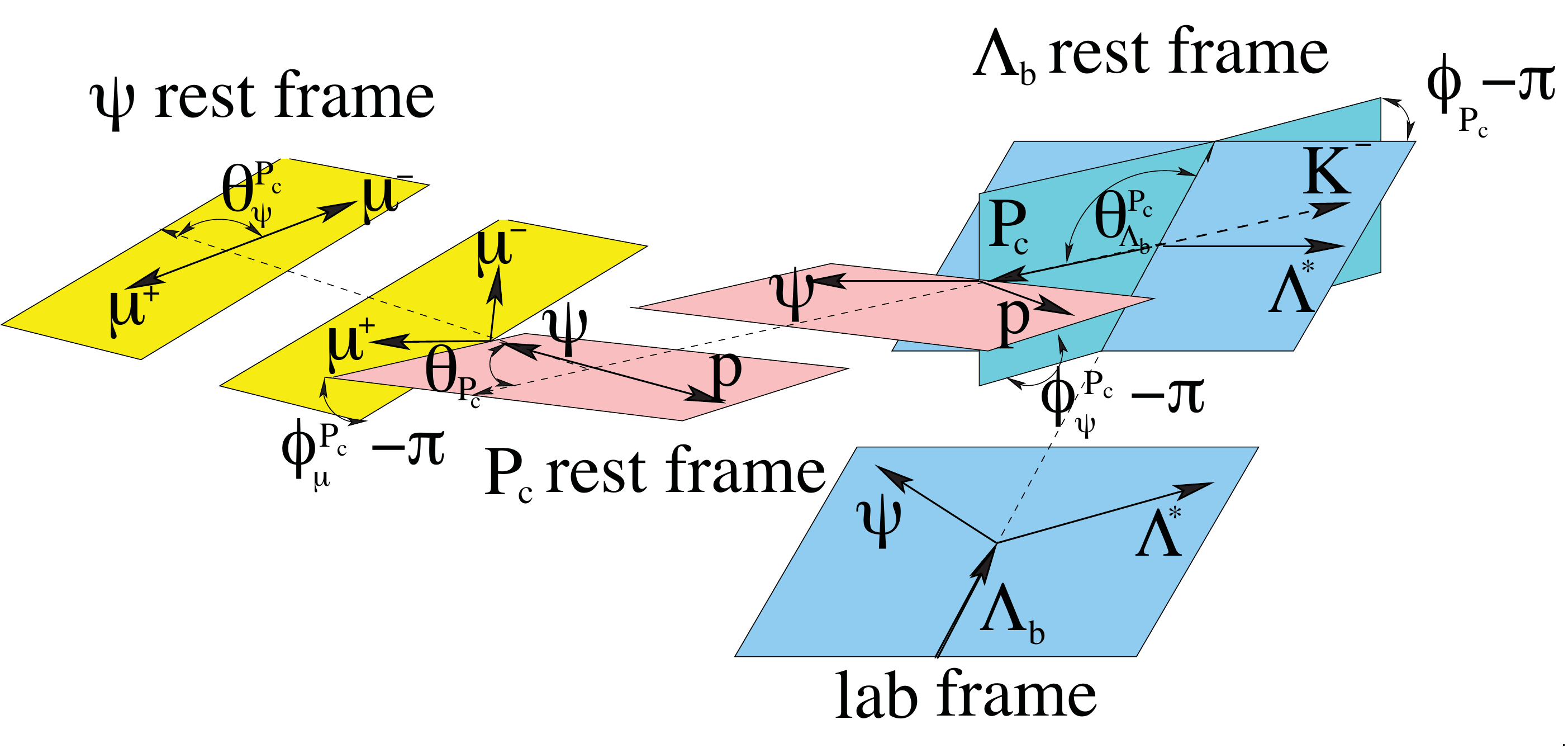}
\end{center}
\vskip -0.8cm
\caption{
Definition of the decay angles in the $\ZP^+$ decay chain (from \cite{Aaij:2015tga}).
}
\label{fig:helicitypc}
\end{figure}

In order to examine whether resonant structure can explain the angular data and the $\jpsi p$ mass spectrum, the matrix element for the decay $\Lb\to\jpsi\ZP^+ K^-$, $\ZP^+\to\jpsi p$ needs to be constructed.
The decay angles for the $P_c^+$ decay sequence are defined in Fig.~\ref{fig:helicitypc}.
The matrix element for the $\ZP^+$ decay chain is given by
\begin{align}
\Mat_{\lambda_{\Lb},\,\lambda_p^{\ZP},\,\Delta\lambda_\mu^{\ZP}}^{\ZP} \equiv &
\sum\limits_{j}
\sum\limits_{\lambda_{\ZP}}
\sum\limits_{\lambda_{\psi}^{\ZP}} 
\,\,
\H^{\Lb\to {\ZP}_j K}_{\lambda_{\ZP},\,0} 
D^{\,\,\frac{1}{2}}_{\lambda_{\Lb},\,\lambda_{\ZP}}(
\phi_{\ZP},\theta_{\Lb}^{\ZP},0)^*\\\nonumber
&\times \H^{{\ZP}_j\to \psi p}_{\lambda_\psi^{\ZP},\lambda_p^{\ZP}} 
D^{\,\,J_{{\ZP}_j}}_{\lambda_{\ZP},\,\lambda_\psi^{\ZP}-\lambda_p^{\ZP}}(
\phi_{\psi},\theta^{\ZP},0)^*
R_{{\ZP}_j}(m_{\psi p}) \\\nonumber
&\times D^{\,\,1}_{\lambda_{\psi}^{\ZP},\,\Delta\lambda_\mu^{\ZP}}(
\phi_{\mu}^{\ZP},\theta_{\psi}^{\ZP},0)^*,
\label{eq:lbtopck}
\end{align}
where the angles and helicity states carry the superscript or subscript $\ZP$ to 
distinguish them from those defined for the $\Lz^*$ decay chain. The sum over $j$ allows for the possibility of contributions from more than one $P_c^+$ resonance.
There are 2 (3) independent helicity couplings 
$\H^{{\ZP}_j\to \psi p}_{\lambda_\psi^{\ZP},\lambda_p^{\ZP}}$ 
for $J_{{\ZP}_j}=\frac{1}{2}$ ($>\frac{1}{2}$), 
and a ratio of the two $\H^{\Lb\to {\ZP}_j K}_{\lambda_{\ZP},\,0}$ couplings, 
to determine from the data.

Before the matrix elements for the two decay sequences can be added coherently,  
the proton and muon helicity states in the $\Lz^*$ decay chain must be 
expressed in the basis of helicities in the $\ZP^+$  decay chain. The appropriate transformation is:
\begin{equation} 
\left| \Mat \right|^2 = 
\sum\limits_{\lambda_{\Lb}}
\sum\limits_{\lambda_{p}}
\sum\limits_{\Delta\lambda_{\mu}}
\left|
\Mat_{\lambda_{\Lb},\,\lambda_p,\,\Delta\lambda_\mu}^{\Lz^*} 
+ 
e^{i\,{\Delta\lambda_\mu}\alpha_{\mu}}\,
\sum\limits_{\lambda_p^{\ZP}} 
d^{\,\,\frac{1}{2}}_{\lambda_p^{\ZP},\,\lambda_p}(\theta_p)\,
\Mat_{\lambda_{\Lb},\,\lambda_p^{\ZP},\,\Delta\lambda_\mu}^{\ZP} 
\right|^2,
\label{eq:matrixelement}
\end{equation}  
where $\theta_p$ is the polar angle in the $p$ rest frame between the boost directions from
the $\Lz^*$ and $\ZP^+$ rest frames, 
and $\alpha_{\mu}$ is the azimuthal angle correcting for the difference between the muon helicity states 
in the two decay chains. 
Note that $m_{\psi p}$,
$\theta_{\Lb}^{\ZP}$, $\phi_{\ZP}$,
$\theta_{\ZP}$, $\phi_{\psi}$, 
$\theta_{\psi}^{\ZP}$, $\phi_{\mu}^{\ZP}$,
$\theta_p$ and $\alpha_\mu$ 
can all be derived from the values of $m_{Kp}$ and $\POmega$, and
thus do not constitute independent dimensions in the $\Lb$ decay phase space. 
 
$\Lb$ production at the LHC is mediated by strong interactions which 
conserve parity and, therefore, cannot produce longitudinal $\Lb$ polarization \cite{Soffer:1991am}. 
Thus, $\lambda_{\Lb}=+1/2$ and $-1/2$ values are equally likely, which is assumed in writing down Eq.~(\ref{eq:matrixelement}). 

\subsubsection{\boldmath{Amplitude fits of $\Lb\to\jpsi K^- p$ allowing $\ZP^+$ states}}
In each fit we minimize  $-2\ln{\cal{L}}$ where ${\cal{L}}$ represents the fit likelihood.
The difference of $\Delta\equiv-2\ln{\cal{L}}$ between different
amplitude models reflects the goodness of fit.
For two models representing separate hypotheses, \eg\, when discriminating between
different $J^P$ values assigned to a $P_c^+$ state, 
the assumption of a $\chi^2$ distribution with one degree of freedom
for $\Delta$ under the disfavored $J^P$ hypothesis 
allows the calculation of a lower limit on the significance of its 
rejection, \ie\, the p-value \cite{james2006statistical}.
Therefore, it is convenient to express  values of $\Delta$ as
${n^2_\sigma}$, where $n_\sigma$ corresponds to the  number of  standard
deviations in the normal distribution with the same p-value. 
When discriminating between models without and with $P_c^+$ states,
$n_\sigma$ overestimates the p-value by a modest amount. Thus, we use
simulations to obtain better estimates of the significance of the
$P_c^+$ states.

\begin{figure}[t]
\begin{center}
\includegraphics[width=0.4\textwidth]{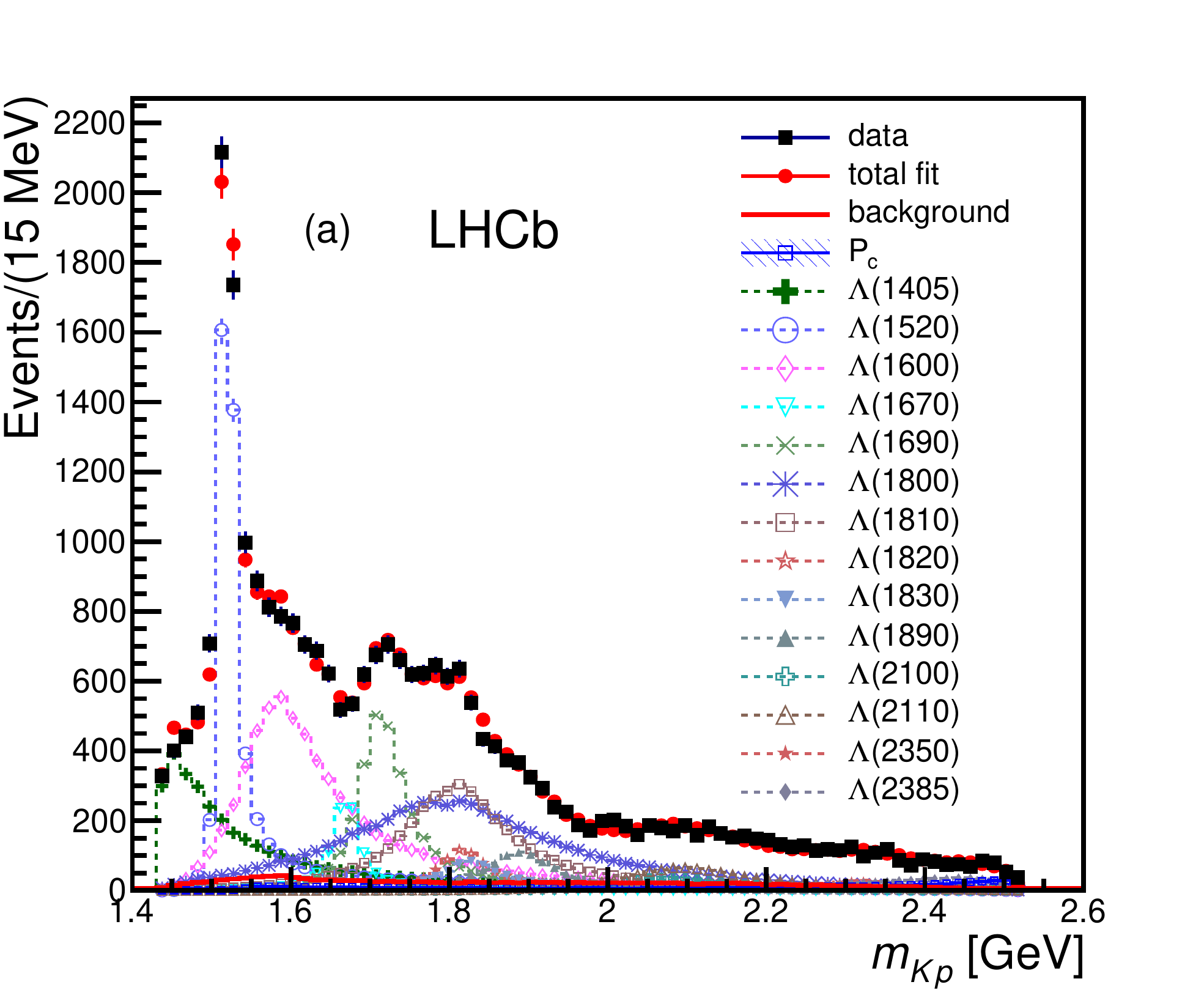}
\includegraphics[width=0.4\textwidth]{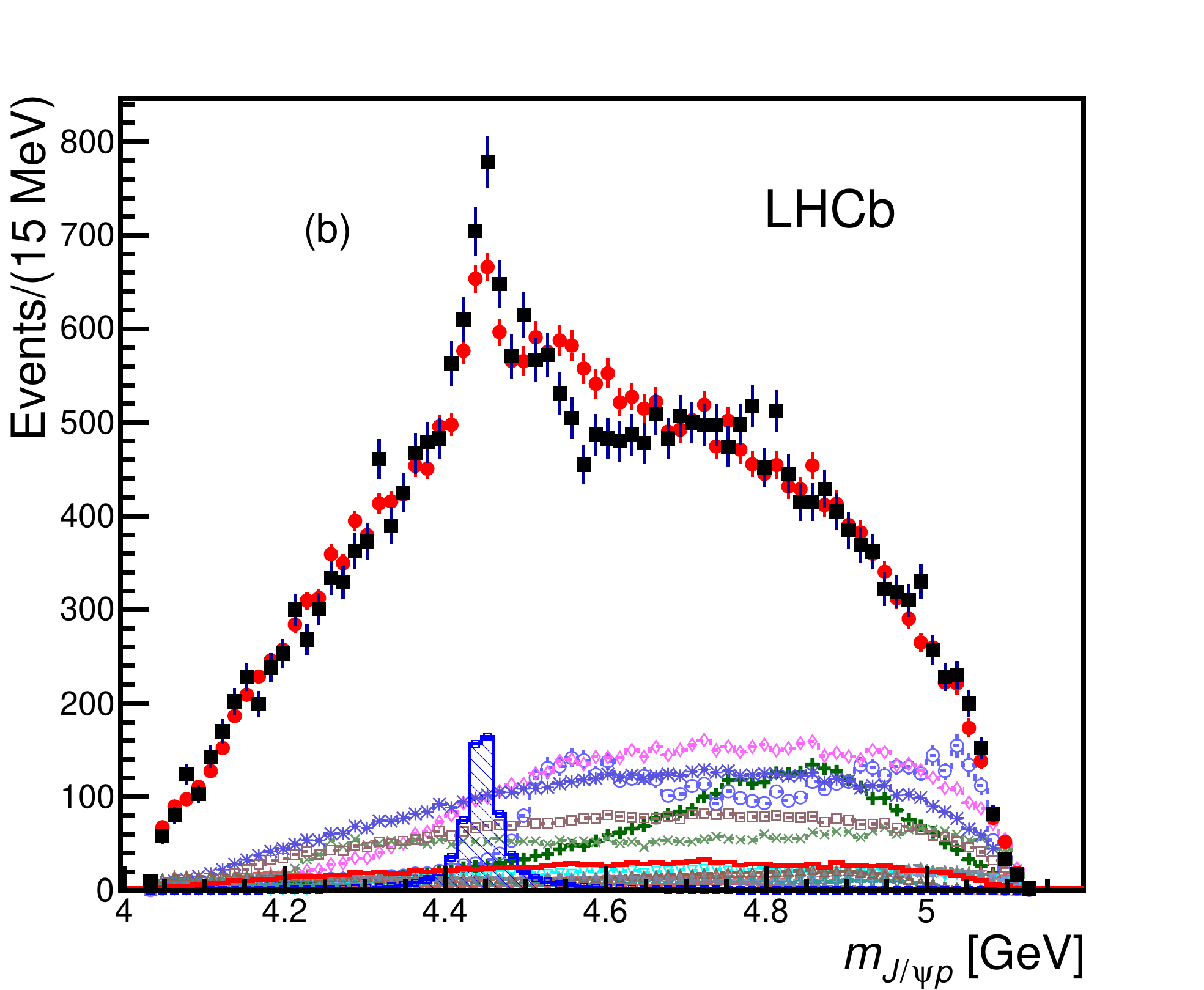}
\end{center}
\vskip -0.5cm
\caption{Results of the fit with one $J^P=5/2^+$ $P_c^+$ candidate. (a) Projection of the invariant mass of $K^-p$ combinations from $\Lb\to\jpsi K^-p$ candidates. The data are shown as (black) squares with error bars, while the  (red) circles show the results of the fit;  (b) the corresponding $\jpsi p$ mass projection. The (blue) shaded plot shows the $P_c^+$ projection, the other curves represent individual  $\Lz^*$ states.  From \cite{Aaij:2015tga}. }
\label{Extended-1Pc}
\end{figure}

We perform separate fits for 
$J^P$ values of $1/2^{\pm}$, $3/2^{\pm}$ and $5/2^{\pm}$. The mass and width of the putative $P_c^+$ state are allowed to vary. The best fit prefers a $5/2^+$ state, which improves $-2\ln {\cal{L}}$ by 14.7$^2$. 
Figure~\ref{Extended-1Pc} shows the projections for this fit.  While the $m_{Kp}$ projection is well described, clear discrepancies in $m_{\jpsi p}$ remain visible. 

\begin{figure}[b!]
\begin{center}
\includegraphics[width=0.4\textwidth]{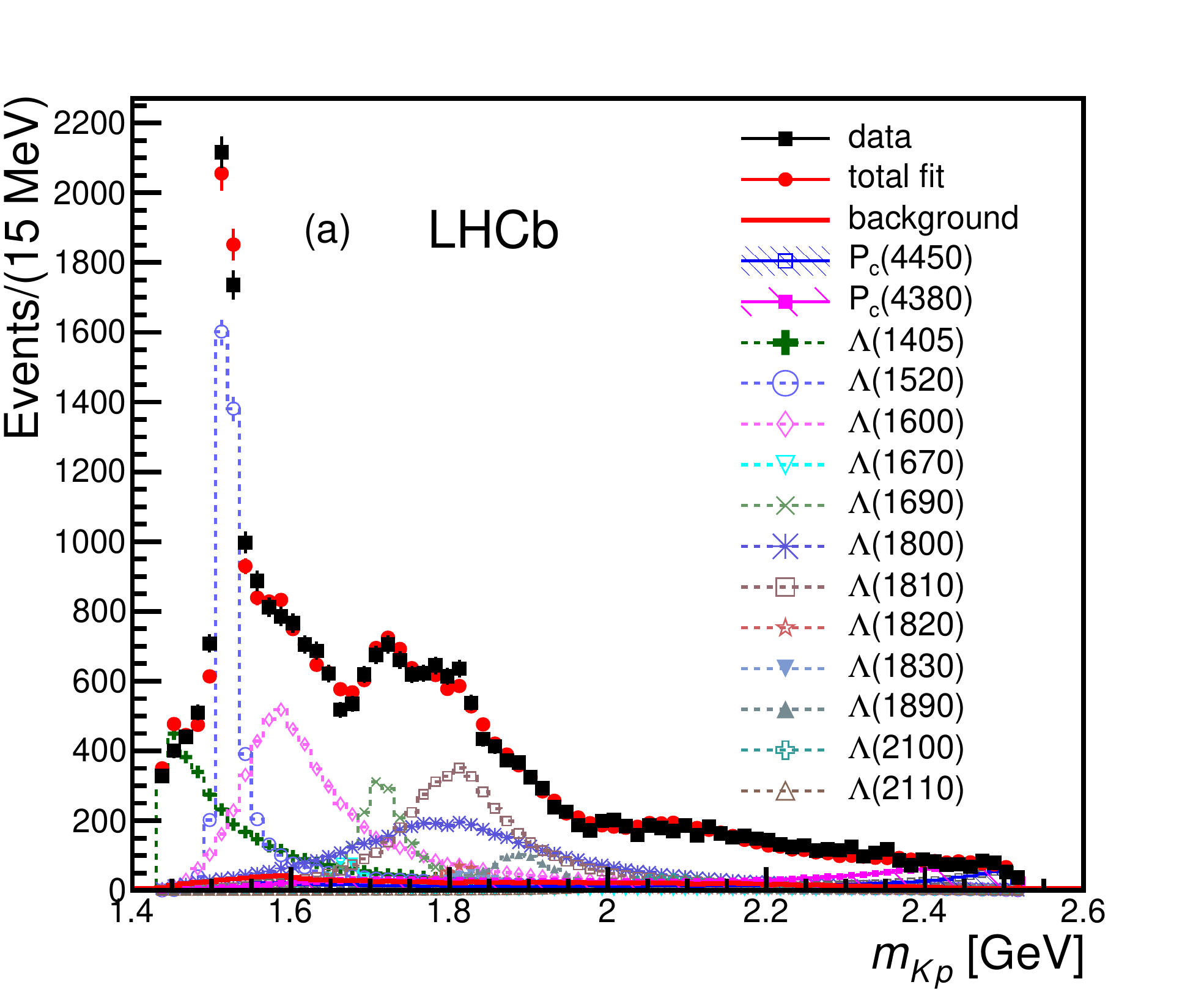}\includegraphics[width=0.4\textwidth]{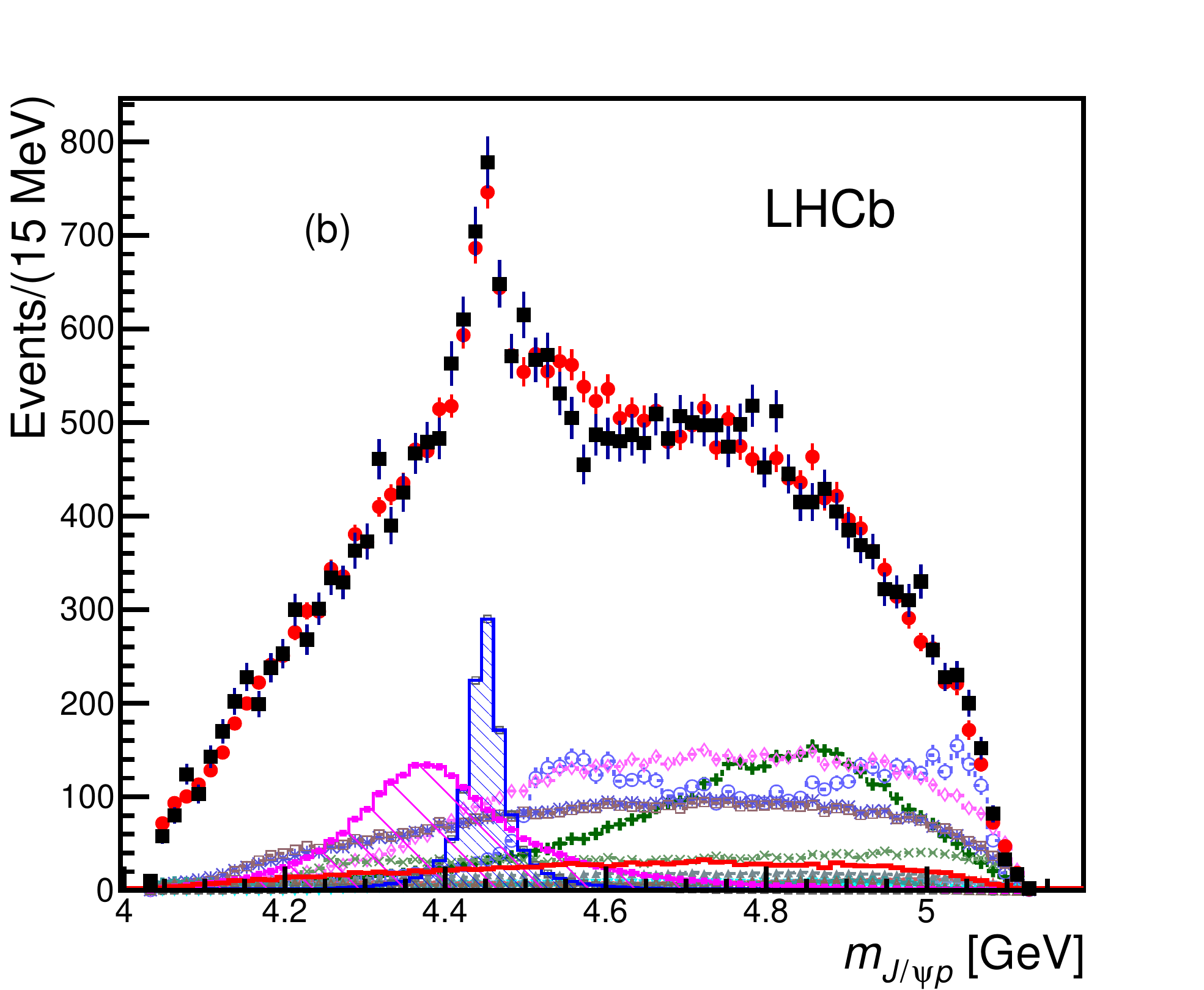}
\end{center}
\vskip -0.1cm
\caption{Fit projections for (a) $m_{Kp}$  and (b) $m_{\jpsi p}$ for the reduced $\Lz^*$ model with two $P_c^+$ states (see Table~\ref{tab:Lstar}). The data are shown as solid (black) squares, while the solid (red) points show the results of the fit.  The solid (red) histogram shows the background distribution. The (blue) open squares with the shaded histogram represent the $P_c(4450)^+$ state, and the shaded histogram topped with (purple) filled squares represents the $P_c(4380)^+$ state. Each $\Lz^*$ component is also shown (from \cite{Aaij:2015tga}).}
\label{Pc2d}
\end{figure}

The next step is to fit with two $P_c^+$ states including their allowed interference. These fits were performed both with the reduced model and the extended model in order to estimate systematic uncertainties.  Toy simulations are done to more accurately evaluate the statistical significances of the two states, resulting in  9 and 12 standard deviations, for  lower mass and higher mass states, using the extended model which gives lower significances. The best  fit projections are shown in Fig.~\ref{Pc2d}. Both  $m_{Kp}$ and the peaking structure in $m_{\jpsi p}$ are reproduced by the fit. The reduced model has 64 free parameters for the $\Lz^*$ rather than 146 and allows for a much more efficient examination of the parameter space and, thus, is used for numerical results.
The two  $P_c^+$ states are found to have masses of  $4380\pm 8\pm 29$~MeV and $4449.8\pm 1.7\pm 2.5$~MeV, with corresponding widths of  $205\pm 18\pm 86$ MeV and $39\pm 5\pm19$ MeV. (Whenever two uncertainties are quoted the first is statistical and the second systematic.)  
The fractions of the total sample due to the lower mass and higher mass states are ($8.4\pm0.7\pm4.2$)\% and ($4.1\pm0.5\pm 1.1)$\%, respectively. The overall branching fraction has recently been determined to be  \cite{Aaij:2015fea}.
\begin{equation}
{\cal{B}}(\Lb\to \jpsi K^- p)= \left(3.04\pm 0.04^{+0.55}_{-0.43}\right)\times 10^{-4},
\end{equation}
where the systematic uncertainty is largely due to the normalization procedure, leading to the product branching fractions:
\begin{align}
{\cal{B}}(\Lb\to P_c(4380)^+ K^- p){\cal{B}}(P_c(4380)^+ \to \jpsi p)= &\left(2.56^{+1.38}_{-1.34}\right)\times 10^{-5} \nonumber \\
{\cal{B}}(\Lb\to P_c(4450)^+ K^- p){\cal{B}}(P_c(4450)^+ \to \jpsi p)= &\left(1.25^{+0.42}_{-0.40}\right)\times 10^{-5},
\end{align} 
where all the uncertainties have been added in quadrature.

\begin{figure}[b!]
\vskip -0.5cm
\begin{center}
\includegraphics[width=0.54\textwidth]{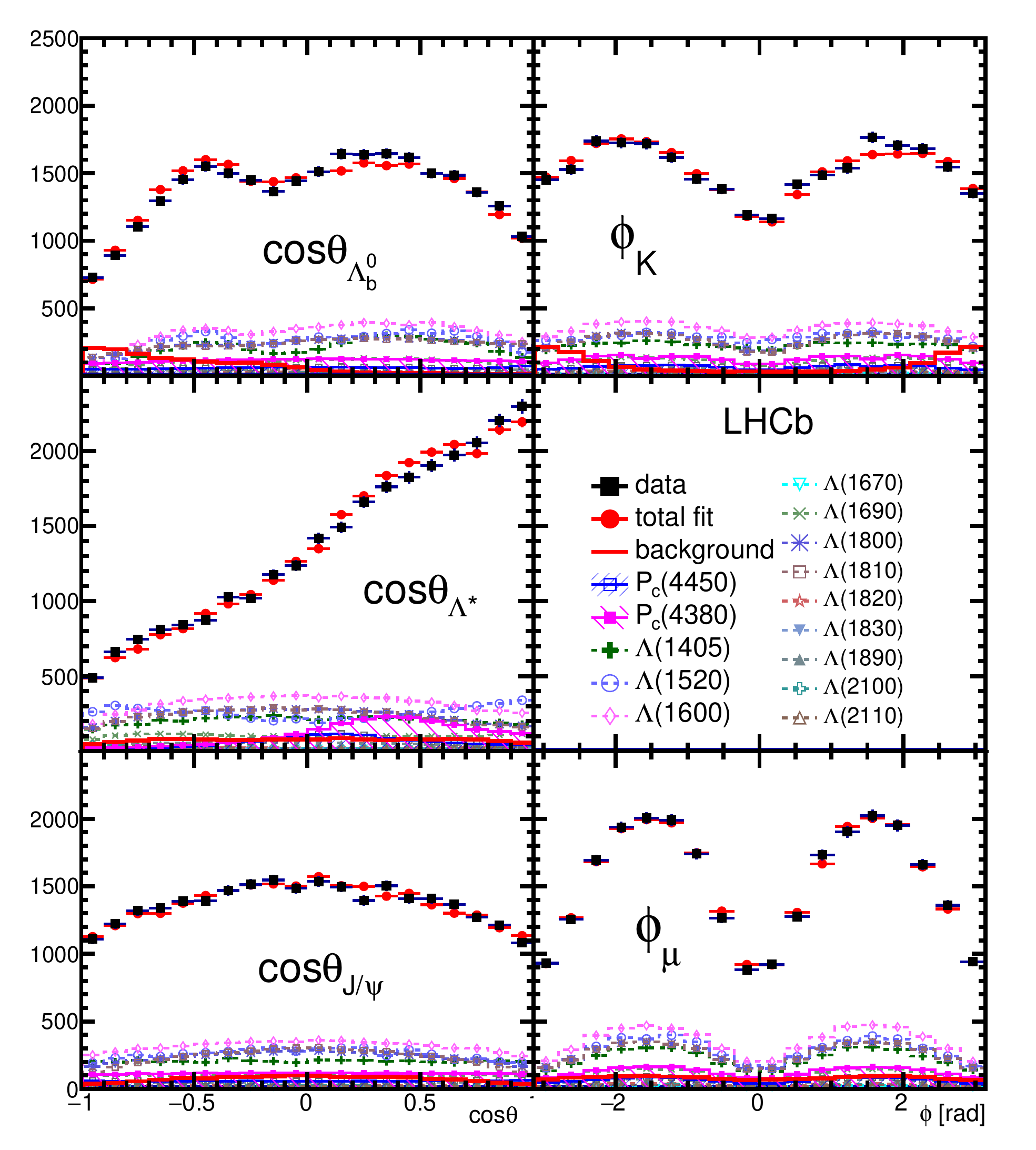}
\end{center}
\vskip -0.5cm
\caption{Various decay angular distributions for the fit with two $P_c^+$ states. The data are shown as (black) squares, while the (red) circles show the results of the fit.  Each fit  component is also shown. The angles are defined in the text (from \cite{Aaij:2015tga}).}
\label{TwoZ-angular-cFit}
\end{figure}

 The best fit solution has spin-parity $J^P$ values of ($3/2^-$, $5/2^+$). Acceptable solutions are  also found for additional cases with opposite parity, either ($3/2^+$, $5/2^-$) or ($5/2^+$, $3/2^-$). 
 The five angular distributions are also well fit as can be seen in Fig.~\ref{TwoZ-angular-cFit}. We note that the concept of dynamically generated resonances led to some predictions of pentaquarks states decaying into $\jpsi p $\cite{Wu:2010jy,*Wu:2010vk}.

\begin{figure}[t!]
\vskip -.5cm
\begin{center}
\includegraphics[width=0.5\textwidth]{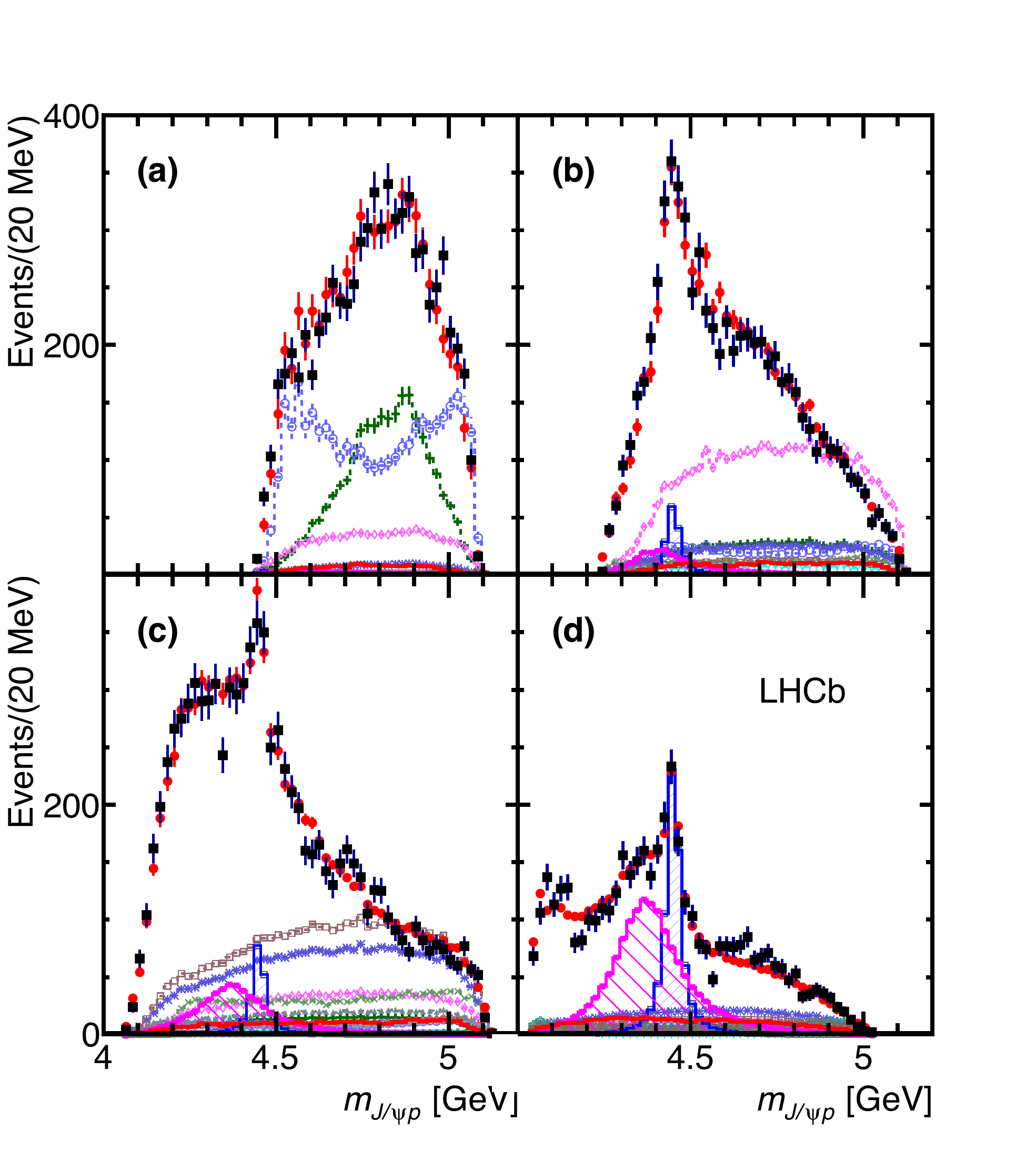}\end{center}
\vskip -0.7cm
\caption{$m_{\jpsi p}$ in various intervals of $m_{Kp}$ for the fit with two $P_c^+$ states: (a) $m_{Kp}<1.55$~GeV, (b) $1.55<m_{Kp}<1.70$~GeV, (c) $1.70<m_{Kp}<2.00$~GeV, and (d) $m_{Kp}>2.00$~GeV.  The data are shown as  (black) squares with error bars, while the (red) circles show the results of the fit. The blue and purple histograms show the two $P_c^+$ states. See Fig.~\ref{TwoZ-angular-cFit} for the legend (from \cite{Aaij:2015tga}).}
\label{TwoZ-mjpsip-bins-cFit}
\end{figure}

\begin{figure}[b!]
\vskip -7mm
\begin{center}
\includegraphics[width=0.45\textwidth]{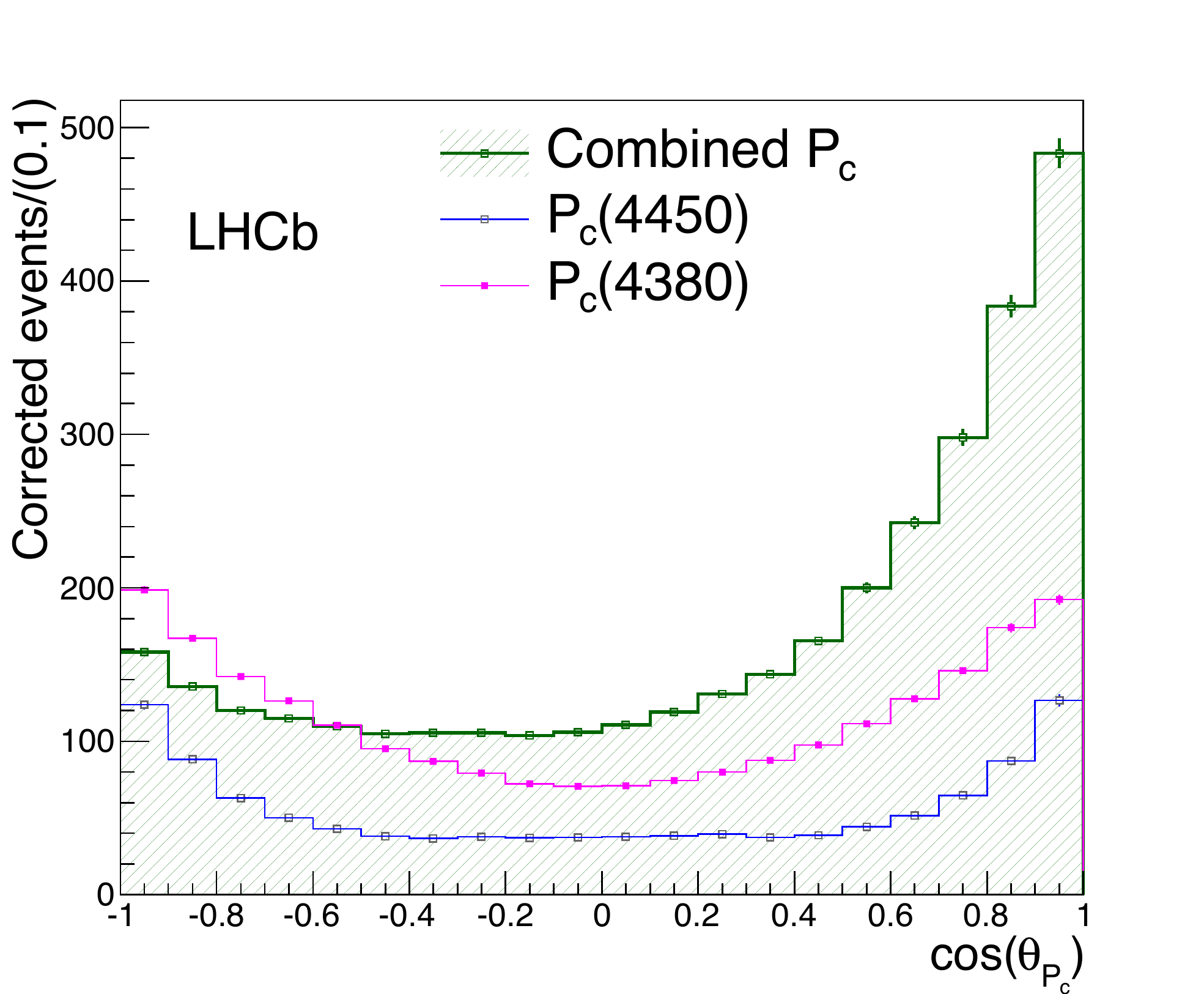}\end{center}
\vskip -0.5cm
\caption{Efficiency corrected and background subtracted fit projections of the decay angular distributions for the two $P_c^+$ states and their sum. Values of $\cos\theta_{P_c}$ near $-1$ are correlated with values of $m_{Kp}$ near threshold, while those near $+1$ are correlated with higher values (from \cite{Stone:2015iba}).}
\label{cosPc_v4}
\end{figure}

The fit projections in different slices of $K^-p$ invariant mass are given in Fig.~\ref{TwoZ-mjpsip-bins-cFit}.  In slice (a) the $P_c^+$ states are not present, nor should they be as they are outside of the Dalitz plot boundary. In slice (d) both $P_c^+$ states form a large part of the mass spectrum; there is also a considerable amount of negative interference between them. This can be seen better by examining the decay angle of the $P_c^+$, $\theta_P$, the angle of the proton in $\jpsi p$ rest frame with respect to the $P_c^+$ direction transformed into its rest frame, shown in Fig.~\ref{cosPc_v4} for the entire $m_{Kp}$ range. The summed fit projections agrees very well with the angular distributions in the data showing that two interfering states are needed to reproduce the asymmetric distribution.\footnote{It can be shown mathematically that the states need to be of opposite parity.}

Systematic uncertainties are evaluated for the masses, widths and fit fractions of the $P_c^+$ states, and for the fit fractions of the two lightest and most significant $\Lz^*$ states. Additional sources of modeling uncertainty that were not considered may affect the fit fractions of the heavier $\Lz^*$ states. The sources of systematic uncertainties  are listed in Table~\ref{tab:syssum}.
They include differences between the results of the extended versus reduced model, varying the $\Lz^*$ masses and widths, uncertainties in the identification requirements for the proton, and restricting its momentum, inclusion of a nonresonant amplitude in the fit, use of separate higher and lower \Lb mass sidebands, alternate $J^P$ fits, varying the Blatt-Weisskopf barrier factor, $d$,  between 1.5 and 4.5~GeV$^{-1}$ in the Breit-Wigner mass shape-function, changing the  angular momentum $L$  by one or two units, and accounting for potential mis-modeling of the efficiencies. For the $\Lz(1405)$ fit fraction we also added an uncertainty for the Flatt\'e couplings, determined by both halving and doubling their ratio, and taking the maximum deviation as the uncertainty. 

The stability of the results are cross-checked by comparing the data recorded in 2011/2012, with the LHCb dipole magnet polarity in up/down configurations, \Lb/\Lbbar decays, and \Lb produced with low/high values of \pt. The fitters were tested on simulated pseudoexperiments and no biases were found. In addition, selection requirements are varied, and the vetoes of \Bsb and \Bdb  are removed and explicit models of those backgrounds added to the fit; all give consistent results.

\begin{table}[t]
\centering
\caption{Summary of systematic uncertainties on $P_c^+$ masses, widths and fit fractions, and $\Lz^*$ fit fractions. 
A fit fraction is the ratio of the phase space integrals of the matrix element squared for a single resonance and for the total amplitude. 
The terms ``low" and ``high" correspond to the lower and higher mass $P_c^+$ states.}
\vspace{0.2cm}
\resizebox{\textwidth}{!}{ 
\begin{tabular}{l|rr|rrrr|cc}
\hline
~~~Source & \multicolumn{2}{c|}{$M_0$~(MeV)} &\multicolumn{4}{c}{$\Gamma_0$~(MeV)} &\multicolumn{2}{c|}{Fit fractions (\%)}\\
 &low& high$\!\!$ & low& high$\!\!$& low& high&$\Lz(1405)$&$\Lz(1520)$\\
\hline
Extended vs. reduced &$21$ &0.2&$54$&10 &$3.14$&0.32&1.37~~~&0.15\\
$\Lz^*$ masses \& widths &7 &0.7 &20&4&0.58&0.37&2.49~~~&2.45\\
Proton ID &2& 0.3 & $1$ & $2$&0.27&0.14&0.20~~~&0.05\\
$10<p_p<100$~GeV &  0&1.2&$1$&$1$&0.09&0.03&0.31~~~&0.01\\
Nonresonant & 3&0.3&$34$&~2 &$2.35$&$0.13$&3.28~~~&0.39\\
Separate sidebands & 0 &~~0&~5&~0&$0.24$&$0.14$&0.02~~~&0.03\\
$J^P$ ($3/2^+$, $5/2^-$) or ($5/2^+$, $3/2^-$) &$10$&$1.2$&34&10&$0.76$&0.44&& \\
$d=1.5-4.5~$GeV$^{-1}$&$9$&0.6&19&$3$&0.29&0.42&0.36~~~&1.91\\
$L_{\Lb}^{{P_c^+}}$ $\Lb\to P_c^+~{\rm (low/high)} K^-$ &6&0.7&4&8&$0.37$&0.16&&\\
$L_{{P_c^+}}$ $P_c^+~{\rm (low/high)}\to\jpsi p$&4&$0.4$&31&7&$0.63$&0.37&&\\
$L_{\Lb}^{\Lz^*_{\!n}}$ $\Lb\to \jpsi \Lz^*$&11&0.3&20&2&0.81&0.53&3.34~~~&2.31\\
Efficiencies &1&0.4&4&0&0.13&0.02&0.26~~~&0.23\\
Change $\Lz(1405)$ coupling&0 &0&0&0&0&0&1.90~~~&~~~0\\
\hline
Overall & 29&2.5&86&19&4.21&1.05&5.82~~~&3.89\\\hline
\end{tabular}
}
\label{tab:syssum}
\end{table}

\begin{figure}[t]
\begin{center}
\includegraphics[width=0.9\textwidth]{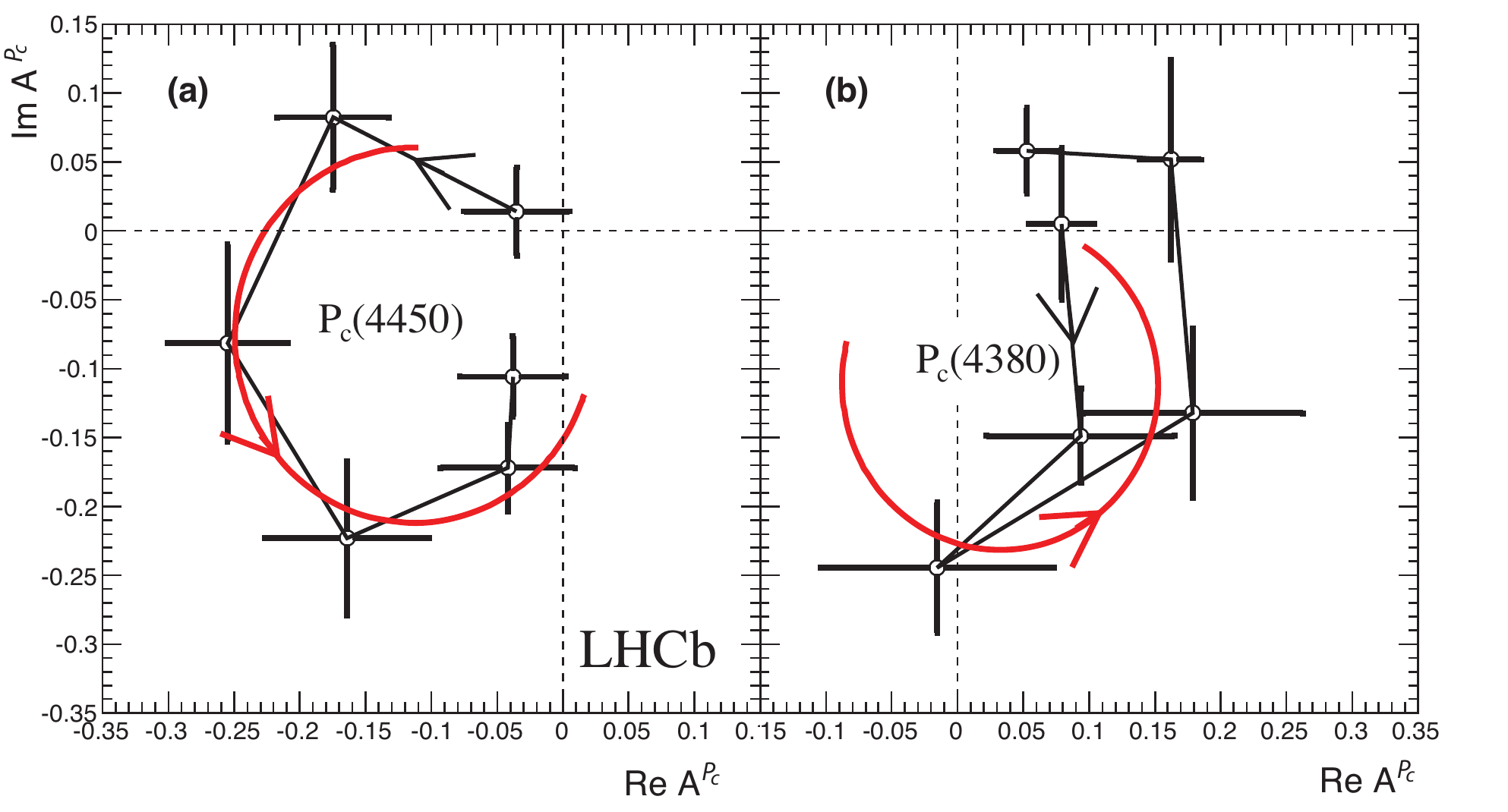}
\vskip -0.2cm
\caption{
Fitted values of the real and imaginary parts of the amplitudes for the baseline ($3/2^-$, $5/2^+$) fit for a) the $P_c(4450)^+$ state and b) the $P_c(4380)^+$ state, each divided into six $m_{\jpsi p}$ bins of equal width between $-\Gamma$ and $+\Gamma$ shown in the Argand diagrams as connected points with error bars ($m_{\jpsi p}$ increases counterclockwise).
The solid (red) curves are the predictions
from the Breit-Wigner formula for the same mass ranges
with  $M$ ($\Gamma$) of
 4450 (39) \mev and 4380 (205) \mev, respectively,
with the phases and magnitudes at the resonance masses set to the
average values between the two points around $M$.  
The phase convention is fixed by the $\Lz(1520)$. Systematic uncertainties are not included (from \cite{Aaij:2015tga}).}
\label{DoubleArgand}
\end{center}
\end{figure}

Further evidence for the resonant character of the higher mass, narrower, $P_c^+$ state is obtained by viewing the evolution of the complex amplitude in the Argand diagram  \cite{pdg}.
In the amplitude fits discussed above, the $P_c(4450)^+$ is represented by a Breit-Wigner amplitude, where
the magnitude and phase vary with $m_{\jpsi p}$ according to
an approximately circular trajectory in the (Re$\,A^{P_c}$, Im$\,A^{P_c}$) plane, where
$A^{P_c}$ is the $m_{\jpsi p}$ dependent part of the $P_c(4450)^+$ amplitude.
An additional fit to the data was performed using the reduced $\Lz^*$ model, in which
we represent the $P_c(4450)^+$ amplitude as the combination of independent
complex amplitudes at six equidistant points from $-\Gamma$ to $\Gamma$ around $M=4449.8\,$MeV as determined in the default fit.
Real and imaginary parts of the amplitude are interpolated in mass between the fitted points.
The resulting Argand diagram, shown in Fig.~\ref{DoubleArgand}(a),
is consistent with a rapid counter-clockwise change of the $P_c(4450)^+$ phase when its magnitude
reaches the maximum, a behavior characteristic of a resonance. A similar study for the wider state is shown in Fig.~\ref{DoubleArgand}(b); although the fit does show a large phase change, the amplitude values are sensitive to the details of the $\Lz^*$ model  and so this latter study is not conclusive. 
\subsection{\boldmath{Consistency check using $\Lb\to\jpsi p \pi^-$ decays}}

\begin{figure}[hbt]
\begin{center}
\includegraphics[width=0.9\textwidth]{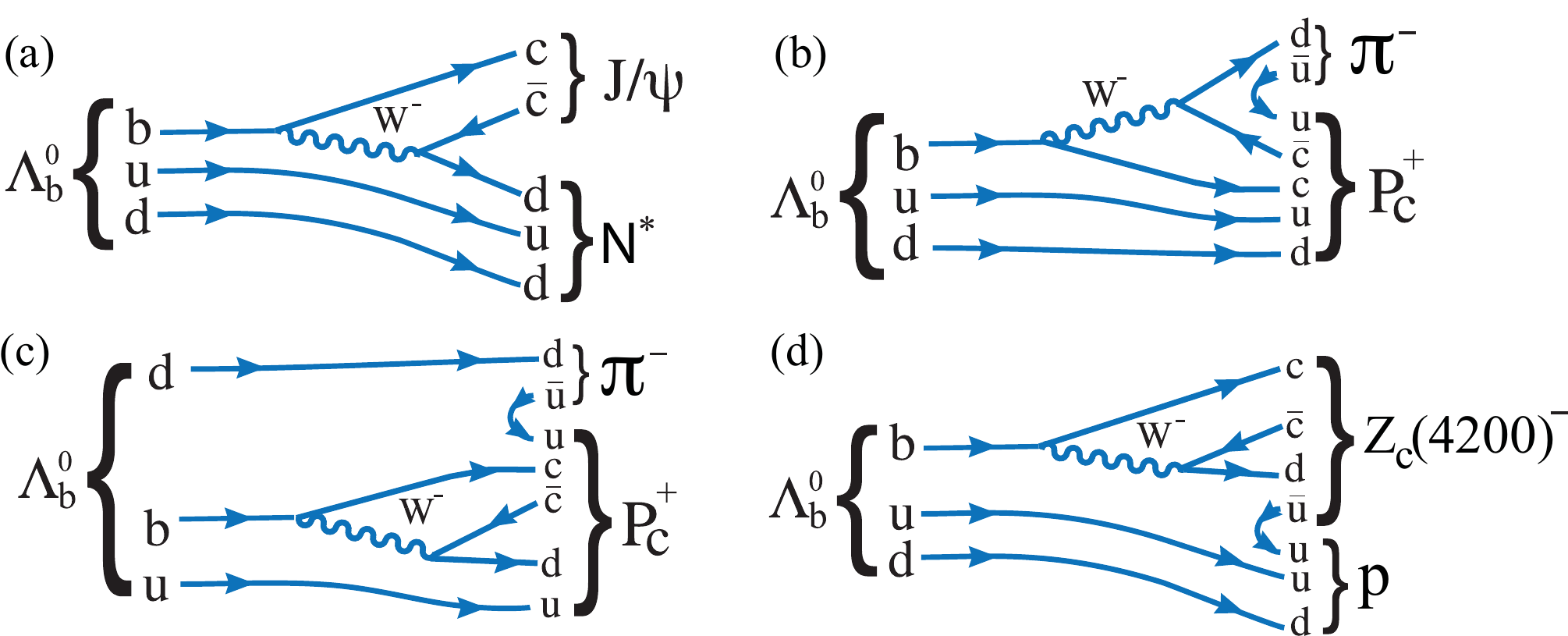}
\end{center}
\vskip -0.3cm
\caption{Feynman diagrams for (a) $\Lb\to \jpsi \Nstar$, (b) $\Lb\to P_c^+ \pi^-$ decay, (c) another diagram for $\Lb\to P_c^+ \pi^-$ decay, and (d) diagram for $\Lb\to Z_c(4200)^- p$ decay.}
\label{Feynman-PC3}
\end{figure}

The $\Lb\to\jpsi p \pi^-$ decay is the Cabibbo suppressed version of the $\Lb\to\jpsi K^+ \pi^-$ decay. Figure~\ref{Feynman-PC3} shows the quark level diagrams for both the decay into (a)  $\jpsi \Nstar$ and (b) $P_c^+\pi^-$. (Compare with Fig.~\ref{Feynman-Pc-both-v5}.) Also note that since there are two $d$ quarks in the final state,  the origin of the two $d$ quarks  can be switched, which corresponds to a different amplitude shown in (c), and these amplitudes can interfere and modify the decay rate \cite{Cheng:2015cca}. The relative branching fractions of the suppressed versus non-suppressed modes has been measured as ${\cal B}(\Lb\to \jpsi p \pi^-)/{\cal B}(\Lb \to \jpsi p K^-)=0.0824\pm0.0024\pm0.0042$~\cite{Aaij:2014zoa}, consistent with the expectation of $\sim\!\!0.05$, however, the crucial question is if we can see the pentaquark states at a rate consistent with this Cabibbo suppression. There is also the possibility of production of the exotic meson state $Z_c(4200)^-$ that decays into $\jpsi\pi^-$; the Feynman diagram is shown in (d). The analogous concern, i.e. of a $\jpsi K^-$ resonant system in the $\Lb\to\jpsi K^+ \pi^-$ decay, was not present because the data showed the absence of such a reasonant state, and because it has not been seen in any other decay.

Event candidates are reconstructed with similar criteria as for the $\jpsi K^- \pi^+$ mode with different particle identification criteria for the pion. Due to the larger backgrounds veto's are employed to get rid of backgrounds caused by proton particle identification failures including the modes $\Bdb \to\jpsi \pip K^-$ or as $\Bsb \to \jpsi K^+ K^-$. Here the protons are interpreted as kaons and if the calculated invariant masses are within $\pm 3\sigma$ of either of the known $B$ masses, the combination is rejected. In addition, $p \pi^-$ combinations within $\pm$ 5 MeV of the \Lz mass are removed. 

\begin{figure}[b!]
\centering
\includegraphics[width=0.7\textwidth]{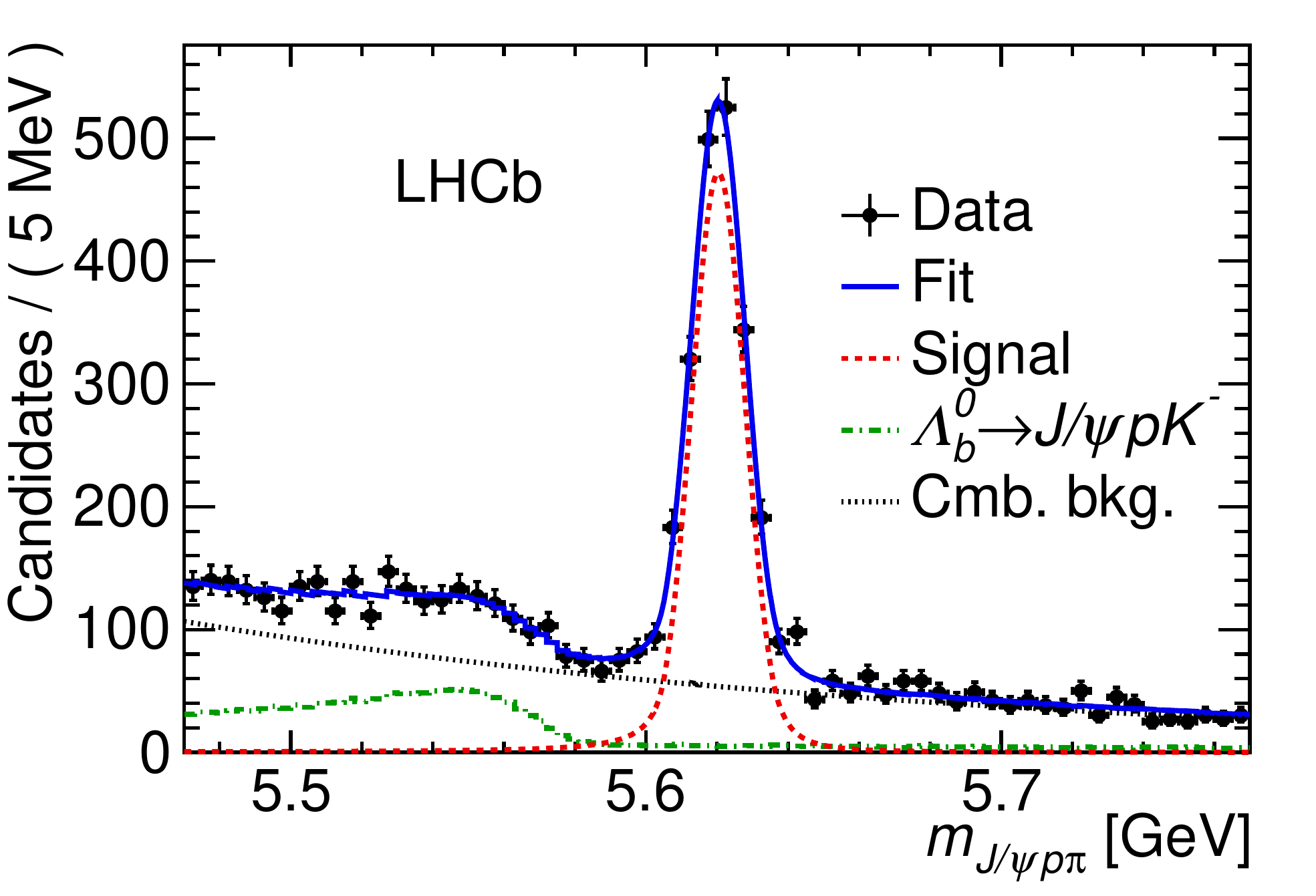}
\vskip -2mm
\caption{Invariant mass spectrum for the selected $\Lb\to \jpsi p \pi^-$ candidates (from ~\cite{Aaij:2014zoa}.} 
\label{fig:MassFit}
\end{figure}

The  invariant mass spectrum of $\Lb$ candidates is shown in Fig.~\ref{fig:MassFit}. The signal yield is  $1885\pm50$. The signal is described by a double-sided Crystal Ball function \cite{Skwarnicki:1986xj}.  
The combinatorial background is modeled by an exponential function. The background shape of $\Lb\to\jpsi p K^-$ events is described by a histogram obtained from simulation; the normalization is not fixed, but allowed to vary in the fit. Candidates are assigned weights using the \sPlot technique \cite{Pivk:2004ty} allowing the signal and background components to be projected out depending on their $\jpsi p \pi^-$ mass.

\begin{table}[h]
\centering
\caption{The $\Nstar$ resonances used in the different fits. 
Parameters are taken from the PDG \cite{pdg}. 
The number of $LS$ couplings is listed in the columns to the right for the two versions Reduced Model (RM)  and  Extended Model (EM). 
To fix overall phase and magnitude conventions, 
the $N(1535)$ complex coupling of lowest $LS$ is set to (1,0).}
\label{tab:Nstar}
\begin{tabular}{lccccc}
\\[-2.5ex] 
State & $J^P$ & Mass (MeV) & Width (MeV)& RM & EM  \\
\hline \\[-2.5ex] 
NR $p\pi$  &1/2$^-$ & - & - & 4  & 4 \\
$N(1440)$ &1/2$^+$ & 1430 & 350 & 3   & 4 \\
$N(1520)$ &3/2$^-$ & 1515 & 115 &3  & 3\\
$N(1535)$ &1/2$^-$ &1535 & 150 &4    & 4\\
$N(1650)$ &1/2$^-$ &  1655 & 140 & 1   & 4\\
$N(1675)$ &5/2$^-$ & 1675 & 150 & 3   & 5\\
$N(1680)$ &5/2$^+$ &  1685 & 130 &-    & 3\\
$N(1700)$ &3/2$^-$ & 1700& 150 &-    & 3\\
$N(1710)$ &1/2$^+$ &1710 & 100 &- & 4\\
$N(1720)$ &3/2$^+$ & 1720 & 250& 3 & 5\\ 
$N(1875)$ &3/2$^-$ & 1875 & 250 & -  & 3\\
$N(1900)$ &3/2$^+$ & 1900 & 200 &-  & 3 \\
$N(2190)$ &7/2$^-$ & 2190& 500 &-  & 3\\  
$N(2300)$ & 1/2$^+$ & 2300 & 340 & -  & 3 \\
$N(2570)$ & 5/2$^-$ & 2570 & 250 & -  & 3 \\ 
\hline
\multicolumn{4}{l}{Free parameters}&  40& 106 \\
\end{tabular}
\end{table}

The  analysis must incorporate all of the amplitudes shown in Fig.~\ref{Feynman-PC3}. The amplitude for diagram (a) is the same as that for $\Lb\to \jpsi K^- p$ but with the final state $\Lz^*$ states being replaced by $\Nstar$ states. These are listed in Table~\ref{tab:Nstar}. Here two models are used, one called the Reduced Model (RM) which uses fewer angular momentum amplitudes and states and the Extended Model (EM).
Table~\ref{tab:Lstar} lists the $\Nstar$ resonances considered in the amplitude model of $p\pi^-$ contributions. There are 15 well-established $\Nstar$ resonances~\cite{pdg}. The high-mass and high-spin states ($9/2$ and $11/2$) are not included,
since they require $L\ge3$ in the $\Lb$ decay and therefore are unlikely to be produced near the upper kinematic limit. Included, however, are two unconfirmed high-mass, low-spin resonances found by the BESIII collaboration, the $N(2300)$ and the $N(2700)$ \cite{Ablikim:2012zk}. 
The amplitudes for diagrams (b) and (c)  differ only by a sign \cite{Cheng:2015cca} and thus are modeled as one amplitude for each of the $P_c^+$ states. Finally, an amplitude is included for the possible $\Lb\to Z_c(4200)^-p$ decay.\footnote{The construction of this amplitude is documented in the supplemental material of Ref.~\cite{Aaij:2015tga}.}

\begin{figure}[b!]
\centering
\includegraphics[width=1.0\textwidth]{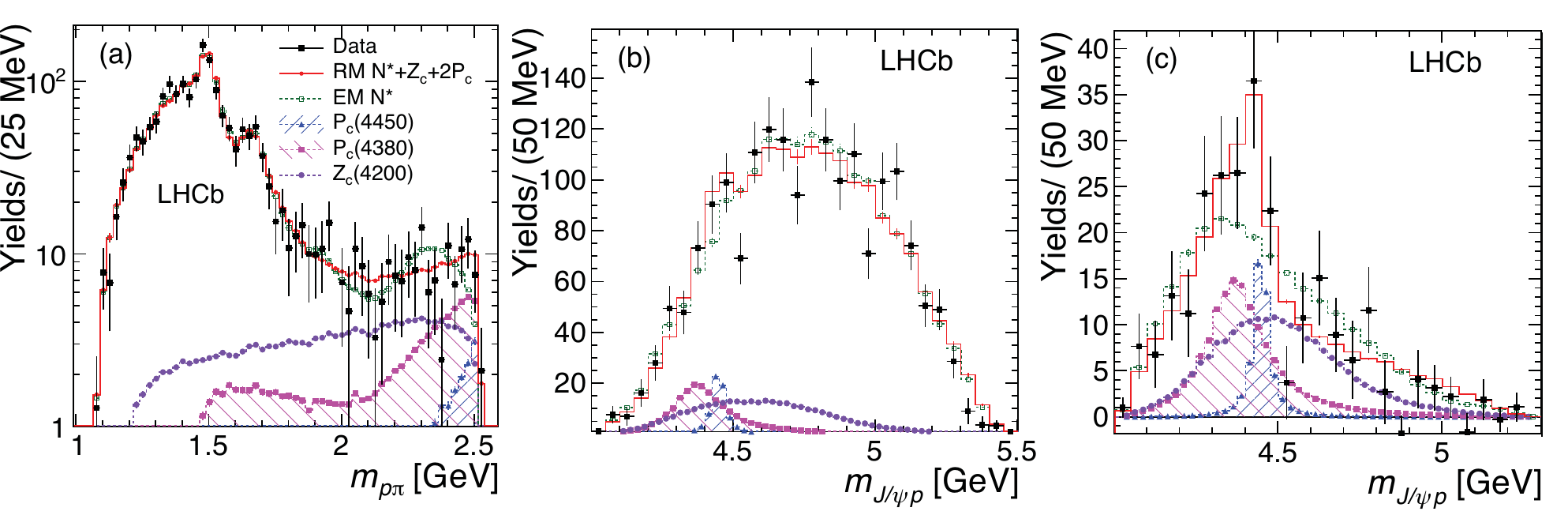}%
\vskip -2mm
\caption{(a) Background-subtracted data and fit projections onto $m_{\jpsi p}$ for all events (log scale),  (b) all events (linear scale),  and (c)  the $m_{p\pi}>1.8$ \gev region. The open circles (green) are for the $\jpsi\Nstar$  model using the EM suite of $\Nstar$ states, while the red histogram includes the RM sample of $\Nstar$ states as well as the two $P_c$ states and the $Z_c$ state; the other curves show the individual components (from ~\cite{Aaij:2014zoa}).}
\label{fig:mppi}
\end{figure}

Amplitude fits are performed by minimizing a 
six-dimensional unbinned negative log-likelihood, $\twolnL$, 
with the background subtracted using the weights described above, and the efficiency folded into the signal probability density function, as discussed in detail in Ref.~\cite{Aaij:2015tga}.
Shown in Fig.~\ref{fig:mppi} are the results of fits to several different amplitude hypotheses to both the $p\pi^-$ and $\jpsi p$ mass spectra. First let us concentrate on the fit where no exotic contributions are considered (amplitude (a) in Fig.~\ref{Feynman-PC3}), the one shown in the green open circles. The fit does describe the $p\pi^-$ and $\jpsi p$ mass projections quite well. However, if one looks at the  $p\pi^-$ mass region above 1.8~GeV, which eliminates the ``background" from the lower mass $\Nstar$ resonances,  there is an excess of events in the $\jpsi p$ mass region between 4.2 and 4.5~GeV. 

\begin{figure}[b!]
\centering
\includegraphics[width=0.35\textwidth]{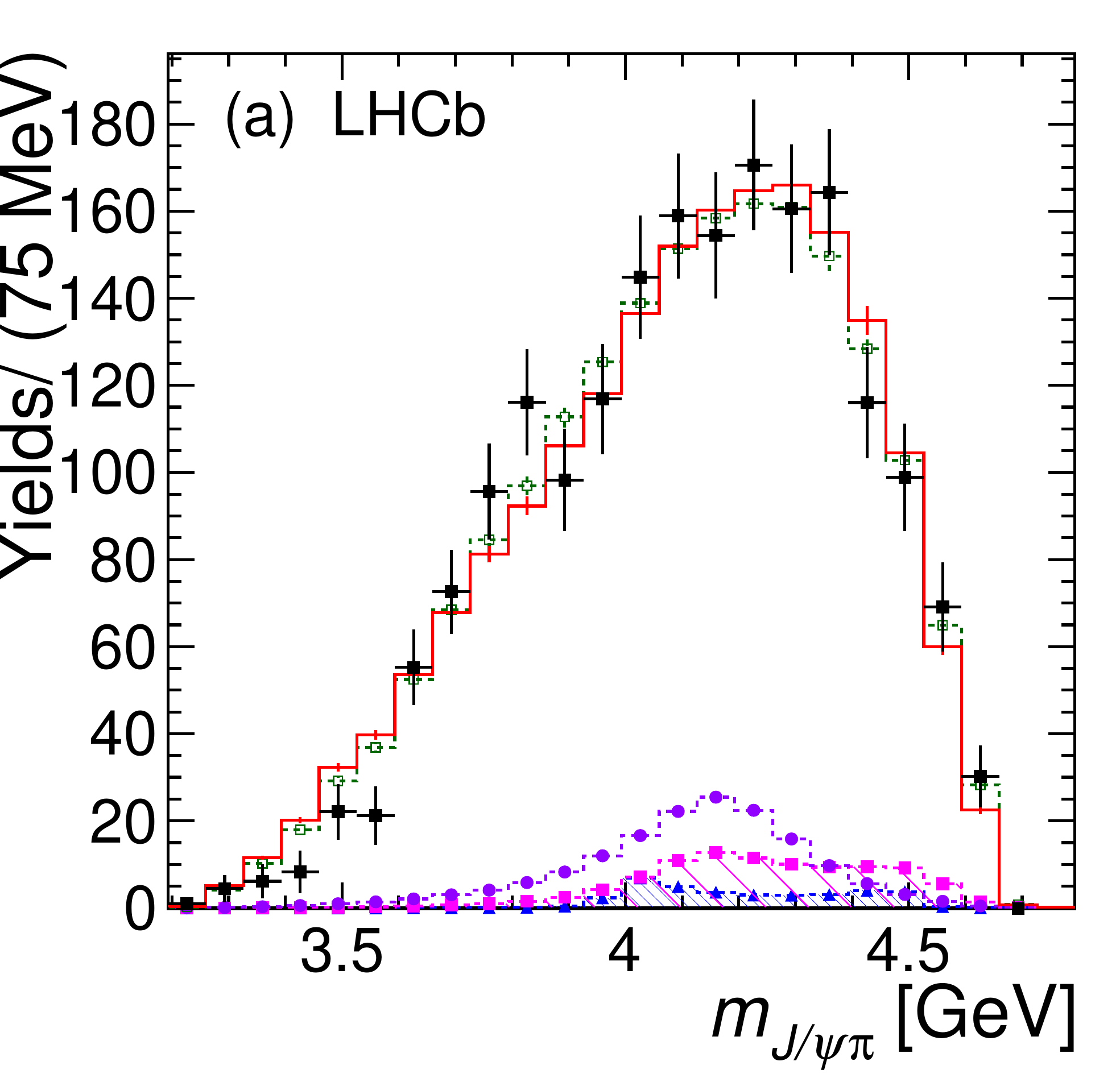}%
\includegraphics[width=0.35\textwidth]{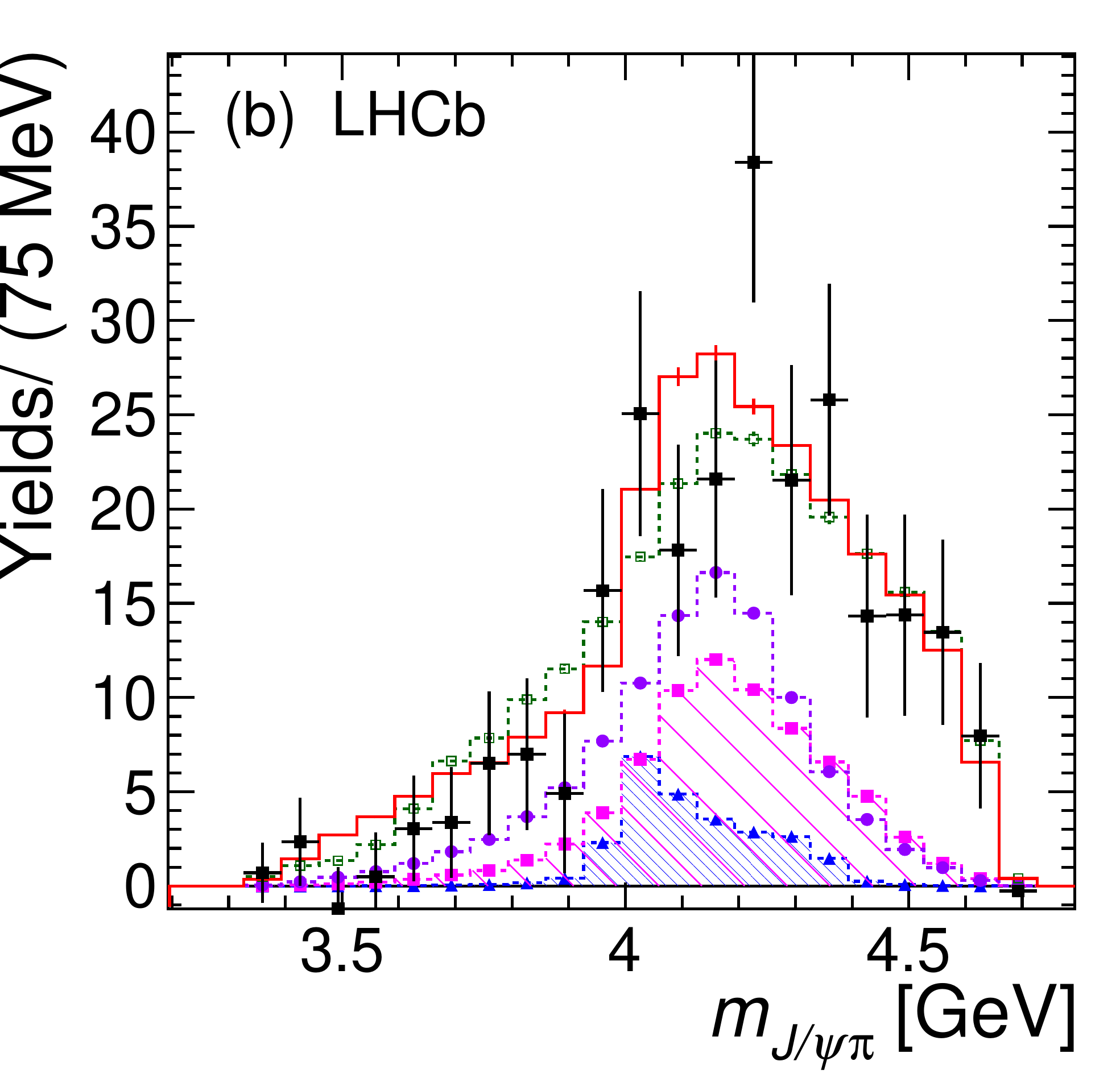}%
\caption{Background-subtracted data and fit projections onto $m_{\jpsi\pi}$ for (a) all events and (b) the $m_{p\pi}>1.8$ \gev region. 
         See the legend and caption of Fig.~\ref{fig:mppi} for a description of the components (from ~\cite{Aaij:2014zoa}).}
\label{fig:mjpsipi}
\end{figure}

It is necessary to also consider the possibility of an exotic $\jpsi \pi^-$ resonance. In Fig.~\ref{fig:mjpsipi} the plot of the $\jpsi \pi^-$ invariant mass for the entire $m_{p\pi}$ and for $m_{p\pi}>1.8$~GeV, shows an excess of events in the 
in the region between 4.1 and 4.3~GeV compared with the fit using only the $\jpsi \Nstar$ model. 

To test these other possibilities full amplitude fits including the decay angles are performed and differences in the $\twolnL$ of fits computed. The baseline model includes the set of RM \Nstar resonances plus all three exotic states, labelled as ``RM \Nstar + $\Z_c$ + 2$P_c$" in the figures. The fit projections of all three exotic states are also shown. The $\jpsi\pi^-$ mass projections are shown in Fig.~\ref{fig:mjpsipi}.

The fit with all three exotic states is superior to that of the only the \Nstar model. Quantitatively, the change in $\twolnL$ ($\Delta(\twolnL)$) corresponds to 3.9$\sigma$. Other models tried included using only the two $P_c$ states, or only the $Z_c$ state. Table~\ref{tab:FitResppi} lists the models, the number of standard deviation (N$\sigma$) changes with respect to the $\Nstar$ EM only model, and the fit fractions.

\begin{table}[hbt]
\centering
\caption{Results of different amplitude model fits to the data. The change N$\sigma$ refers to the improvement in the fit with respect to the EM $\Nstar$ model. The fit fractions are the percentages of each of the final states as determined by the fit.}
\label{tab:FitResppi}
\begin{tabular}{lcccc}
\\[-2.5ex] 
Model &N$\sigma$& \multicolumn{3}{c}{Fit Fractions (\%)}\\
& & $P_c(4380)^+$ &  $P_c(4450)^+$ & $Z_c(4200)^-$ \\
\hline \\[-2.5ex] 
$\Nstar+P_c+Z_c$  & 3.9 & $5.1\pm1.5^{+2.6}_{-1.6}$ & $1.6^{+0.8+0.6}_{-0.6-0.5}$& $7.7\pm 2.8^{+3.4}_{-4.0}$\\ 
$\Nstar+P_c$ & 3.3 & $\approx 5.1$ &$\approx 1.6$ &--\\
$\Nstar+Z_c$ &3.2 &  --&--& $17.2\pm 3.5$\\
\hline
\end{tabular}
\end{table}

The fit with either one of the two $P_c$ states, or only the $Z_c$ state can adequately describe the data, although the $Z_c$ only fit has a seemingly rather large fit fractions of $(17\pm 3.5)$\%. Adding the $Z_c$ state hardly changes the $P_c$ fit fractions. These correspond to the ratio of branching fractions ${\cal{B}}(\Lb\to P_c(4380)^+\pi^-)/{\cal{B}}(\Lb\to P_c(4380)^+K^-)=0.050\pm 0.016^{+0.026}_{-0.016}\pm 0.025$ and ${\cal{B}}(\Lb\to P_c(4450)^+\pi^-)/{\cal{B}}(\Lb\to P_c(4450)^+K^-)=0.033^{+0.016+0.011}_{-0.014-0.010}\pm 0.009$, where the first uncertainty is statistical, the second systematic and the third due to the error on the fit fractions in the $\jpsi K^- p$ mode.  These are consistent with the expectations for Cabibbo suppressed rates within the large uncertainties \cite{Cheng:2015cca}. 

To summarize this section, the $\jpsi \pi^- p$ mode shows that the data are inconsistent with not having exotic components, are consistent with the expectations for the rates of the Cabibbo suppressed $P_c$ states, but also could be explained by an exotic $\Z_c$ or both exotic contributions.

\subsection{\boldmath{Independent confirmation of the $P_c(4450)^+$ using a model independent $\Lz^*$ description}}

While a convincing case has been made for the existence of the two $P_c$ states, there is an alternative method of examining the consistency of the data with the presence of exotic states.  The poorly understood nature of the $\Lz^*$ states \cite{pdg} and the difficulty of knowing exactly which amplitudes are present can be circumvented by using moments related to the $\Lz^*$ decay angle that can be calculated without knowledge of what $\Lz^*$ states are present. The only assumption made is that very high spin states cannot be present in $\Lb\to\jpsi\Lz^*$ decays especially near the largest possible allowed $\Lz^*$ masses. This approach was developed by the BaBar collaboration for their studies of \cite{z4430_babar_1} $\Bzb\to\jpsi \pi^- K^+$ decays, where they attempted to explain the $Z(4430)^-\to \jpsi\pi^-$ as the sum of interfering $K^*$ resonances. The same method was adapted by LHCb with opposite conclusions \cite{Aaij:2015zxa}.

In principle, after integrating over the decay angular variables the decay amplitude can be written in terms of $m_{Kp}$ and $m_{\jpsi p}$, or equivalently $m_{Kp}$ and $\cos\theta_{\Lz^*}$, where the latter is the helicity decay angle of the ${K^-p}$ system. For each bin in $m_{Kp}$ the $\cos\theta_{\Lz^*}$ distribution can be expressed as a sum over Legendre polynomials:
\begin{equation} 
dN/d\cos\theta_{\Lz^*}=\sum_{\ell=0}^{\ell_{\rm max}} {\langle P_{\ell}^U \rangle P_{\ell}(\cos\theta_{\Lz^*})},
\end{equation}
where $N$ is the efficiency-corrected and background-subtracted signal yield, $\ell_{\rm max}$ is the maximum allowed angular momentum value, and $\langle P_{\ell}^U \rangle$ 
is an unnormalized Legendre moment of rank $\ell$, 
\begin{equation}
\langle P_{\ell}^U \rangle= \int_{-1}^{+1} d\cos\theta_{\Lz^*}\, P_{\ell}(\cos\theta_{\Lz^*}) \,
dN/d\cos\theta_{\Lz^*}.
\end{equation}
To ensure the validity of this procedure, the event sample must be background subtracted and efficiency corrected over the full range of angular variables in the decay amplitude.

\begin{figure}[tb]
   \vskip-0.8cm
  \begin{center}
  \includegraphics*[width=0.7\textwidth]{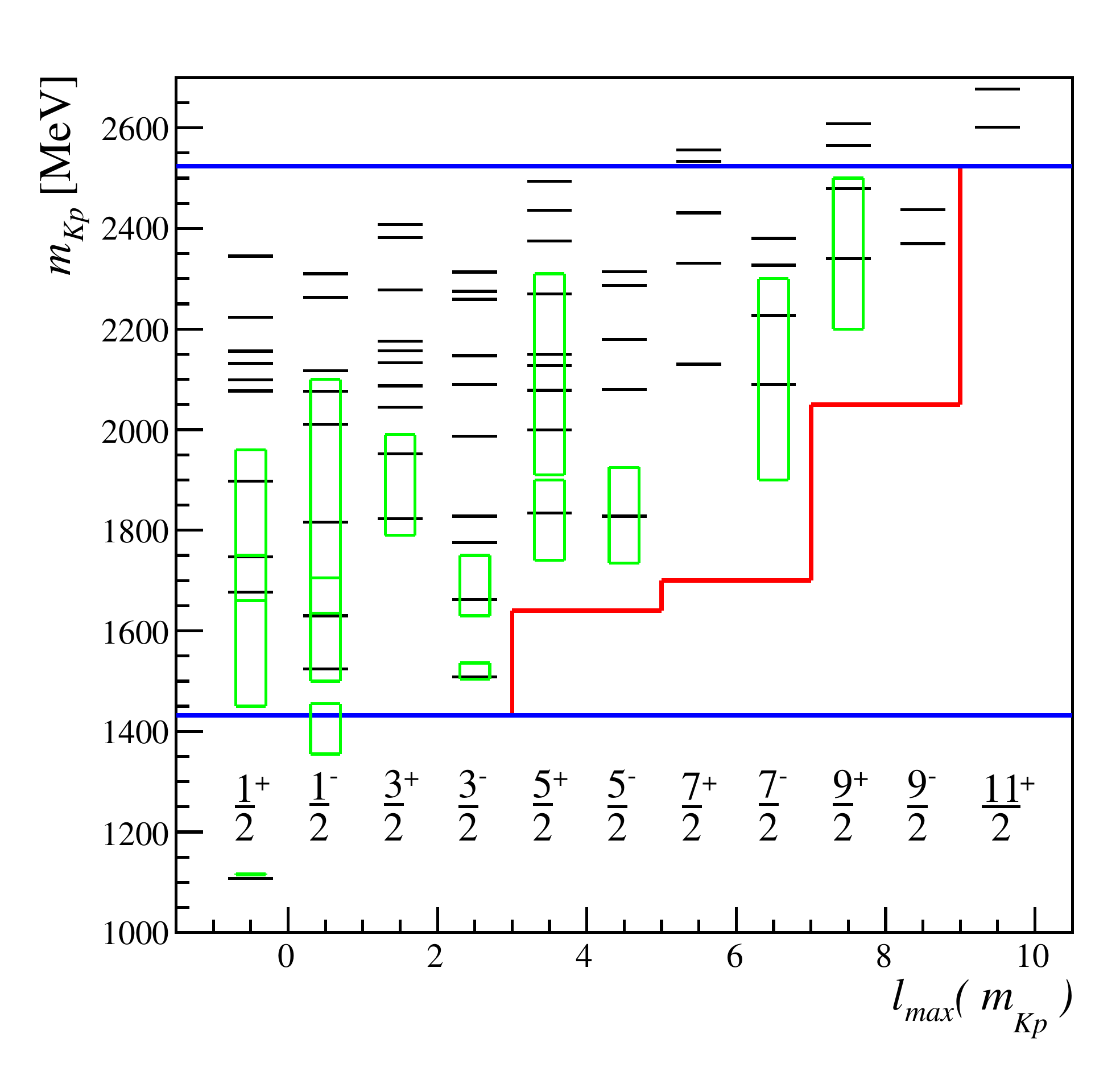}
  \end{center}
  \vskip-1.1cm
\caption{
   Excitations of the $\Lz$ baryon. States predicted 
   in ref.~\cite{Loring:2001kx} are shown as short 
   horizontal lines (black)
   and experimentally well-established $\Lz^*$ states 
   are shown as green
   boxes covering the mass ranges from 
   $M_0-\Gamma_0$ to $M_0+\Gamma_0$.
   The $m_{Kp}$ mass range probed in $\Lb\to\jpsi p \Km$ decays
   is shown by long horizontal lines (blue).
   The $\ell_{\rm max}$ upper limit as a function of $m_{Kp}$ is shown as a stepped line (red).
   All contributions from $\Lz^*$ states with 
   $J^P$ values to the left of the red line 
   are accepted by the filter (from \cite{Aaij:2015zxa}).
   \label{fig:lmax}
  }
\end{figure}

In this analysis \cite{Aaij:2015zxa} two hypotheses are considered, one where the $P_c^+$ states are absent, denoted as $H_0$, and another where they are present, called $H_1$. For $H_0$ the maximum allowed value of $\ell_{\rm max}$ is determined from the angular momentum of the $\Lz^*$. Since the $\Lz^*$ decays into a spinless kaon and a spin 1/2 proton, $\ell_{\rm max}$  can only be as large as twice the spin of any particular $\Lz^*$ resonance.\footnote{The angular distribution of a spin $\ell/2$ resonance is described by a Legendre polynomial of order $\ell-1$ to which one more unit of additional angular momentum is possible due to the spin 1/2 nature of the proton.} The maximum spin of these resonances are limited by the $\Lz^*$ mass, with higher masses having larger allowed spins. This rather mild hypothesis is based on both experimental measurement and theoretical predictions as illustrated in Fig.~\ref{fig:lmax}.

\begin{figure}[b!]
  \begin{center}
  \includegraphics*[width=0.65\textwidth]{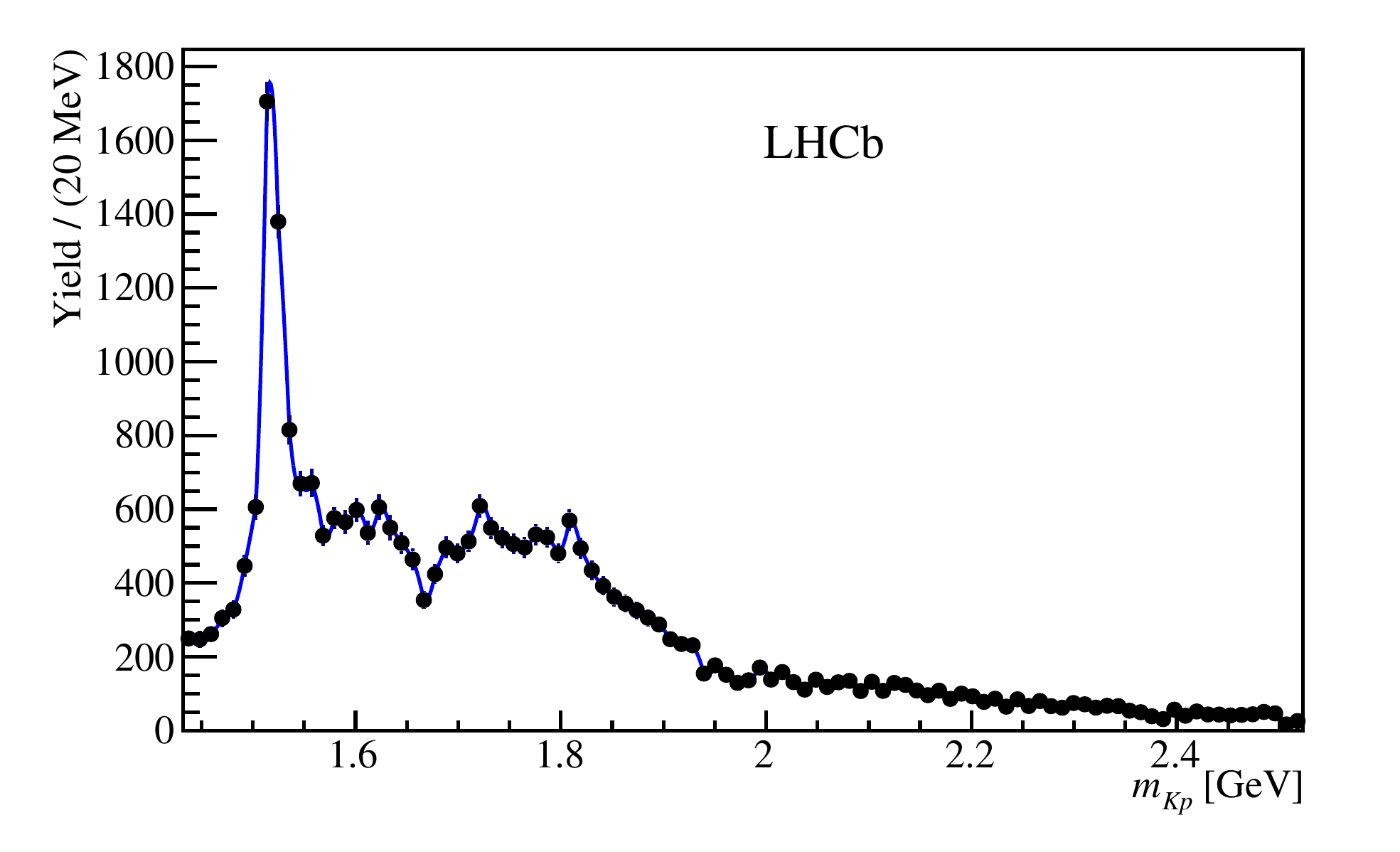}
  \end{center}
  \vskip-0.7cm\caption{
    Efficiency-corrected and background-subtracted $m_{Kp}$ distribution 
    of the data (black points with error bars),
    with $\PDF(m_{Kp}|H_0)$ superimposed (solid blue line). $\PDF(m_{Kp}|H_0)$ fits 
    the data by construction (from~\cite{Aaij:2015zxa}).
   \label{fig:mkp}
  }
\end{figure}

This analysis takes advantage of the limited $\ell_{\rm max}(m_{Kp}$) by allowing only moments above the red line in Fig.~\ref{fig:lmax}. Thus the largest $\ell_{\rm max}$ to be considered is 9 and that is only for $m_{Kp}>2.050$~GeV. The background subtracted and efficiency corrected data projected in $m_{Kp}$ are shown in Fig.~\ref{fig:mkp}. 

A probability density function based on
the  $\Lz^*$ helicity distribution is then formed via linear interpolation between the $m_{Kp}$ bins, labeled as $k$, as  
\begin{equation}
\label{eq:PDF1}
\PDF(\cos\theta_{\Lz^*}|H_0,{m_{Kp}}^k)=\sum_{l=0}^{\ell_{\rm max}({m_{Kp}}^k)} 
\langle P_{\ell}^N \rangle^k P_{\ell}(\cos\theta_{\Lz^*}).
\end{equation}
Here the Legendre moments $\langle P_{\ell}^N \rangle^k$
are normalized by the yield in the corresponding $m_{Kp}$ bin.
These data are used to determine
\begin{equation}
\langle P_{\ell}^U \rangle^k = \sum_{i=1}^{{n_{\rm cand}}^k} (w_i/\epsilon_i)  P_{\ell}(\cos\theta_{\Lz^*}^i). 
\end{equation}
Here the index $i$ runs over selected $\jpsi p K^-$ candidates in 
the signal and sideband regions for the $k^{th}$ bin of $m_{Kp}$  
(${n_{\rm cand}}^k$ is their total number),
$\epsilon_i=\epsilon( {m_{Kp}}^i,\cos{\theta_{\Lz^*}}^i,{\Omega_a}^i)$ is the efficiency correction, 
and $w_i$ is the background subtraction weight. 

\begin{figure}[b!]
  \begin{center}
  \includegraphics*[width=0.7\textwidth]{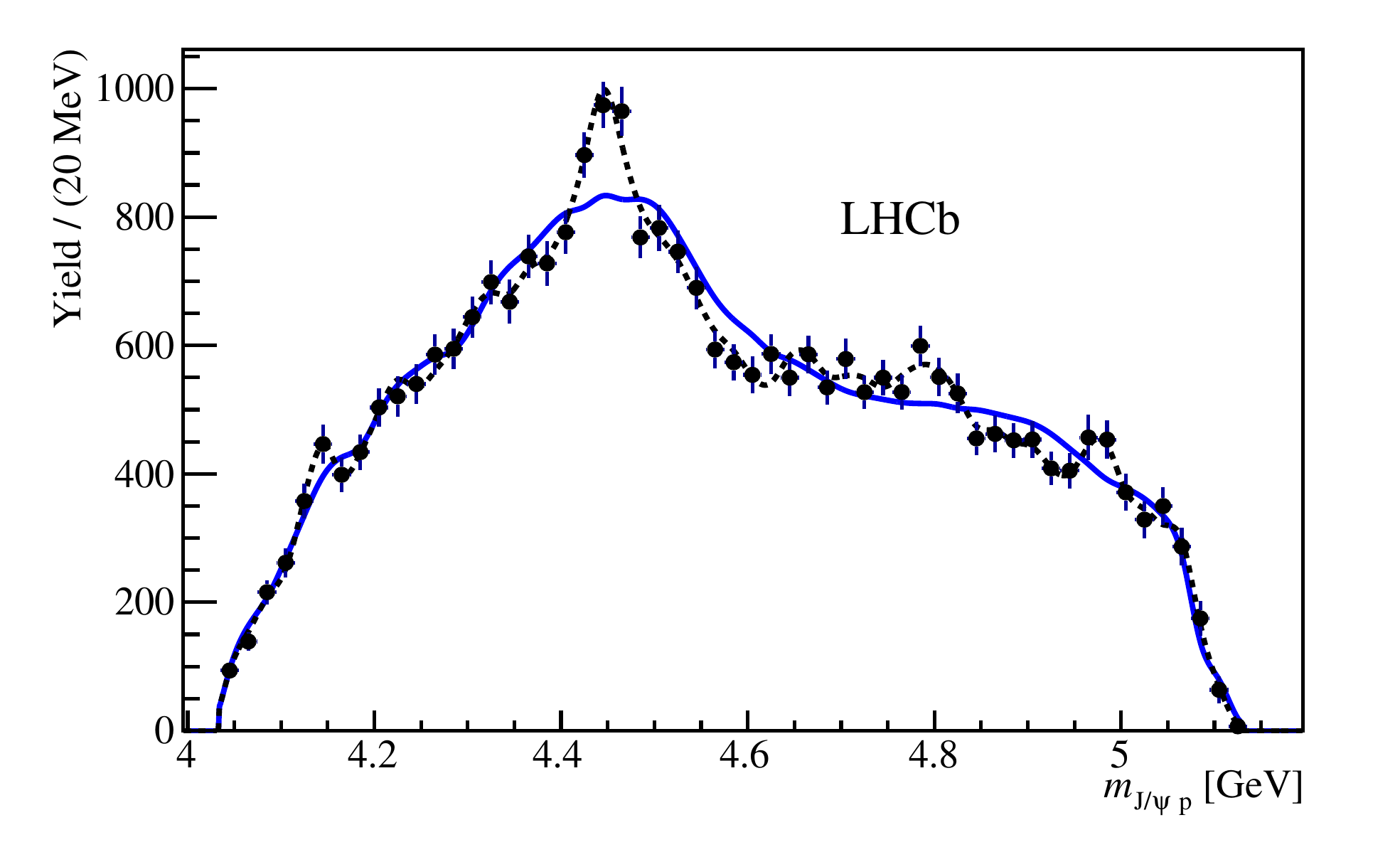}
  \end{center}
  \vskip-0.9cm\caption{
    Efficiency-corrected and background-subtracted $m_{\jpsi p}$
    distribution of the data (black points with error bars),  
    with $\PDF(m_{\jpsi p}|H_0)$ (solid blue line) and 
    $\PDF(m_{\jpsi p}|H_1)$ (dashed black line)
    superimposed (from ~\cite{Aaij:2015zxa}).
   \label{fig:mjpsip}
  }
\end{figure}

Now that the moments are computed, under the hypothesis that the $\Lz^*$ states indeed follow the $\ell_{\rm max}$ limit, they need to compared to the data to see if they explain it. The best comparison is achieved using the $m_{\jpsi p}$ distribution.  After some mathematical gymnastics the  $\PDF({m_{\jpsi p}}|H_0)$ distribution is formed and compared to the directly obtained efficiency-corrected 
and background-subtracted $m_{\jpsi p}$ distribution in the data  in Fig.~\ref{fig:mjpsip}.

While it is clear to the naked eye that the $H_0$ hypothesis is a bad fit to the data, mostly due the structure near 4450~MeV, a quantitative method of distinguishing between this and other models is needed. Consider the specific alternative hypothesis that includes the two pentaquark states, $H_1$.  To evaluate $\PDF(\cos\theta_{\Lz^*}|H_1,{m_{Kp}}^k)$, similar to the determination of $H_0$ in Eq.~\ref{eq:PDF1}, $\ell_{\rm max}$ is increased to allow for the effects of pentaquark states on the higher moments. The actual choice of the new $\ell_{\rm max}$ is a matter of judgement: a very large increase would describe the data better, but at the expense of a weakened test due to adding possible additional fluctuations. In this analysis the maximum value somewhat arbitrarily was chosen as 31. 

A good metric to use in order to test for the best hypothesis is the likelihood ratio 
\begin{equation}
\label{eq:test}
{\Delta(\!-2\ln L\!)}=\sum_{i=1}^{n} 
w_i \ln \frac{ \PDF({m_{\jpsi p}}^i|H_0)/I_{H_{0}} }{ \PDF({m_{\jpsi p}}^i|H_1)/I_{H_{1}} } ,
\end{equation}
where $n$ is the number of signal and background events, $w_i$ the background subtraction weight, and  the $I_{H}$'s  are normalization factors.

\begin{figure}[b!]
  \vskip-0.3cm
  \begin{center}
  \includegraphics*[width=0.5\textwidth]{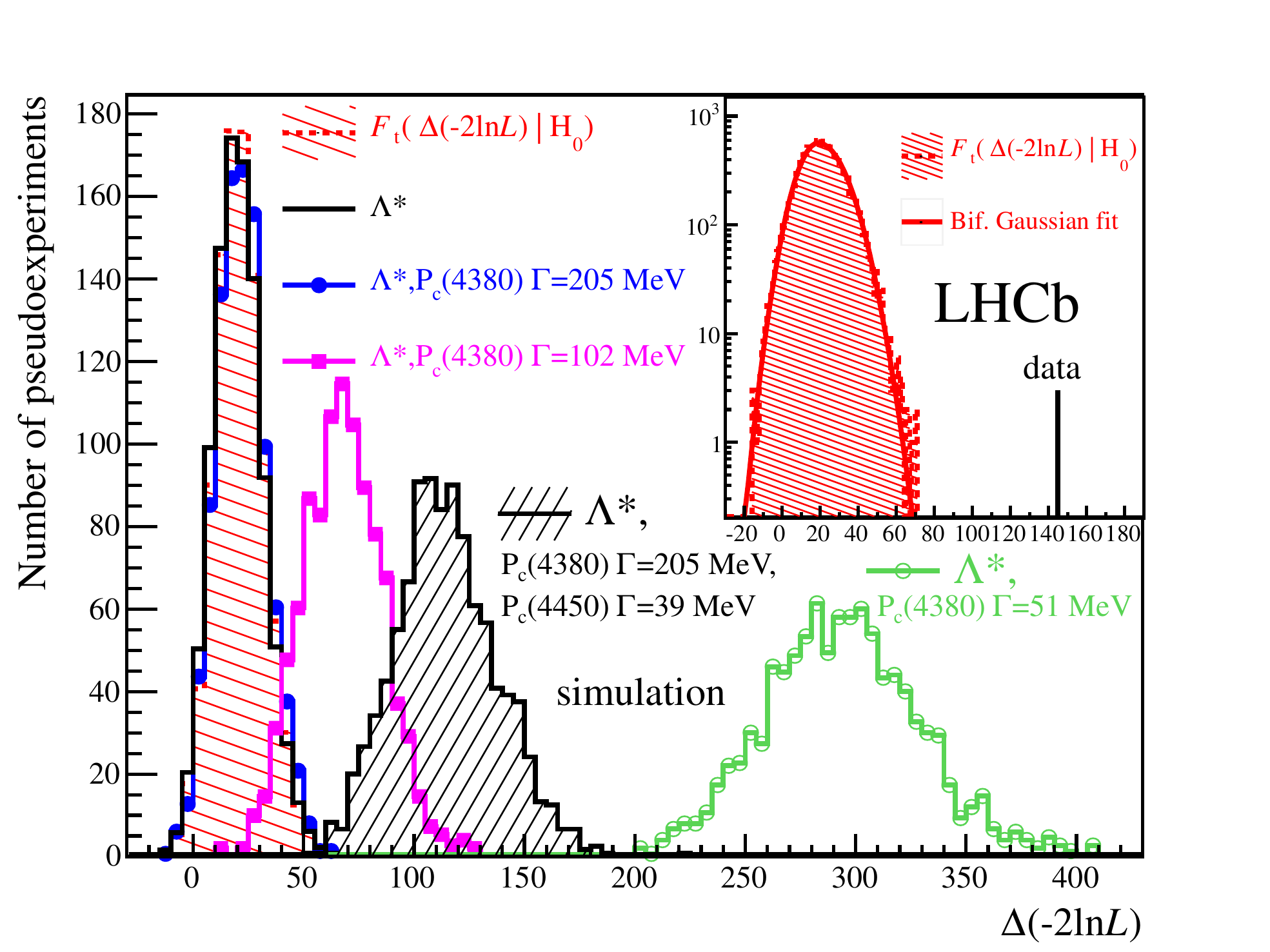} 
  \end{center}
  \vskip-0.3cm\caption{
   Distributions of ${\Delta(\!-2\ln L\!)}$
   in the model-independent pseudo-experiments 
   corresponding to $H_0$ (red falling hatched)  
   compared to the distributions for pseudo-experiments 
   generated from various amplitude models and, in the inset,
   to the bifurcated Gaussian fit function (solid line) and
   the value obtained for the data (vertical bar) (from \cite{Aaij:2015zxa}).
   \label{fig:dll}
  }
   \vskip-0.2cm
\end{figure}

It is necessary to ascertain the distribution of  the test variable for the $H_0$ hypothesis, ${\PDFt}(m_{\jpsi p}^i|H_0)$. This is done by simulating many pseudo-experiments, varying the measured parameters, but keeping the number of signal and background events fixed. Signal events are generated using the  $\PDF(m_{\jpsi p}^i|H_0)$ function, while the background is generated using the parameters determined in the amplitude analysis in Ref.~\cite{Aaij:2015fea}.  The distribution is shown in the Fig.~\ref{fig:dll}, and in the inset in logarithmic scale.

It is interesting to see how adding $P_c^+$ resonances into the simulation would change the likelihood function. When the $P_c(4380)^+$ with 205~MeV total width ($\Gamma$) is added the likelihood distribution hardly changes from ${\PDFt}(m_{\jpsi p}^i|H_0)$,  (see Fig.~\ref{fig:dll}) showing that this technique cannot detect its presence. Other simulations show that if its width were 102~MeV, or better yet 51~MeV its presence would be clear. The predictions for  $\Gamma= 205$~MeV
 (for $P_c(4380)^+$) and  $\Gamma=39$~MeV (for $P_c(4450)^+$)  show a large separation and this is confirmed by the value of ${\Delta(\!-2\ln L\!)}$ from the data, shown in the insert.  The $p$-value obtained from Eq.~[\ref{eq:test}] corresponds to 9 standard deviations and confirms the presence of the $P_c(4450)^+$.

Searches for other pentaquark states have started. Signals have been observed in the modes  $\Lambda_b^0 \to \chi_{c1,c2} p  K^-$  \cite{Aaij:2017awb}, and $\varXi^{-}_{b}\to J/\psi\varLambda K^{-}$ \cite{Aaij:2017bef}, but the number of events are too small currently to allow for Dalitz plot analyses.

\subsection{\boldmath{Impending studies using photoproduction of $\jpsi p$ resonances}}
\begin{figure}[t!]
  \begin{center}
  \includegraphics*[width=0.8\textwidth]{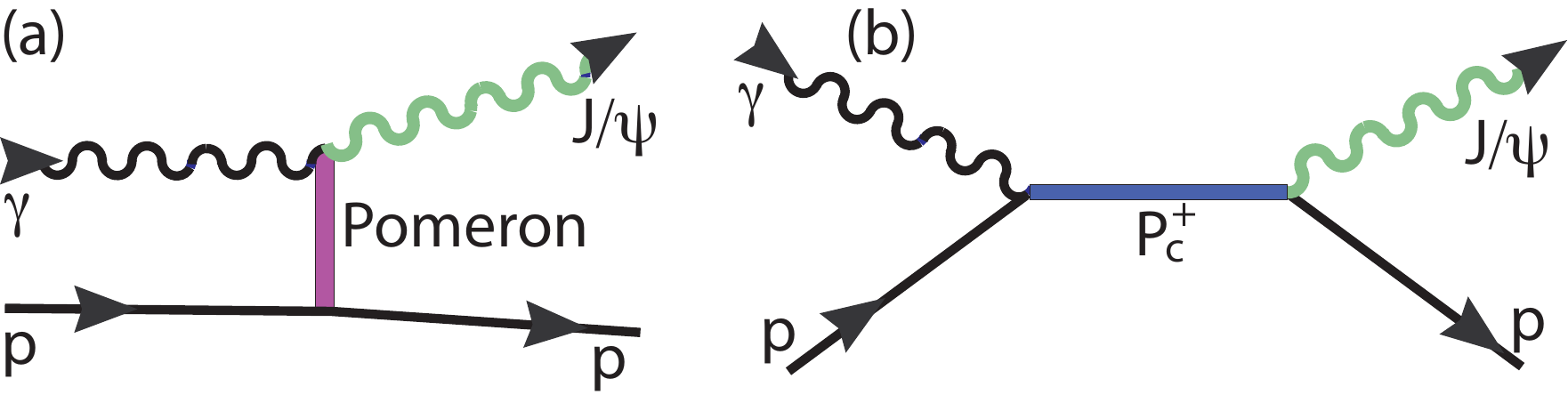} 
  \end{center}
    \vskip-0.3cm
\caption{Feynman diagrams for $\gamma p\to \jpsi p$ via (a)  the diffractive (Pomeron) process 
and (b)  mediated by the  $P_c^+$ states.
   \label{gammaP}
  }
   \vskip-0.2cm
\end{figure}

While LHCb has collected more data in 2015 and 2016, about doubling their sample of \Lb baryons, and will collect even more data in the future, from which more information on the $P_c^+$ states will be extracted, it is very important to have confirmation of the two $P_c^+$ states from another experiment. A very promising avenue was suggested soon after the pentaquark discovery was announced involving $\jpsi$ photoproduction via the reaction $\gamma p\to \jpsi p$ \cite{Wang:2015jsa, Kubarovsky:2015aaa, Karliner:2015voa}.  Since the photon has the same quantum numbers as the $\jpsi$ in the field of a particle it can transform temporarily into a $\jpsi$, or for that matter any vector meson. This process is called ``vector dominance" \cite{Sakurai:1960ju}. The virtual $\jpsi$ then interacts with the proton. This process has been studied experimentally  \cite{pdg}, and is thought to proceed via the diffractive diagram shown in Fig.~\ref{gammaP}(a) where the gluon exchanges are summarized as a ``Pomeron." The resonant diagram is shown in  Fig.~\ref{gammaP}(b).

\begin{figure}[h]
  \begin{center}
  \includegraphics*[width=0.95\textwidth]{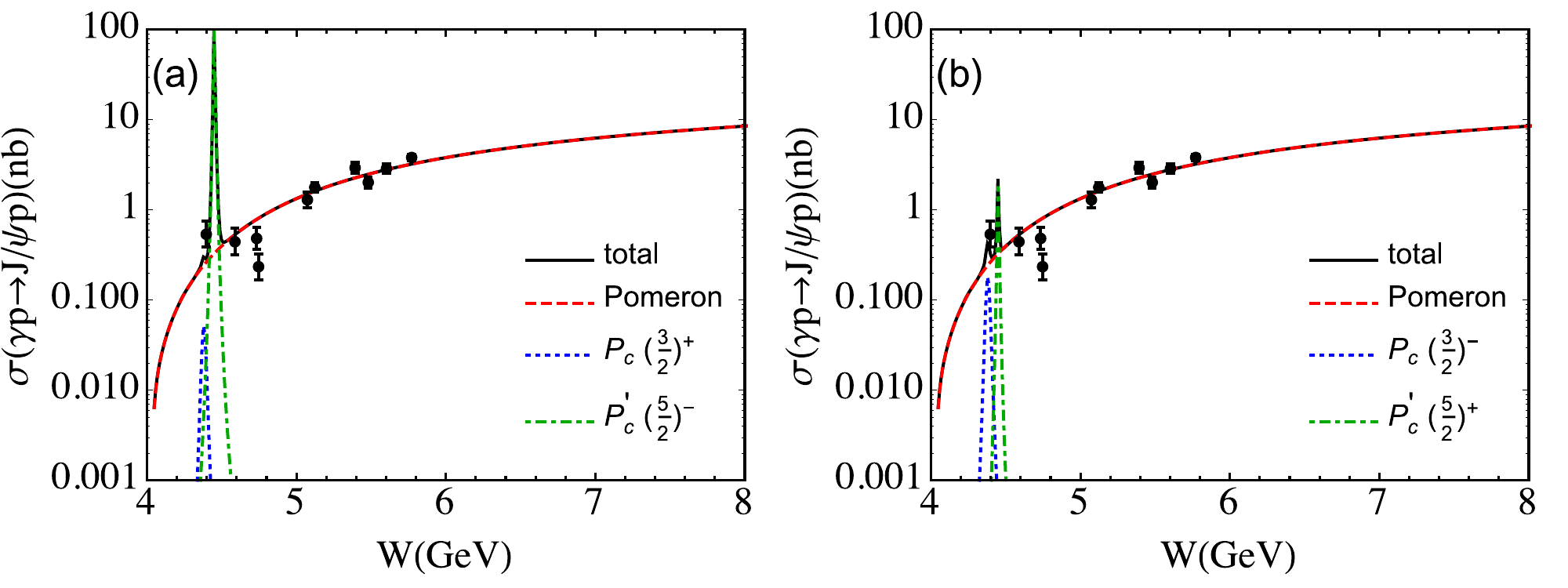} 
  \end{center}
    \vskip-0.6cm
\caption{Measurements of the $\gamma p\to \jpsi p$ cross-section as a function of the total center-of-mass energy W for the diffractive (Pomeron) process and the two observed $P_c^+$ states, under the assumption that ${\cal{B}}(P_c^+\to \jpsi p)$ is 5\%, for two different parity assignments as indicated (from \cite{Wang:2015jsa}). 
   \label{Pcxsect}
  }
   \vskip-0.2cm
\end{figure}

The
presence of resonant $P_c^+$ states could be observed if ${\cal{B}}(P_c^+\to\jpsi p)$ is large enough.  Figure~\ref{Pcxsect} shows the expected $\gamma p\to \jpsi p$ cross-section as a function of energy, assuming hat ${\cal{B}}(P_c^+\to \jpsi p)$ is 5\%. Measurements in the region where the $P_c$ states could be observed are scarce and easily could have missed any resonance structure because of acceptance and resolution issues.  Experiments have been approved at Jefferson lab using a 12~GeV photon beam \cite{Blin:2016dlf,Kubarovsky:2016whd,Meziani:2016lhg} and we eagerly await the results.

%% file: Sec-Experimental-evidence-for-tetraquarks-rev.tex
\section{Experimental evidence for tetraquarks}
\label{sec:Experimental-evidence-for-tetraquarks}
\input{x3872-rev}
\input{x3940}

\input{y4260-rev}
\input{z4430}

\input{z3900-rev}
\input{zb-rev}

%% file: x3872-rev.tex
\subsection{\boldmath{The $X(3872)$ state}}

\label{cx3872}

Results of the E705 experiment from interactions of a 300~GeV/c momentum $\pi^{\pm}$ beam on a nuclear $Li$ target  reported in 1993
indicated a state just above 3.8~GeV in mass decaying into $\jpsi\pi^+\pi^-$ \cite{Antoniazzi:1993jz}.
The first observation was made by Belle \cite{x3872belle}
in a data set of 152$\times$10$^6$ $B$$\bar{B}$ events
with a large statistical significance of 9.4$\sigma$
in the decay $B^{\pm}\rightarrow K^{\pm}X(3872)(\rightarrow \jpsi\pi^+\pi^-)$. 
Since an assignment of a charmonium state was a priori not possible, 
the state was called $X$(3872). 
Soon afterwards it was confirmed by 
\BaBar \cite{Aubert:2004ns,Aubert:2008gu}, 
CDF II \cite{Acosta:2003zx,Abulencia:2005zc,Aaltonen:2009vj}, 
D0 \cite{Abazov:2004kp}, 
LHCb \cite{x3872lhcb} and CMS \cite{x3872cms}.

\begin{figure}[t!]
\centerline{\includegraphics[width=\textwidth,height=4.0cm]{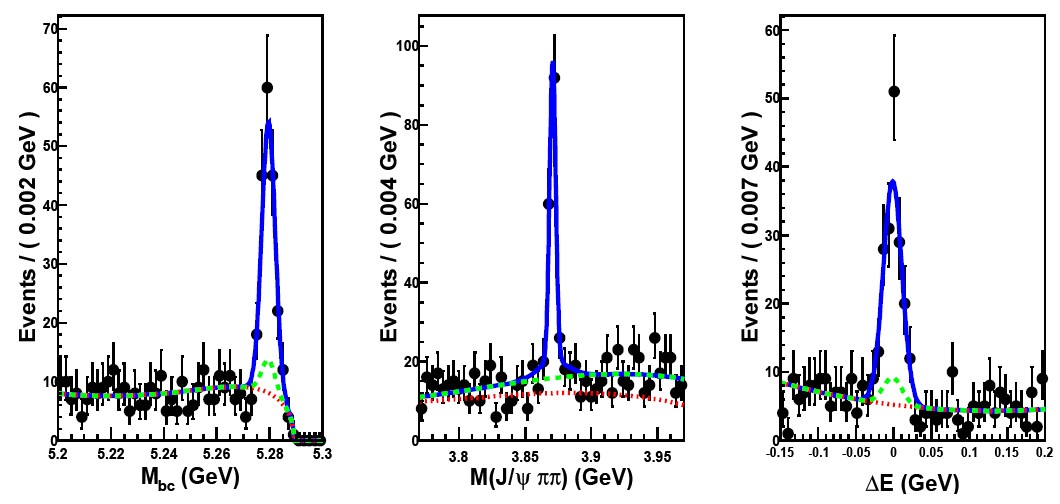}}
\centerline{\includegraphics[width=0.7\textwidth,height=5.5cm]{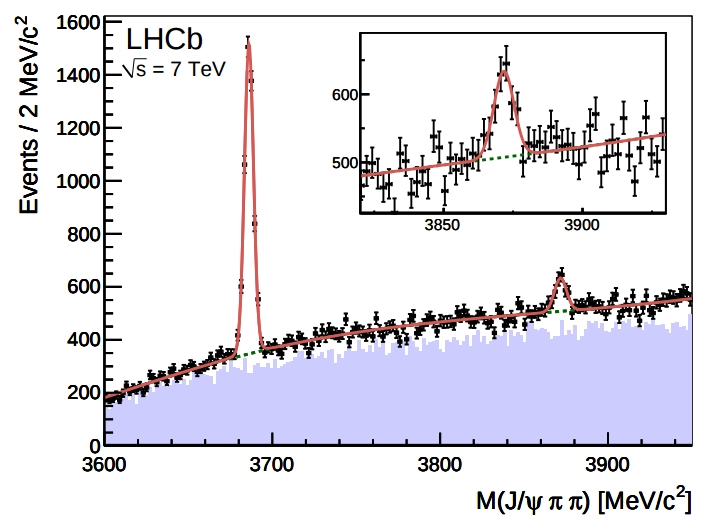}}
\caption{{\it Top:} Beam constraint mass $M_{\rm bc}$ (left), 
invariant mass $m(\jpsi\pi^+\pi^-)$ (center)
and $\Delta$$E$ (right) for $B^+$$\rightarrow$$K^+$$X$(3872)($\rightarrow$$\jpsi$$\pi^+$$\pi^-$)
from Belle \cite{x3872belle_width_2011}.
The blue line indicates signal, and the dashed green line background.  {\it Bottom:} Invariant mass $m$($\jpsi$$\pi^+$$\pi^-$)
in inclusive production in $p$$p$ collision at $\sqrt{s}$=7~TeV from LHCb \cite{x3872lhcb}.}
\label{fx3872_signal}
\end{figure}

Two measurements serve as examples of two different production mechanisms. 
First of all, in $B$ mesons decays, Belle performed an analysis 
using the complete data set of 711~fb$^{-1}$
collected at the $\Upsilon$(4$S$) resonance \cite{x3872belle_width_2011}. 
The decays $B^+$$\rightarrow$$K^+$$X$(3872) 
and $B^0$$\rightarrow$$K^0$($\rightarrow$$\pi^+$$\pi^-$)$X$(3872)
were investigated. 
For the determination of the mass and the width (see below) 
of the $X$(3872) in the mode $X(3872)\rightarrow \jpsi\pi^+\pi^-$,
a 3-dimensional fit was performed using the three variables:
beam constraint mass $M_{\rm bc}$=$\sqrt{(E_{beam}^{cms})^2-(p_B^{cms})^2}$ 
(with the energy in the center-of-mass system $E_{beam}^{cms}$, and the momentum
of the $B$ meson in the center-of-mass system $p_B^{cms}$), 
the invariant mass $m(\jpsi\pi^+\pi^-)$
and the energy difference $\Delta$$E$=$E_B^{cms}$$-$$E_{beam}^{cms}$ 
(with the energy of the $B$ meson in the center-of-mass system $E^{cms}_B$).
For the mass measurement, in a first step, the fit was performed for 
the reference channel $\psi'\rightarrow\jpsi\pi^+\pi^-$,
and the resolution parameters 
(i.e.\ the widths of a core Gaussian and a tail Gaussian)
were then fixed for the fit of the $X$(3872).
Figure~\ref{fx3872_signal} (top) shows the data and the fits for the $X$(3872) 
in the projections of the three variables as defined above.
The yield is 151$\pm$15 events for $B^+$ decays
and 21.0$\pm$5.7 events for $B^0$ decays.
Secondly,
LHCb observed the inclusive production of the $X$(3872) in $p$$p$ collisions at $\sqrt{s}$=7~TeV
with an integrated luminosity of 34.7 pb$^{−1}$ \cite{x3872lhcb}.
Figure~\ref{fx3872_signal} (bottom) shows the invariant mass $m(\jpsi\pi^+\pi^-)$
with a fitted signal yield of 565$\pm$62 events for the $X$(3872).
In an updated analysis \cite{x3872lhcb_quantum_number} using increased integrated luminosity of 1~fb$^{-1}$,
but a tighter selection, a signal yield 313$\pm$26 was reported. 
{\color{revcolor} As will be discussed below in Sec.~\ref{cy4260}, the $X$(3872) has also been observed
in radiative transitions at $\sqrt{s}$=4.26~GeV by BESIII \cite{Ablikim:2013dyn}.}

\subsubsection{\boldmath{Mass of the $X$(3872)}}

\label{cx3872_mass}

The world average mass of the $X$(3872) is 3871.69$\pm$0.17~MeV \cite{pdg}.
As will be discussed below,
the $X$(3872) does not fit into potential model predictions,
and is discussed in literature as a possible $S$-wave $D^{*0}$$\overline{D}^0$
molecular state, in particular because its mass is close
to the $D^{*0}$$\overline{D}^0$ threshold.
In this case, the binding energy $E_b$ is given
by the mass difference $m$(X)$-$$m$($D^{*0}$)$-$$m$($D^0$),
yielding $E_b$=0.01$\pm$0.18~MeV using present world average masses of the $X$, the $D^0$ and the $D^{*0}$ \cite{pdg}.
For an $S$-wave near-threshold resonance with an assumed positive
scattering length, $E_b$ is inversely proportional to the squared
scattering length $a$ according to $E_b$=$\hbar^2$/2$\mu$$a^2$ \cite{Braaten:2007dw,Braaten:2007ft,Braaten:2003he}
using the reduced mass $\mu$.
The average radius can be approximated in first order by $<$$r$$>$=$a$/$\sqrt{2}$,
which would lead to a very large value of $<$$r$$>$$\geq$31.7$^{+\infty}_{-24.5}$~fm.
{\color{revcolor} From theory point of view, if requiring isospin $I$=0 for the X(3872),
both neutral $D^{*0}$$\overline{D}^0$ and charged $D^\pm$$D^{*\mp}$ should be present in the molecule \cite{Gamermann:2009uq}.
For the charged threshold, the binding energy would be $\simeq$8 MeV instead of 0.01$\pm$0.18~MeV,
and the question arises about the validity of the above size estimate.
On the one hand, the charged component could appear e.g.\ in a $D^{+}$$D^{-}$$\gamma$ final state,
but has not been observed experimentally yet.
On the other hand, with two thresholds in the system, the question is non-trivial, which one to select for the binding energy.
A bound state wave function at large distances scales as $\exp(-\gamma r)/r$,
where $r$ is the distance between the constituents and $\gamma$ denotes the typical momentum scale defined via
$\gamma = \sqrt{2 \mu E_b}$ using the reduced mass $\mu$.
Thus, even in case of two thresholds, the closer one largely dominates the long-range part \cite{Guo:2017jvc},
and the above estimate of the size of the object remains valid.}

\subsubsection{\boldmath{Width of the $X$(3872)}}

Belle used the above mentioned 3-dimensional fit to attempt to measure the width.  The 3-dimensional fits are more sensitive to the natural width
than the mass resolution provided by the detector
because of the constraints which enter from additional kinematic parameters.
In the first step, the fit was performed for 
the reference channel $\psi'$$\rightarrow$$\jpsi$$\pi^+$$\pi^-$,
and the resolution parameters 
(i.e.\ the widths of a core Gaussian and a tail Gaussian)
were then fixed for the fit of the $X$(3872).
With this 3-dimensional fit, a new precise measurement of the width 
of the $X$(3872) was performed.
This method of determining 
the width was tested using the $\psi'$ as reference providing 
a result of $\Gamma_{\psi'}^{\rm measured}$=0.52$\pm$0.11~MeV larger than
the world average of $\Gamma_{\psi'}^{PDG}$=0.304$\pm$0.009~MeV \cite{pdg},
indicating a bias in the Belle measurement of $\Delta$$\Gamma$=$+$0.23$\pm$0.11~MeV.
The procedure for the determination of the upper limit  on  $\Gamma_{X(3872)}$ is:
for a given fixed width $\Gamma$ the number of signal events
and the number of peaking background events is kept floating in the 3-dimensional fit, and the 
likelihood is calculated. Then the 90\% likelihood interval is 
determined by finding $w_{90\%}$ for the integral $\int_0^{w_{90\%}}$$\Gamma$$d\Gamma$=0.9.
This procedure gives $w_{90\%}$=0.95~MeV, to which the bias has to be added,
so that the final result is $\Gamma_{X(3872)}$$<$1.2~MeV at 90\% C.L. 
This upper limit is a factor of $\simeq$2 smaller than the previous upper limit.
Higher precision might be difficult to achieve using this method, 
however, the width of the $X$(3872) can be measured by a resonance scan 
using a cooled anti-proton beam \cite{Galuska:2012cr}, \cite{panda_x3872_charm13}.
The narrow width of the $X$(3872) is quite remarkable, as it is about two orders of magnitude smaller
than potential model predictions for the $\chi_{c1}'$, a predicted charmonium state
with nearby mass and identical quantum numbers. One of the proposed explanations
is isospin violation (see below).

\subsubsection{\boldmath{Decays of the $X$(3872)}}

The $X$(3872) is one of the few among the new charmonium-like states which has been observed 
in more than one decay channel, clearly proving that it is not a threshold effect. 
The decay channels are
$X$(3872)$\rightarrow$$\jpsi$$\pi^+$$\pi^-$ \cite{x3872belle} \cite{Aubert:2004ns,Aubert:2008gu} \cite{Acosta:2003zx,Abulencia:2005zc,Aaltonen:2009vj} \cite{Abazov:2004kp},  
$X$(3872)$\rightarrow$$\jpsi$$\gamma$ \cite{x3872jpsigamma_belle} \cite{x3872jpsigamma_babar},
$X$(3872)$\rightarrow$$\psi'$$\gamma$ \cite{x3872jpsigamma_babar} \cite{x3872lhcb_radiative},
$X$(3872)$\rightarrow$$\jpsi$$\pi^+$$\pi^-$$\pi^0$ \cite{x3872jpsipipipi},
$X$(3872)$\rightarrow$$D^0$$\overline{D}^0$$\pi^0$ \cite{x3872ddpi_belle_1} \cite{x3872ddpi_belle_2}, and 
$X$(3872)$\rightarrow$$D^0$$\overline{D}^0$$\gamma$ \cite{x3872ddpi_belle_2}.

\paragraph{\boldmath{Decays of $X$(3872) to $\jpsi\pi\pi$}}

The decay $X$(3872) to $\jpsi\pi^+\pi^-$ was the first observed decay mode 
of the $X$(3872) \cite{x3872belle}.
The $\pi^+$$\pi^-$ invariant mass shows evidence
that the decay actually proceeds through the sequential decay
$X$(3872)$\rightarrow$$\jpsi$$\rho$ with $\rho$$\rightarrow$$\pi^+$$\pi^-$ 
(see below). 
For the case of non-resonant $\pi^+$$\pi^-$, the isospin could be $I$=0 or $I$=1.
However, for the case of an intermediate $\rho$ the isospin is fixed to $I$=1.
Therefore the decay $X$(3872)$\rightarrow$$\jpsi$$\rho$ violates 
strong isospin conservation.
There are only two additional isospin violating transitions known 
in the charmonium system \cite{pdg}, namely
$\psi'$$\rightarrow$$\jpsi$$\pi^0$ 
(${\cal B}$=1.3$\pm$0.1$\times$10$^{-3}$ \cite{pdg}, 
${\cal B}$=1.26$\pm$0.02$\pm$0.03$\times$10$^{-3}$ \cite{bes3_isospin})
and $\psi'$$\rightarrow$$h_c$$\pi^0$ 
(${\cal B}$=8.4$\pm$1.6$\times$10$^{-4}$ \cite{pdg}).
For the $X$(3872) the branching fraction of isospin violating 
transition is (among the known decays) $>$2.6\% \cite{pdg}
and thus seems to be largely enhanced.
There are several possible mechanisms for the isospin violation:

\begin{itemize}

\item The $u$/$d$ quark mass difference in strong interaction
induces isospin violation. 
Such a difference is immediately provided here, as the $X$(3872) 
can only decay to $D^{0*}$$\overline{D}^0$ (containing
$u$-type quarks), but not into $D^{+*}$$D^-$ (containing
$d$-type quarks), for which the threshold is $\simeq$8~MeV
higher. 

\item The $u$/$d$ quark charge difference in electromagnetic interactions (EM) 
induces isospin violation. 
Isospin should only be conserved in strong interaction, 
but not in EM interaction. 
Thus, one of the possible explanation is that the decay
$X$(3872)$\rightarrow$$\jpsi$$\rho$($\rightarrow$$\pi^+$$\pi^-$)
  is proceeding via EM interaction, i.e.,
the $\rho$ is created not by gluons but by a virtual photon. The
branching ratio would be then comparable to radiative decays (see below).

\item {\color{revcolor} As the $\rho$ has a large width of 149.1$\pm$0.8~MeV \cite{pdg}, 
mixing of $\rho$ and $\omega$ may induce isospin violation \cite{terasaki_rhoomegamixing}.
In fact, the experimentally observed large ratio of 
${\cal B}$$X$(3872)$\rightarrow$$\jpsi$$\omega$/
${\cal B}$$X$(3872)$\rightarrow$$\jpsi$$\rho$ 
(see below in Sec.~\ref{cjpsiomega})
is supporting evidence.

\item The difference in hadronic loops 
$X$(3872)$\rightarrow$
$D^{0*}$$\overline{D}^0$$\rightarrow$
$D^{+*}$$D^-$$\rightarrow$$\jpsi$$\rho$
vs.\ 
$X$(3872)$\rightarrow$
$D^{0}$$\overline{D}^0$$\gamma$$\rightarrow$
$D^{+}$$D^-$$\gamma$$\rightarrow$$\jpsi$$\rho$
may induce isospin violation as well.
Although being a small effect due to the small mass difference between the charged and the neutral channel,
the effect can be amplified by the relatively large mass of the $\rho$, since the phase space is severely restricted
\cite{Gamermann:2009uq,Kalashnikova:2005ui,Voloshin:2006pz,Danilkin:2009hr,Barnes:2007xu,Pennington:2007xr,Meng:2007cx}.} 

\end{itemize}

\paragraph{\boldmath{Decays of $X$(3872) to $D^{(*)}\overline{D}^{(*)}$}}

\begin{figure}[t!]
\centerline{\includegraphics[width=0.7\textwidth]{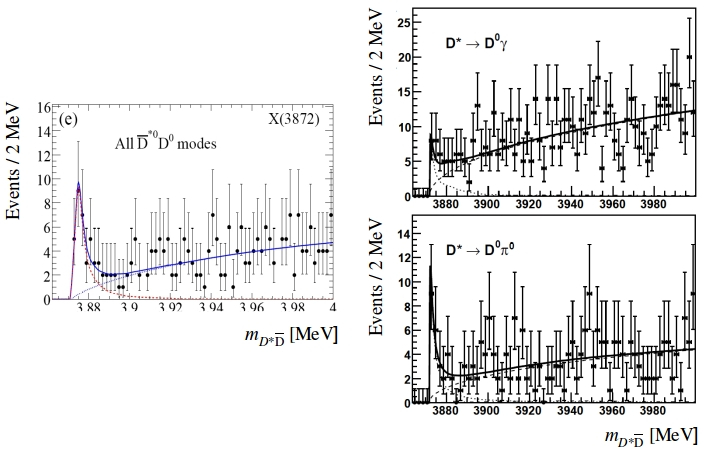}}
\caption{$D^{0*}$$\bar{D}^0$ invariant mass for the decay
$X$(3872)$\rightarrow$$D^{0*}$$\bar{D}^0$ from 
\BaBar \cite{x3872ddpi_babar} (left, for both 
$D^{*0}$$\rightarrow$$D^{0}$$\gamma$ and $D^{*0}$$\rightarrow$$D^{0}$$\pi^0$) 
and from Belle \cite{x3872ddpi_belle_1} \cite{x3872ddpi_belle_2} 
(right top for $D^{*0}$$\rightarrow$$D^{0}$$\gamma$ and 
right bottom for $D^{*0}$$\rightarrow$$D^{0}$$\pi^0$).
\label{fx3872_ddpi}}
\vskip -3mm
\end{figure}

As the mass of the $X$(3872) is very close to the 
$D^0$$\overline{D}^{0*}$ threshold, the investigation 
of open charm decays is very important.
The decay into $D^0$$\bar{D}^{0*}$ is a strong decay and 
among the so far observed decays it represents the dominant one, 
i.e.\ the branching fraction is a factor $\simeq$9
higher than for the decay into $\jpsi$$\pi^+$$\pi^-$.
In this decay channel, \BaBar 
measured surprisingly a high mass 
of the $X$(3872) as
$m$=3875.1$^{+0.7}_{-0.5}$(stat.)$\pm$0.5(syst.)~MeV 
\cite{x3872ddpi_babar} (see Fig.~\ref{fx3872_ddpi}, left), suggesting that there might be two different states,
namely $X$(3872) and $X(3876)$ , which would fit to a tetraquark hypothesis \cite{x3872_tetraquark_maiani}
for two states $[$$c$$\overline{c}$$u$$\overline{u}$$]$
and $[$$c$$\overline{c}$$d$$\overline{d}$$]$.
On the other hand, Belle measured the mass in the same decay channel  as 
$m$=3872.9$^{+0.6}_{-0.4}$(stat.)$^{+0.4}_{-0.5}$(syst.)~MeV \cite{x3872ddpi_belle_2}
(see Fig.~\ref{fx3872_ddpi}, right) 
which is consistent with the world average \cite{pdg}.
A possible explanation of the discrepancy is the 
difficulty of performing fits to signals close to threshold.
In fact, the two experiments used two very different approaches:

\begin{itemize}

\item \BaBar used a 1-dimensional binned maximum likelihood fit \cite{x3872ddpi_babar}
with the $D^*D$ invariant mass as the only variable,  
where the signal probability density function (p.d.f.) was extracted 
from MC simulations assuming a $S$=1 resonance produced with $L$=0. 
An exponential function $a(m-m_0)^b\times\exp(m-m_0)$ 
with a threshold mass $m_0$ and 
parameters $a$, $b$, $c$ was used for the background parametrization. 

\item Belle used an 2-dimensional unbinned maximum likelihood fit \cite{x3872ddpi_belle_2} 
to the beam constrained mass, with a Gaussian signal function and an Argus function \cite{Albrecht:1990am} for the background, and the 
$D^*D$ invariant mass using a Breit-Wigner signal function and
$a$$\sqrt{m-m_0}$ for the background parametrization. 

\end{itemize}

The lineshape in Fig.~\ref{fx3872_ddpi} was found to be compatible with 
both the $X(3872)$ being a virtual or a bound state \cite{Kang:2016jxw}. Also,
following a suggestion from \cite{flatte_hanhart}
the fit in the Belle case was checked with a Flatt\'{e} p.d.f.
instead of a Breit-Wigner p.d.f., in particular to take into account the contribution
of $D^{+*}$$D^{-}$ to the tail shape. Results were consistent.

An important question is if the decay proceeds resonant
in the $D^*$ (i.e.\ $X$(3872)$\rightarrow$$D^0$$\overline{D}^{0*}$)
or non-resonant
(i.e.\ $X$(3872)$\rightarrow$$D^0$$\overline{D}^0$$\pi^0$).
As the $D^{*0}$ ($D^0$) carries $J$=1 ($J$=0), 
the first case would correspond to $L$=0,
the latter case to $L$=1.
In the analyses, a mass cut on the $D^*$ can distinguish
between the resonant and non-resonant case. 
The product branching fractions for the sum of the resonant and non-resonant case
is 
(1.22$\pm$0.31$^{+0.23}_{-0.30}$)$\times$10$^{-4}$ \cite{x3872ddpi_belle_1}, 
for the resonant case alone 
(0.80$\pm$0.20$\pm$0.10)$\times$10$^{-4}$ \cite{x3872ddpi_belle_2} 
and thus corresponding to $\simeq$65\%. 
Note that the decay of $D^{0*}$$\rightarrow$$D^0$$\gamma$ 
(Fig.~\ref{fx3872_ddpi}, right top) only contributes 
to the resonant case. So far no evidence for non-resonant
$X$(3872)$\rightarrow$$D^0$$\overline{D}^0$$\gamma$ has been
reported. 
The charged decay $X$(3872)$\rightarrow$$D^+$$D^{*-}$
was not observed, as the mass of the $X$(3872) is 8.1~MeV 
below the threshold and consequently is kinematically forbidden.

\paragraph{\boldmath{Radiative Decays of the $X$(3872)}}

\begin{figure}[b!]
\centerline{
\centerline{\includegraphics[width=\textwidth]{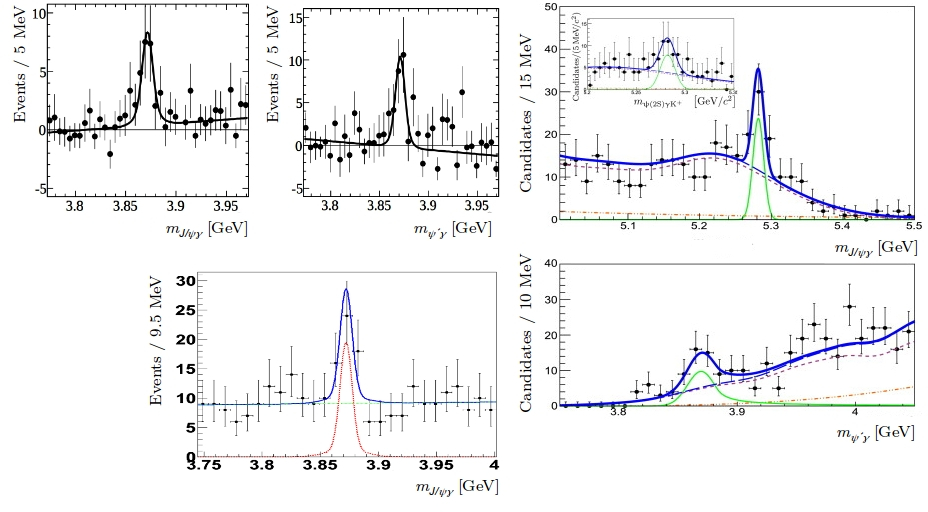}}
}
\caption{$\jpsi$$\gamma$ and $\psi'$$\gamma$ invariant mass
for the decays $X$(3872)$\rightarrow$$\jpsi$$\gamma$ and 
$X$(3872)$\rightarrow$$\psi'$$\gamma$ 
from \BaBar \cite{x3872jpsigamma_babar} (top left, top center), 
from Belle \cite{x3872belle_radiative_2011} (bottom left, only $\jpsi$$\gamma$ observed)
and from LHCb \cite{x3872lhcb_radiative} (top right, bottom right).
\label{fx3872_jpsigamma}}
\end{figure}

The decay $X(3872)\rightarrow \jpsi\gamma$
represents a decay into two eigenstates of C-parity.
Therefore its observation implies that the charge conjugation 
of the $X(3872)$ must be C=$+$1.
The observation of this decay was reported by Belle
with a data set of 256~fb$^{-1}$, 
a yield of 13.6$\pm$4.4 events
and a statistical significance of 4.0$\sigma$ \cite{x3872jpsigamma_belle}.
The combined branching ratio was measured to
${\cal B}$($B^{\pm}$$\rightarrow$$X$$K^{\pm}$,
$X$$\rightarrow$$\gamma$$\jpsi$)=
(1.8$\pm$0.6$\pm$0.1)$\times$10$^{-6}$),
i.e.\ the branching fraction of the rare decay $X$(3872)$\rightarrow$$\jpsi$$\gamma$ is 
a factor $\simeq$6 smaller than the one for $X$(3872)$\rightarrow$$\jpsi$$\pi^+$$\pi^-$.
\BaBar confirmed the observation 
with a data set of 260~fb$^{-1}$, 
a yield of 19.4$\pm$5.7 events
and a statistical significance of 3.4$\sigma$ \cite{x3872jpsigamma_babar}
(see Fig.~\ref{fx3872_jpsigamma}, bottom).
The combined branching ratio was measured to
${\cal B}$($B^{\pm}$$\rightarrow$$X$$K^{\pm}$,
$X$$\rightarrow$$\gamma$$\jpsi$)=
(3.4$\pm$1.0$\pm$0.3)$\times$10$^{-6}$),
i.e. about a factor of two higher than Belle.

\BaBar found evidence for the decay 
$X$(3872)$\rightarrow$$\psi'$$\gamma$ \cite{Aubert:2004ns}
with 424~fb$^{-1}$ and 25.4$\pm$7.4 signal events. 
The ratio of the branching fractions 
$R_{\psi\gamma}$=${\cal B}$(X(3872)$\rightarrow$$\psi$(2S)$\gamma$)/${\cal B}$(X(3872)$\rightarrow$$\jpsi$$\gamma$)
was measured as $R_{\psi\gamma}$=3.4$\pm$1.4.
indicating the surprising fact that the transition of the $X$(3872) to the $n$=2 state 
is significantly stronger than the transition to the $n$=1 state.
In the case of $X$(3872)$\rightarrow$$\jpsi$$\gamma$
the photon energy is $E_{\gamma}$=775~MeV, and thus 
due to vector meson dominance $\rho$ and $\omega$ can contribute
to the amplitudes. However, in the case of 
$X$(3872)$\rightarrow$$\psi'$$\gamma$
with the smaller $E_{\gamma}$=186~MeV
the transition can only proceed through light quark 
annihilation with an expected small amplitude.

The ratio $R_{\Ppsi\g}$ is predicted to be in the range
$( 3 - 4 )\times10^{-3}$ for a~$\D\Dstarb$~molecule~\cite{Swanson,FerrettiGalata},
$1.2 - 15$ for a pure charmonium state~\cite{BarnesGodfreySwanson,BarnesGodfrey,LiChao,Lahde,Simonov,identifyX1,identifyX2}
and $0.5 - 5$ for a~molecule-charmonium mixture~\cite{Simonov,EichtenLaneQuigg,DongFaeser}.
An updated measurement by Belle of both radiative channels 
was based upon a data set of 711~fb$^{-1}$ \cite{x3872belle_radiative_2011}.
The background was studied in MC simulations
and revealed peaking behavior in some background components
close to the signal region.
The signal $X$(3872)$\rightarrow$$\jpsi$$\gamma$ was clearly re-established
(see Fig.~\ref{fx3872_jpsigamma}, top)
with 
30.0$^{+8.2}_{-7.4}$ signal events 
(4.9$\sigma$ significance)
for $B^+$$\rightarrow$$K^+$$X$(3872)
and 
5.7$^{+3.5}_{-2.8}$ signal events  
(2.4$\sigma$ significance)
for $B^0$$\rightarrow$$K^0$$X$(3872).
For $X$(3872)$\rightarrow$$\psi'$$\gamma$, 
the shape of the $\psi' K^*$ and $\psi' K$ background,
and in particular the peaking structures, 
was modeled as a sum of bifurcated Gaussians
using a large MC sample. 
The signal yields were determined as
5.0$^{+11.9}_{-11.0}$ signal events 
(0.4$\sigma$ significance)
for $B^+$$\rightarrow$$K^+$$X$(3872)
and 
1.5$^{+4.8}_{-3.9}$ signal events  
(0.2$\sigma$ significance)
for $B^0$$\rightarrow$$K^0$$X$(3872).
With an upper limit of $R_{\psi\gamma}$$<$2.1 (90\% CL), 
Belle was not able to confirm the large
$R_{\psi\gamma}$ value by \BaBar. 
Finally, with an integrated luminosity of 3 fb$^{-1}$, LHCb 
confirmed the evidence of $X$(3872)$\rightarrow$$\psi'$$\gamma$ 
with a higher signal yield of 36.4$\pm$9.0 \cite{x3872lhcb_radiative},
corresponding to a significance of 4.4$\sigma$.
A large ratio of $R_{\psi\gamma}$= 2.46$\pm$0.64$\pm$0.29
was measured. 
{\color{revcolor} 
The large value supports the charmonium interpretation, or the interpretation of a mixture of a molecule and a charmonium state, but disfavors a pure molecule ~\cite{Swanson,FerrettiGalata}. 
However, there is recent evidence that the radiative transition is controlled by short range dynamics \cite{Guo:2017jvc}
and therefore any information about the long-range, molecular contribution is difficult to be extracted by the $R_{\psi\gamma}$ ratio.
Instead, there are proposals to use different ratios such as
${\cal B}$(X(3872)$\rightarrow$$\jpsi\gamma$)/${\cal B}$(X(3872)$\rightarrow$$\jpsi$$\pi^+\pi^-$) \cite{DongFaeser} \cite{Aceti:2012cb}.}

\paragraph{\boldmath{Decays of $X$(3872) into $\jpsi\pi\pi\pi$}}

\label{cjpsiomega}

As one of the observed proposed explanations for the isospin violation
(see above) is $\rho-\omega$ mixing \cite{terasaki_rhoomegamixing}, 
the search for the decay 
$X$(3872)$\rightarrow$$\jpsi$$\omega$($\rightarrow$$\pi^+$$\pi^-$$\pi^0$)
is of importance.
The difficulty here is the nearby $Y(3940)$ state, 
which is also known to decay into the same final state. 
Belle observed a signal 
for $X$(3872)$\rightarrow$$\jpsi$$\omega$($\rightarrow$$\pi^+$$\pi^-$$\pi^0$),
based upon a data set of 256~fb$^{-1}$ \cite{x3872jpsipipipi}.
A cut on the $\omega$ meson in the 3-pion mass from 0.750 GeV to 0.775  GeV was used
in the analysis. The observed yield was 12.4$\pm$4.1 events, corresponding
to a significance of 4.3$\sigma$.
The measured efficiency corrected ratio of 
$X$(3872)$\rightarrow$$\jpsi$$\pi^+$$\pi^-$$\pi^0$/
$X$(3872)$\rightarrow$$\jpsi$$\pi^+$$\pi^-$=
1.0$\pm$0.4(stat.)$\pm$0.3(syst.) indicates isospin violation in a decay of the $X$(3872), 
as the additional $\pi^0$ carries $I$=1. 

In an analysis by \BaBar \cite{y3940_babar_1}, using a three-pion mass from 0.7695 GeV to 0.7965 GeV
as a cut for the $\omega$, no evidence for $X$(3872)$\rightarrow$$\jpsi$$\omega$($\rightarrow$$\pi^+$$\pi^-$$\pi^0$)
was found. 
In a re-analysis with 426~fb$^{-1}$ by \BaBar \cite{x3872_babar_2010}
the lower boundary of the kinematical search window for the three-pion mass
was extended from 0.5~GeV to 0.9~GeV. 
Here, the Belle signal was confirmed
with a significance of 4.0$\sigma$. 
Also the large isospin violation is suggested 
by the measurement of the ratio
$X$(3872)$\rightarrow$$\jpsi$$\pi^+$$\pi^-$$\pi^0$/
$X$(3872)$\rightarrow$$\jpsi$$\pi^+$$\pi^-$ 
as 0.7$\pm$0.3(stat.) and 1.7$\pm$1.3(stat.) 
for $B^+$ and $B^0$ decays, respectively.

For the three decays with the largest branching fraction,
numbers are given in Table~\ref{tx3872br}.
The product branching fraction for the $\jpsi$$\pi^+\pi^-$
final state was derived from \cite{x3872jpsigamma_belle} 
by using published branching fractions for the $\psi$(2$S$)
in \cite{pdg}.
The relative rates for
$\jpsi$$\pi^+\pi^-$ : $D^{*0}$$\overline{D}^0$ : $\jpsi$$\gamma$
are 1 : 0.18 : 0.025.

\begin{table}[htb]
\centering
\caption{Branching fractions of decays of the $X$(3872).}
\vspace{0.2cm}
\begin{tabular}{lcc}
\hline\\[-2.5ex]
Product branching fraction & & \\
\hline\\[-2.5ex]
${\cal B}$($B$$\rightarrow$$K$X) 
$\times$ 
${\cal B}$($X$(3872)$\rightarrow$$\jpsi$$\pi^+\pi^-$) &
(1.31$\pm$0.24$\pm$0.13)$\times$10$^{-5}$ & \cite{x3872belle} \\
& (1.28$\pm$0.41)$\times$10$^{-5}$ & \cite{Aubert:2004ns,Aubert:2008gu} \\
\hline
${\cal B}$($B$$\rightarrow$$K$X)
$\times$
${\cal B}$($X$(3872)$\rightarrow$$D^{*0}$$\overline{D}^0$) &
(0.73$\pm$0.17$\pm$0.08)$\times$$10^{-4}$ & 
\cite{x3872ddpi_belle_1}
\cite{x3872ddpi_belle_2} \\
 & 
(1.41$\pm$0.30$\pm$0.22)$\times$$10^{-4}$ & 
\cite{x3872ddpi_babar} \\
\hline
${\cal B}$($B$$\rightarrow$$K$X)
$\times$
${\cal B}$($X$(3872)$\rightarrow$$\jpsi$$\gamma$) &
(1.8$\pm$0.6$\pm$0.1)$\times$10$^{-6}$ &
\cite{x3872jpsigamma_belle} \\
 & 
(3.3$\pm$1.0$\pm$0.3)$\times$10$^{-6}$ &
\cite{x3872jpsigamma_babar} \\
\hline
\end{tabular}
\label{tx3872br}
\end{table}

\subsubsection{\boldmath{Production of the $X$(3872)}}

The $X$(3872) was discovered in three different $B$ meson decays
$B^0$$\rightarrow$$K^0$($\rightarrow$$\pi^+$$\pi^-$)$X$(3872),
$B^{\pm} \rightarrow K^{\pm}X$(3872), and 
$B^0\rightarrow K^{0*}(\rightarrow K^+ \pi^-)X(3872)$. 
It has also been observed as an inclusive signal 
in direct $p$$\overline{p}$ collisions at
CDF II \cite{Acosta:2003zx,Abulencia:2005zc,Aaltonen:2009vj}, 
CMS \cite{x3872cms}, 
D0 \cite{Abazov:2004kp} 
and LHCb \cite{x3872lhcb} (see Fig.~\ref{fx3872_signal}). 
However, so far, there was no indication of direct $X$(3872) production
in $e^+$$e^-$ collisions, $\gamma$$\gamma$ collisions 
or production in initial state radiation.
For the observation in $B$ meson decays, 
it is important to note that the $K^0$, $K^{\pm}$
have $J$=0, while the $K^*$ has $J$=1.
Thus, the $X$(3872) has been observed in $B$ meson decays
both with a pseudoscalar or a vector particle accompanied.
No evidence of the $X$(3872) was found in 
$\gamma$$\gamma$$\rightarrow$$\jpsi$$\omega$
\cite{jpsiomega_gammagamma_belle} \cite{jpsiomega_gammagamma_babar} as expected, 
because it should not be observable for $J$=1 due to the Landau-Young theorem.

\paragraph{\boldmath{Production of $X$(3872) in $B$$\rightarrow$$K$$\pi$$X$(3872)}}

All the above analyses of $X$(3872) were performed in the decays
$B$$\rightarrow$$K$$X$(3872). However, it was also observed in
$B$$\rightarrow$$K$$\pi$$X$(3872) with an additional $\pi^{\mp}$.
$B$$\rightarrow$$K$$\pi$$X$(3872) can contain resonant ($K^*$) 
and non-resonant amplitudes in the $K^{\pm}$$\pi^{\mp}$ system,
while $B\rightarrow K X(3872)$ is by definition only resonant. 
The non-resonant part was clearly observed and a product
branching fraction of 
${\cal B}(B^0\rightarrow X[K^{\pm}\pi^{\mp}_{\rm non-resonant})$
$\times {\cal B}(X\rightarrow \jpsi\pi^+\pi^-) =
(8.1\pm 2.0 ^{+1.1}_{-1.4})\times 10^{-6}$ was measured \cite{x3872belle_charged_neutral},
corresponding to $N$$\simeq$90 observed $X$(3872) events. 
This is surprisingly as large as
${\cal B}$($B^0$$\rightarrow$X$K$)
$\times$ ${\cal B}$($X$$\rightarrow$$\jpsi$$\pi^+$$\pi^-$) =
(8.10$\pm$0.92$\pm$0.66)$\times$10$^{-6}$ \cite{x3872belle_charged_neutral}
and 
(8.4$\pm$1.5$\pm$0.7)$\times$10$^{-6}$ \cite{Aubert:2004ns,Aubert:2008gu},
although the phase space is smaller. 
Another surprise is that the resonant part is very small,
and only an upper limit 
$<$3.4$\times$10$^{-6}$ \cite{x3872belle_charged_neutral}
could be given. 
This behavior is very different from other charmonium channels, 
e.g.\ $B$$\rightarrow$$K$$\pi$$\psi'$ \cite{x3872belle_charged_neutral}, 
$B$$\rightarrow$$K$$\pi$$\jpsi$ \cite{x3872_kpi_crossref_jpsi}, or 
$B$$\rightarrow$$K$$\pi$$\chi_{c1}'$ \cite{x3872_kpi_crossref_chic1}:
in all cases the resonant $K^*$(892) and $K^*(1430)$ are dominating
these decays almost completely. This observation might support 
the indication that the $X$(3872) does not represent conventional 
charmonium.

\paragraph{\boldmath{Production of $X$(3872) in charged and neutral $B$ decays}}

Production of the $X$(3872) in $B^+$ and $B^0$ decays can be quite different
from each other. 
According to \cite{swanson_new_mesons} there are two different Feynman graphs
for $B$$\rightarrow$$K$$D^*$$\overline{D}$:

\begin{itemize}

\item In case of external $W$ emission the process is color enhanced, 
as the color is decoupled from the $B$$\rightarrow$$K$ transition.
This process is possible for
any of the transistions
$B^0$$\rightarrow$$K^+$, 
$B^0$$\rightarrow$$K^0$, 
$B^+$$\rightarrow$$K^+$, and 
$B^+$$\rightarrow$$K^0$.

\item In case of internal $W$ emission the color is locked 
by the spectator quark, which is in the same loop 
with the $\overline{D}$ and the $D^*$. 
This process is only possible for  
$B^+$$\rightarrow$$K^+$ and
$B^0$$\rightarrow$$K^0$, 
i.e.\ a change of the charge 
for the $B$$\rightarrow$$K$ transition is not possible.
The charge sign flips by the $W$ boson, and then flips back again, 
when the loop is closed.
This process is color suppressed. 

\end{itemize}

In a factorization ansatz \cite{Braaten:2007dw,Braaten:2007ft,Braaten:2003he}, the total amplitude
for $B^+$$\rightarrow$$K^+$ has three contributions:
{\it (1)} $B^+\rightarrow \overline{D}^{*0} V$, $V\rightarrow[D^0K^+]$, 
{\it (2)} $B^+\rightarrow \overline{D}^{0}V$, $V\rightarrow[D^{*0}K^+]$, 
and {\it (3)} $B^+\rightarrow K^+V$, $V\rightarrow [\overline{D}^{*0}D^0]$. All three 
involve the Cabibbo-allowed $\bar{b} \to \bar{c} W^+$, followed by $W^+ \to c\bar{s}$,
and a $q\bar{q}$ vacuum excitation, with (1) and (2)  color-connected and  (3) 
Fierz transformed.
In contrast, the total amplitude for $B^0\rightarrow K^0$ has only one amplitude
$B^0\rightarrow K^0 V$ with $V\rightarrow [\overline{D}^{*0}D^0]$.
According to a detailed calculation \cite{Braaten:2007dw,Braaten:2007ft,Braaten:2003he}, 
the ratio $\Gamma(B^0\rightarrow K^0 X)/\Gamma(B^+\rightarrow K^+X)$ 
should be $\simeq$1 for a charm meson molecular state.
The most recent measurements of 
0.41$\pm$0.24$\pm$0.05 by \BaBar \cite{Aubert:2004ns,Aubert:2008gu}
and 0.50$\pm$0.14$\pm$0.04 by Belle \cite{x3872belle_width_2011}
{\color{revcolor} may disfavour the molecule interpretation, if only the neutral $D^0$$\overline{D}^{0*}$ component is used in the molecular model.
However, the charged component (although, as mentioned above, not experimentally observed so far)
may change the conclusion [125]. In addition, involved uncertainties are large.}

\paragraph{\boldmath{Quantum numbers of the $X$(3872)}}

\paragraph{\boldmath{Quantum Numbers from the 2$\pi$ and 3$\pi$ invariant mass distributions}}

Early analyses \cite{angular_choi_olsen} \cite{angular_cdf2}
tried to employ the shape of the $\pi^+$$\pi^-$ mass distribution in the decay
$X$(3872)$\rightarrow$$\jpsi$$\pi^+$$\pi^-$.
If the relative angular momentum is fixed, then conclusions about the quantum numbers
of the $X$(3872) can be drawn. 
For $S$-wave the spectrum should show a dependence $\sim$$q^*$($\jpsi$)
for $P$-wave $\sim$$q^*$($\jpsi$)$^2$, where $q^*$($\jpsi$) denotes the 
momentum of the $\jpsi$ in the rest frame of the $X$(3872).
If the $X$(3872) decays into two $J^P$=$1^-$ particles 
(i.e.\ the $\jpsi$ and the $\rho$) with an $S$-wave, 
this implies that $J^P$=$0^-$, $1^+$, $2^-$, ... 
However, it was proven later by Belle \cite{x3872belle_width_2011}, that $\rho-\omega$
interference has an significant impact on the shape, and $S$-wave and $P$-wave fits
turned out to be both compatible with the observed distribution.  
In the analysis of the $X$(3872)$\rightarrow$3$\pi$ by \BaBar \cite{x3872_babar_2010}, 
the shape of the 3$\pi$ mass distribution was investigated 
in order to constrain the quantum number of the $X$(3872).
It was found that, different from the case of the 2$\pi$ mass, 
the shape of the 3$\pi$ mass distribution seems to indicate that $P$-wave is preferred.
Surprisingly, this implies $J^{PC}$=2$^{-+}$ for the $X$(3872), which was later
proven to be incorrect by other analyses. 

\paragraph{\boldmath{Quantum numbers from the decay into $\jpsi\gamma$}}

As mentioned above, the branching fraction of the the radiative
decay $X$(3872)$\rightarrow$$\jpsi$$\gamma$ is about one order
of magnitude smaller than $X$(3872)$\rightarrow$$\jpsi$$\pi^+$$\pi^-$
and about two orders of magnitude smaller than $X$(3872) decaying
into open charm.
Although rare, this decay channel is very important, as 
its observation implies the decay into two particles, which 
are C=$-$1, as they are identical with their anti-particles, 
and clearly establishes a $C$=+1 charge parity 
assignment to the $X$(3872).
Consequently, the $C$=+1 assignment importantly implies that the decays 
$X$(3872)$\rightarrow$$\jpsi$$\pi^0$$\pi^0$, $\jpsi$$\pi^0$, 
$\jpsi$$\eta$, $\chi_{c}$$\gamma$ or $\eta_c$$\gamma$ are 
forbidden. None of them has been observed.

The $S$-wave assignment from the $\pi^+$$\pi^-$ invariant mass 
and the $C$=+1 assignment are consistent.
As for mesons the total wave function must be symmetric,
isospin $I$=1 and $J^{P}$=$1^-$ for the case of the $\rho$
in combination of an $S$-wave does require $C$=+1.
In addition, the search for a partner of the $X$(3872) 
with negative $C$-parity in the decay 
$X(3872)\rightarrow \jpsi\gamma\gamma$ was 
negative \cite{D-wave_belle} with a product branching fraction 
of ${\cal {B}}<1.9\times 10^{-4}$.

\paragraph{Quantum numbers from angular analysis}

In order to apply further constraints on the quantum number assignment
of the $X$(3872), angular analyses have been performed.
Initial studies by Belle \cite{angular_choi_olsen} were based upon two assumptions:

\begin{itemize}

\item The decay is assumed to be $X$(3872)$\rightarrow$$J/\psi$$\rho$,
  rather than $X$(3872)$\rightarrow$$J/\psi$$\pi^+$$\pi^-$,
  i.e.,\ a two-particle decay and not a three-particle decay.

\item The polarisation of the $\jpsi$ is orthogonal 
to the axis of its decays to $e^+$$e^-$ in its rest frame.

\end{itemize}

All the amplitudes are calculated in the $\jpsi$ rest frame.
For the definition of the coordinate systems see \cite{angular_rosner} and \cite{angular_bugg}.
$J^{PC}$=0$^{-+}$ and $J^{PC}$=0$^{++}$ were disfavoured by this analysis.
The test for $J^{PC}$=1$^{+-}$ was disfavoured by an analysis based upon the $\jpsi$ helicity angle,
i.e. the angle between the $\jpsi$ and the $B$ meson in the $X$(3872) rest frame.

In an updated analysis by Belle \cite{x3872belle_width_2011}  
a test with the full  data set was performed to distinguish
in particular these two assignments 
($J^{PC}$=$1^{++}$ or $J^{PC}$=$2^{-+}$)
using an angular analysis.
For this purpose, it was assumed that the decay $X$(3872)$\rightarrow$$\jpsi$$\pi^+$$\pi^-$
proceeds via $X$(3872)$\rightarrow$$\jpsi$$\rho$($\rightarrow$$\pi^+$$\pi^-$) in the kinematic
limit, i.e.\ both particles are at rest in the $X$(3872) rest frame. Due to
$m_{X(3872)}$$\simeq$$m_{\rho}$+$m_{J/\psi}$ this is a valid assumption and
it also implies that any higher partial waves can be neglected. 
For $J^{PC}$=1$^{++}$, there is only one amplitude with $L$=0 and $S$=1, 
where $L$ is the total orbital angular momentum between the particles, and $S$ the total spin 
constructed from the $\rho$ and the $\jpsi$. 
For $J^{PC}$=2$^{-+}$, there are two amplitudes with $L$=1 and
$S$=1 or $S$=2. 
These two amplitudes can be mixed using a parameter $\alpha$, which is a complex number.
The angular reference frame follows the definition in \cite{angular_rosner}. 
The angle $\theta_X$ is chosen as the angle between the $\jpsi$ and
the kaon direction in the $X$(3872) rest frame.
The angular distributions $d$$\Gamma$/$d$$\cos$($\theta_X$) for the different quantum numbers are: 
constant for $J^{PC}$=$1^{++}$,
$\sin$$^2$($\theta_X$) for $J^{PC}$=2$^{-+}$ for the case of $\alpha$=0,
and 1+3$\cos$$^2$($\theta_X$) for $J^{PC}$=$2^{-+}$ in case of $\alpha$=1.

Two additional angles are defined as follows:
the $xy$-plane is spanned by the kaon direction and the $\pi^+$ and $\pi^-$ (back-to-back) directions
in the $X$(3872) rest frame.
The $x$-axis is chosen to be along the kaon direction.
The $z$-axis is constructed perpendicular to the $xy$-plane.
The angle $\chi$ is chosen between the $x$-axis and the $\pi^+$ direction.
The angle $\theta_{\mu}$ is chosen between the $\mu^+$ direction 
and the $z$-axis.
A simultaneous fit for all three angles was performed.
The $\chi^2$ values are listed in Table~\ref{tchi2}. For the case of $J^{PC}$=2$^{-+}$, 
the values in Table~\ref{tchi2} are given for a complex amplitude $\alpha$=0.69$\times$$\exp$($i$23$^o$),
which was found in a grid search and 
which is the only value which gives a confidence level $>$0.1 for all three angles.
With the available statistics, the two quantum numbers could not be distinguished; 
however $J^{PC}$=1$^{++}$ was slightly preferable. 

\begin{table}[tb]
\centering
\caption{$\chi^2$ values for the fit of the angular distributions.
See text for the definitions of the angles.}
\vspace{0.2cm}
\begin{tabular}{lcccc}
\hline\\[-2.5ex]
Angle & $\chi^2$/n.d.f. & C.L. & $\chi^2$/n.d.f. & C.L. \\
\hline\\[-2.5ex]
 & \multicolumn{2}{c}{$J^{PC}$=$1^{++}$} & \multicolumn{2}{c}{$J^{PC}$=$2^{-+}$} \\
\hline\\[-2.5ex]
$\chi$ & 1.76/4 & 0.78 & 4.60/4 & 0.33 \\
$\theta_{lepton}$ & 0.56/4 & 0.97 & 5.24/4 & 0.26 \\
$\theta_X$ & 3.82/4 & 0.51 & 4.72/4 & 0.32 \\
\hline
\end{tabular}
\label{tchi2}
\end{table}

In an analysis by LHCb \cite{x3872lhcb_quantum_number} with a data set of 1.0~fb$^{-1}$, 
the likelihood with respect to the parameter $\alpha$ 
was tested, while each event was weighted according to the mass difference 
$m$($\jpsi$$\pi^+$$\pi^-$)-$m$($\jpsi$). 
The likelihood was calculated using five angles, i.e.\ three helicity angles
and two decay plane angles:
the polar angle between the $X$(3872) and the $\jpsi$ ($\vartheta_X$), 
the polar angle between the $\pi^+$ and the $\pi^+\pi^-$ system ($\vartheta_{\pi\pi}$), 
the polar angle between the $\jpsi$ and the $\mu^+$ ($\vartheta_{J/\psi}$), 
the azimuthal angle between the plane spanned by the $\pi^+$ and the $\pi^-$
and the plane spanned by the $X$ and the $\jpsi$ ($\Delta$$\phi_{X,\pi\pi}$), 
and the azimuthal angle between the plane spanned by the $\mu^+$ and the $\mu^-$
and the plane spanned by the $X$ and the $\jpsi$ ($\Delta$$\Phi_{X,J/\psi}$).
The result strongly favours $J^{PC}$=1$^{++}$
with a confidence level of 34\%, while $J^{PC}$=2$^{-+}$ is disfavoured by 8.2$\sigma$, 
in contrast to the result from the \BaBar analysis of the $\jpsi$$\pi$$\pi$$\pi$
decay \cite{x3872_babar_2010} (see above).

The $J^{PC}$=$1^{++}$ assignment has implications for the production process.
The decay $B$$\rightarrow$$K$$X$(3872) would be $0^-$$\rightarrow$$0^-$$1^+$.
This means parity ($-$1) on the left hand side and parity 
($-$1)$\times$($+$1)$\times$($-$1)$^L$ for the right hand side. Creating $J$=1 for the $X$(3872) 
would require $L$=1, but this implies parity $+1$ for the right hand side.
Consequently it represents a parity violating weak decay. 

\subsubsection{\boldmath{Interpretation of the $X$(3872)}}

\paragraph{Interpretation as a charmonium state}

If the $X$(3872) is a conventional charmonium state, the $J^{PC}$=1$^{++}$ assignment
leaves as the only candidate the $\chi_{c1}'$, a $^3P_1$ state.
The predicted mass by potential models is $m$=3953~MeV, thus $\simeq$70~MeV
higher than the observed $X$(3872) mass.
This would be a $n$=2 radial excitation, and the quantum numbers are favoured 
by angular analyses \cite{angular_cdf2} \cite{angular_choi_olsen}.
However, there are three arguments against the assignment as the $\chi_{c1}'$ \cite{Barnes:2005pb}: 

\begin{itemize}
\item Potential models predict that the mass 
should be theoretically higher by $\simeq$70~MeV.
For almost all of the other known charmonium states
deviations of the predicted values for the masses 
and for the widths are $<$10~MeV.
\item The width should be larger 
with $\simeq$130 MeV, compared 
to the experimental upper limit  $\leq$2.3~MeV for the $X$(3872), 
as mentioned above.
\item The observed ratio of
$X$(3872)$\rightarrow$$\jpsi$$\gamma$ to $X$(3872)$\rightarrow$$\jpsi$$\pi^+$$\pi^-$
is $\simeq$0.18 (see above). 
However, for the $\chi_{c1}'$ as a $P$-wave state,
radiative decays are expected to be dominant,
namely the ratio should be $\geq$40.
\end{itemize}

A notable implication for the potential model is,
that the $LS$ term is $-$2$<$1/$r^3$$>$ for the $1^{++}$ state, 
while it would be zero for a $2^{-+}$ state. 
If the $X$(3872) is the $\chi_{c1}'$, the decay to $\chi_{c1}$$\pi^+$$\pi^-$ should be observable. However, a search by Belle \cite{Bhardwaj:2015rju}
with 711 fb$^{-1}$ was negative. The upper limit on the product branching fraction 
${\cal B}$($B^+$$\rightarrow$$K^+$$X$(3872)) $\times$ ${\cal B}$($X$(3872)$\rightarrow$$\chi_{c1}'$$\pi^+$$\pi^-$) $<$
1.5 $\times$ 10$^{−6}$ (90\% C.L.)
is already a factor $\simeq$5$-$6 smaller than 
${\cal B}$($B^+$$\rightarrow$$K^+$$X$(3872)) $\times$ ${\cal B}$($X$(3872)$\rightarrow$$\jpsi$$\pi^+$$\pi^-$) =
(8.6$\pm$0.8)$\times$10$^{-6}$. 
The same argument would apply, if the $X$(3872) is not pure charmonium, but an admixture of a tetraquark or molecule
with identical quantum numbers, which is disfavored by the small measured upper limit.

An important test for future experiments would be, that 
there should be a second state nearby,
which should exhibit the same mass shift as the $X$(3872).
If the $X$(3872) is a $J^{PC}$=$1^{++}$, then it would be $n=2$($^3P_1$).
Then there must be the $h_c'$ ($2^1P_1$, $J^{PC}$=1$^{+-}$) nearby.\footnote{T. Burns private communication.}

\paragraph{Interpretation as a molecule}

As mentioned above in Sec.~\ref{cx3872_mass},
an intriguing feature of the $X$(3872) is that its measured mass is very
close to the sum of the masses of the $D^0$ and $D^{*0}$ mesons
with a mass difference $m(X)-m(D^{*0})-m(D^0)$ of $-0.01\pm$0.18~MeV.
This correspondence has led to considerable
speculation that the $X$(3872) {\color{revcolor} is a molecule-like bound state of a
$D^0$ and $\overline{D}^{*0}$, with an admixture of $D^\pm$ and $D^{\mp *}$ if one requires isospin $I$=0.}

A $J^{PC}$=1$^{++}$ quantum number assignment for the $X$(3872) implies that $S$-wave couplings
of the $X$ to $D^0$$\overline{D}^{*0}$ is permitted, and these result in a strong coupling
between the $X$ and the two mesons. This strong coupling can produce a bound
state with a molecular structure just below the two-particle threshold. 
There are quite a number of arguments in favour of the molecule
interpretation, as pointed out by \Tornqvist \cite{Tornqvist:2004qy}:

\begin{itemize}
\item The molecule states should be $J^{PC}$=$0^{-+}$ or $J^{PC}$=$1^{++}$. 
For other quantum numbers pion exchange is repulsive, or so weak that bound 
states are not expected.
\item No $D$$\overline{D}$ are expected since the three pseudoscalar coupling
vanishes because of parity.
\item If isospin were exact, the $X$(3872) as a molecule would be a pure isosinglet
with a mass very close to the $D$$\overline{D}^*$ threshold, consistent
with observation. For isovector states pion
exchange is generally one third weaker than for isoscalar states. 
\item For a state with small binding energy (for comparison, for the deuteron it is
$\Delta E_B=2.2$~MeV) the state should be large in spatial size. It should then
have a very narrow width since annihilation of this loosely bound $D\overline{D}^*$
state to other hadrons is expected to be small, although states containing
the $\jpsi$  are favoured compared to states with only light hadrons due to
the OZI rule.
\end{itemize}

Even the surprising isospin violation can be explained in the molecule picture; 
the $D$$\overline{D}^*$ molecule wave function is expected to contain
an admixture of $\rho$$\jpsi$ and $\omega$$\jpsi$. 
As in a meson-meson molecular state the long-range parts of the wave function
would naively be enhanced, the total width of the state could be larger. 
An interesting measurement would be, if the partial width of 
$\varGamma$($X$(3872)$\rightarrow$$\jpsi$$\pi^+$$\pi^-$) is of the order 
of $\simeq$40~keV (as is predicted for the $\chi_{c1}'$ charmonium state) \cite{chen_ma}
or $\simeq$202$-$237~keV (as predicted for for $D^{0*}$$\overline{D}^{0}$ molecular state) \cite{chen_ma}.
In any case the total width of a molecule must be larger 
than the width of its constituent $D^{*0}$, which is 82.3$\pm$1.2$\pm$1.4~keV \cite{babar_width_D0star}.

An idea has been proposed 
\cite{Braaten:2003he} to test the molecule hypothesis
by comparison of the $X$(3872) production yields
for $B^0$ and $B^+$ decays.
Based upon factorization, heavy-quark and isospin symmetries, 
it was predicted that the neutral/charged ratio has a value

\begin{equation}
\frac{{\cal B} ( B^0 \rightarrow K^0 {X}(3872) ) }
{{\cal B} ( B^+ \rightarrow K^+ {X}(3872) ) } \leq 0.1 \ .
\end{equation}

In a simplified picture, the reason is 
that the $B^0$ and the $K^0$ mesons contain $d$ quarks,
but the $B^+$, the $K^+$ and the $D^0$ and $\overline{D}^{*0}$
mesons contain $u$ quarks.
This ratio is expected to be unity for charmonium
as well as for hybrids ($ccg$) and glueballs ($gg$).
Belle measured the ratio \cite{x3872belle_charged_neutral} to 
0.82$\pm$0.22$\pm$0.05 
{\color{revcolor} which seems to contradict the prediction for molecules in \cite{Braaten:2003he},
which however was based on neutral molecular components only.
Inclusion of charged components change the prediction
(see \cite{Braaten:2007ft} for a detailed discussion).}

\paragraph{Interpretation as a tetraquark}

Following a suggestion of \cite{tetraquark_x3872} 
the $X$(3872) might represent a tetraquark, in particular 
in the form with two coloured di-quark pairs. 
As an indication in favour of the tetraquark interpretation,
a mass difference was observed in the two different decay channels
\begin{center}
\begin{tabular}{ll}
$m$($\jpsi$$\pi^+$$\pi^-$)=3871.2$\pm$0.5~MeV & \cite{pdg}\\
$m$($D^0$$\overline{D}^0$$\pi^0$)=3875.4$\pm$0.7$^{+1.2}_{-2.0}$~MeV & \cite{x3872ddpi_belle_1}\\
$m$($D^0$$\overline{D}^0$$\pi^0$)=3875.6$\pm$0.7$^{+1.4}_{-1.5}$~MeV & \cite{x3872ddpi_babar}\\
\end{tabular}
\end{center}
which could be regarded as an indication of the possible existence of two different
states $X$(3872) and $X(3876)$.
As pointed out in \cite{x3872_tetraquark_maiani},
in a tetraquark model these two states could be identified 
with $X$(3872)=$[cu][\overline{c}\overline{u}]$ and 
$X$(3876)=$[cd][\overline{c}\overline{d}]$.
In a different approach \cite{terasaki_tetraquark}, 
it was proposed to identify
the two states $X$(3872) and $X(3876)$ with the two opposite $G$-parity states
$X(3872)=([cq][\overline{c}\overline{q}]-[cq][\overline{c}\overline{q}])_{I=0}$ and 
$X(3876)=([cq][\overline{c}\overline{q}]+[cq][\overline{c}\overline{q}])_{I=0}$,
where $q$ represents a $u$ or a $d$ quark. 
The states $[$$c$$\overline{c}$$u$$\overline{u}$$]$ and $[$$c$$\overline{c}$$d$$\overline{d}$$]$ 
are contributing with identical weight.

If the $X$(3872) is a tetraquark, then also charged partners 
of the form $c$$\overline{c}$$u$$\overline{d}$ and 
$c$$\overline{c}$$d$$\overline{u}$ should possibly exist.
\BaBar performed a search in the decay
$B\rightarrow K X^{\pm}$, $X^{\pm}$$\rightarrow$$\jpsi$$\pi^{\pm}$$\pi^0$
\cite{x3872_charged_partner_babar} with 234$\times$$10^6$ $B$$\overline{B}$ events, 
however with a negative result.

{\color{revcolor} Searches for a $c$$\overline{c}$$s$$\overline{s}$ state
in the decay $B \to \jpsi \eta K$ 
have also been performed by \BaBar \cite{x3872_jpsieta_babar}
with 90$\times$$10^6$ $B$$\overline{B}$ events
and by Belle \cite{x3872_jpsieta_belle} with 772$\times$$10^6$ $B$$\overline{B}$ events.
With an $s$$\bar{s}$ component in the wave function of the $\eta$ meson,
the $\jpsi \eta$ final state provides sensitivity to $c$$\overline{c}$$s$$\overline{s}$
(and $c$$\overline{c}$$q$$\overline{q}$). 
Both searches yielded also a negative result, i.e.\ no $X$(3872) signal
was observed in this channel.}

\paragraph{Interpretation as a threshold effect}

The constituent quark model as e.g.\ used in the Cornell potential \cite{Eichten:1978tg}
assumes that the hadronic interaction in the final state 
is not a significant effect. However, there can be dynamics 
leading to a modified observed pole position of a resonance.
Examples are attraction or repulsion of two resonances 
or of a resonance and a threshold. A mechanism to generate
such a dynamics is the coupled channel approach.
As the $X$(3872) is thus close to the $D^0$$\overline{D}^{0*}$ threshold,
it was suggested that it is a Wigner-cusp \cite{Wigner:1948zz}. 
A quantitative
estimate \cite{cusp_bugg} shows that the $X$(3872) is too narrow to be a pure cusp,
i.e.,\ the upper limit of the width $\leq$1.2~MeV (see above) must be
compared to an expected width of the cusp of $\simeq$15~MeV.\footnote{D. Bugg private communication.}
However, it is possible that the cusp in the real part 
of the amplitude captures the resonance at the threshold \cite{cusp_bugg}, an effect which 
seems to be observed in the light meson sector as well, 
e.g.\ the $f_0$(980) at the $K$$K$ threshold, or the $f_2$(1565) 
at the $\omega$$\omega$ threshold.
The mixing of the $D^0$$\overline{D}^{0*}$ threshold and 
the resonance (decaying into $D^0$$\overline{D}^{0*}$) 
could be provided by coupling via loop diagrams. 

A test scenario was proposed by searching for the partner states.
 When a $c$$\overline{c}$
state mixes with $D^0$$\overline{D}^{0*}$, formally two eigenstates
are produced: one gets pulled down by the mixing, but the 
other is pushed up by level repulsion. The upper one is usually invisible
because of its large width. In the case of the $X$(3872), 
the second state could be the $Y(3940)$.
As the $X$(3872) and the $Y(3940)$ have a common decay channel, 
i.e.\ $\jpsi$$\omega$, the proposal is intriguing. 
A disadvantage is that coupling constants for the charmonium states 
to $D^0$$\overline{D}^{0*}$ are not at all well known.

A recent unquenched lattice QCD calculation
\cite{lattice_cc_new_regensburg} interestingly 
predicts a 1$^{++}$ molecule,
but no evidence for 1$^{-+}$ or 1$^{--}$ molecules.
However, the calculated binding energy is 88 MeV and thus too large.

As an additional note, the $X$(3872) has implications for the spectrum
of conventional charmonium states. 
Recently the $X(3820)$ \cite{D-wave_belle}
has been observed, probably representing the conventional $1^3D_2$ state
with $J^{PC}$=$2^{--}$. 
As $J^{PC}$=$2^{-+}$ is excluded for the X(3872),
it can be concluded that a yet unobserved $1^1D_2$ state 
with $J^{PC}$=$2^{-+}$ should be close to the X(3820) \footnote{T.~D.~Burns private communication.}, 
and possibly to be observed at future experiments. 


%% file: x3940.tex
\subsection{\boldmath{The $X(3940)$  state}}

\label{cx3940}

\begin{figure}[t!]
\centerline{\includegraphics[width=.9\textwidth]{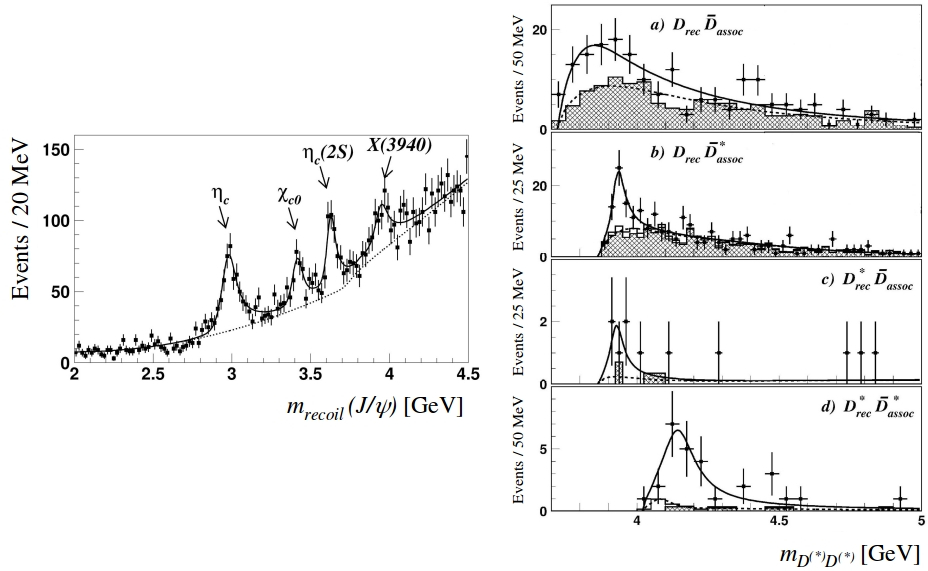}}
\caption{{\it Left:} Recoil mass for double charmonium production from Belle \cite{Abe:2007jna}.
The new state $X(3940)$  is visible, along with the known states
$\eta_c$(1$S$), the $\chi_{c0}$ and the $\eta_c'$(2$S$).
{\it Right:} $D$$\bar{D}$ {\it (a)},  
$D$$\bar{D}^*$ {\it (b)},
$D^*$$\bar{D}$ {\it (c)} and 
$D^*$$\bar{D}^*$ {\it (d)} invariant mass distributions
of mesons recoiling against a $\jpsi$ 
in double charmonium production from Belle \cite{x3940_2}.
In {\it (d)}, the new state $X(4140)$ is visible, 
which has a preferred charge conjugation assignment of $C$=$+$1.\label{fx3940}
}
\end{figure}

Another new state was observed by Belle in 
double charmonium production 
$e^+$$e^-$$\rightarrow$$c$$\bar{c}$$c$$\bar{c}$.
In the reconstruction of the final state a $\jpsi$ is found, and then the recoil mass 
against the $\jpsi$ is calculated.
In this particular production mechanism interestingly the $C$-parity of the recoil (possibly being an
$XYZ$ state) is fixed to $C$=+1. 
Fig.~\ref{fx3940} (left) shows the recoil mass for a data set 
of 350~fb$^{-1}$.
A new state was observed at a mass of $m$=3.940$\pm$0.011~GeV
with $N$=148$\pm$33~events, corresponding to a statistical significance 
of 4.5$\sigma$.
The other signals in the recoil mass spectrum can be attributed
to the conventional charmonium states $\eta_c$(1$S$), the $\chi_{c0}$, and the $\eta_c'$(2$S$).
It is remarkable that only $J$=0 states are observed.
As can be seen from Fig.~\ref{fx3940} (left), the width is surprisingly 
narrow, smaller or comparable to the resolution of 32~MeV.
The background curve is given by a second order polynomial
plus a threshold term for $D^{(*)}$$\overline{D}^{(*)}$$\jpsi$.
It is interesting to note that no signal of X(3872)
is observed.
As in addition to the new $X(3940)$, three known states with $J$=0 
are observed in the recoil mass spectrum, implying  that the new state might also have $J$=0.
There are three possible candidate charmonium states:

\begin{itemize}

\item The $\eta_c''$(3$S$) (3$^1$S$_0$, $J^{PC}$=0$^{-+}$)
has a $\simeq$100~MeV higher predicted mass of $m$=4064~MeV \cite{Barnes:2005pb}.
Predictions of static potential models should be more accurate
in particular for the case $S$=0 and $L$=0.

\item The $h_c´$ (2$^1$P$_1$, $J^{PC}$=1$^{+-}$) is expected 
around 3934-3956~MeV \cite{Barnes:2005pb}. 
However, states with negative $C$-parity (such as the lower $h_c$)
are not visible. 

\item The identification as the $\chi_{c0}'$ is not preferred either,
as the $\chi_{c2}'$ was observed at $\simeq$3930, 
and due to the spin-spin forces the $J=0$ state should be 
63 MeV lower than the $J$=2 state.

\end{itemize}

The tentative conclusion is that the $X(3940)$  is probably not a charmonium state.
An interesting additional question is, if the $X$(3940) and the $X$(3915) may be the same state, 
as the latter one is presently a strong candidate for the conventional charmonium state $\chi_{c0}'$. 
For the $X$(3915) an upper limit of 
${\cal B}(B\rightarrow X(3915)K)\times {\cal B}(X(3915)\rightarrow D^{*0}\overline{D}^0)<0.67\times$10$^{-4}$ at 90\% CL was measured \cite{x3872ddpi_belle_2}.
By averaging the branching fractions of
\cite{y3940_belle_2}~and~\cite{y3940_babar_1}, one obtains 
${\cal B}(B\rightarrow X(3915)K)\times {\cal B} (X(3915)\rightarrow \omega \jpsi)=(0.51\pm 0.11)\times 10^{-4}$.
Combining the two numbers, one gets the ratio
${\cal B}(X(3915)\rightarrow \omega\jpsi$) / 
${\cal B}(X(3915)\rightarrow D^{*0}\overline D^0)>0.71$ at\ 90\%\ CL. 
This must be compared with the 90\% CL limits from  
${\cal B}$($X$(3940)$\rightarrow$$\omega$$\jpsi$)$<$0.26 and ${\cal B}$($X$(3940)$\rightarrow$$D^{*0}$$\bar D^0$)$>$0.45,
and thus the ratio of the two branching ratios
 ${\cal B}$($X$(3940)$\rightarrow\omega$$\jpsi$) / 
${\cal B}$($X$(3940)$\rightarrow$$D^{*0}$$\bar D^0$)$<$0.58 at\ 90\%\ CL \cite{Abe:2007jna}.
This incompatibility suggests that the $X(3940)$ and the $X$(3915) are different states.

In a further study \cite{x3940_2}, 
the analysis was extended to additional
specification and identification of the recoil particles,
using a data set of 693~fb$^{-1}$.
The technique uses one fully reconstructed $D^{(*)}$ meson,
which is denoted as $D_{rec}$ in Fig.~\ref{fx3940} (right).
The $D^*$ mesons are reconstructed via their decays $D$$\pi$.
The $D$ mesons are reconstructed in the five decay channels 
$K^-$$\pi^+$, 
$K^-$$K^+$, 
$K^-$$\pi^-$$\pi^+$$\pi^+$, 
$K_s^0$($\rightarrow$$\pi^+$$\pi^-$)$\pi^+$$\pi^-$ and 
$K^-$$\pi^+$$\pi^0$($\rightarrow$$\gamma$$\gamma$).
The second $D^{(*)}$ meson is then identified by using the recoil 
mass against the $\jpsi$$D_{rec}$ system.
All of the observed final states seem to exhibit 
an $S$-wave enhancement. 

Fig.~\ref{fx3940} (right, (a)) shows the $D$$\bar{D}$ invariant mass
recoiling against a $\jpsi$ in double charmonium production.
The  $X(3880)$  resonance is visible, although only with a 3.8$\sigma$ statistical significance. 
As the fit is not stable under the variation of background parametrization,
no observation was claimed.
Fig.~\ref{fx3940} (right, (b) and (c)) show the $\bar{D}$$D^*$
invariant mass recoiling against a $\jpsi$ in double charmonium production,
in the two different cases of either the $\bar{D}$ or the $D^*$ side  reconstructed.
The $X(3940)$  (mentioned above)
is visible with a statistical significance of 6.0$\sigma$.
The mass and the width were determined as 3942$^{+7}_{-6}$$\pm$6~MeV and 37$^{+26}_{-15}$$\pm$8~MeV, respectively. 
Note that the $X(3940)$  is only observed here in 
$D^*$$\bar{D}$ and not in $D$$\bar{D}$, giving preliminary preference
to an assignment of $J$=1 (because of the $D^*$) instead of $J$=0.
In addition, a decay to 
$\jpsi$$\omega$ was checked and not observed, importantly indicating 
that the $X(3940)$  and the $Y(3940)$ (see Sec.~\ref{cx3940}) are not the same.

Fig.~\ref{fx3940} (right, (d)) shows the $D^*$$\bar{D}^*$ invariant mass 
recoiling against a $\jpsi$ in double charmonium production.
Another state, tentatively called $X(4140)$, is visible with 
a statistical significance of 5.5$\sigma$ ($N$=24$\pm$5~events).
From the fit, a mass of 4156$^{+25}_{-20}$$\pm$15~MeV and a width of 139$^{+111}_{-61}$$\pm$21~MeV
were determined.
A calculation of $D^*$$\bar{D}^*$ molecules predicts 
a tensor state $J^{PC}$=$2^{++}$ state at this mass.\footnote{E. Oset private communication.}
This hypothesis could be tested by searching for
vector vector decays such as $K^*$$\overline{K}^*$, 
$\omega$$\omega$ or $\phi$$\phi$ which are predicted
to have a significant branching,
although these decays would be SU(3) violating.

Concerning the quantum numbers of the new states, 
in the case of no other final state particle
(i.e.\ exclusive $\jpsi D^* \bar{D}^*$), 
the charge conjugation $C$=$+$1 value is preferred. 
$C$=$+$1 charmonium states are interesting, as they
cannot annihilate to a virtual photon. Thus
they cannot decay to $e^+$$e^-$ or $\mu^+$$\mu^-$,
but only to $\gamma$$\gamma$ or two gluons. 
Interestingly, the decay widths such as 
for the decay $^3$P$_0$$\rightarrow$$\gamma$$\gamma$
are proportional not to the squared wave function 
at the origin, but to the squared {\it derivative}
of the wave function
$|$$\partial$$\psi$/$\partial$$r$($r$=0)$|^2$.
Thus $C$=$+$1 states represent 
a precise tool to study the exact wave function
behavior for $r$$\rightarrow$0.

The relevant threshold for the new state is given by
$m(D^{*+})+m(D^{*-})\simeq$4020~MeV.
Thus, the new state has a positive mass difference 
of $\Delta$$m$$\simeq$120~MeV, or, in other words,
the binding energy has the wrong sign.
The dynamics, which could create a state 
with such a large $\Delta$$m$ is not clear yet.
The two new states $X(3940)$ and $X(4140)$ have not\footnote{\color{revcolor} Note that the $X(4140)$ seen in the $D^*$$\bar{D}^*$ final state,
  with the name assigned by Belle, and the $X(4140)$ in the $\jpsi \phi$ final state (see Sec.~\ref{4140}),
  with the name assigned by the Particle Data Group, are not necessarily
  identical.} been confirmed yet by any other experiment in the same decay modes. 


The observed states can be compared to predictions for molecular 
states based upon a one-pion exchange model \cite{Tornqvist:2004qy}
which are listed in Table~\ref{tDDtornqvist}. The observed masses are
higher than the predicted masses, e.g.\ by $\simeq$125~MeV for the $X(4140)$
if compared to the hypothetical 0$^{++}$ state with positive charge conjugation.
As the difference is positive, it can not be interpreted as a binding energy.
Instead, if the molecular hypothesis is correct, the states would be virtual
states above threshold. 

\begin{table}
\centering
\caption{Possible molecular $D^{(*)}$$\overline{D}^{(*)}$ states as predicted in Ref. \cite{Tornqvist:2004qy}.}
\vspace{0.2cm}
\begin{tabular}{lcc}
\hline\\[-2.5ex]
Constituents & $J^{PC}$ & Mass (MeV) \\
\hline\\[-2.5ex]
$D$$\overline{D}^*$ & 0$^{-+}$ & $\simeq$3870 \\
$D$$\overline{D}^*$ & 1$^{++}$ & $\simeq$3870 \\
\hline
$D^*$$\overline{D}^*$ & 0$^{++}$ & $\simeq$4015 \\
$D^*$$\overline{D}^*$ & 0$^{-+}$ & $\simeq$4015 \\
$D^*$$\overline{D}^*$ & 1$^{+-}$ & $\simeq$4015 \\
$D^*$$\overline{D}^*$ & 2$^{++}$ & $\simeq$4015 \\
\hline
\end{tabular}
\label{tDDtornqvist}
\end{table}


%% file: y4260-rev.tex
\subsection{\boldmath{$Y$ states}}
\label{cy4260}

Another new charmonium-like state was discovered by \BaBar \cite{y4260_babar_1} with a data set of 211 fb$^{-1}$.
The production process is given by initial state radiation 
$e^+$$e^-$$\rightarrow$$\gamma_{ISR}$$\jpsi$$\pi^+$$\pi^-$.
A photon is radiated by either the $e^+$ or the $e^-$ in the initial
state, lowering $\sqrt{s}$ and producing the $Y$(4260).§
The state was observed in the invariant mass $m(\jpsi\pi^+\pi^-)$.
Based upon this production process, in which $e^+e^-\to \gamma_{ISR}\gamma_V$, $\gamma_V\to Y(4260)$,
the quantum numbers are $J^{PC}$=$1^{--}$. The tentative name $Y$ was assigned
\footnote{\color{revcolor}Although the Particle Data Group \cite{pdg} changed the names of all  the vector states discussed in this section  from $Y$ to $X$, we keep the historic nomenclature.}. 
which is typical for vector states and resembles similarity to $\Upsilon$ or $\psi$.
The mass was measured  as
4259$\pm$8$^{+2}_{-6}$~MeV, $\simeq$500~MeV above the 
$D$$\overline{D}$ threshold.
The width was determined as 88$\pm$23$^{+6}_{-4}$~MeV.
Thus, the state is significantly broader than the $X$(3872),
but still surprisingly narrow considering its high mass. 
The state was confirmed by Belle with a data set of 553.2~fb$^{-1}$ \cite{y4260_belle_1}.
However, the mass was measured to be $\simeq$2.5$\sigma$ higher than 
the mass measured by \BaBar$\!$, and the width $\simeq$50\% 
wider than in the case of \BaBar$\!$.
Later both experiments updated their results \cite{y4260_belle_1,y4260_babar_2,y4260_babar_3}.

\begin{figure}[t!]
\centerline{\includegraphics[width=0.8\textwidth]{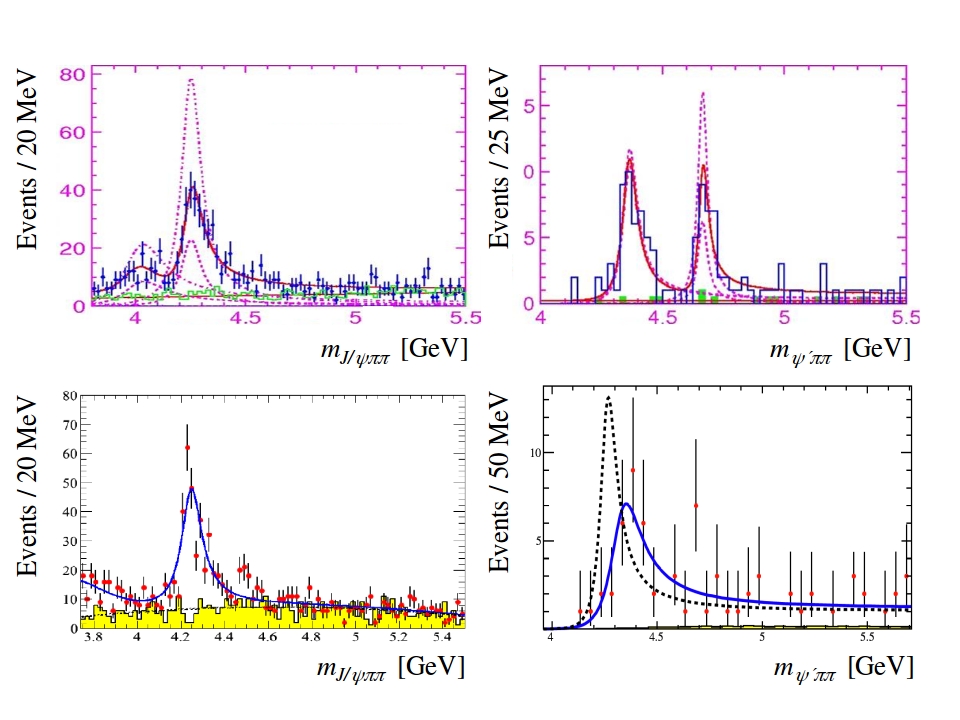}}
\caption{Summary of the $Y$ states observed in ISR.
  {Top left:} Analysis of $e^+$$e^-$$\rightarrow$$\gamma_{ISR}$$\jpsi$$\pi^+$$\pi^-$ with a data set of 550 fb$^{-1}$
  from Belle \cite{y4260_belle_2}. 
  {Top right:} Analysis of $e^+$$e^-$$\rightarrow$$\gamma_{ISR}$$\psi'$$\pi^+$$\pi^-$ with a data set of 670 fb$^{-1}$
  from Belle \cite{y4350_belle}. 
  {Bottom left:} Analysis of $e^+$$e^-$$\rightarrow$$\gamma_{ISR}$$J$$\psi$$\pi^+$$\pi^-$ with a data set of 454 fb$^{-1}$
  from \BaBar \cite{y4260_babar_2}. 
  {Bottom right:} Analysis of $e^+$$e^-$$\rightarrow$$\gamma_{ISR}$$\psi'$$\pi^+$$\pi^-$ with a data set of 298 fb$^{-1}$
  from \BaBar \cite{y4350_babar}.
  Dotted and dash-dotted lines indicate solutions of the fits including and excluding interference.
\label{fy4260} 
}
\end{figure}

Figure~\ref{fy4260} shows $Y$(4260) signals for Belle and \BaBar .
As can be seen in the invariant mass $m$($\jpsi$$\pi^+$$\pi^-$) spectra,
the background shapes are somewhat different
between the experiments.
The reason is the different design of the interaction
regions in the two experiments, namely \BaBar had collisions 
head-on, which required a dipole magnet very close to the
interaction point generating additional background. 
Belle used a steering angle between both beams,
not requiring magnets close to the interaction point.
As can also be seen in Fig.~\ref{fy4260},
the lineshape of the $Y$(4260) shows a long range tail at high masses.
This particular lineshape results not from radiative effects, but from a strong 
dependence of the reconstruction efficiency as a function of the 
invariant mass $m$($\jpsi$$\pi^+$$\pi^-$).  
The fit function applied in e.g.\ \cite{y4260_belle_1} is a Breit-Wigner function folded 
with a phase space\footnote{Note that the energy dependent width 
$\Gamma (s)$ corresponds to a phase space increase,
however the $Y$(4260) is narrow, so $\Gamma$ is
chosen independent of $s$. Thus, in the Belle analyses of the Y(4260) \cite{y4260_belle_1}
an additional phase space term is 
applied.} term and efficiency $\varepsilon$, which is parametrised as 
$\varepsilon$=$a$($m$-$m_0$)+$b$ with $a$=7.4$\pm$1.3~GeV$^{-1}$
and $b$=9.31$\pm$0.07, i.e.,\ the efficiency changes by a factor $\simeq$2 
over the peak region.

In an independent additional measurement, 
the $Y$(4260) was confirmed by CLEO-c \cite{y4260_cleo-c} and by BESIII \cite{Ablikim:2016qzw}
using a scan technique, i.e.\ by variation of the beam energies.
{\color{revcolor} Surprisingly, BESIII reports a mass which is $\simeq$25~MeV lower and a width which is a factor $\simeq$2 narrower
than the average obtained from all other experiments \cite{pdg}.} 
The measurements of all experiments are summarized in Table~\ref{ty4260}.
\begin{table}
\centering
\caption{Summary of the mass and width measurements of the $Y$(4260).  The measured masses $m$, and widths, $\Gamma$, are given in units of MeV; ${\cal S}$ denotes the significance.}
\vspace{0.2cm}
\begin{tabular}{llllllll}
\hline\\[-2.5ex]
 & 
\BaBar & 
CLEO-c & 
Belle & 
Belle &
\BaBar &
\BaBar &
{\color{revcolor} BESIII} \\
 & 
\cite{y4260_babar_1} & 
\cite{y4260_cleo-c} & 
\cite{y4260_belle_1} & 
\cite{y4260_belle_2}  &
\cite{y4260_babar_2} &
\cite{y4260_babar_3} &
\cite{Ablikim:2016qzw}\\
\hline\\[-2.5ex]
${\cal L}$ &
211~fb$^{-1}$ &
13.3~fb$^{-1}$ &
553~fb$^{-1}$ &
548~fb$^{-1}$ &
454~fb$^{-1}$ &
454~fb$^{-1}$ &
{\color{revcolor} 9~fb$^{-1}$} \\
N & 
125$\pm$23 & 
14.1$^{+5.2}_{-4.2}$ & 
165$\pm$24 & 
324$\pm$21 &
344$\pm$39 &
$-$ &
{\color{revcolor} 3853$\pm$68}\\
${\cal S}$ & 
$\simeq$8$\sigma$ & 
$\simeq$4.9$\sigma$ & 
$\geq$7$\sigma$ & 
$\geq$15$\sigma$ &
$-$ &
$-$ &
{\color{revcolor} 7.6$\sigma$} \\ 
$m$  & 
4259$\pm$8$^{+2}_{-6}$ & 
4283$^{+17}_{-16}$$\pm$4 & 
4295$\pm$10$^{+10}_{-3}$ & 
4247$\pm$12$^{+17}_{-32}$ &
4252$\pm$6$^{+2}_{-3}$ &
4244$\pm$5$\pm$4 &
{\color{revcolor} 4222.0$\pm$3.1$\pm$1.4}\\
$\Gamma$ & 
88$\pm$23$^{+6}_{-4}$ & 
70$^{+40}_{-25}$ & 
133$\pm$26$^{+13}_{-6}$ & 
108$\pm$19$\pm$10 &
105$\pm$18$^{+4}_{-6}$ &
114$^{+16}_{-15}$$\pm$7 &
{\color{revcolor} 44.1$\pm$4.3$\pm$2.0}\\
\hline
\end{tabular}
\label{ty4260}
\end{table}
CLEO-c also reported a 5.1$\sigma$ evidence for a second decay channel
$Y$(4260)$\rightarrow$$\jpsi$$\pi^0$$\pi^0$ \cite{y4260_cleo_kk},
which was confirmed by Belle \cite{y4260_belle_pi0pi0}.
The ratio of the two branching fractions was determined as\\
${\cal B}(\jpsi\pi^0\pi^0)/{\cal B}(\jpsi\pi^+\pi^-)\simeq 0.5$.
This leads to the important conclusion, based upon the Clebsch-Gordan coefficients 
in isospin space, that the isospin of the $\pi$$\pi$ system must be zero, 
i.e., $I^G=0^+$.
By symmetry arguments, this leads to $J^{PC}$=0$^{++}$ or 2$^{++}$
for the $\pi$$\pi$ system. 
\BaBar investigated the $\pi^+$$\pi^-$ invariant mass distribution \cite{y4260_babar_1}.
The observed angular distribution 
is consistent with J$^{PC}$=0$^{++}$, i.e.\ consistent with $\pi^+$$\pi^-$ phase space 
for $S$-wave $\jpsi$$\pi^+$$\pi^-$.
Note, in particular, that this is completely different from the $X(3872)$, 
where the $\pi^+$$\pi^-$ system is in a J$^{PC}$=1$^{--}$ state with $I$=1. 
The $Y$(4260) does not violate isospin in the decay.

In addition, CLEO-c reported a 3.7$\sigma$ evidence for a third decay channel, 
namely $Y$(4260)$\rightarrow$$\jpsi$$K^+$$K^-$ \cite{y4260_cleo_kk},
which unfortunately could not be confirmed by Belle \cite{y4260_belle_kk}.
Searches were also performed for the $Y$(4260) 
in $B$ meson decays \cite{y4260_babar_4}, however 
the phase space is small and the results are compatible with 
statistical fluctuations.

The $\pi^+$$\pi^-$ mass distribution in $Y$(4260)$\rightarrow$$J$$\psi$$\pi^+$$\pi^-$
exhibits an $f_0$(980) signal \cite{y4260_babar_3} with $J^{PC}$=0$^{++}$ (positive parity).
This is clearly different from 
the $X(3872)$, in which the $\pi^+$$\pi^-$ system exhibits
a $\rho$ signal (see Sec.~\ref{cx3872}) 
with a $J^{PC}$=1$^{--}$ (negative parity),
and is a consequence of the different quantum numbers 
of the $Y$(4260) (negative parity) and the $X(3872)$ (positive parity)
under the assumption of $S$-wave decays. 

{\color{revcolor} Recently, BESIII reported a state decaying into $h_c \pi^+ \pi^-$ \cite{BESIII:2016adj},
  which may be identical to the $Y$(4260).
Since the $\jpsi$ in $\jpsi \pi^+\pi^-$ is $S$=1 and the $h_c$ in $h_c \pi^+ \pi^-$ is $S$=0,
the latter decay involves a spin-flip of the heavy quark system.
Resonance parameters of this state were measured with the mass being 4218.4$^{+5.5}_{-4.5}$$\pm$0.9~MeV, about 30 MeV lower
than the average mass of 4251$\pm$9 MeV of the $Y(4260)$ obtained from all other experiments \cite{pdg}, 
and the width being 66.0$^{+12.3}_{-8.3}$$\pm$0.4~MeV, a factor $\simeq$2
narrower than the present PDG average width of the $Y(4260)$ of 120$\pm$12 MeV \cite{pdg}.
Puzzlingly,  both the mass and the width are consistent with the recent measurements
of the $Y$(4260) parameters by BESIII \cite{Ablikim:2016qzw}. This may point to be a systematic discrepancy
between resonance parameters extracted from the ISR measurements at \BaBar and Belle and
from the scan in $e^+$$e^-$ direct production at BESIII.}

Belle reported a second state with $\simeq$250 MeV lower mass: the $Y$(4008) \cite{y4260_belle_1}.
The width is reported to be about a factor $\simeq$2 larger
than that of the $Y$(4260).
{\color{revcolor} Supporting evidence for this state was reported later by BESIII \cite{Ablikim:2016qzw},
  however multiple stable fit solutions were found. In one particular solution,
  the product of the coupling to $e^+$$e^-$ and the branching fraction
  $\Gamma_{e^+e^-}$ $\times$ ${\cal B}$($Y(4008) \to \jpsi \pi^+ \pi^-$) is a factor 3.5 higher
  than for the $Y$(4260). However, a fit with an exponential instead of a Breit-Wigner parametrization
  was found to describe the data equally well. Thus, presently a firm conclusion about the existence of the $Y$(4008) is difficult to make.}
Another important result of the Belle analysis is that
no evidence for any higher mass state up to 7 GeV was found. 
There has also been the attempt to quantify the contribution
of higher $\psi$ resonances ($\psi$(3D), $\psi$(5$S$), $\psi$(4$D$), $\psi$(6$S$) and $\psi$(5$D$))
to the $\jpsi$$\pi^+º$$\pi^-$ spectrum \cite{beveren_higher_psi}.

In 2007 initial state radiation processes with a $\psi'$ 
instead of a $\jpsi$ were investigated by \BaBar$\!$, and 
another new state was found in the $\psi'$$\pi^+$$\pi^-$
invariant mass (see Fig.~\ref{fy4260})
in a data sample of 298 fb$^{-1}$.
The new resonance, tentatively called $Y$(4350)\footnote{\color{revcolor} Historically,
  the name $Y$(4350) originates from a fit with one resonance only. Later,
the name was changed to $Y$(4360), as fits with two resonances and interference yielded a slightly higher mass (see below).}, has 
a  peak cross section of $\simeq$80~pb$^{-1}$,  almost as large as the $Y$(4260). However,
the peak position at 4324$\pm$24~MeV is significantly 
different from the mass of the $Y$(4260)
The width was measured to be 172$\pm$33~MeV.
Belle was able to confirm the $Y$(4350) \cite{y4350_belle}
(Fig.~\ref{fy4260}).
The mass and the width measurements are summarized in Table~\ref{ty4350}.
Note that the experimentally measured widths differ by a factor $\simeq$2.

{\color{revcolor}
Recently a state $Y$(4320) was observed by BESIII \cite{Ablikim:2016qzw} with parameters very similar to the $Y$(4350), 
but decaying into $\jpsi \pi^+ \pi^-$ instead of $\psi' \pi^+ \pi^-$.
The measured mass of 4320.0$\pm$10.4$\pm$7.0~MeV is consistent with the measurement of \BaBar (Table~\ref{ty4350}), 
the width of 101.41$^{+25.3}_{19.7}$$\pm$10.2~MeV is consistent with the measurement of Belle (Table~\ref{ty4350}).
However, one should keep in mind, that the $\jpsi \pi^+ \pi^-$ final state would then be common for the $Y$(4260) and
for that $Y$(4350) candidate state. For the latter one, sitting on the tail of the $Y$(4260), the  interference can be significant,
leading to difficulties in extracting the resonance parameters. 
In the BESIII analysis, the $Y$(4008), the $Y$(4260) and the $Y$(4320) were fitted simultaneously,
and four solutions with very different extracted yields were found, all describing the data well.
The present world average for the couplings are 
$\Gamma_{e^+e^-}$$\times$${\cal B}$($Y(4260) \to \jpsi \pi^+ \pi^-$)=9.2$\pm$1.0~eV \cite{pdg} and
$\Gamma_{e^+e^-}$$\times$${\cal B}$($Y(4350) \to \psi' \pi^+ \pi^-$)=9.15$\pm$3.15~eV \cite{pdg}.
In the BESIII fits, the possible range is determined as 
1.1$\leq$$\Gamma_{e^+e^-}$$\times$${\cal B}$($Y(4320) \to \jpsi \pi^+ \pi^-$)$\leq$21.1~eV,
making it difficult to draw a solid conclusion.
In fact, the fit is driven by only one single data point, providing a statistical significance of 7.6$\sigma$. 
Assuming the most conservative number of 1.1~eV would imply a suppression of the $Y$(4350)
in $\jpsi \pi^+ \pi^-$ by a factor 5.5$-$11.1 compared to $\psi' \pi^+ \pi^-$.
In any case, the coupling can be considered small, as they are about two orders of magnitude smaller than 
$\Gamma_{e^+e^-}$$\times$${\cal B}$($\psi' \to \jpsi \pi^+ \pi^-$)=807.1$\pm$20.8~eV \cite{pdg}.
There are also preliminary results by BESIII on $e^+ e^- \to \pi^+ D^0 D^{*-} + c.c.$ \cite{talk_changzheng_2017}
indicating two structures. While the $Y$(4260) parameters are consistent with the parameters extracted from
$\jpsi \pi^+ \pi^-$, the structure at the $Y$(4320) seems to have a factor $\simeq$2 larger width,
thus indicating that the $Y$(4350) and the $Y$(4320) are probably two different states. 
Signals may have also been observed in $e^+ e^- \to \chi_{c0} \omega$ \cite{Ablikim:2015uix},
however with limited statistics. Born cross sections are smaller by factors 3$-$6 compared to $\jpsi \pi^+ \pi^-$. 
For an attempted combined fit involving all observed final states see \cite{Gao:2017sqa}.
}

\begin{table}
\centering
\caption{Summary of the mass and width measurements of the $Y$(4350).}
\vspace{0.2cm}
\begin{tabular}{llll}
\hline\\[-2.5ex]
 &
Integrated luminosity &
$m$ (MeV) &
$\Gamma$ (MeV) \\
\hline\\[-2.5ex]
\BaBar \cite{y4350_babar} &
298~fb$^{-1}$ &
4324$\pm$24 &
172$\pm$33\\
Belle \cite{y4350_belle} &
673~fb$^{-1}$ &
4361$\pm$9$\pm$9 &
74$\pm$15$\pm$10\\
{\color{revcolor} \BaBar \cite{Lees:2012pv}} &
{\color{revcolor} 520~fb$^{-1}$} &
{\color{revcolor} 4340$\pm$16$\pm$9} &
{\color{revcolor} 94$\pm$32$\pm$13}\\
{\color{revcolor} Belle \cite{Wang:2014hta}} &
{\color{revcolor} 980~fb$^{-1}$} &
{\color{revcolor} 4347$\pm$6$\pm$3} & 
{\color{revcolor} 103$\pm$9$\pm$5}\\
\hline
\end{tabular}
\label{ty4350}
\end{table}

In the analysis of the $Y$(4350) Belle used a factor $\simeq$3 larger data sample
of 670 fb$^{-1}$, which also revealed evidence for another
new state $Y$(4660) with a statistical significance of 5.8$\sigma$.
The measured mass and width are 4664$\pm$11$\pm$5~MeV and
48$\pm$15$\pm$3~MeV, respectively.
The $Y$(4660) is the heavyist charmonium-like state ever observed,
with a mass of $\simeq$1.5~GeV above the charmonium ground state.
{\color{revcolor} Later, measurements were updated with full data sets by both \BaBar \cite{Lees:2012pv} and Belle \cite{Wang:2014hta}.}

\begin{figure}[htb]
\centerline{\includegraphics[width=0.75\textwidth]{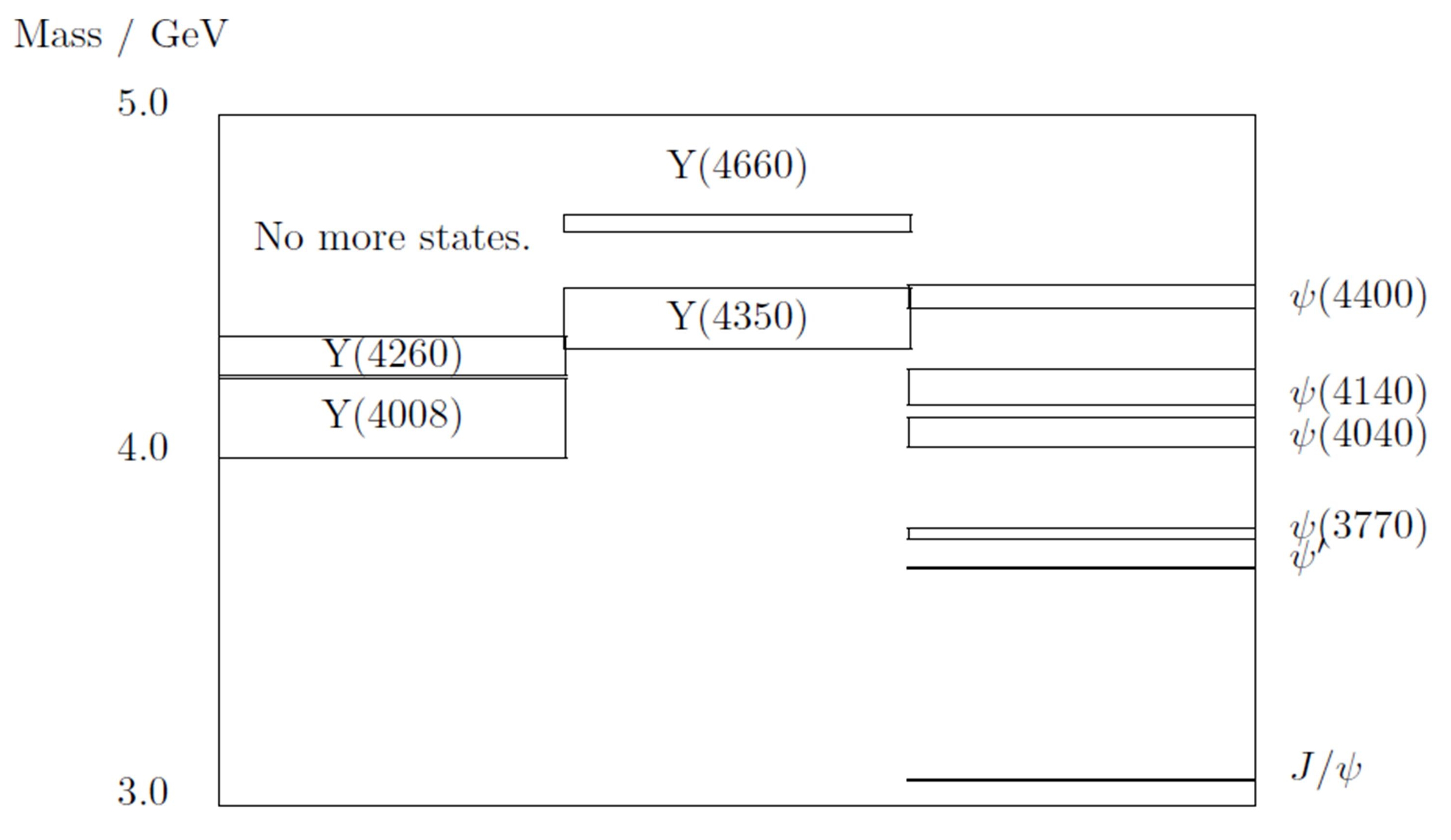}}
\caption{Mass level scheme of the observed $Y$ states 
in comparison to the known excited $\psi$($n$$S$) states
All of them carry the quantum numbers $J^P$=1$^{--}$, 
but due to the experimental observation they are separated in three different columns:
states observed in ISR which decay into $\jpsi$$\pi^+$$\pi^-$ (left), 
states observed in ISR which decay into $\psi'$$\pi^+$$\pi^-$ (center) and the
known $\psi$ states (right). The vertical sizes of the boxes indicate
their widths. \label{fy4260_term}}
\end{figure}

\paragraph{Interpretation as a charmonium state}

All the $Y$ states have to carry the quantum numbers
$J^{PC}$=$1^{--}$, due to their observation in an ISR process.
There are five known and assigned $J^P$=$1^{--}$ charmonium states: $\jpsi$, $\psi$(2$S$), $\psi$(4040), $\psi$(4160) and $\psi$(4415). As we will see there is a clear overpopulation of 1$^{--}$ states 
in the $m$$\geq$4~GeV region. Somewhat surprisingly, although they are partially overlapping due to their widths,  
apparently there seems to be no mixing: 

\begin{itemize}

\item the $Y$(4008) candidate state and the $Y$(4260) decay to $\jpsi$$\pi^+$$\pi^-$, 
the $Y$(4350) and the $Y$(4660) decay to $\psi'$$\pi^+$$\pi^-$. 
{\color{revcolor} In the ISR measurements, neither of one has been observed in the other channel yet. 
In the scan experiments by BESIII, there seems to be preliminary indication of the $Y$(4350)
in $\jpsi \pi^+ \pi^-$ \cite{Ablikim:2016qzw}.
However, as explained above, interference can be significant and leads to multiple fit solutions.
In any case, a suppression by a factor 5.5$-$11.1 compared to $\psi' \pi^+ \pi^-$
can be calculated, using the most conservative result of the fit.} 

\item no mixing of the  $\psi$ states with the $Y$ states has been observed so far. 

\end{itemize}

On the other hand, there is evidence of destructive interference
from the analysis of the known $\psi$ states 
in the region 4.2$\leq$$m$$\leq$4.4~GeV \cite{bes_psi404041604415}. 
In particular, the $e^+e^-$ cross section shows a local minimum 
in the $Y$(4260) mass region, pointing to 
destructive interference of the $\psi$ states 
with the $Y$(4260).
The pattern of the $Y$ states appears non-trivial (see Fig.~\ref{fy4260_term}): 
two non-mixing doublets without parity flip and without charge flip. 
It remains unclear what the underlying symmetry is.
In addition, there is no obvious pattern so far, as to
how the masses of the $\psi$ states 
and the masses of the $Y$ states might be related.

Intriguingly, the $Y(4260)$ state is again close to a threshold \cite{swanson_new_mesons}.
The $\overline{D}^0$$D_1$(2420) threshold is at 4285.63~MeV
with a summed narrow width of the constituent mesons of $\simeq$27~MeV,
and thus the $Y$(4260) has been discussed as a molecule \cite{y4260_hanhart}.
The quantum numbers of $J^P$=0$^-$ and $J^P$=1$^+$ can be combined correctly 
in a relative $S$-wave to the observed $J^P$=1$^-$. 
{\color{revcolor} The threshold would be located in the tail of the $Y$(4260), 
but the only indication of a modified line shape so far comes from BESIII \cite{Ablikim:2016qzw}.}
A second nearby threshold is $D^{0}$$\overline{D}_2^{*0}$(2460) 
at 4327.5~MeV with a summed width of the constituent mesons of $\simeq$49~MeV.
However, a molecule would not be an option for the $Y$(4260)
due to the $J^P$=$2^+$ of the $\overline{D}_2^{*0}$(2460), i.e.\ 
an $S$-wave coupling would lead to $J^P$=$2^-$, but not $J^P$=$1^-$.
{\color{revcolor} The $D_s^*$$\overline{D_s^*}$ threshold is located at 4224.2~MeV, and thus very close to the recent measurement
of the mass of the $Y$(4260) by BESIII (see Table~\ref{ty4260}). However, except the marginal evidence
for a decay to $\jpsi K^+ K^-$ by CLEO-c \cite{y4260_cleo_kk}, no decays involving strangeness have been found.}

{\color{revcolor} BESIII observed the final state $\gamma$$X$(3872) at $\sqrt{s}$=4.26~GeV \cite{Ablikim:2013dyn}
with a significance of 6.3$\sigma$.
Although this is not conclusive evidence that the $Y$(4260) is the initial state, 
there is evidence of a peaking behavior of the cross section at this center of mass energy.
The measurement is based upon only 20.0$\pm$4.6 events, however the implication is strong:
this seems to represent a transition between the $Y$ and the $X$,
thus pointing the a common nature of these states. 
Interestingly, the branching fraction ${\cal B}$($Y(4260) \to \gamma X(3872)$) is a factor $\leq$50 higher
than for an E1 charmonium transition, assuming typical scaling behavior $\propto$$E_{\gamma}^{2-3}$ for pure
charmonium states. When interpreting the $Y$(4260) as a $D$$\overline{D}_1$ molecule and the $X$(3872)
as a $D$$\overline{D}^*$ molecule, this branching fraction was predicted to be enhanced \cite{Guo:2013nza}
due to $D_1 \to D^* \gamma$ transitions inside the molecule, while keeping the molecule intact.}

The $Y$(4260) has also been discussed in literature as a hybrid $[c\overline{c}_8g]$
with a color octet $c$$\overline{c}$ pair bound to a valence type gluon.
However, recently there is evidence that the $Y$(4260) decays also to
$h_c \pi^+ \pi^-$ \cite{BESIII:2016adj},
which would imply a spin flip of the heavy quark system.
If this decay is confirmed by another measurement,
an interpretation of the $Y$(4260) as a hybrid would be strongly disfavoured.  As discussed in the section on
theoretical models, a similar pattern is observed in the decays $\Upsilon(10860) \to \Upsilon(nS) \pi^+ \pi^-$, 
$nS=1S,2S,3S$, 
and $\Upsilon(10860) \to (h_b(1P), h_b(2P)) \pi^+ \pi^-$, which are dominated by the intermediate $Z_b(10650)^+ \pi^-$ and
$Z_b(10610)^+ \pi^-$ states. In the tetraquark interpretation, the two $Z_b$ states have both $b\bar{b}$ (spin-0) and
$b\bar{b}$ (spin-1) components in their Fock space. Presumably, the same phenomenon is at work in the decays of
$Y(4260)$.

\begin{figure}[t!]
\centerline{\includegraphics[width=0.48\textwidth]{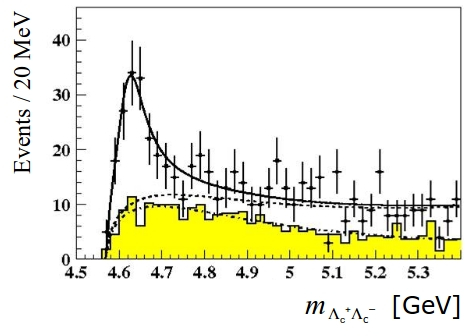}}
\vspace*{-4mm}
\caption{Invariant mass $m$($\Lambda_c^+$$\Lambda_c^-$)
for the process 
$e^+$$e^-$$\rightarrow$$\gamma_{ISR}$$\Lambda_c^+$$\Lambda_c^-$
from Belle \cite{x4630_belle} showing the signal for the $X$(4630).
\label{fx4630}}
\end{figure}

A state probably identical to the $Y$(4660) 
has also been observed at Belle \cite{x4630_belle} in the ISR process
using a data set of 670~fb$^{-1}$,
but in a different decay channel, i.e.\ the signal was observed in
$e^+$$e^-$$\rightarrow$$\gamma_{ISR}$$\Lambda_c^+$$\Lambda_c^-$. 
The state is usually referred to as the $X$(4630).
The $\Lambda_c^+$ is reconstructed in the final states
$p$$K^0_s$($\rightarrow$$\pi^+$$\pi^-$),
$p$$K^-$$\pi^+$, and 
$\Lambda$($\rightarrow$$p$$\pi^-$)$\pi^+$.
For the $\Lambda_c^-$ only partial reconstruction is used: 
The recoil mass to [$\Lambda_c^+$$\gamma$] is investigated while requiring
an anti-proton (from the $\Lambda_c^-$ decay) as a tag and then a cut
around the $\Lambda_c^-$ mass is applied. 
The measured mass is 4634$^{+8}_{-7}$$^{+5}_{-8}$~MeV and 
the measured width is 92$^{+40}_{-24}$$^{+10}_{-21}$~MeV.
Figure~\ref{fx4630} shows the invariant mass 
$m$($\Lambda_c^+$$\Lambda_c^-$).
A signal with a statistical significance 
of 8.2$\sigma$ is observed.
The observation of this state is remarkable because of 
two reasons: it represents the highest charmonium-like state observed so far 
(along with the $Y$(4660) of almost same mass, but decaying into $\jpsi$$\pi^+$$\pi^-$, and 
it is the only $XYZ$ state so far observed decaying into baryons.

%% file: z4430.tex

\subsection{\boldmath{${Z}$ states of type I in the charmonium mass regime}}

\label{cz4430}

\begin{figure}[H]
\centerline{\includegraphics[width=0.72\textwidth]{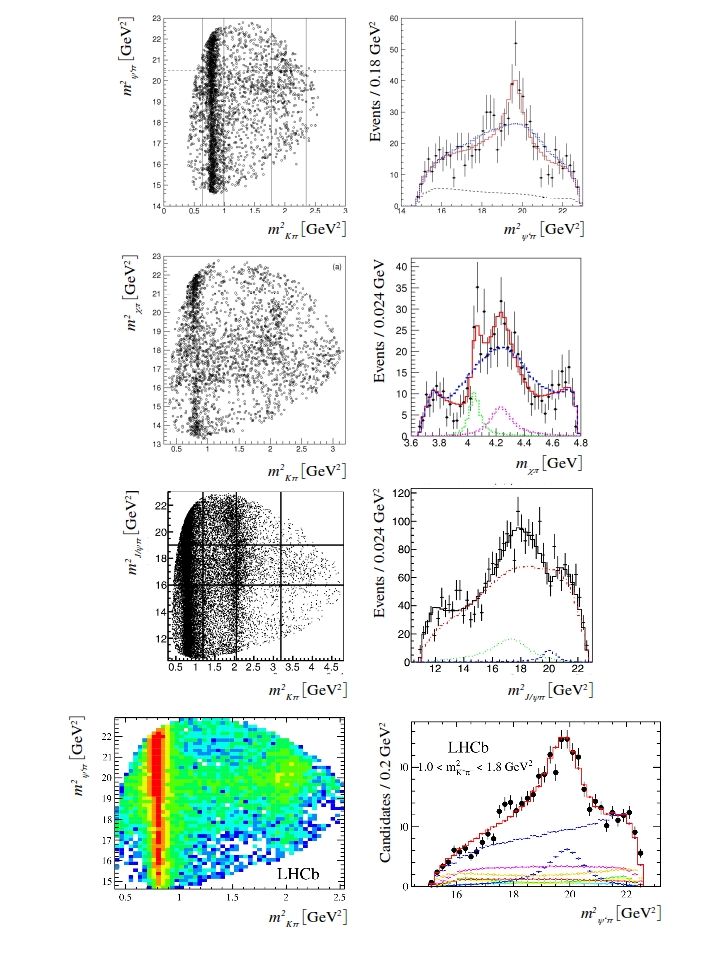}}
\vspace{-4mm}
\caption{
Dalitz plots (left) and mass squared projections for charmonium plus a single charged pion for different final states or experiments.
(Top row) $B^0\rightarrow \psi' K^{\pm}\pi^{\mp}$ from Belle \cite{z4430_belle_1,z4430_belle_2}; $m^2_{\psi^\prime \pi^+}$ is shown with the mass regions around the $K^{*0}(890)$ and $K^{*0}(1400)$ being vetoed.
(Second row) $B^0\rightarrow \chi_{c1}\pi^{\mp}$ from Belle \cite{z4430_belle_3};  $m^2_{\chi_{c1}\pi^{+}}$  is shown for a slice in $m^2(K^-\pi^+)$ between 1.0 and 1.75 GeV$^2$. The dotted (blue) line shows the fit result without any $\chi_{c1}\pi^{+}$ resonant contributions, and the solid (red) line shows the fit including two resonances. The signal shapes are shown by the dotted (green) line for the $Z_{c,1}(4051)$, and the dotted (magenta) line for the $Z_{c,2}(4248)$.
(Third row) $B^0\rightarrow K^{\pm}\jpsi\pi^{\mp}$ from Belle \cite{z4200_belle}; $m^2_{\jpsi\pi^{+}}$ is shown for $2.05<m^2(K^-\pi^+)<3.2$~GeV$^2$, the solid (black) line is the result of the total fit, the dashed (red) line is the fit without $Z_c$ states, the dotted (green) line shows the $Z_c(4200)^+$ contribution, and the dashed (blue) line shows the $Z_c(4430)^+$.
(Bottom row) $B^0\rightarrow \psi' K^{\pm}\pi^{\mp}$ from LHCb \cite{Stone:2015iba}.
The upper red line indicates to total fit, the upper blue curve indicates the fit without the $Z_c$(4430).
The other lines indicate fit contributions from the $Z_c$(4430) (lower blue curve),
the $K^*$(892) (dark yellow)
the $K \pi$ $S$-wave (magenta),
the $K_s^*$(1430) (green)
the $K^*$(1680) (light yellow)
the $K^*$(1410) (light blue) and additional other background sources (lower red curve). 
For details of the analysis see \cite{z4430_lhcb}.
\label{fz4430}}
\end{figure}

The first of the charged charmonium-like states was first observed by Belle in the decay channel
$B^0\rightarrow \psi'K^{\pm}\pi^{\mp}$.
We will refer to this state, and similar states observed in $B$ meson decays, as $Z$ states of type I.
Since it was observed in the charmonium mass regime, and decays to final states with open or hidden charm,
we will also use the short form $Z_c^+$.
A Dalitz plot of this decay channel is shown along with the 
$\psi'$$\pi^\mp$ squared invariant mass projection in Fig.~\ref{fz4430} (top).
The peak in the mass spectrum is due to the $Z_c^+(4430)$.
The existence of this resonance is remarkable, as it is charged, while a charmonium state must be neutral.
In the analysis it is important to take into account resonances in the $K^{\pm}$$\pi^{\mp}$ 
channel, in order to solely extract the dynamics in the 
$\psi'$$\pi^{\mp}$ channel. In the Dalitz plot in Fig.~\ref{fz4430} resonances in 
$\!\!\!\!\!\!\!\!\!$the $K^{\pm}\pi^{\mp}$ channel are seen as vertical bands, 
resonances in $\psi'$$\pi^{\mp}$ as horizontal bands.
Two vertical bands, corresponding to the $K^*$(892) and the $K_2^*$(1430)
are visible and expected.
In Fig.~\ref{fz4430} (top right), a veto on both $K^*$ resonances
is applied. The first observation of this state \cite{z4430_belle_1}
with a data set of 605~fb$^{-1}$ showed a statistical significance of 6.5$\sigma$.

The resonance parameters were determined
from a fit to $\psi'\pi^+$ mass spectrum giving
a mass of 4433$\pm$4$\pm$2~MeV
and a width of 45$^{+18}_{-13}$$^{+30}_{-13}$~MeV.
This is quite narrow for a state with such a high mass.
\BaBar found contrary evidence for the $Z_c^+$(4430)
with a data set of 413 fb$^{-1}$ \cite{z4430_babar_1}.
The argument was based upon the observation 
that the kinematic variables in this final state
are strongly correlated. 
Particular importance was attributed to cos~$\theta_K$, the normalized dot-product between 
the $K$$\pi$ three-momentum vector in the parent-$B$ rest frame 
and the kaon three-momentum vector 
after a Lorentz transformation from the $B$ meson rest frame 
to the $K$$\pi$ rest frame. 
This parameter cos~$\theta_K$ is correlated with $m$($K^{\pm}$$\pi^{\mp}$) \cite{z4430_babar_1}. 
A forward/backward asymmetry is observed, i.e.\ cos~$\theta_K$$<$0 
is preferred for the $K^*$(892) and the $K_2^*$(1430).  
This effect can produce structures in the $m^2$($\psi'\pi^{\mp}$)
invariant mass in the high mass region, depending 
on applied $K^*$ cuts in the analysis.
The \BaBar analysis yields only a 2.7$\sigma$ statistical significance
for the $Z_c^+$(4430).

In a re-analysis \cite{z4430_belle_2} 
of the original Belle observation in \cite{z4430_belle_1}
the $Z_c^+$(4430) was confirmed using a 2-dimensional fit of the 
Dalitz plot (Fig.~\ref{fz4430}, top left).
The resulting statistical significance is 6.4$\sigma$.
Note that  both charged states $Z_c^+$ and $Z_c^-$ are observed.
If these $Z$ states would turn out to carry isospin,
this would have the important implication of the $Z^{\pm}$ being an iso-doublet. 

In a  3~fb$^{-1}$ data set recorded at $\sqrt{s}$=7 and 8 TeV,
LHCb finally confirmed this state with a very large significance $\geq$13.9$\sigma$ \cite{z4430_lhcb}.
The Dalitz plot and the projected squared mass distribution are shown in Fig.~\ref{fz4430} (bottom). 
The full amplitude analysis used a 4-dimensional Dalitz fit of two squared masses $m_{K\pi}^2$ and $m_{\psi'\pi}^2$, 
and two angles $\theta_{\psi'}$ (the $\psi'$ helicity angle, defined as the angle between the
momenta of the ($K^+\pi^-$) system and the $\mu^-$ in the $\psi'$ rest frame), and $\phi$
(the angle between the planes defined by the $\mu^+\mu^-$ and $K^\pm\pi^\mp$ systems). 
The quantum numbers of the $Z_c^{\pm}$(4430) were unambiguously established as $J^P$=$1^+$, 
excluding $0^-$, $1^-$, $2^+$, and $2^-$ by 9.7$\sigma$, 15.8$\sigma$, 16.1$\sigma$ and 14.6$\sigma$. 
The measured mass and width are consistent with the Belle determination.

A second $Z$ state with a significance 6$\sigma$ was also observed
with a mass of 4239$\pm$18$^{+45}_{-10}$~MeV
and a width of 220$\pm$47$^{+108}_{-74}$~MeV. 
In the fit, 
$J^P$=$0^-$ was preferred over the $J^P$=$1^-$, $2^-$, and $2^+$ by 8$\sigma$.
The $0^+$ interpretation  was preferred over the $1^+$ by only 1$\sigma$,
however, in a fit with $J^P$=$1^+$ the width of $Z_c$(4430) would also increase by 660 MeV,
not being consistent anymore with the Belle measurement.

The observation of a charged state in the decay $B^0\to\jpsi \pi^+K^-$,
tentatively called the $Z_c$(4200), 
was made by Belle using the full data set of 711 fb$^{-1}$ \cite{z4200_belle},
with a measured mass of 4196$^{+31}_{-29}$$^{+17}_{-13}$~MeV 
and a width of 370$^{+70}_{-70}$$^{+~70}_{-132}$~MeV.
The Dalitz plot and projected squared mass distribution are shown in Fig.~\ref{fz4430} (second row). 
The significance was 6.2$\sigma$.
Again, $J^P$=$1^+$ was preferred
with an exclusion level of 6.1$\sigma$ , 7.4$\sigma$, 4.4$\sigma$, 7.0$\sigma$
for $J^P$=$0^-$, $1^+$, and $2^-$, $2^+$.  Note, $0^+$ is forbidden by parity conservation.

Another very important result of this analysis was the observation of 
a second decay mode for the $Z_c$(4430) with $Z_c$(4430)$\rightarrow$$J$$\psi$$\pi^{\pm}$,
established with a significance of 4.0$\sigma$. The
mass and width of the $Z_c$(4430) were fixed in the fit. 
The measured branching fraction of the $Z_c$(4430) was a factor $\simeq$4
smaller than for the $Z_c$(4260), presumably due to the smaller phase space. 
This observation, in addition to the resonant behaviour shown in the Argand diagram in \cite{z4430_lhcb} (see below), 
proves that the $Z_c$(4430) is not a kinematic effect such as induced by a triangle singularity. 

\begin{figure}[t!]
\centerline{\includegraphics[width=0.8\textwidth]{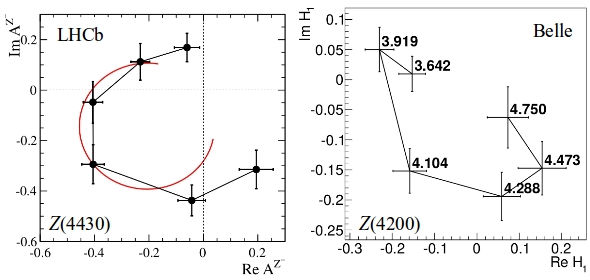}}
\caption{
Argand plots for the $Z_c$(4430) (left) from LHCb \cite{z4430_lhcb} and the $Z_c$(4200) (right) from Belle \cite{z4200_belle},
proving the resonant nature. For details see text.
\label{fz_argand}}
\end{figure}

To plot the Argand diagrams, the Breit-Wigner shape in the signal regions was replaced by six complex amplitudes in consecutive intervals in the invariant mass, that are allowed to float in the fit. Then, the real and imaginary parts of these amplitudes are plotted in a two-dimensions. 
Figure~\ref{fz_argand} shows  these  Argand plots. 
In the case of $Z_c$(4200) \cite{z4200_belle}, two independent complex $Z$ helicity couplings,
${\cal H}_{\lambda}$ for $\lambda$=0,+1, were allowed to float in the fit.
Note that parity conservation requires ${\cal H}_{-1}$=${\cal H}_{+1}$.
Figure~\ref{fz_argand} (right) shows only the real and imaginary parts of the ${\cal H}_{1}$ amplitude,
which is more conclusive than ${\cal H}_{0}$.
In the case of $Z_c$(4430) \cite{z4430_lhcb}, 
the $D$-wave contribution is found to be insignificant when allowed in the fit. 
Consequently, a pure $S$-wave $Z_c$ decay was assumed, implying ${\cal H}_{+1}$=${\cal H}_{0}$$\equiv$${\cal A}^{Z}$,
for which the real and imaginary parts of the amplitude are shown in Fig.~\ref{fz_argand} (left). In both cases the phase motion of the amplitude is consistent with that of a Breit-Wigner resonance.

\afterpage{\clearpage}

In an additional channel, the decay 
$B^0\rightarrow \chi_{c1}K^+ \pi^-$ was observed to contain 
two new states called the $Z_{c,1}$(4051) and $Z_{c,2}$(4248) decaying into
$\chi_{c1}\pi^-$, using  a data set of 605~fb$^{-1}$ \cite{z4430_belle_3}.
Figure~\ref{fz4430} (second row) shows the 
$\chi_{c1}\pi^{\mp}$ invariant mass after
applying $K^*(892)$ and $K^*(1430)$ veto cuts.
The $Z_{c,1}$ and the $Z_{c,2}$  are clearly visible. 
The fit shows that the addition of two resonances 
are preferred by 13.2$\sigma$, 
and two resonances are preferable to one by 5.7$\sigma$.
A fit using Breit-Wigner shapes for both states
gives for the $Z_{c,1}$ a mass of 4051$\pm$14$^{+20}_{-41}$~MeV, 
and a width of 82$^{+21}_{-17}$$^{+47}_{-22}$~MeV, 
and for the $Z_{c,2}$ a mass of  $4248^{+44}_{-29}$$^{+180}_{-~35}$~MeV and a width of 177$^{+54}_{-39}$$^{+316}_{-~61}$~MeV.
For the fits, different assumptions
for the quantum number assignment combinations of 
$J_{1,2}$=0 or $J_{1,2}$=1 were tested, however, the $\chi^2$ of the fit doesn't change significantly.
An important point in these analyses by Belle is
the interference (i.e.\ mixing between the states and
interference of the states with the background) was taken into account in the Dalitz fit model. 
This can be seen in Fig.~\ref{fz4430} in the mass regions $m_{\chi_{c1}\pi^{\pm}}$$\simeq$3.9~GeV
and $m_{\chi_{c1}\pi^{\pm}}$$\simeq$4.5~GeV, in which
the red line (describing the fit with the $Z_{c,1}$ and the $Z_{c,2}$) 
falls below the dashed blue line (describing the fit without
any $Z$ resonances). It can be seen that in these mass
regions the interference is significantly destructive. 
The $Z_{c,1}$ and $Z_{c,2}$ were searched for in an analysis by \BaBar \cite{z4430_babar_2}
in a data set of 429~fb$^{-1}$, but similar to the case of the $Z_c$(4430)
no evidence was found and upper limits were assigned. 

For a charged state in the charmonium mass regime,
the minimal quark content of the $Z_c^+$(4430) 
must be $[$$c$$\bar{c}$$u$$\bar{d}$$]$,
and thus it cannot be a charmonium state.
An interpretation as a $D$$\bar{D}$ molecule is possible,
as the mass is close to the sum of the masses of
$D^*$(2010) and $D_1$(2420), both having narrow 
widths of 96~keV and 20.4~MeV, respectively.
In the case of the $Z_{c,1}$, an interpretation 
as a molecule is also possible, 
as the sum of the masses $m$($D^{*0}$)+$m$($D^{*+}$)$\simeq$4017~MeV
is close to the observed mass of $Z_{c,1}$ of 4051~MeV.
However, for the $Z_{c,2}$, there are no narrow excited $D^{(*)}$ mesons
which would fit.
The mass and width of the $Z_2$ are consistent with the $J^P$=$0^-$ $\psi'$$\pi^{\pm}$ state observed by LHCb \cite{z4430_lhcb}, 
but a strong decay $0^-$$\rightarrow$$1^+$$0^-$ is not allowed due to conservation of parity.
Thus, there are two different charged states with two different decays, but almost identical mass.  

%% file: z3900-rev.tex
\subsection{\boldmath{$Z$ states of type II in the charmonium mass regime }}

\label{cz3900}

\begin{figure}[t!]
\centerline{\includegraphics[width=0.84\textwidth]{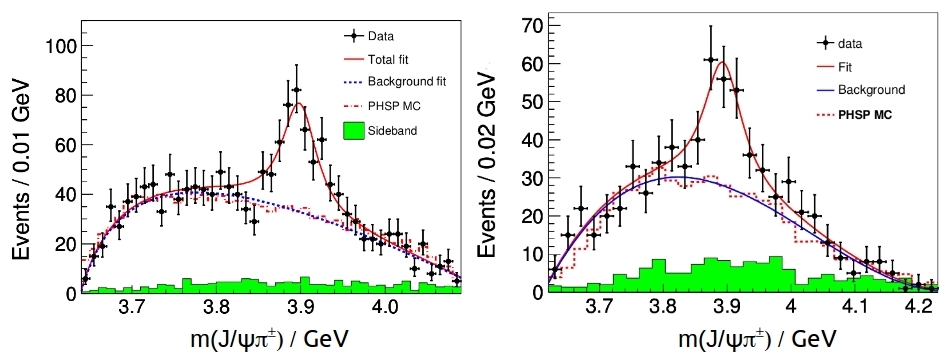}}
\caption{$J$/$\psi$$\pi^{\mp}$ invariant mass in Y(4260) decays, 
indicating the $Z_c^+$(3900) signal from BESIII (left) \cite{bes3_z3900_charged} 
and from Belle (right) \cite{zc3900_belle}. 
\label{fzc3900}}
\end{figure}

Recently, more charged charmonium-like states have been discovered
in production mechanisms different from those found in $B$ meson decays.
In the following, we will refer to them as $Z$ states of type II. 
As the first in this class of new states, the $Z_c^+$(3900) was observed by 
BESIII \cite{bes3_z3900_charged} in the decay  $Y$(4260)$\rightarrow$$Z_c^+$(3900)$\pi^{-}$
in a data set of 525~pb$^{-1}$ collected at $\sqrt{s}$=4.26~GeV.
The $Z_c^+$(3900) was reconstructed in the decay to $J$/$\psi$$\pi^{\pm}$.
Figure~\ref{fzc3900} (left) shows the observed signal, which has a statistical 
significance of $>$8$\sigma$. From the two charged pions, the one is used  
which gives the larger $\jpsi\pi^{\pm}$ invariant mass,
in order to remove combinatorical background from the pion 
of the $Y$(4260) transition to the new state. 
The measured mass is 3899.0$\pm$3.6$\pm$4.9~MeV and the measured
width 46$\pm$10$\pm$20~MeV.
The observation of the decay $Y$$\rightarrow$$Z^{+}_c\pi^-$ is remarkable, 
as it provides for the first time a connection between $Z$ states and $Y$ states, possibly
pointing to the same interpretation of their nature. 

Only a few days later, the state was confirmed by Belle \cite{zc3900_belle} in the same decay channel 
and also in $Y$(4260) decays, while in the Belle case the $Y$(4260) was produced 
in the ISR process $\Upsilon(nS)\rightarrow \gamma_{ISR}Y(4260)$ (see also Sec.~\ref{cy4260}). 
The mass of 3894.5$\pm$6.6$\pm$4.5~MeV and the width of 63$\pm$24$\pm$26~MeV are 
both consistent with the BESIII measurement. 
Figure~\ref{fzc3900} (right) shows the observed signal, which has a statistical 
significance of $>$8$\sigma$ in a data set of 967~fb$^{-1}$.
Although the yield of produced $Z_c$(3900) is comparable, 
Fig.~\ref{fzc3900} (left) corresponds to about 4 weeks of data taking
at BESIII, while Fig.~\ref{fzc3900} (right) corresponds to about 10 years data taking
at Belle. 
Again, as in the case of the $Z_c^+$(4430), the state was observed in both positive and negative charged states  
with about the same yield \cite{bes3_z3900_charged}, indicating a doublet. 

Concerning the quantum numbers, remarkably the isospin must be $I$=1, 
(as the isospin of the pion is $I$=1), if we assume $I$=0 for the $Y$(4260). 
If the heavy meson pair is assumed to be in the $S$-wave, 
the spin-parity of the state is uniquely determined as $J^P$=$1^+$.
$C$-parity $(-1)^{L+S}$ is only defined for neutral particles, 
thus there can only be a $G$-parity assignment to the $Z_c^+$(3900). 
The $G$-parity $(-1)^{L+S+I}$ with $L$=0, $S$=1 and $I$=1 thus gives G=+.
As $G$-parity should be preserved in strong decays, 
this assignment, due to the negative $G$-parity of the pion, has the 
interesting implication that the $Y$(4260) would have $G$=$-$.
This would be compatible with an $I$=0 isosinglet assignment for the $Y$(4260),
which would imply that there is no charged partner of the $Y$(4260). 
{\color{revcolor} Using a partial wave analysis, the spin and parity were finally determined to $1^+$
with a statistical preference of more than 7$\sigma$ over all other quantum numbers \cite{Ablikim:2017hyd}.}

A similar structure to the $Z_c$(3900) was observed in an analysis of CLEO-c data \cite{zc3900_cleo}
however in data recorded at $\sqrt{s}$=4.17~GeV.
The fitted mass of 3885$\pm$5$\pm$1~MeV and width of 34$\pm$12$\pm$4~MeV of an observed 
state are consistent with the $Z_c$(3900). 
The fitted yield is 81$\pm$20 events, corresponding to a statistical significance of 6.1$\sigma$. 
However, the lower center-of-mass energy would imply that this state is not produced in $Y$(4260) decays,
but directly via $e^+$$e^-$$\rightarrow$$Z_c^{\pm}$$\pi^{\mp}$.

\begin{figure}[t!]
\centerline{
\includegraphics[width=0.32\textwidth]{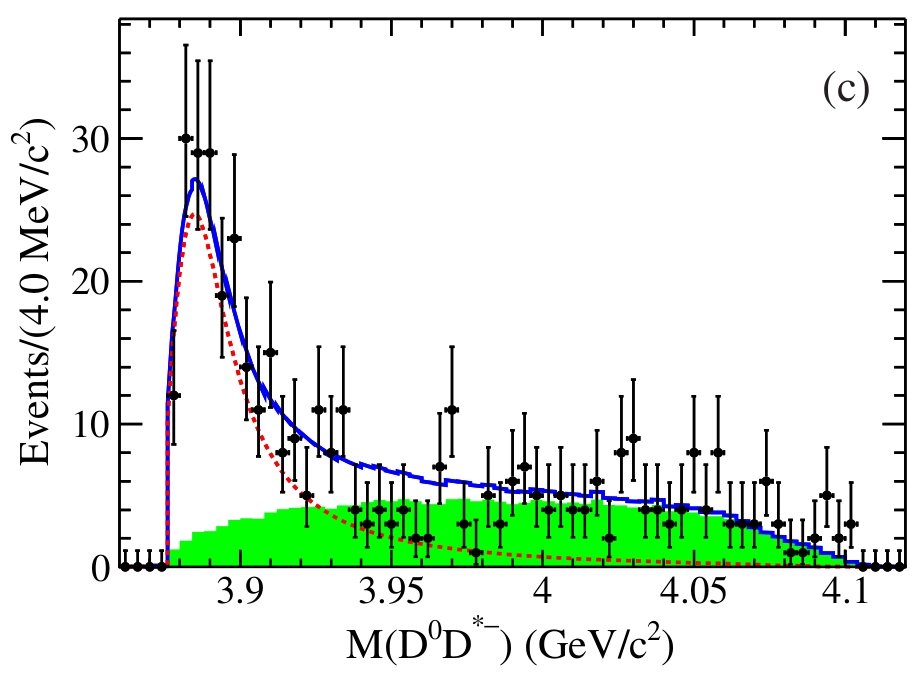}
\includegraphics[width=0.25\textwidth]{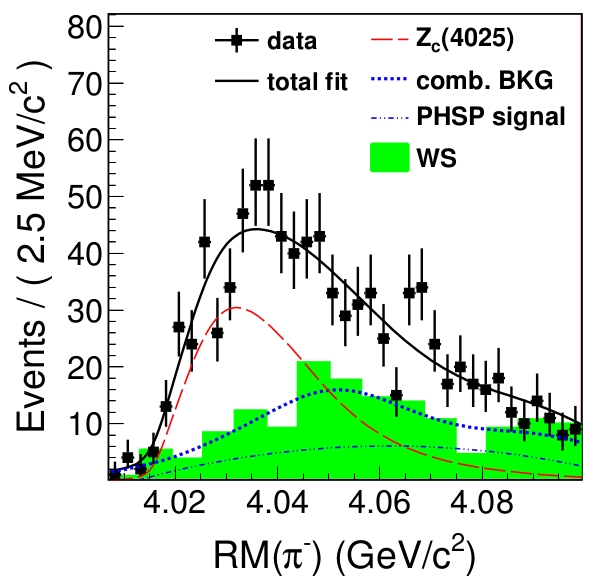}
\includegraphics[width=0.32\textwidth]{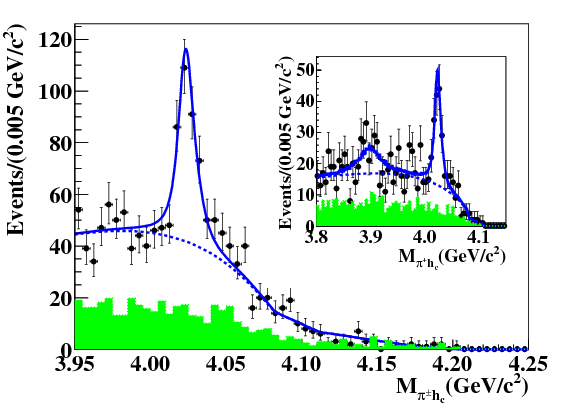}
}
\caption{Invariant mass distributions for type II charged states. The solid (green) represents the backgrounds.
{\it Left:} $D^0$$D^{*-}$ mass at $\sqrt{s}$=4.26~GeV  showing the signal for the $Z_c$(3885) from BESIII \cite{bes3_z3885_charged_double_tag}
The  solid line (blue) shows the combined fit to background and signal, the dotted line (red) shows the signal,
and the background is fitted with a phase space probability density function.  
{\it Center:}  recoil mass from $\pi^{\mp}$  for $e^+e^-\rightarrow(D^*\overline{D}^*)^{\pm}\pi^{\mp}$ at
$\sqrt{s}$=4.26~GeV from BESIII \cite{bes3_z4025_charged}. The black line represents an unbinned maximum likelihood fit. 
The dashed-line  (red) shows the fitted signal contribution of the $Z_c$(4025), 
the dotted-line (blue) the combinatorial background, and
the dash-dotted line (blue) the result of an MC simulation according to a non-resonant phase space distribution. 
{\it Right:} $h_c\pi^{\pm}$ invariant mass for $e^+e^-\rightarrow h_c\pi^{\pm}$
at $\sqrt{s}$=4.23, 4.26 and 4.36~GeV with the fit for the $Z_c$(4020) from BESIII \cite{bes3_z4020_charged}.
The dotted curves show the fitted background.
The inset shows the combined fit for the $Z_c$(4020) and the $Z_c$(3900); for the latter no significant signal was observed. 
\label{f_z3885_z4020_z4025}}
\end{figure}

If the $Z_c$(3900) represents a charged partner of the $X$(3872), one of the important 
questions is: does it decay not only into a final state with closed charm (i.e.\ $J$/$\psi$),
but also in states with open charm (i.e.\ $D^{(*)}$$\overline{D}^{(*)}$). 
In fact, subsequently another type II state was then observed at
 BESIII \cite{bes3_z3885_charged_double_tag} in 
$e^+e^-\rightarrow\pi^+(D\overline{D}^*)^-$ at $\sqrt{s}$=4.26~GeV using a data set 
of 525~pb$^{-1}$. The mass and width were determined as 3883.9$\pm$1.5$\pm$4.2~MeV and 24.8$\pm$3.3$\pm$11.0~MeV,
consistent with the mass and the width of the $Z_c$(3900) as measured by BESIII \cite{bes3_z3900_charged}.
Figure~\ref{f_z3885_z4020_z4025} (left) shows the $Z_c$(3885) signal from an additional analysis by
 BESIII \cite{bes3_z3885_charged_double_tag}
with double charmed meson tag, which confirmed the $Z_c$(3885). 
The angular analysis of the $\pi$$Z_c$(3885) system leads to a preference 
for a $J^P$=1$^+$ assignment
(while disfavoring 1$^-$ or 0$^-$), which would also be consistent with the $Z_c$(3900) (see above). 

The above states obviously are close to the $D^{*}$$\overline{D}$ threshold.
A second class of observed states is close to the $D^*$$\overline{D}^*$ threshold.
The $Z_c$(4025) was observed at BESIII \cite{bes3_z4025_charged} in the reaction 
$e^+$$e^-$$\rightarrow$($D^*$$\overline{D}^*$)$^{\pm}$$\pi^{\mp}$
at $\sqrt{s}$=4.26~GeV in a data set of 827~pb$^{-1}$. 
It was observed by a recoil mass technique
with a mass of 4026.3$\pm$2.6$\pm$3.7~MeV
and a width of 24.8$\pm$5.6$\pm$7.7~MeV. 
Figure~\ref{f_z3885_z4020_z4025} (center) shows the $\pi^{\mp}$ recoil mass.
The $Z_c$(4025) signal is observed with a statistical significance of $\geq$13$\sigma$.

Another type II state, denoted $Z_c$(4020) \cite{bes3_z4020_charged}, which may be identical 
to the $Z_c$(4025), has been observed in a different decay. 
The data set is comprised of several center-or-mass energies $3.90\leq\sqrt{s}\leq 4.42$~GeV, 
i.e.\ not only $Y$(4260) decays. It was observed in the $h_c$$\pi^{\pm}$ invariant 
mass with a mass of $4022.9\pm 0.8 \pm 2.7$~MeV and a width of $7.9\pm 2.7 \pm 2.6$~MeV.
Figure~\ref{f_z3885_z4020_z4025} (right) shows the fitted signal for a fit of only the $Z_c$(4020) and for a combined
fit for the $Z_c$(4020) and the $Z_c$(3900), for which no significant signal was observed. 
The $h_c$ was reconstructed in the decay $h_c$$\rightarrow$$\gamma$$\eta_c$ 
with the recontruction of the $\eta_c$ in the 16 decays 
$p$$\overline{p}$, 
2($\pi^+$$\pi^-$), 
2($K^+$$K^-$), 
$K^+$$K^-$$\pi^+$$\pi^-$, 
$p$$\overline{p}$$\pi^+$$\pi^-$,  
3($\pi^+$$\pi^-$), 
$K^+$$K^-$2($\pi^+$$\pi^-$), 
$K_s$$K^{\pm}$$\pi^{\mp}$, 
$K_s$$K^{\pm}$$\pi^{\mp}$$\pi^{\pm}$$\pi^{\mp}$, 
$K^+$$K^-$$\pi^0$, 
$p$$\overline{p}$$\pi^0$, 
$\pi^+$$\pi^-$$\eta$, 
$K^+$$K^-$$\eta$, 
2($\pi^+$$\pi^-$)$\eta$, 
$\pi^+$$\pi^-$$\pi^0$$\pi^0$, 
2($\pi^+$$\pi^-$)$\pi^0$$\pi^0$. 
$K_s$ mesons are reconstructed from their decay into $\pi⁺$$\pi^-$, 
$\pi^0$ and $\eta$ mesons are reconstructed from their decay into $\gamma$$\gamma$. 

{\color{revcolor} A charged state decaying to $\psi' \pi^+$ has been observed with a significance of 9.2$\sigma$
at a mass of 4032.1$\pm$2.4~MeV with a width of 26.1$\pm$5.3~MeV. 
This is again close to the $D^*$$\overline{D}^*$ threshold, and the state may thus be identical to the $Z_c$(4020),
seen in $h_c \pi$, and the $Z_c$(4025), seen in $D^*$$D^*$.
The relative ratio if the branching fractions, not measured so far, would give an important
hint for the interpretation as molecules.
Remarkably, no significant signal in the $\jpsi \pi$ final state has been observed, yet.
Future measurements have to show, if the decay to the charmonium ground state is suppressed
compared to the decay to the excited $\psi'$, although the phase space being larger,
which would be an striking observation.
In the analysis, $J^P$=$1^+$ is assumed and found to be consistent with the observation. 
This would be the identical assignment as the $Z_c$(3900), which in fact decays dominantly into $\jpsi \pi^+$.}

{\it Interpretation} --- As the $Z_c^+$(3900) is charged, it can neither be a 
charmonium state nor a hybrid state. 
The vicinity to the $D^+$$\overline{D}^{0*}$ ($D^0$$\overline{D}^{+*}$)
and the $D^{0*}$$\overline{D}^{\pm *}$ thresholds obviously make the
states candidates for a charmed meson molecules, similar to the $X(3872)$.
Table~\ref{tzc} lists the measured masses, the nearby thresholds, 
and the mass differences $\Delta$$m$ between them. 
There are two important differences with respect to the $X(3872)$.
On the one hand (as can be seen in Table~\ref{tzc}), 
in all the cases the masses seem to be higher than the threshold,
and thus the ``binding'' energy would be positive and the state
would be a virtual state. For the $ X(3872)$, the measured mass is 
within $\simeq$1~MeV of the threshold and the binding energy is negative,
although small. 
On the other hand, the measured width of the $Z_c$(3900) 
with $\geq$10~MeV is much larger than in case of the $X$(3872)
with $\leq$1~MeV.
An interpretation as a tetraquark $[$$c$$d$$]_{\bar{3}}$$[$$\overline{c}$$\overline{u}$$]_3$ 
is another option and may 
explain the isospin $I$=1 e.g.\ with a direct coupling between the light quarks.
In fact, even prior to the observation of the $Z_c$(3900), two 
nearby tetraquarks with J$^{PC}$=1$^{+-}$ were predicted \cite{zb_interpretation_ali_hambrock}
with masses of 3.752 and 3.882 GeV, respectively. The heavier predicted state
may represent the $Z_c$(3900). 

These states are presently subject to intensive study in order to clarify their nature. Further data taking at BESIII may largely improve the understanding of these states in the near future.
In particular, it may be important to investigate the combined pattern of the three (and possibly more yet to be discovered) states:

\begin{itemize}

\item Are the $Z_c$(4020) and the $Z_c$(4025) identical or different states?
One possible way to investigate this question is to assign the quantum numbers using angular distributions. 

\item Is the $Z_c$(3900) only generated in the decay of the $Y(4260)$, and thus possibly the decay 
of an exotic state into another exotic state, or also in direct production at center-of-mass energies below? 

\item In the analysis of the $Z_c$(4020) \cite{bes3_z4020_charged} no significant signal for the $Z_c$(3900) was observed.
Is this a result of different quantum numbers, insufficient statistics or does it maybe point to a different nature?

\end{itemize}

\begin{table}[hbt]
\centering
\caption{Masses, nearby thresholds, and mass differences for charged $Z$ states. 
For all the states, the masses are higher than the thresholds and thus the states
may be virtual states.}
\vspace{0.2cm}
\begin{tabular}{lcccc}
\hline\\[-2.5ex]
State & $m$ (MeV) & Threshold & $\Delta$$m$ (MeV) \\
\hline\\[-2.5ex]
$Z_c$(3900) & 
3899.0$\pm$3.6$\pm$4.9 & 
$D^+$$\overline{D}^{0*}$ & 
+22.4 &
\cite{bes3_z3900_charged} \\
$Z_c$(3900) & 
3899.0$\pm$3.6$\pm$4.9 & 
$D^0$$\overline{D}^{+*}$ &
+23.9 &
\cite{bes3_z3900_charged} \\
$Z_c$(3900) & 
3894.5$\pm$6.6$\pm$4.5 &
$D^+$$\overline{D}^{0*}$ &
+17.9 &
\cite{zc3900_belle} \\
$Z_c$(3900) & 
3894.5$\pm$6.6$\pm$4.5 &
$D^0$$\overline{D}^{+*}$ & 
+19.4 &
\cite{zc3900_belle} \\
$Z_c$(3900) & 
3885$\pm$5$\pm$1 &
$D^+$$\overline{D}^{0*}$ &
+8.4 &
\cite{zc3900_cleo} \\
$Z_c$(3900) & 
3885$\pm$5$\pm$1~MeV &
$D^0$$\overline{D}^{+*}$ &
+9.9 &
\cite{zc3900_cleo} \\
$Z_c$(3885) & 
3883.9$\pm$1.5$\pm$4.2 &
$D^+$$\overline{D}^{0*}$ & 
+7.4 &
\cite{bes3_z3885_charged_single_tag} \\
$Z_c$(3885) & 
3883.9$\pm$1.5$\pm$4.2 &
$D^0$$\overline{D}^{+*}$ &
+8.8 &
\cite{bes3_z3885_charged_single_tag} \\
$Z_c$(4020) & 
4022.9$\pm$0.8$\pm$2.7 & 
$D^{0*}$$\overline{D}^{\pm *}$ & 
+5.6 &
\cite{bes3_z4020_charged} \\
$Z_c$(4025) & 
4026.3$\pm$2.6$\pm$3.7 &
$D^{0*}$$\overline{D}^{\pm *}$ & 
+9.0 &
\cite{bes3_z4025_charged} \\
{\color{revcolor} $Z_c$(4032)$^+$} &
{\color{revcolor}$\simeq 4032.1\pm$2.4} &
{\color{revcolor} $D^{0*}$$\overline{D}^{\pm *}$} &
{\color{revcolor} +15.0} & 
\cite{Ablikim:2017oaf} \\
\hline
\end{tabular}
\label{tzc}
\end{table}

\begin{table}[hbt]
\centering
\caption{Masses and widths in comparison for the charged and neutral $Z$ states observed at BESIII.}
\vspace{0.2cm}
\begin{tabular}{lcccc}
\hline\\[-2.5ex]
State & $m$ (MeV) & Width (MeV) & Decay & \\
\hline\\[-2.5ex]
$Z_c$(3900)$^+$ & 3899.0$\pm$3.6$\pm$4.9 & 46$\pm$10$\pm$20 & $J$/$\psi\pi^+$ & \cite{bes3_z3900_charged} \\
$Z_c$(3900)$^0$ & 3894.8$\pm$2.3$\pm$2.7 & 29.6$\pm$8.2$\pm$8.2 & $J$/$\psi\pi^0$ & \cite{bes3_z3900_neutral} \\
\hline
$Z_c$(3885)$^+$ & 3883.9$\pm$1.5$\pm$4.2 & 24.8$\pm$3.3$\pm$1.0 & ($DD^*$)$^+$ & 
\cite{bes3_z4025_charged} \\
$Z_c$(3885)$^0$ & 3885.7$_{-5.7}^{+4.3}$$\pm$8.4 & 35$_{-12}^{+11}$$\pm$15 & ($DD^*$)$^0$ & \cite{bes3_z3885_neutral} \\
\hline
$Z_c$(4020)$^+$ & 4022.9$\pm$0.8$\pm$2.7 & 7.9$\pm$2.7$\pm$2.6 & $h_c\pi^+$ & \cite{bes3_z4020_charged} \\
$Z_c$(4020)$^0$ & 4023.8$\pm$2.2$\pm$3.8 & Fixed to 7.9 & $h_c\pi^0$ & \cite{bes3_z4020_neutral} \\
\hline 
$Z_c$(4025)$^+$ & 4026.3$\pm$2.6$\pm$3.7 & 24.8$\pm$5.6$\pm$7.7 & $(D^*D^*)^+$ & 
\cite{bes3_z4025_charged} \\
$Z_c$(4025)$^0$ & 4025.5$_{-4.7}^{+2.0}$$\pm$3.1 & 23.0$\pm$6.0$\pm$1.0 & $(D^*D^*)^0$ & \cite{bes3_z4025_neutral} \\
\hline
    {\color{revcolor} $Z_c$(4032)$^+$} &
    {\color{revcolor} $\simeq 4032.1\pm$2.4} &
    {\color{revcolor} $\simeq 26.1\pm$5.3} &
    {\color{revcolor} $\psi' \pi^+$} &
    \cite{Ablikim:2017oaf} \\
    {\color{revcolor} $Z_c$(4032)$^0$} &
    \multicolumn{4}{c}{\color{revcolor} not observed yet} \\
\hline
\end{tabular}
\label{t_z_charged_neutral}
\end{table}

Soon after the discovery of the charged $Z_c^+$(3900) \cite{bes3_z3900_charged}
the neutral partner state, decaying into $J$/$\psi$ $\pi^0$
was discovered by BESIII \cite{bes3_z3900_neutral}.
On the one hand, this observation may lead to the tempting conclusion of isospin triplets,
i.e.\ that the $Z^{0,\pm}$ may carry isospin $I$=1.
This conclusion is supported by the observation, that the ratio of the cross sections of $e^+$$e^-$$\rightarrow$$Z^0\pi^0$
and $e^+$$e^-$$\rightarrow$$Z^\pm\pi^\mp$ is found to be close to unity for the $Z_c$(3885) \cite{bes3_z3885_neutral},
the $Z_c$(4020) \cite{bes3_z4020_neutral} and the $Z_c$(4025) \cite{bes3_z4025_neutral}. 
On the other hand, the observation of both charged and neutral states points to different 4-quark contents:
$[c\overline{c}u\overline{d}]$ or 
$[c\overline{c}d\overline{u}]$ for the charged $Z$, 
$[c\overline{c}u\overline{u}]$ or 
$[c\overline{c}d\overline{d}]$ for the neutral $Z$. 
In such a case, the masses may be different. 
Meanwhile, for all four observed charged $Z$ states 
\cite{bes3_z3900_charged} \cite{bes3_z3885_charged_single_tag} \cite{bes3_z4020_charged} 
\cite{bes3_z4025_charged},
corresponding neutral partners were identified
\cite{bes3_z3900_neutral} \cite{bes3_z3885_neutral} \cite{bes3_z4020_neutral} \cite{bes3_z4025_neutral}. 
Table~\ref{t_z_charged_neutral} shows the comparison of the masses and the widths
of corresponding charged and neutral $Z$ states. 

If, in fact, if $Z$ states of type II are isospin triplets,
the interesting question would be: where is the $I$=0 partner?
Let us consider the $Z_c$(3900) as an example.
The $X$(3872) is a tempting candidate, even though $\simeq$10~MeV below (as the $Z$ states are all slightly above threshold)
and significantly more narrow ($\leq$1.2~MeV instead of a few tens MeV). 
In the case of the $Z_c^{0,\pm}$(3900), from the decay to $J$/$\psi$ $\pi$ and assuming $L$=0 due to the vicinity of the threshold,
the quantum number assignment would be $J^P$=$1^+$, identical to the $J^P$ of the $X$(3872). 
However, the $X$(3872) carries $C$=+1. Under the above assumptions, the $Z^0$(3900) may only carry $C$=$-$1.
Note that in the tetraquark interpretation, the $Z$(3900)$^0$ will carry a charge conjugation quantum number
in any case, as with a $[c\overline{c}q\overline{q}]$ the $Z$would be its own antiparticle.
The only question is, if the content is $[c\overline{c}d\overline{d}]$ or $[c\overline{c}u\overline{u}]$,
however not changing the quantum number assignment. 

An interesting observation is, that the branching fractions of decays into open charm meson pairs are large
compared to branching fractions into charmonium, which may indicate molecular contributions in the wave function. 

\begin{itemize}

\item Assuming, that the $Z_c$(3900) and the $Z_c$(3885) are the same state,
the ratio of the branching fractions of the decay into $(DD^*)^{\pm}$ and
into $J$/$\psi$ $\pi^+$ is 6.2$\pm$1.1$\pm$2.7 \cite{bes3_z3885_charged_single_tag}.

\item Assuming, that the $Z_c$(4020) and the $Z_c$(4025) are the same state,
the ratio of the branching fractions of the decay into $(DD^*)^{\pm}$ and
into $h_c$$\pi^+$ is 12$\pm$5 \cite{Bian:2014vfa}.

\end{itemize}

We emphasize that this is very different from $Y$(4260) decays, for which no open charm decay has ever been found (see Sec.~\ref{cy4260} .
However, there is strong evidence that all the $Z$ states of type II are produced in decays of the $Y$(4260). 
If $Y$ states and $Z$ states of type II are of the same nature, the decay pattern seems to be quite different. 
The ratio can be compared to the other cases:

\begin{itemize}
\item For the $X$(3872), the ratio \cite{pdg} is 

\begin{equation}
  \frac{ {\cal B} ( X(3872) \rightarrow D^0 D^{0*} ) }
       { {\cal B} ( X(3872) \rightarrow J\psi \pi^+ \pi^- ) }
       = \frac{>24\%}{>2.6\%}. 
\end{equation}

\item For the conventional charmonium states $\psi$(3770),
the ratios of decays into $D^*\overline{D}^*$ and $J$/$\psi$$\pi^+$$\pi^-$
is very large with 482$\pm$84 \cite{pdg}.
In a simplified point of view, it seems that open charm decays of $Z$ states are blocked by a yet unknown mechanism.  

\end{itemize}

Summarizing, we find the following differences for the two classes of $Z$ states. 
$Z_c$ states of type I are observed in $B$ meson decays, 
not obviously related to any nearby threshold,
have large widths (partially $\geq$100 MeV or more) and
are so far only observed as charged states. 
$Z_c$ states of type II
are observed in $e^+$$e^-$ and ISR production, 
are found within a few MeV of a nearby threshold
(and, perhaps as a hint to the nature of these states, even a few MeV above threshold),
are narrow with a maximum width of a few tens MeV, and 
are observed as charged and neutral states
(within the errors of the measurement with degenerate masses and identical widths), 
and thus probably representing isospin triplets. 
No common member of type I and type II has been found yet. Thus, at present, it must be assumed
that these are two different classes of $Z$ states.
As a striking fact, the quantum numbers are 1$^+$ for any state of type I and any state of type II,
wherever tested experimentally.

%% file: zb-rev.tex
\subsection{\boldmath{$Z$ states in the bottomonium mass regime}}

\label{czb}

\begin{figure}[b]
\centerline{\includegraphics[width=\textwidth]{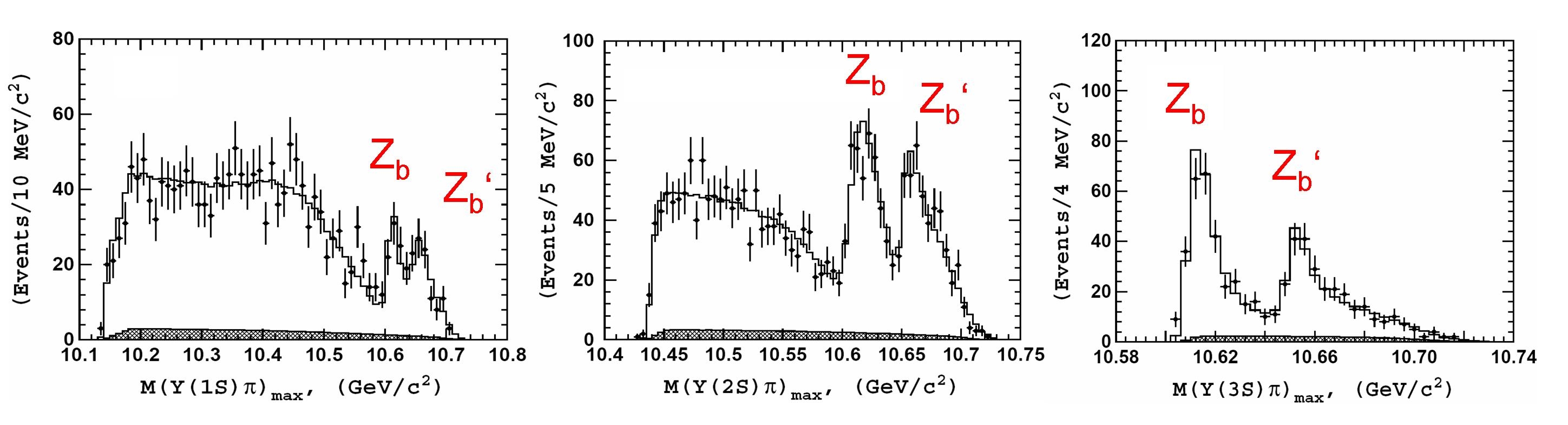}}
\caption{Charged pion recoil mass for 
$\Upsilon$(5$S$)$\rightarrow$$\Upsilon$(1$S$)$\pi^+$$\pi^-$ (left), 
$\Upsilon$(5$S$)$\rightarrow$$\Upsilon$(2$S$)$\pi^+$$\pi^-$ (center) and 
$\Upsilon$(5$S$)$\rightarrow$$\Upsilon$(3$S$)$\pi^+$$\pi^-$ (left) from Belle \cite{zb_observation}. 
The $Z_b$ and $Z_b'$ states are labelled.
\label{fzb}}
\end{figure}

In $\Upsilon(10860)$ decays, Belle observed two new states
with masses of 
$m=9898.3\pm 1.1^{+1.0}_{-1.1}$~MeV and
$m= 10259.8\pm 0.6^{+1.4}_{-1.0}$~MeV, respectively \cite{Adachi:2011ji}. 
These new states are widely accepted to represent the conventional bottomonium 
states $h_b$ ($1^1P_1$, 1$^{+-}$) and $h_b'$ ($2^1P_1$, 1$^{+-}$), 
as their masses were found within 2.7 MeV and 1.2~MeV of relativistic potential model predictions \cite{Ebert:2002pp}.
In a second step, the observation of the $h_b$(1$P$) and the $h_b$(2$P$)
also enabled the study of their specific production mechanism in $\Upsilon$(5$S$) decays,
i.e. are they produced according to phase space or are there any intermediate resonances.
Surprisingly,  both  the $h_b$(1$P$)$\pi^+$$\pi^-$ and 
$h_b$(2$P$)$\pi^+$$\pi^-$ final states contain a large fraction of $h_b$($n$P)$\pi^{\pm}$ resonances.
In the analysis \cite{zb_observation}, instead of directly plotting the $h_b$($n$P)$\pi^{\pm}$ mass, 
the recoil mass off of the bachelor charged pion is plotted,
because its relatively low momentum leads to excellent mass resolution.
In addition to the $h_b$($n$P)$\pi^{\pm}$ systems, the $\Upsilon$($n$$S$)$\pi^{\pm}$ systems
were investigated. In fact, all five systems show two intermediate resonances \cite{zb_observation},
which similar to the $Z$ states in charmonium (Sec.~\ref{cz4430}), were given the names
$Z_b$ (or $Z_b$(10610)) and $Z_b'$ (or $Z_b$(10650)). 
Figure~\ref{fzb} shows the recoil mass for
$\Upsilon$(1$S$)$\pi^{\pm}$, $\Upsilon$(2$S$)$\pi^{\pm}$ and $\Upsilon$(3$S$)$\pi^{\pm}$. 
As these two resonances are charged, they cannot be
bottomonium states.
Fits were performed using two Breit-Wigner shapes
with different masses and widths. 
For $\Upsilon$(5$S$)$\rightarrow$$\Upsilon$($n$$S$)$\pi^+$$\pi^-$ an
$S$-wave Breit-Wigner shape was taken, as the $\Upsilon$ states
have the same quantum numbers. 
For $\Upsilon$(5$S$)$\rightarrow$$h_b$($n$$P$)$\pi^+$$\pi^-$  a
$P$-wave Breit-Wigner shape was taken 
because of $\Delta$$S$=1 for the $b$$\overline{b}$ transition.
Phases $\phi_i$ were included into the fit functions
by $\exp(i\phi_i$) terms for the different signals $i$.
Table~\ref{tzb} shows the fitted masses, widths and statistical
significances. As the observation of the same two 
resonances is made in five different final states,
the total significance is very high. 
Note that, as a result from fitting with two Breit-Wigner shapes
with a relative phase, the result is that the phase in $\Upsilon$($n$$S$) 
and $h_b$($n$$P$) final states seems to be shifted by 180$^{\circ}$. 
Interestingly, the $Z_b$ is very close to the $B^{0*}$$\overline{B}^{\pm}$ threshold
and the $Z_b'$ to the $B^{0*}$$\overline{B}^{*\pm}$ threshold.
Mass differences with respect to the thresholds are only +2.6~MeV and +2.0~MeV, respectively.
Note the surprising fact, that both mass differences are positive
and thus indicate no binding energy in the system, although $\Delta$$m$
and the errors in the mass determination of 2.0~MeV and 1.5~MeV
are of the same order of magnitude, and thus some caution is still advised before drawing a conclusion.
If this behavior is confirmed, the states would be very similar to the $Z$ states of type II
in the charmonium mass regime.
Precise measurements of the pole positions, if available, would also provide a way to esitmate the weight
of the $B$$\overline{B}$ components in the wavefunctions \cite{Kang:2016ezb}.

\begin{table}
\centering
\caption{Measured masses and width of the charged $Z_b$ and $Z_b'$ states.}
\vspace{0.2cm}
\begin{tabular}{llcccc}
\hline\\[-2.5ex]
\quad & 
$\Upsilon$(1$S$)$\pi^+$$\pi^-$ &
$\Upsilon$(1$S$)(2$S$)$\pi^+$$\pi^-$ &
$\Upsilon$(1$S$)(3$S$)$\pi^+$$\pi^-$ &
$h_b$(1$P$)$\pi^+$$\pi^-$ &
$h_b$(2$P$)$\pi^+$$\pi^-$ \\
\hline\\[-2.5ex]
$m$(Z$_b$(10610)) (MeV) &
10611$\pm$4$\pm$3 &
10609$\pm$2$\pm$3 &
10608$\pm$2$\pm$3 &
10605$\pm$2$^{+3}_{-1}$ & 
10599$^{+6}_{-3}$$^{+5}_{-4}$ \\
$\Gamma$(Z$_b$(10610)) (MeV) &
22.3$\pm$7.7$^{+3.0}_{-4.0}$ & 
24.2$\pm$3.1$^{+2.0}_{-3.0}$ & 
17.6$\pm$3.0$\pm$3.0 &
11.4$^{+4.5}_{-3.9}$$^{+2.1}_{-1.2}$ &  
13.0$^{+10}_{-8}$$^{+9}_{-7}$ \\
$m$(Z$_b$(10650)) (MeV) &
10657$\pm$6$\pm$3 &
10651$\pm$2$\pm$3 &
10652$\pm$1$\pm$2 &
10654$\pm$3$^{+1}_{-2}$ & 
10651$^{+2}_{-3}$$^{+3}_{-2}$ \\
$\Gamma$(Z$_b$(10650)) (MeV) &
16.3$\pm$9.8$^{+6.0}_{-2.0}$ & 
13.3$\pm$3.3$^{+4.0}_{-3.0}$ & 
8.4$\pm$2.0$\pm$2.0 & 
20.9$^{+5.4}_{-4.7}$$^{+2.1}_{-5.7}$ & 
19$\pm$7$^{+11}_{-7}$ \\
\hline
\end{tabular}
\label{tzb}
\end{table}

An angular analysis was performed by Belle as well. 
In particular {\it (a)} the angle between the charged pion $\pi_1$ and the $e^+$ from the $\Upsilon$ decay
and {\it (b)} the angle between  the plane ($\pi_i$,$e^+$) and the plane ($\pi_1$,$\pi_2$) turned out
to be useful. For example, in case of $J^P$=$1^+$ both distributions are approximately flat, while for $1^-$, $2^+$ and $2^-$
they indicate parabolic shapes. Note that the quantum numbers $0^+$ and $0^-$ are forbidden by parity conservation.
All distributions turned out to be consistent with $J^P$=$1^+$, while all other quantum numbers were disfavoured
at typically $\geq$3$\sigma$ level. Thus, it is likely that the $Z_b$ and the $Z_b'$ carry the same spin and parity as the $X$(3872). Note that $C$-parity is only defined for neutral states, therefore only $G$-parity could be 
assigned to the $Z$ states, but has not been investigated yet. 

There are numerous attempts to explain the $Z_b$ states e.g. coupled channel effects \cite{zb_interpretation_danilkin}, 
a cusp effect \cite{zb_interpretation_bugg} or tetraquarks 
\cite{zb_interpretation_ali_hambrock} \cite{zb_interpretation_karliner_lipkin}. 
A particular attempt was made \cite{zb_interpretation_bondar} 
to explain the states along with 
the anomalous observations in $h_b$ production. 
The ansatz is to interpret the new resonances 
as $B^{0*}\overline{B}^{\pm}$ and $B^{0*}\overline{B}^{*\pm}$ 
molecular states, and to form 1$^{+}$ states based 
upon the quantum number from the angular distribution tests (see above).
The $Z_b$ and $Z_b'$ states are constructed as orthogonal states
in the limit of large $m_b$ quark mass. In this limit, 
their heavy quark spin structure should be the same as of the following
pairs,
\begin{eqnarray}
\ket{Z_b} & = & \frac{1}{\sqrt{2}}
( 0_{b\overline{b}}^- \otimes 1_{q\overline{q}}^-  \quad + \quad 
1_{b\overline{b}}^- \otimes 0_{q\overline{q}}^- ) \nonumber\\
\ket{Z_b'} & = & \frac{1}{\sqrt{2}} 
( 0_{b\overline{b}}^- \otimes 1_{q\overline{q} }^-  \quad - \quad 
1_{b\overline{b}}^- \otimes 0_{q\overline{q}}^- ). 
\label{eq:molecule-Zb}
\end{eqnarray}
Here $0^-$ and $1^-$ stand for para- and ortho-states with negative parity, and
 both $Z_b$ and $Z_b^\prime$ are assumed to have $J^P$=$1^+$.  The above representation 
 in its Fierz transformed form implies that the decays of the type $Z_b^\prime \to B^* \bar{B}^*$
are forbidden, despite being allowed by phase space.
The other consequences of this model are:

\begin{enumerate}

\item It would be expected, that the two states are degenerate in the large $m_b$ limit.
Therefore the widths should be equal, and this might explain why 
the $h_b$$\pi^+$$\pi^-$ final state is  not suppressed relative 
to $\Upsilon$($n$$S$)$\pi^+$$\pi^-$.

\item The relative phase of the coupling
of these two resonances to the ortho-bottomonium 
(i.e.\ the $b$$\overline{b}$ part) is opposite 
to that for the para-bottomonium.
This would explain the relative phase difference of 180$^{\circ}$.

\item It also would explain the similarity of the mass differences of
$m$($Z_b'$)$-$$m$($Z_b$)$\simeq$50~MeV and 
$m(B^*)-m(B)\simeq$46~MeV. 

\end{enumerate}

For further details see \cite{zb_interpretation_bondar}. These features can equally well be explained in the tetraquark
interpretation of the two $Z_b$ states, irrespective of whether they are the decay products of the $\Upsilon(10860)$, or of
the $Y_b(10890)$, as discussed in the section on theoretical models. 

The hidden $b \bar{b}$ state
$Y_b(10890)$ with  $J^{\rm P}=1^{--}$  was discovered by Belle in 2007~\cite{Abe:2007tk} in the
 process $e^+e^- \to Y_b(10890) \to (\Upsilon(1S), \Upsilon(2S), \Upsilon(3S)) \pi^+ \pi^-$ just above the
$\Upsilon(10860)$. More on this state later.

Similar to the case of the type II $Z_c$ states (see Sec.~\ref{cz3900}) it was found,
that decays to meson pairs seem to dominate, i.e.\
${\cal{B}}(Z_b(10610) \to B\bar{B}^*+c.c.)/{\cal{B}}(Z_b(10610) \to {\rm bottomonium}) = 4.76\pm0.64\pm0.75$
and
${\cal{B}}(Z_b(10650) \to B^*\bar{B}^*)/{\cal{B}}(Z_b(10650) \to {\rm bottomonium}) = 2.40\pm0.44\pm0.50$ \cite{Garmash:2015rfd}.
This may indicate molecular contributions in the wave function.
For the $Z_b$(10610), a neutral partner has been observed, pointing to an isospin triplet
similar to the type II $Z_c$ states (see Sec.~\ref{cz3900}).
Principally, in case of sufficiently precise experimental mass resolution, this would
allow tests of different quark contents, e.g.\
$[b\overline{b}u\overline{d}]$ or
$[b\overline{b}d\overline{u}]$ for the charged $Z_b$ and
$[b\overline{b}u\overline{u}]$ or
$[b\overline{b}d\overline{d}]$ for the neutral $Z_b$.
However, so far the measured masses are compatible within the errors,
with 10609$\pm$4$\pm$4~MeV for the neutral $Z_b$ \cite{Krokovny:2013mgx}
and 10607.2$\pm$2.0 for the charged $Z_b$ \cite{Belle:2011aa}.
{\color{revcolor} The isospin $I$=1 assignment has implications for the molecular interpretation:
one pion exchange is not allowed by OZI rule and the interaction should be based upon e.g.\ 
two light meson exchange or vector meson exchange \cite{Dias:2014pva}.}


%% file: Sub-sec-4140.tex
\subsection{\boldmath{Another curious system: $\jpsi\phi$ resonances}}

\label{4140}

Charmonium states are composed of $c\overline{c}$ quarks. The charged states described above decaying into a $\jpsi\pi^{\pm}$ or $\psi'\pi^{\pm}$ cannot be composed of only $c\overline{c}$. Resonant states decaying into $\jpsi\phi$ could be charmonium states or have a more exotic quark content. In 2009 the CDF collaboration found  58$\pm$10 signal events in the channel $B^-\to\jpsi\phi K^-$ \cite{Aaltonen:2009tz}. The $\jpsi\phi$ invariant mass spectrum showed a spike near 4140~MeV, close to the threshold of about 4116~MeV, with a very narrow width of about 15~MeV. 

\begin{figure}[b!]
\begin{center}
\vskip -2.6cm
    \includegraphics[width=0.6\textwidth]{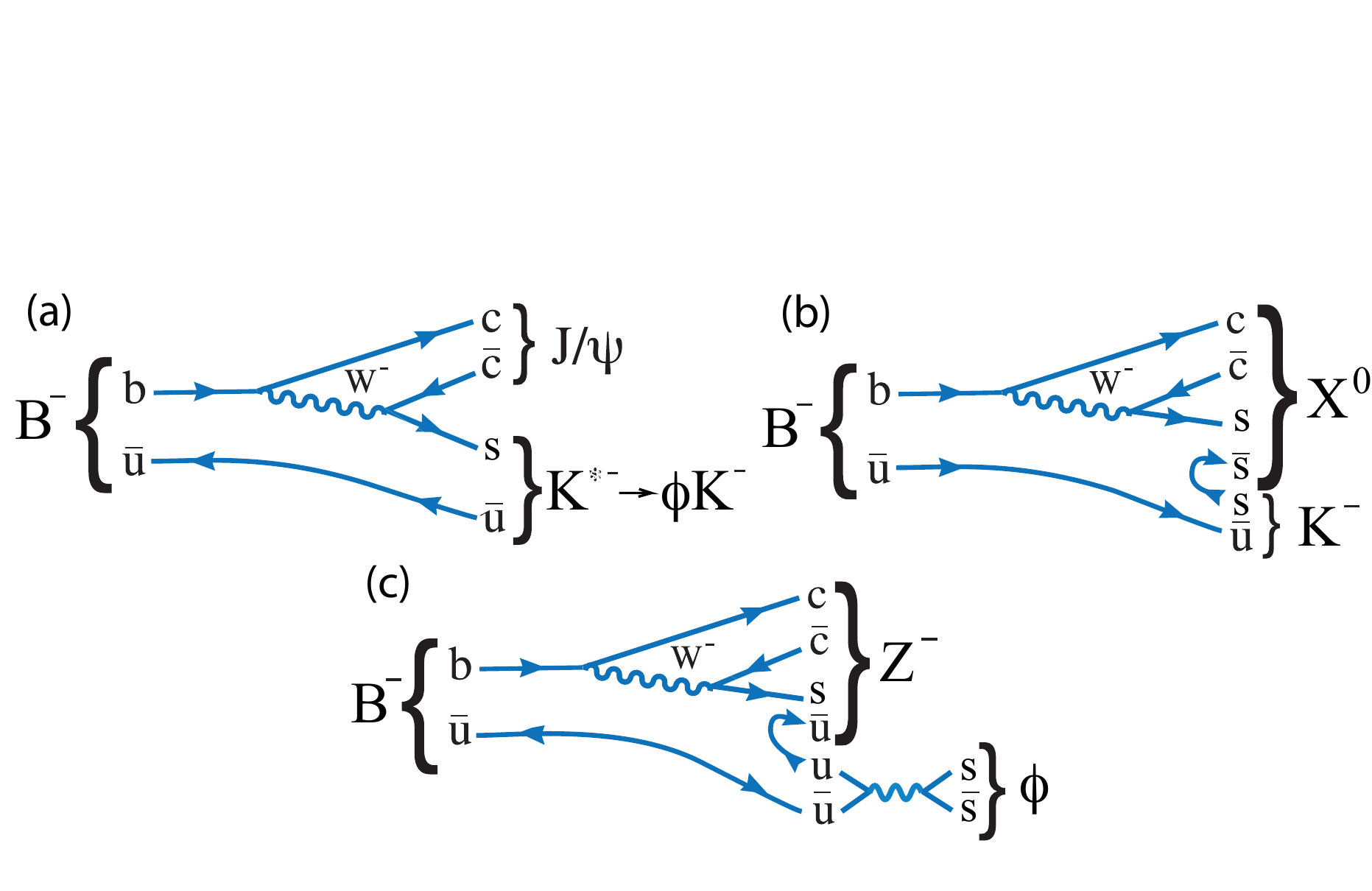} \end{center}
\vskip -0.7cm
\caption{Diagrams for $B^-\to\jpsi\phi K^-$ proceeding via three possible intermediate states. a) The $\phi K^-$ states are produced by an intermediate $K^{*-}$. (b) Neutral exotic states are produced by the merger of the $c\overline{c}s\overline{s}$ quarks. (c) There is an exotic $Z^-$ resonance formed from $c\overline{c}s\overline{u}$ quarks. For this to happen the $u\overline{u}$ must make a $\phi$ meson, which dominantly has an $s\overline{s}$ wavefunction.}
\label{Feyn-4140-pub}
\end{figure}

Feynman diagrams for the $B^-$ decay into $\jpsi\phi K^-$ are shown in Fig.~\ref{Feyn-4140-pub}. One possibility is that the decay reflects simply the well known process $B^-\to\jpsi K^{*-}$ with  $K^{*-}\to \phi K^-$ instead of $K^-\pi^0$ or $K^0\pi^-$. Another, implied by the narrowness of the observed state, is that the $c\overline{c}s\overline{s}$ produced in the decay form an object called the $X(4140)$ plus an additional $K^-$. Yet another process involves the production of an exotic $Z$ like state that decays into $\jpsi K^-$; here the $u\overline{u}$ produced in the decay must transform into a $\phi$, which makes this process very unlikely.

\begin{figure}[t!]
\begin{center}
    \includegraphics[width=0.98\textwidth]{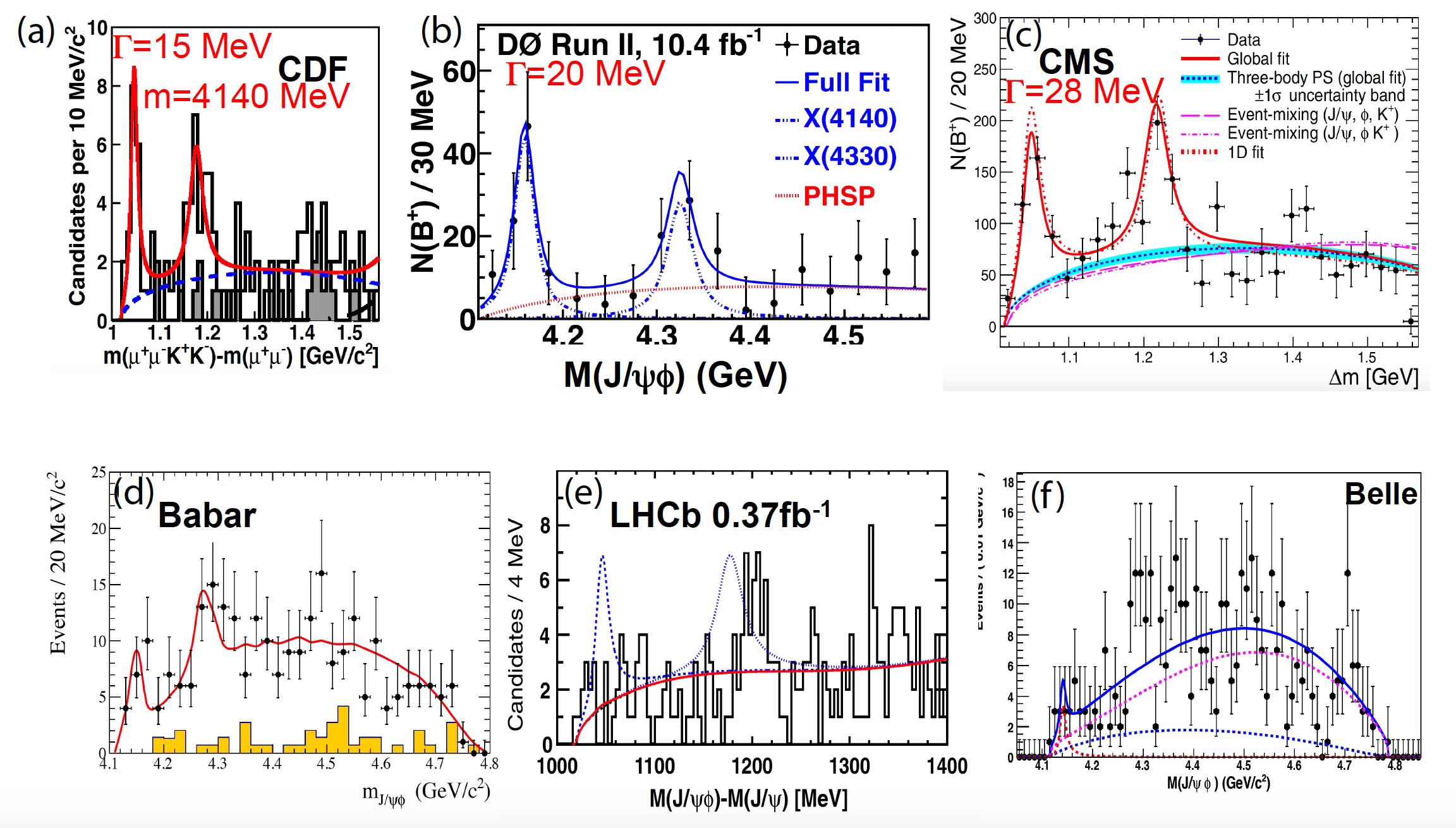} \end{center}
\vskip -.5cm
\caption{Experimentally determined $\jpsi\phi$ mass spectra for  $B^-\to\jpsi\phi K^-$ decay. From (a) CDF \cite{Aaltonen:2009tz}, (b) D0 \cite{Abazov:2013xda},  (c)  CMS \cite{Chatrchyan:2013dma}, (d) \BaBar \cite{Lees:2014lra}, (e) LHCb \cite{Aaij:2012pz,Aaij:2016iza}, and (f) Belle \cite{ChengPing:2009vu}. The graphs in the variable $\Delta m$ can be compared to the others by adding  the $\jpsi$ mass of  3097~MeV.}
\label{X4140-all}
\end{figure}

The CDF data can be seen in Fig.~\ref{X4140-all}(a). The fit is to an $S$-wave relativistic Breit-Wigner convolved with the experimental mass resolution for the peak near 4140~MeV plus a three-body phase space background. The significance of the peak is 5$\sigma$. Another peak near 4.3~GeV is also present in the fit and gives a 3.1$\sigma$ significance. 

Other experiments followed. D0 observed a similar structure at 4159~MeV with a somewhat larger width of 19.9~MeV \cite{Abazov:2013xda}. (All mass and width values are listed in Table~\ref{tab:jpsiphih1}.) The second structure was seen but the statistics were insufficient to determine either the mass or width [Fig.~\ref{X4140-all}(b)]. CMS followed with clear observations of both structures with the lower mass peak at 4148~MeV and a width of 28~MeV \cite{Chatrchyan:2013dma};  they claimed evidence but not observation of the higher mass structure [Fig.~\ref{X4140-all}(c)].  There were also experiments that did not observe the peak near 4140~MeV including Babar [Fig.~\ref{X4140-all}(d)] \cite{Lees:2014lra}, LHCb with a 0.37~pb$^{-1}$ sample  [Fig.~\ref{X4140-all}(e)] \cite{Aaij:2012pz}, and Belle  [Fig.~\ref{X4140-all}(f)] \cite{ChengPing:2009vu}. While the $X(4140)$ state remained controversial, the data from the different experiments seems to have a significant enhancement near 4300~MeV, although this was not emphasized, and indeed no statistically significant observation in any experiment had been made.

\begin{table}[htp]
\centering
\caption{History of  $X(4140)$ and $X(4274)$ observations in $B^-\to\jpsi\phi K^-$ prior to 2016.  The term ``fix" means the values were not taken from the data but fixed to previous measurements. The number of standard deviation significances ($\sigma$) are given when quoted by the experiments. The u.l. in the LHCb row for $X(4140)$ indicates an upper limit that is in contradiction with the CDF result by 2.4$\sigma$.}
\vspace{0.2cm}
\begin{tabular}{llrccccc}
\hline
\multicolumn{8}{c}{$X(4140)$}\\
Year & Exp. & fb$^{-1}$& \# $B^-$& Mass (MeV) & $\Gamma$  (MeV) &$\sigma$\\
\hline
\rule{0pt}{2.5ex}2008 & CDF \cite{Aaltonen:2009tz}& 2.7 & ~58$\pm10$ &$4143.0\pm 2.9\pm1.2$ & $11.7^{+8.3}_{-5.0}\pm 3.7$  & 3.8 \\
2009 & Belle \cite{ChengPing:2009vu}& 840 & 325$\pm21$ &4143 fix & 11.7 fix & 1.9 \\
2011 & CDF \cite{Aaltonen:2011at} & 6.0 & 115$\pm12$ &$4143.4^{+2.9}_{-3.0}\pm0.6$ & $15.3^{+10.4}_{-6.1}\pm 2.5$  & 5.0 \\
2011 & LHCb \cite{Aaij:2012pz}& 0.37 & 346$\pm20$ &4143.4 fix& 15.3 fix  & u.l. \\
2013 & CMS \cite{Chatrchyan:2013dma}& 5.2 &2480$\pm160$ &$4148.0\pm 2.4\pm 6.3$ & $28.0^{+15}_{-11}\pm 19$  & 5.0 \\
2013 & D0 \cite{Abazov:2013xda}& 10.4 & 215$\pm37$ &$4159.0\pm 4.3 \pm 6.6$ & $19.9\pm 12.6^{+1.0}_{-8.0}$ & 3.1 \\
2014 & \BaBar \cite{Lees:2014lra}& 422 &189$\pm14$ &4143.4 fix & 15.3 fix  & 1.6 \\
\hline 
\multicolumn{8}{c}{$X(4274)$}\\
Year & Exp. & fb$^{-1}$& \# $B^-$& Mass (MeV)& $\Gamma$  (MeV) &$\sigma$ &\\
\hline
\rule{0pt}{2.5ex}2008 & CDF \cite{Aaltonen:2009tz}& 2.7 & ~58$\pm10$ &&& \\
2009 & Belle \cite{ChengPing:2009vu} & 840 & 325$\pm21$  &&&\\
2011 & CDF \cite{Aaltonen:2011at} & 6.0 & 115$\pm12$  &$4274.4^{+8.4}_{-6.7}\pm1.9$ &$32.3^{+21.9}_{-15.3}\pm7.6$&3.1\\
2011 & LHCb \cite{Aaij:2012pz}& 0.37 & 346$\pm20$&4274.4 fix&32.3 fix&\\
2013 & CMS\cite{Chatrchyan:2013dma} & 5.2 &2480$\pm160$ &$4313.8\pm 5.3\pm 7.3$ &$38^{+30}_{-15}\pm 16$&\\
2013 & D0 \cite{Abazov:2013xda}& 10.4 & 215$\pm37$ &4328.5&30 fix&\\
2014 & \BaBar \cite{Lees:2014lra}& 422 &189$\pm14$  &4274.4 fix&32.2 fix&1.2\\
\hline 
\end{tabular}
\label{tab:jpsiphih1}
\end{table}

\begin{figure}[b!]
\begin{center}
\vskip -0.4cm
\includegraphics[width=0.5\textwidth]{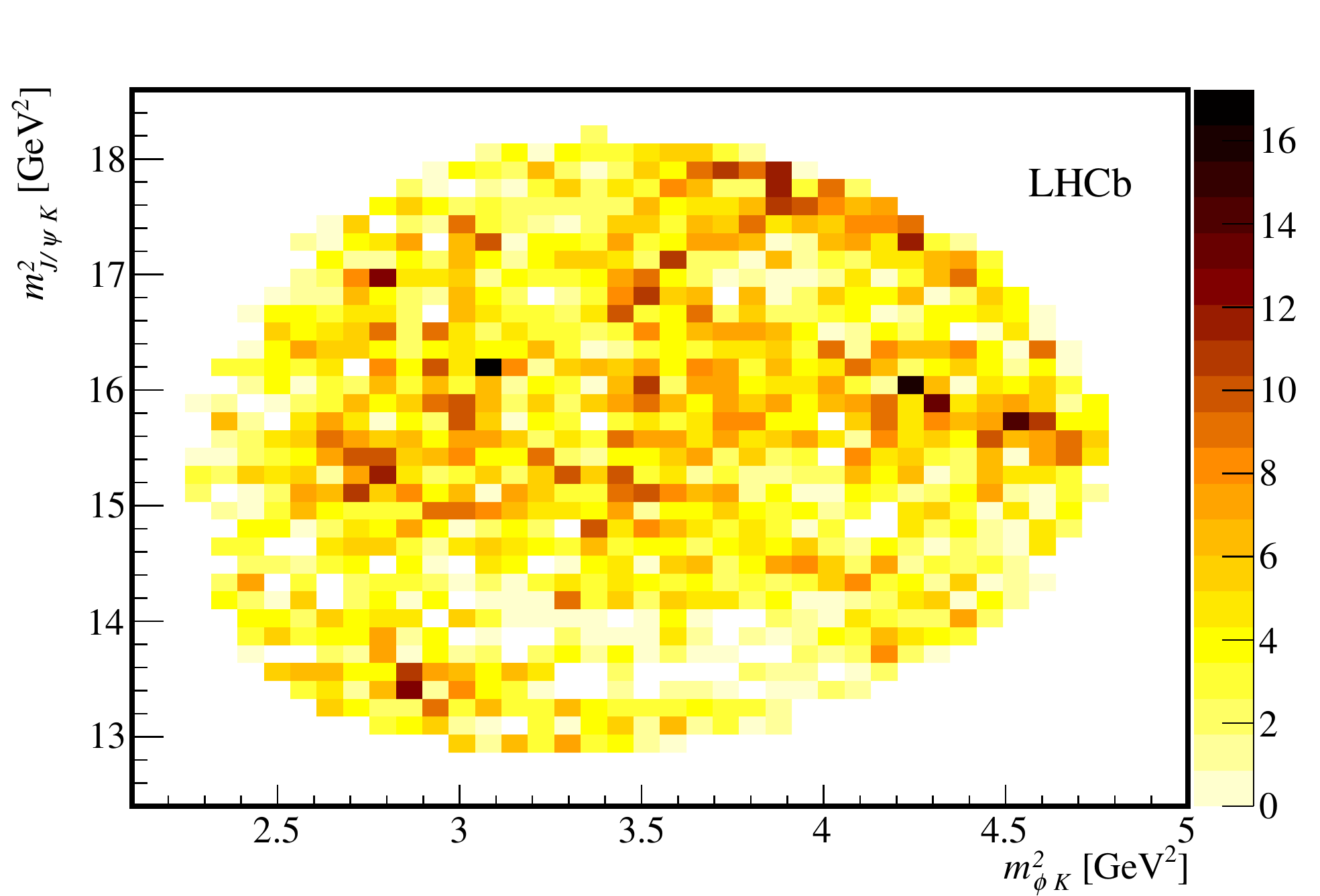}
 \end{center}
\vskip -0.4cm
\caption{Dalitz plot for $B^-\to\jpsi\phi K^-$ decays, background subtracted and efficiency corrected (from \cite{Aaij:2016nsc}).}
\label{Dalitzjpsiphi}
\end{figure}

In 2016 LHCb performed a full amplitude analysis of this final state \cite{Aaij:2016nsc}. Since the final state is three-body, a Dalitz plot can be constructed and is shown in Fig.~\ref{Dalitzjpsiphi} \cite{Dalitz:1953cp}.\footnote{Dalitz plots for three spinless particles in the final state reveal directly the structure of the matrix element. For non-zero spin they are also instructive.}  Note that there are structures evident in both the $\jpsi\phi$ and $\phi K^-$ masses. The amplitude formalism used is very similar to that used for analysis of the pentaquark states described in Section~\ref{sec:LcKp} \cite{Aaij:2015tga}. For the pentaquark states the first decay mode that was analyzed was $\Lb\to\jpsi K^- p$, a similar three-body final state to the one considered here with the $\jpsi$ decay into muons being a common feature. Of course the spins of the particles are different and the amplitudes must reflect this, giving different decay angular distributions for the different resonant final states. Here all three amplitudes shown in Fig.~\ref{Feyn-4140-pub} are allowed to interfere. (It turns out the one describing the possible $Z^-$ has no effect on the results and is used only to set part of the systematic uncertainty.)

While resonant structures in most analyses have relied on seeing Breit-Wigner shaped peaks in the mass spectrum, the absence of such evidence does not mean that several resonances are not present, merely that they can be wide and therefore washed out. Conversely, there have been arguments that seeing such peaks may be evidence of the rescattering of intermediate particles rather than resonance structures. We will return to this discussion later. 

\begin{figure}[t!]
\begin{center}
\includegraphics[width=0.45\textwidth]{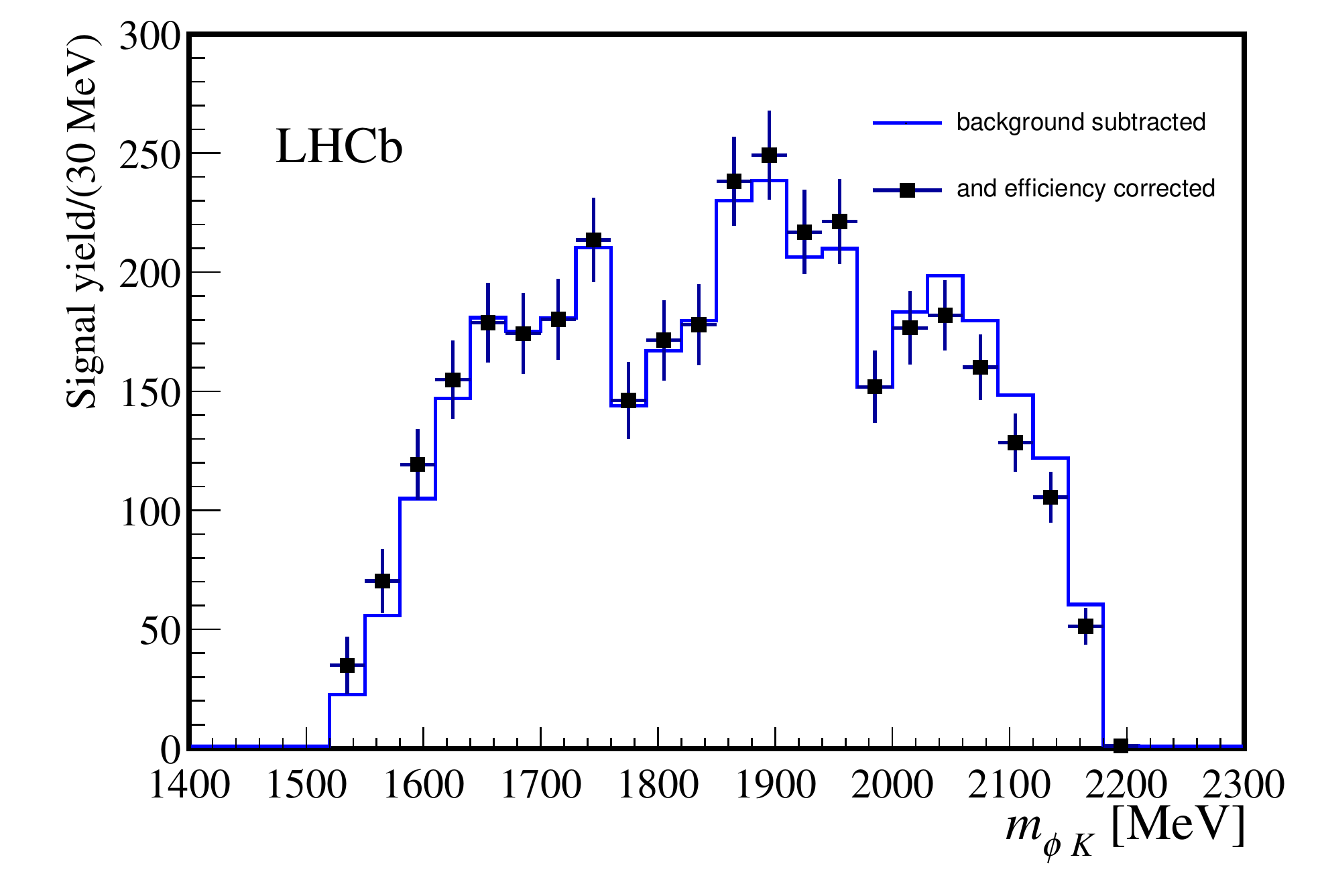}\includegraphics[width=0.45\textwidth]{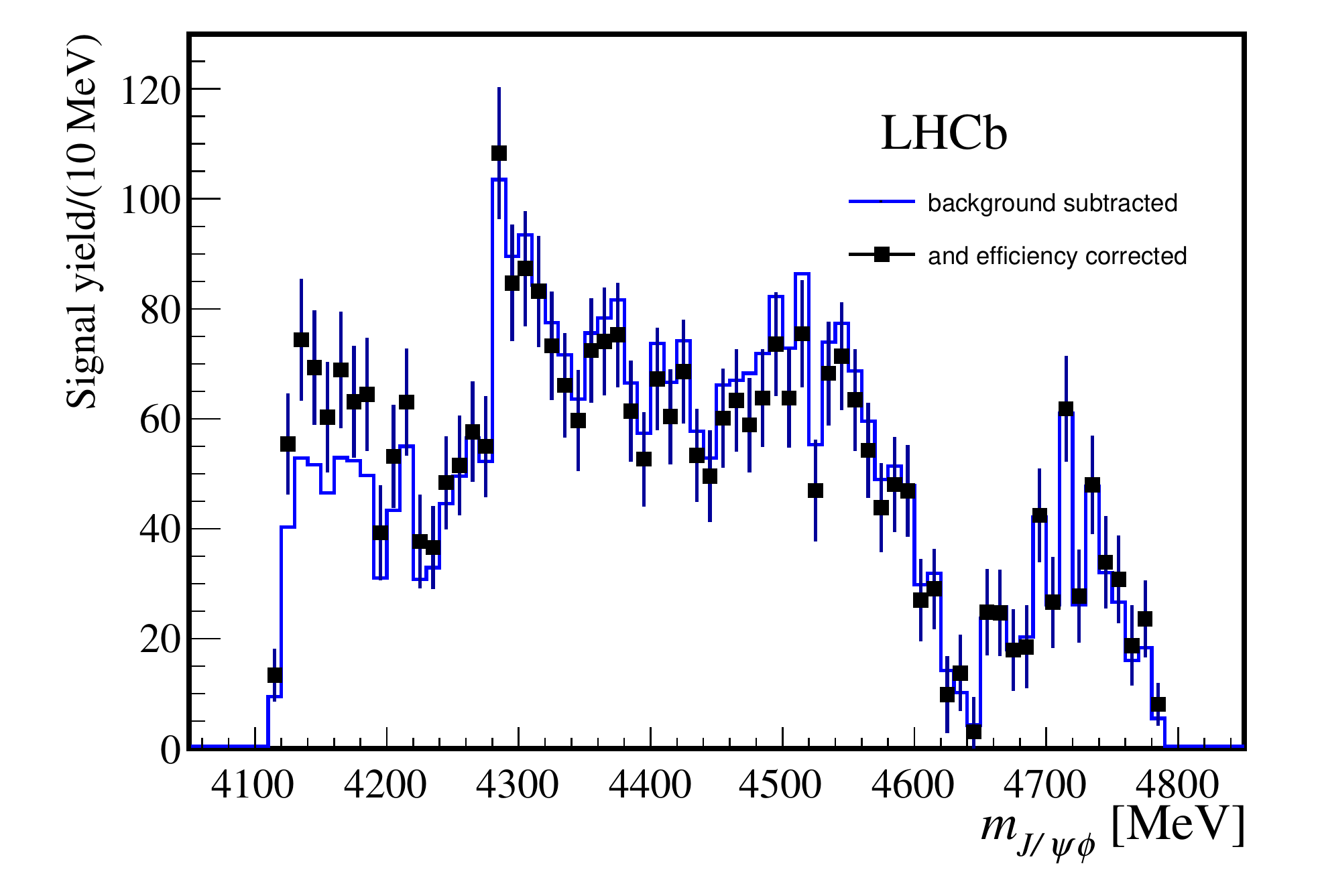}
 \end{center}
\vskip -0.7cm
\caption{Invariant mass distributions of  $\phi K^-$ (left) and $\jpsi\phi$ (right), background subtracted and efficiency corrected (from \cite{Aaij:2016nsc}). The data are shown once with background subtraction only and then again efficiency corrected, so as to emphasize the relatively small nature of the efficiency corrections. }
\label{m-jpsiphi}
\end{figure}

\begin{figure}[h]
\begin{center}
\includegraphics[width=0.5\textwidth]{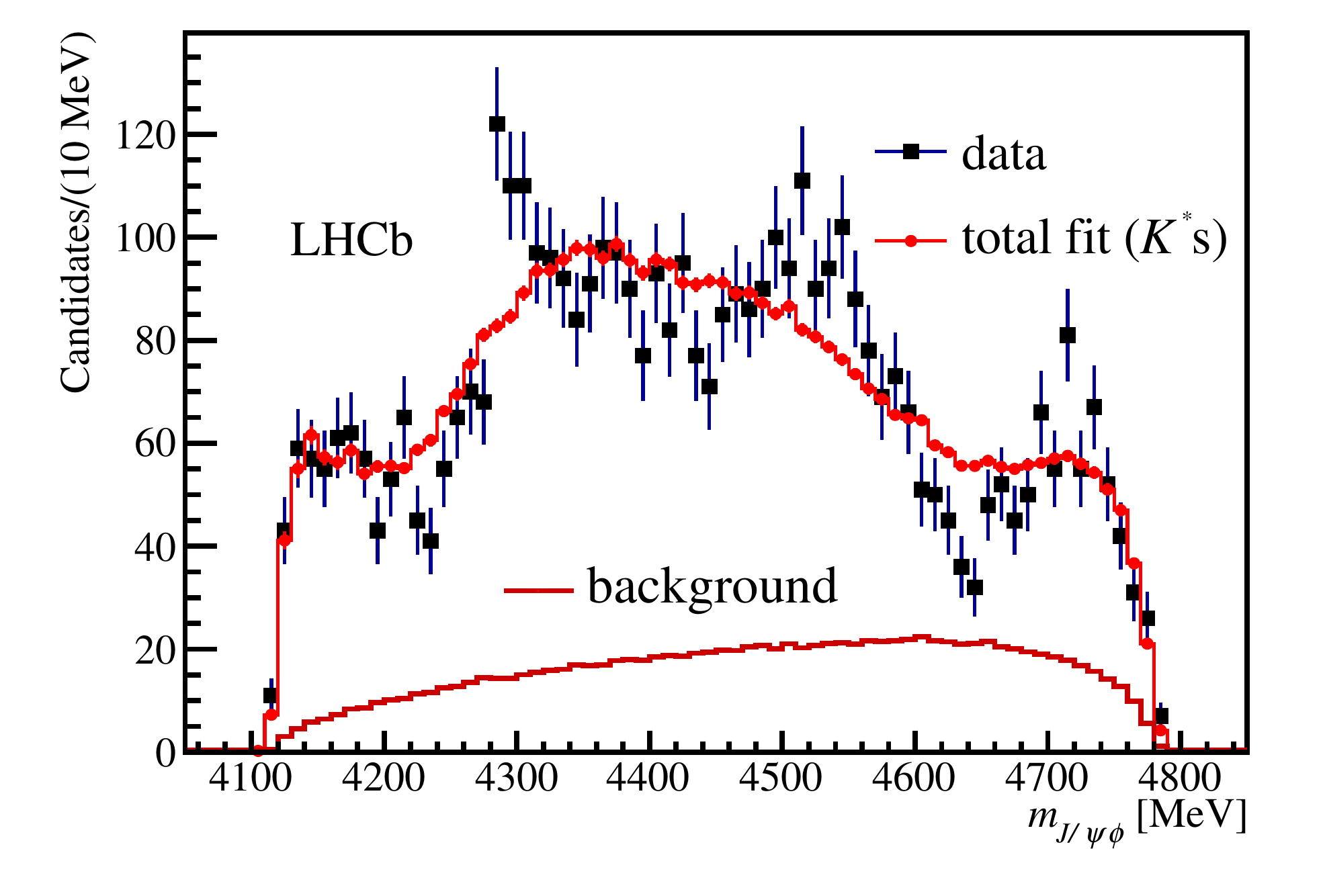}
 \end{center}
\vskip -1.0cm
\caption{Best fit of $\jpsi\phi$ mass distribution using only $K^{*-}$ resonances (from \cite{Aaij:2016nsc}).}
\label{Fit-bad-jpsiphi}
\end{figure}

The efficiency corrected and background subtracted projections of the Dalitz plot are shown in Fig.~\ref{m-jpsiphi}. The $\phi K^-$ mass spectra seems devoid of resonant activity, while the $\jpsi\phi$ shows evidence of structures. 
LHCb first tried to fit the data with only $K^{*-}$ contributions. While the $\phi K^-$ mass distribution was adequately described, the peaks in the $\jpsi\phi$ distribution could not be satisfactorily reproduced (see Fig.~\ref{Fit-bad-jpsiphi}). There is always some art in Dalitz plot analyses in choosing the number of such resonances to use. Generally those that are not significant are dropped.\footnote{At least one author has given some thought to these matters \cite{Weisser:2016cnc}.}  The amplitude analysis performed by LHCb reveals the presence of several resonant $K^{*-}\to \phi K^-$ states as well as four $\jpsi\phi$ structures.

\begin{figure}[hbt]
\begin{center}
\includegraphics[width=0.99\textwidth]{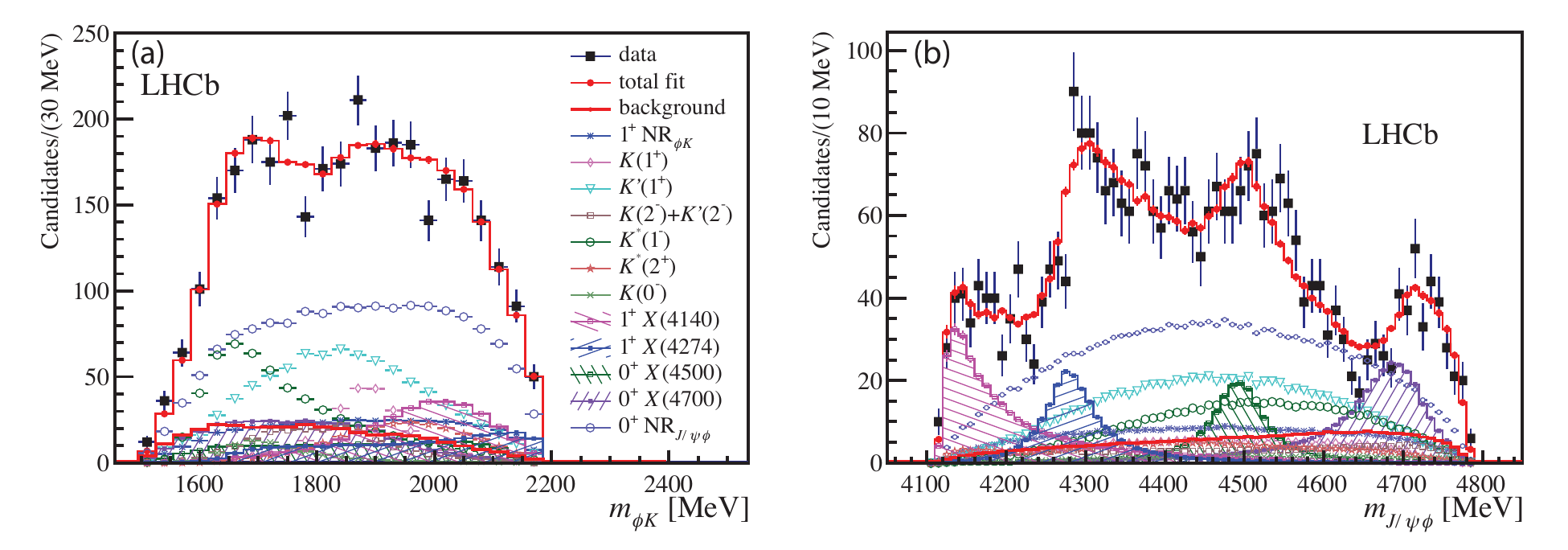}
 \end{center}
\vskip -0.5cm
\caption{Fit projections of the mass distributions for  (a) $\phi K^-$ and  (b) $\jpsi\phi$ (right). The separate resonance contributions are also shown (from \cite{Aaij:2016nsc}).}
\label{Fit-phiK-jpsiphi}
\end{figure}

\begin{figure}[hbt]
\begin{center}
\includegraphics[width=0.45\textwidth]{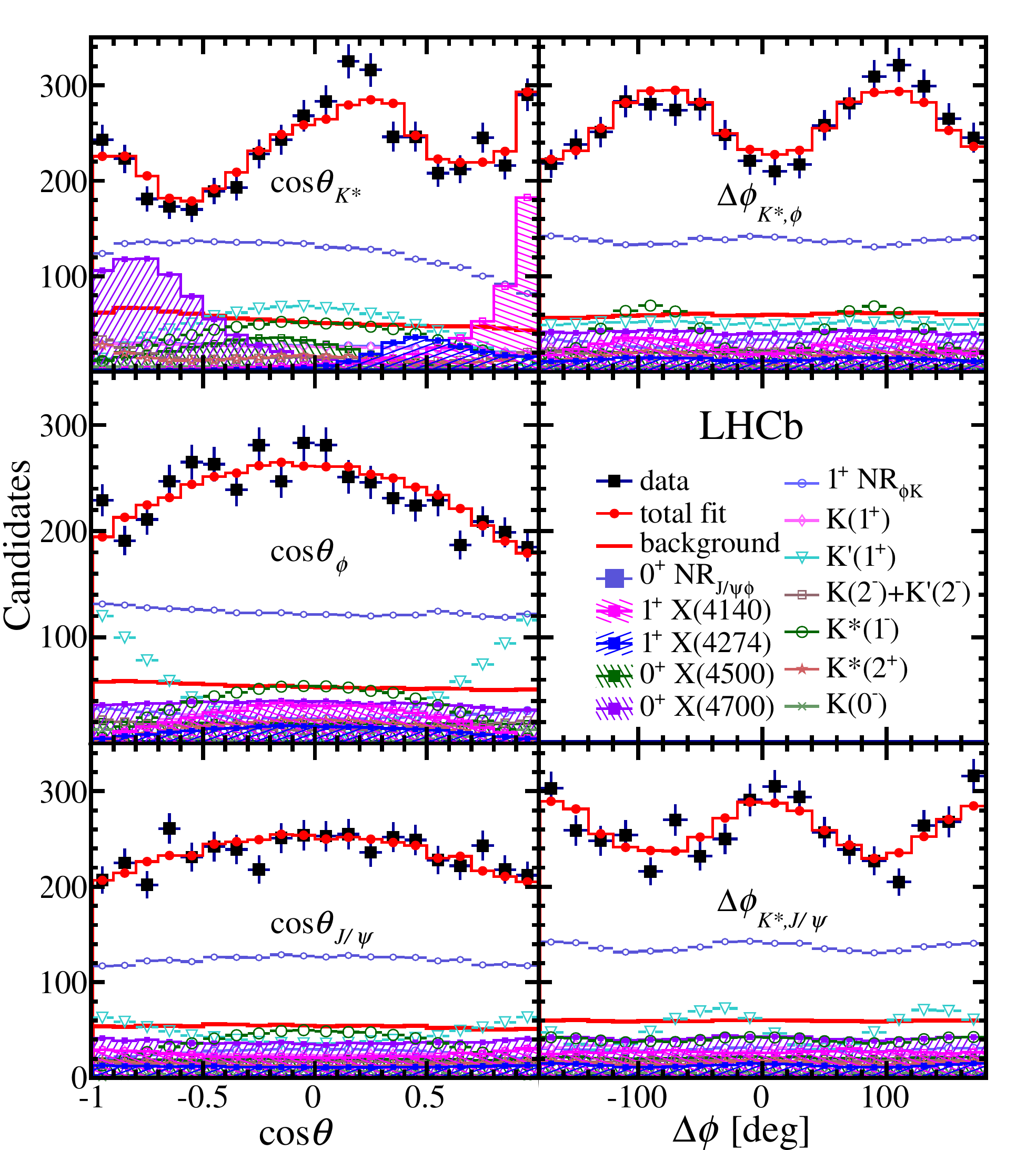}\includegraphics[width=0.45\textwidth]{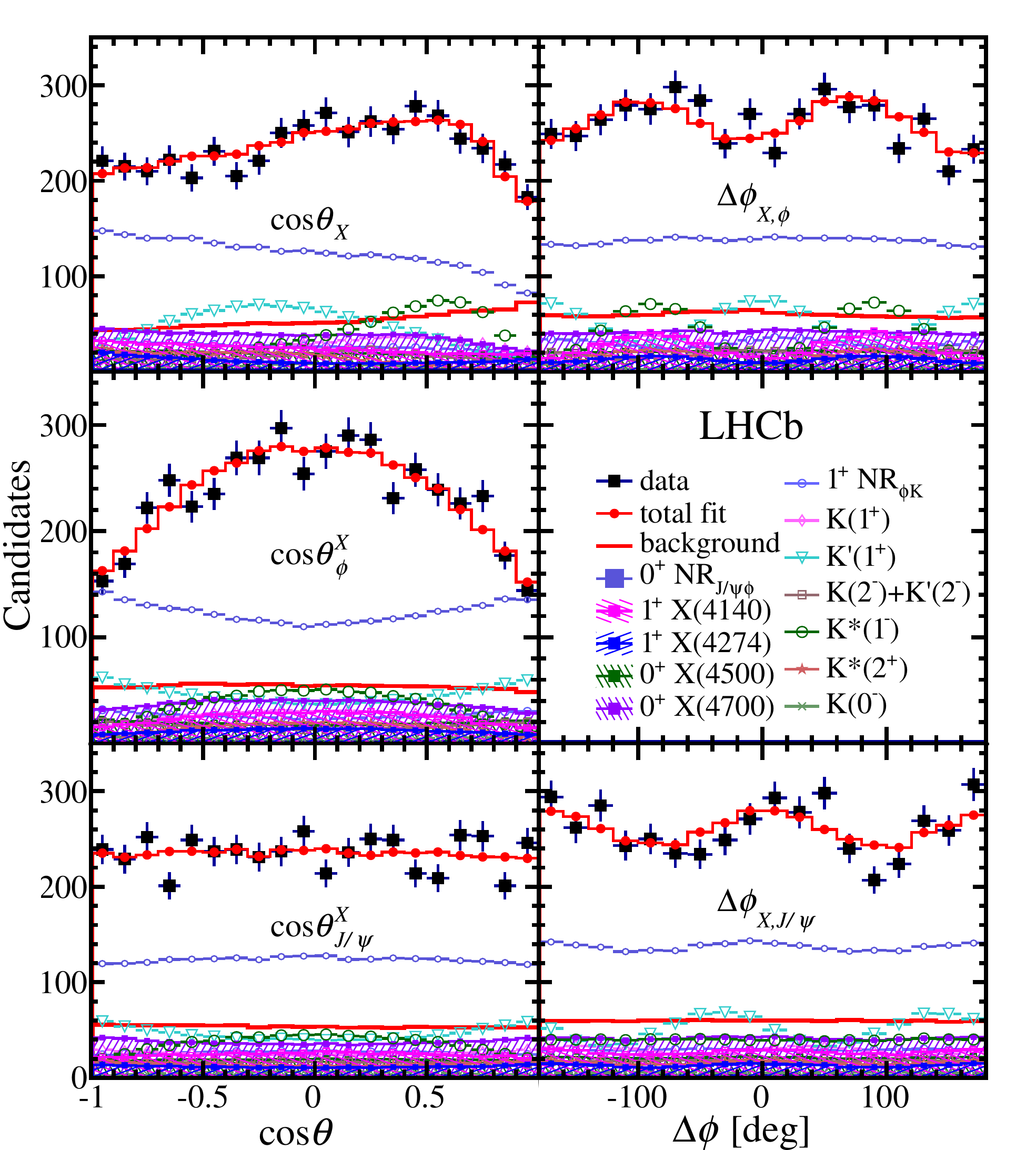}
 \end{center}
\caption{Fit projections of the angular distributions from the $K^{*-}$ decay sequence (left) and $\jpsi\phi$ chain (right). The separate resonance contributions are also shown (from \cite{Aaij:2016nsc}).}
\label{Fit-Ksangles}
\end{figure}

The projections of the fit are shown in Fig. ~\ref{Fit-phiK-jpsiphi}. The amplitude analysis allows not only their masses and widths to be determined, but also their $J^P$.   Table~\ref{tab:jpsiphires} gives the five putative resonant structures found in the fit, while Table~\ref{tab:phiK} lists the $K^{*-}$ resonances. Figure~\ref{Fit-Ksangles} shows the fit projections in terms of the angular variables. (While a complete discussion of the $K^*$ states found here is outside of the scope of this review, these results add importantly to our knowledge of these resonances.) 

{\renewcommand{\arraystretch}{1.2}
\begin{table}[htp]
\centering
\caption{Properties of resonant $\jpsi\phi$ states found in the LHCb amplitude analysis, and the non-resonant (NR) fraction.}
\vspace{0.2cm}
\begin{tabular}{lcccrc}
\hline
\rule{0pt}{2.5ex}Particle & $J^P$ & $\sigma$ & Mass & Width~~~~ & Fit~~~~~~~~~\\
               &            &                   & (MeV) & (MeV)~~~~ & Fraction(\%)~~\\
\hline
\rule{0pt}{2.5ex}$X(4140)$ & $1^+$ & 8.4 &\,\,$4146\pm 4.5^{+4.6}_{-2.8}$&$83\pm21^{+21}_{-14}$&$13.0\pm3.2^{+4.8}_{-2.0}$\\
$X(4274)$ & $1^+$ & 6.0& $4273.3\pm 8.3^{+17.2}_{-~3.6}$&$56\pm11^{+~8}_{-11}$&~$7.1\pm2.5^{+3.5}_{-2.4}$\\
 $X(4500)$ & $0^+$ & 5.6 & $4506\pm 11^{+12}_{-15}$&$92\pm21^{+21}_{-14}$&~\,$6.6\pm2.4^{+3.5}_{-2.3}$\\
 $X(4700)$ & $0^+$ & 5.6 & $4704\pm 10^{+14}_{-24}$&$120\pm31^{+42}_{-33}$&\!\!\!\!$12\pm5^{+9}_{-5}$\\
NR & $0^+$&6.4&&&$\,46\pm11^{+11}_{-21}$\\
\hline 
\end{tabular}
\label{tab:jpsiphires}
\end{table}

{\renewcommand{\arraystretch}{1.0}

 Finding four states was quite unexpected and surprising. The lowest mass state near 4140~MeV is consistent with the mass of the state claimed by CDF, but its width is substantially higher. In fact, a narrow mass peak is not present in the LHCb data, showing inconsistency with the findings of CDF, D0 and CMS. This state is revealed in the LHCb data only through the amplitude analysis. The state at 4274~MeV was put on a firm foundation and the data are consistent among the experiments. The higher mass states were not seen by other experiments; this is perhaps an acceptance issue. 
 
Another possibility is that these are not resonance states but a manifestation of rescattering.  The basic idea is that a basic process such as $B^-\to D_s^{*-}D_s^+K^-$ occurs and the $D_s^{*-}D_s^+$ rescatter into $\jpsi\phi$. Since the mass of the $D_s^{*-}$ plus $D_s^+$ of 4080~MeV is just below 4140~MeV, such a cusp could cause behavior similar to that of a resonance \cite{Swanson:2015bsa,Liu:2009iw}. LHCb did fit their data to a variant of the Swanson model for the 4140~MeV state  \cite{Swanson:2015bsa}. The fit was marginally better than for a Breit-Wigner resonance. The other states were found not be describable by this model.

Prior to the LHCb analysis there had been other proposals explaining the $X(4140)$ including a molecular
state \cite{Karliner:2016ith,Chen:2015fdn,Wang:2009ry,Liu:2009pu,Zhang:2009st,Ding:2009vd,Albuquerque:2009ak,Branz:2009yt,Liu:2009ei}, or a tetraquark state \cite{Lebed:2016yvr,Anisovich:2015caa,Wang:2015pea,Drenska:2009cd, Stancu:2009ka}, hadrocharmonium \cite{Dubynskiy:2008mq}, or a hybrid state \cite{ Mahajan:2009pj,Wang:2009ue}. It remains to be seen which of these can explain the data. We note, however, that molecular states or re-scattering models cannot account for the $1^+$ nature of the $X(4274)$.  

It is difficult to understand the observed ($1^+$, $1^+$, $0^+$, $0^+$) pattern seen here and is suggestive that not all of them have the same nature. Many explanations of these states have appeared subsequent to the LHCb publication.  One interesting suggestion starts with taking the underlying structure of these states as $c\overline{c}s\overline{s}$ tetraquarks, that these can be further broken down into $cs$ and $\overline{c}\overline{s}$ diquarks, and then they assume that the spin-spin interactions inside
the diquark are dominant  \cite{Maiani:2016wlq}.  They predict the masses of states with this tetraquark content depending on the $J^{PC}$ of the resonance. They suggest that the state near 4274~MeV is not $1^+$ but the superposition of a $0^{++}$ state with a $2^{++}$ state. Further analysis will be needed to see if this is required by the data. In the hadrocharmonium model \cite{Dubynskiy:2008mq}, the $X(4140)$ can be described as a $\jpsi \phi$ state and would correctly give the observed $1^+$ in $S$-wave. Although the $s \overline{s}$ component does not resemble a light quark cloud (see Sec.~\ref{sec:Theoretical-models-for-tetraquarks} below), the mass of the $\phi$ is still within the range of the model. However, it would be a virtual state, as the mass is $\simeq$20~MeV above the $\jpsi \phi$ threshold at 4116~MeV.

Another state near 4350~MeV was found by Belle in $\gamma \gamma \to \phi\jpsi$ collisions with 3.2$\sigma$ significance.   \cite{Shen:2009vs}. It needs confirmation. Clearly it is not seen in the $B^-$ decay investigated by LHCb. 

\begin{table}[htp]
\centering
\caption{Properties of resonant $K^{*-}$ states found in the LHCb amplitude analysis, and the non-resonant (NR) fraction taken as a $1^+$ contribution. Predicted masses are from Godfrey and Isgur \cite{Godfrey:1985xj} The fraction of longitudinal and transverse polarizations are available in \cite{Aaij:2016nsc}.}
\vspace{0.2cm}
\begin{tabular}{lcccccc}
\hline    
             &                                        &                   & \multicolumn{3}{c}{Measured} &{Predicted}\\
Particle & $n^{2S+1}L_J$ & $\sigma$ & Mass & Width~~~~ & Fit~~~~~~~~~&Mass \\
               &                              &                          &(MeV) & (MeV)~~~~ & Fraction(\%)~~&(MeV)  \\
\hline
\rule{0pt}{2.5ex}$K(1^+)$  & $2^1P_1$ & 7.6 & $1793\pm59^{+153}_{-101}$&$365\pm 157^{+157}_{-215}$ & $12\pm 10^{+17}_{-~6}$&1900\\
$K'(1^+)$  & $2^3P_1$ & 1.9 & $1968\pm65^{+~\,70}_{-172}$&$396\pm 170^{+174}_{-178}$ & $23\pm 20^{+31}_{-29}$&1930\\
NR &  & &&&$42\pm8^{+5}_{-9}$&\\
$K(2^-)$  & $1^1D_2$ & 5.0 & $1777\pm35^{+122}_{-~77}$&$217\pm 116^{+221}_{-154}$ & $11\pm 3^{+2\dagger}_{-5}$&1780\\
$K'(2^-)$  & $1^1D_2$ & 3.0 & \!\!$1853\pm27^{+18}_{-35}$&$167\pm 58^{+83}_{-72}$ & {$\dagger$Shared with $K(2^-)$}&1810\\
$K^*(1^-)$  & $1^3D_2$ & 8.5 & $1722\pm20^{+~\,33}_{-109}$&$354\pm 75^{+140}_{-181}$ & $6.7\pm1.9^{+3.2}_{-3.9}$&1780\\
$K*(2^+)$  & $2^3P_2$ & 5.4 & $2073\pm94^{+245}_{-240}$&$678\pm 311^{+1153}_{-~\,559}$ & $2.9\pm0.8^{+1.7}_{-0.7}$&1940\\
$K(0^-)$  & $3^1S_0$ & 3.5 & $1874\pm43^{+~\,59}_{-115}$&$168\pm 90^{+280}_{-104}$ & $2.6\pm1.1^{+2.3}_{-1.8}$&2020\\
\hline 
\end{tabular}
\label{tab:phiK}
\end{table}

%% file: Sec-Theoretical-models-for-tetraquarks-rev.tex
\section{Theoretical models for tetraquarks}
\label{sec:Theoretical-models-for-tetraquarks}
The exotic particles discussed in the preceding sections have either a hidden $c\bar{c}$ or a
$b\bar{b}$ pair in their Fock-space. No exotic hadron with a single $c$ or a single $b$ quark has been 
found yet. No doubly-charged exotic hadron has been seen so far either.   
However, the nonet of the lightest scalar mesons in the Particle Data Group,
called $\sigma$ or $f_0(500)$, $\kappa(800)$, $a_0(980)$ and $f_0(980)$, 
have been argued to be tetraquark candidates due to their inverted mass hierarchy compared to the 
light pseudoscalar and vector hadrons,
with the $(I=1)$ $f_0(980)$ heavier than the $I=1/2$ $\kappa(800)$, and the isosinglet $f_0(500)$ being 
the lightest \cite{Hooft:2008we,Fariborz:2008bd}. The  dynamics of the light scalar mesons is influenced by the
infrared sector of QCD. In particular, instanton effects play a crucial role and this was recognized already in the early
QCD epoch \cite{Belavin:1975fg,Shifman:1979uw,tHooft:1986ooh,tHooft:1999cta,Black:1999yz}. 
We shall not discuss the light scalar nonet here and will restrict our discussion to the charmonium-like
and bottomonium-like exotics in which, due to the heavy quark constituents, instanton-induced effects are anticipated
to be small. For a detailed review of the scalar nonet, covering various dynamical frameworks, such as dispersion
relations and chiral lagrangians, we refer to the reviews \cite{Pelaez:2015qba,Oller:2000ma}. 
Several explicit kinematic and dynamical mechanisms have been devised to work out the spectroscopy of
the quarkonia-like exotics, which we have already mentioned earlier.
 They go by the names: cusps, hadroquarkonia, hybrids, hadron molecules, and
compact diquarks. Theoretical details are, however, still rather sketchy, and an underlying organizational
principle is either lacking or not yet properly formulated. In particular, production cross-sections of the
multiquark hadrons are not yet calculable. Despite this, some characteristic features specific to each
of the theoretical framework can be defined and will be shortly discussed below.   

\subsection{Tetraquarks as cusps}
The cusp approach is used
to explain, for example, the origin of the charged states $Z_c(3900)[D\bar{D}^*]$, $Z_c(4025)[D^*\bar{D}^* ]$, 
$Z_b(1610) [B\bar{B}^* ]$,  and $Z_b(10650) [B^*\bar{B}^* ]$, as their masses lie just above the indicated
thresholds.  For tetraquarks~\cite{Swanson:2015bsa,Swanson:2014tra}, it is
assumed that threshold re-scatterings  are enough to describe the data, and as such
there is no need for poles in the scattering matrix.  Discussed long ago by Wigner~\cite{Wigner:1948zz} in the context of the non-relativistic two-body scattering
 theory, and resuscitated more recently by T\"{o}rnqvist~\cite{Tornqvist:1995kr} in an attempt to
understand the low-lying scalar meson $q\bar{q}$ nonet, and in a broader sweep
by Bugg~\cite{Bugg:2008wu}  interpreting the resonances as synchronized artefacts, this effect has to do with
 the behavior of  the scattering cross-sections
$\sigma(E)$ near thresholds, say $E=E_0$.  As one approaches $E_0$ from above or below, the cross-section remains  finite but the slope $d\sigma(E)/dE \to -\infty$, indicating a discontinuity, which can result in a
cusp, as  $\sigma(E)$ is continued below $E_0$. This can be illustrated by mapping the  two-particle scattering amplitude 
to a two-point function (self-energy), thereby relating the opening channel singularity to the  self-energy threshold singularity. The imaginary part of the self-energy ${\rm Im}\Pi(s)$ is zero for
$\sqrt{s}$ below the threshold, and it turns on rapidly once the threshold is
crossed.\footnote{Here $E$ and $\sqrt{s}$ are used interchangeably.} The resulting enhancements by cusps can mimic genuine S-matrix poles (resonances). However, the two can be
distinguished by studying the phase motion of the amplitudes. Representing a resonance by a Breit-Wigner
amplitude, $f(s)=\frac{\Gamma}{2}/(M-\sqrt{s} -i\Gamma/2)$, the magnitude and phase of this amplitude vary with $\sqrt{s}$, according to a circular trajectory in the Argand diagram. Cusps, on the other hand, have characteristically different
dependence on  $\sqrt{s}$. In terms of the variable $z=c (m_A + m_B -\sqrt{s})$, where $A$ and $B$ are the intermediate states and $c$ is a normalizing constant, one can show that the imaginary part of a cusp amplitude is zero for positive $z$ (i.e., below threshold) and turns on rapidly as the threshold is crossed, reflecting essentially the function  ${\rm erfc }(\sqrt{z})$,
 which governs ${\rm Im}(\Pi(s))$. This phase motion differs, in principle,  from that of a genuine Breit-Wigner.
 In the experimental analysis of some of the exotic mesons this phase motion is not measured, and in those cases a cusp-interpretation remains a logical, though by no means a unique, option. However, in at least three cases, $Z_c(4430)$, $Z_c(4200)$ and $P_c(4450)$, data analysis produced  Argand diagram; the first two are shown in Fig.~\ref{fz_argand} and the third in Fig.~\ref{DoubleArgand},
 in complete agreement with the characteristic Breit-Wigner motion 
 for a genuine resonance.

\subsection{Tetraquarks as hadroquarkonia}
This mechanism is motivated by analogy with the hydrogen atom.
 In the hadroquarkonium model,  a $Q\bar{Q}$ $(Q=c, b)$ pair forms the hard core surrounded by light matter (light $q\bar{q}$  in the case of tetraquarks and $qqq$ for pentaquarks),
with the two systems bound by  the QCD analog of a residual van der Waals type force.  For example, the hadrocharmonium core may consist of the $J/\psi, \psi^\prime$ or $\chi_c$,
 and the light $q\bar{q}$ degrees of freedom can be combined to accommodate the observed
 hadrons~\cite{Dubynskiy:2008mq}. 
 The effective Hamiltonian for the interaction can be written using the QCD multipole expansion
$H_{\rm eff} =-1/2 \alpha^{(\psi_1 \psi_2)} E_i^{a} E_i^{a} $, where $E_i^{a}$ is a chromoelectric field,
and $\alpha^{(\psi_1 \psi_2)} $ is the chromo-electric polarizability, which can be measured from the dipionic
transitions involving two heavy quarkonia $Q\bar{Q}$ states $\psi_1 \to \psi_2\;\pi^+ \pi^-$. Decays into heavy flavor mesons in this picture are suppressed, as this requires the splitting of the  $Q\bar{Q}$ core by
means of the soft gluons, present in the cloud. 
A variation on this theme is that the hard core quarkonium
could be in a color-adjoint representation \cite{Braaten:2013boa}, in which case the light degrees of freedom are also a
color-octet to form an overall singlet. 
 Hadroquarkonium models have conceptual problems:  if the binding force is weak, the question is why the system remains stable for long enough a time to be identified
as a distinct state. If the force is strong, it is not clear why the $Q\bar{Q}$ core and the light degrees
of freedom don't rearrange themselves as a pair of heavy mesons ($D\bar{D}^*, B\bar{B}^*$ etc.). 
While this does seem to happen in the decays of the $Z_b(10610)$ and $Z_b(10650)$, generally this is
not the case, as discussed in the previous section. This would have suppressed  the appearance of the states $(J/\psi,h_c) \pi \pi$ in their decays, which
in fact, are in many cases the discovery modes of such exotic multiquark states. 

\subsection{Tetraquarks as hybrids}
The hybrid models for exotic hadrons are   based on the QCD-inspired flux-tubes, which predict
exotic $J^{\rm PC}$ states of both the light and heavy quarks~\cite{Close:1994hc}.
 Hybrids are hadrons formed from the valence quarks and gluons, for example, consisting of $Q\bar{Q}g$, following the color algebra
\begin{equation}
3_c \otimes \bar{3}_c \otimes 8_c= (8_c \oplus 1_c) \otimes 8_c= 27_c \oplus \bar{10}_c \oplus 10_c 
\oplus 8_c \oplus 8_c \oplus 8_c \oplus 1_c,
\end{equation}
resulting in a color-singet hybrid hadron. States dominated by gluons
form glueballs, which are firm predictions of QCD, but have proven to be so far elusive experimentally.
Current lattice-QCD computations~\cite{Dudek:2011bn} suggest that non-perturbative
gluons, the object of interest in constructing the hybrids, are quasiparticles having $J^{PC}=1^{+-}$
with an excitation energy of approximately 1 GeV. This would put the lightest charmonium hybrid multiplets
at around 4200 MeV. Extensive studies of such hybrids have been carried out on the lattice by the Hadron
Spectrum Collaboration~\cite{Ryan:2016lml}, though for a heavy pion mass, $m_\pi  \sim 400 $ MeV and at fixed lattice 
spacing. More recently, simulations are also undertaken at a lower mass, and the results for   $m_\pi  \sim 240$ MeV
from Ref. \cite{Cheung:2016bym} are shown in Fig. \ref{fig:hsc}.  
Several states in this computation are identified as charmonium hybrid multiplets, having the quantum
numbers $J^{PC}= 0^{-+}, 1^{--},2^{-+},1^{-+}$, with their masses ($M$) estimated to lie in the range $M - M_{\eta_c} \simeq 
1200 - 1400$ MeV. Very much along the same lines, but much earlier, a hybrid interpretation 
was advanced  for the $J^{\rm PC}=1^{--}$ state $Y(4260)$, which has a small $e^+e^-$
annihilation cross section~\cite{Close:2005iz,Kou:2005gt,Zhu:2005hp}. In the meanwhile, hybrids have been offered as
templates for other exotic hadrons as well~\cite{Meyer:2015eta,Pennington:2015pda}. They have
been put on firmer theoretical footings in the framework of effective field theories \cite{Berwein:2015vca}. Despite all these theoretical
advances, which are impressive and may eventually provide reliable quantitative predictions, an unambiguous hybrid candidate has yet to be identified in the current experiments. Advances in lattice QCD techniques, enabling a firm phenomenological profile of the glueballs and the $Q\bar{Q}g$ hybrids, and dedicated experiments, such as the GlueX \cite{Ghoul:2015ifw} and 
${\rm \overline{P}ANDA}$ \cite{Lutz:2009ff}, may change this picture dramatically.

\subsection{Tetraquarks as hadron molecules}
This very popular approach assumes that the tetraquarks  and pentaquarks are  meson-meson and
meson-baryon bound states, respectively,  formed by
an attractive residual van der Waals force \footnote{This applies to the central part of the force, as there is also
a tensor part, decreasing as $1/r^2$.} generated by mesonic exchanges~\cite{Tornqvist:1993ng,Cleven:2011gp,Braaten:2003he,Close:2003sg,Swanson:2006st,Cleven:2013sq,Aceti:2014uea,Karliner:2015ina}.  
This hypothesis is in part supported by the closeness of the observed exotic hadron masses to their respective
 two-particle thresholds in many cases leading to a very small binding energy, which imparts the exotic hadrons  with very large hadronic radii, following from the Heisenberg uncertainty principle.
 A good illustration is given by the  $X(3872)$,  which has an $S$-wave coupling to $D^* \bar{D}$ (and its conjugate) and  has a binding energy  ${\cal E}_X=M_{X(3872)}-M_{D^{*0}} -M_{\bar{D}^0}=+ 0.01\pm 0.18$ MeV.
As mentioned in Sec.~\ref{cx3872_mass}, such a hadron molecule will have a large mean square separation of the constituents
$<$$r$$>$$\geq$31.7$^{+\infty}_{-24.5}$~fm.
This would lead to small production cross-sections in hadronic collisions~\cite{Bignamini:2009sk}, contrary to
what has been observed in a number of experiments at the Tevatron and the LHC. In some theoretical constructs, this problem is mitigated by invoking a hard (point-like) core for  the hadron molecules. In that sense, such models resemble hadroquarkonium models, discussed above.
In yet others, rescattering effects are invoked to substantially increase the cross-sections~\cite{Artoisenet:2009wk}.
A variation on this theme is to invoke that the $D\bar{D}^*$ state in question is a coherent mixture of the neutral and
charged components, with the latter components bound by 7 MeV \cite{Gamermann:2009uq}. A  consequence of this assumption would be the (almost) on-shell decays $X(3872) \to D^{\pm} D^{*\mp}$. 
 Experimentally, there is no
trace of such decays, and hence, in our opinion, there is no easy way out to invoke the charged component to increase the
binding energy, thereby substantially enhancing the hadronic production cross section, yet cancel out the charged $D\bar{D}^*$
decays from the decays of the $X(3872)$. 
 A crucial test is the $p_T$-spectrum of the exotic hadrons in question in
prompt production processes at the LHC. This spectrum has been measured for the $X(3872)$ by the CMS collaboration \cite{x3872cms}
over a $p_T$-range of $10-50$ GeV, and compared to the corresponding spectrum of the $ \psi^\prime$. The ratio
of the two $p_T$-spectra is found to be constant over this range, within experimental errors. This strongly suggests
that the $X(3872)$ is a compact hadron. 

The hadron molecular picture is plausible in explaining some other aspects of
the current data, namely the lack of experimental evidence of a quartet of exotic states, almost degenerate in
mass with the $X(3872)$, containing a light quark-antiquark pair $q\bar{q}$, $q=u,d$, leading to the formation of $I=1$ and $I=0$ multiplets. These multiplets are anticipated in the diquark picture, discussed below. 
However, in the molecular picture, due to the exchange of a pion providing the main binding, and pion being an isospin $I$=1 meson, not all isospin configurations will bind. 
 In line with this, no resonant structure is seen near the $D^0 \bar{D}^0$ threshold, consistent with the inadmissibility of a strong interaction coupling  of three pseudo-scalars $D^0 \bar{D}^0 \pi^0$ which violates parity conservation. More data is needed to observe or rule out the isospin partners of the $X(3872)$.
 On the other hand, the case for hadron molecules is less compelling for those exotics whose
masses are well above the respective thresholds. For example, the $Z_c(3900)^+$ is a case in point
whose  mass lies 20 MeV above the $D\bar{D}^*$ threshold, which incidentally is also its main
decay mode. This is hard to accommodate in the hadron molecular picture. 
Theoretical interest in hadron molecules has remained unabated, and 
 there exists a vast and growing literature on this topic with ever increasing sophistication, a sampling of which is referenced here~\cite{Gutsche:2014zda,Cleven:2014qka,Guo:2016bjq,Artoisenet:2010va,Hanhart:2011jz,Meng:2014ota,Barnes:2014csa}. 

\subsection{Tetraquarks as compact diquark-antidiquark mesons}

Last on this list are QCD-based interpretations in which tetraquarks and pentaquarks are genuinely new hadron species in which a color-nonsinglet diquark is the
essential building block~\cite{Maiani:2004uc,Maiani:2004vq,Brodsky:2014xia}.
In the large $N_c$ limit of QCD, tetraquarks, treated as diquark-antidiquark mesons, 
have been shown to exist~\cite{Weinberg:2013cfa,Knecht:2013yqa,Rossi:2016szw} as poles in the S-matrix.
They may have narrow widths in this approximation, and hence they are reasonable candidates for
 multiquark states. First attempts to study multiquark states using Lattice QCD have been
 undertaken~\cite{Padmanath:2015era,Hosaka:2016pey,DeTar:2015orc,Chen:2014afa,Ikeda:2016zwx} in which
correlations involving four-quark operators are studied numerically. 
Evidence of tetraquark states in the sense of  S-matrix poles using these methods
is still lacking. Establishing the signal of a resonance requires good control of the
background. In the lattice QCD simulations of multiquark states, this is currently not the case. This may be
traced back to the presence of a number of nearby hadronic thresholds and to lattice-specific issues,
such as an unrealistic pion mass. More powerful analytic and computational techniques are needed to
draw firm conclusions.

In the absence of reliable first principle calculations, approximate phenomenological methods are 
 the way forward. In that spirit,  an effective Hamiltonian approach has been often
 used~\cite{Maiani:2004uc,Maiani:2004vq,Maiani:2014aja,Ali:2009pi,Ali:2009es,AHS:2010}, 
in which tetraquarks are assumed to be  diquark-antidiquark objects, bound by gluonic exchanges
 (pentaquarks are diquark-diquark-antiquark objects). The diquarks are bound by the spin-spin interaction
 between the two quarks of a diquark (or between two antiquarks of an  anti-diquark). Motivated by the
  phenomenologically successful constituent quark model, the constituent diquark model allows one to work out the
spectroscopy and some aspects of tetraquark decays. Heavy quark symmetry is a help in that it can be used for the heavy-light diquarks relating the charmonia-like states to the bottomonium-like counterparts, and also in the
characteristics of $b$-baryon decays leading to pentaquarks.
As detailed below, diquark models anticipate a very rich spectroscopy of tetraquarks
and pentaquarks, only a small part of which has been possibly observed experimentally. Hence,
 diquark models are in dire need of dynamical selection rules to restrict the number of
observable states. The underlying multiquark dynamics is complex and
the effective Hamiltonian framework, in which the parameters are assumed to subsume the
dynamics, is obviously inadequate. Salient features of the phenomenology of the diquark picture 
are discussed below to test how far such models go in describing the observed features of the exotic hadrons  measured in current experiments. 

 For recent in-depth reviews of all the models discussed above and the theoretical techniques employed see 
\cite{Esposito:2016noz,Briceno:2015rlt,Chen:2016qju,Lebed:2016hpi,Guo:2017jvc}.

\subsection{The diquark model}

\begin{figure}[t]
\centerline{
\centerline{\includegraphics[width=5cm]{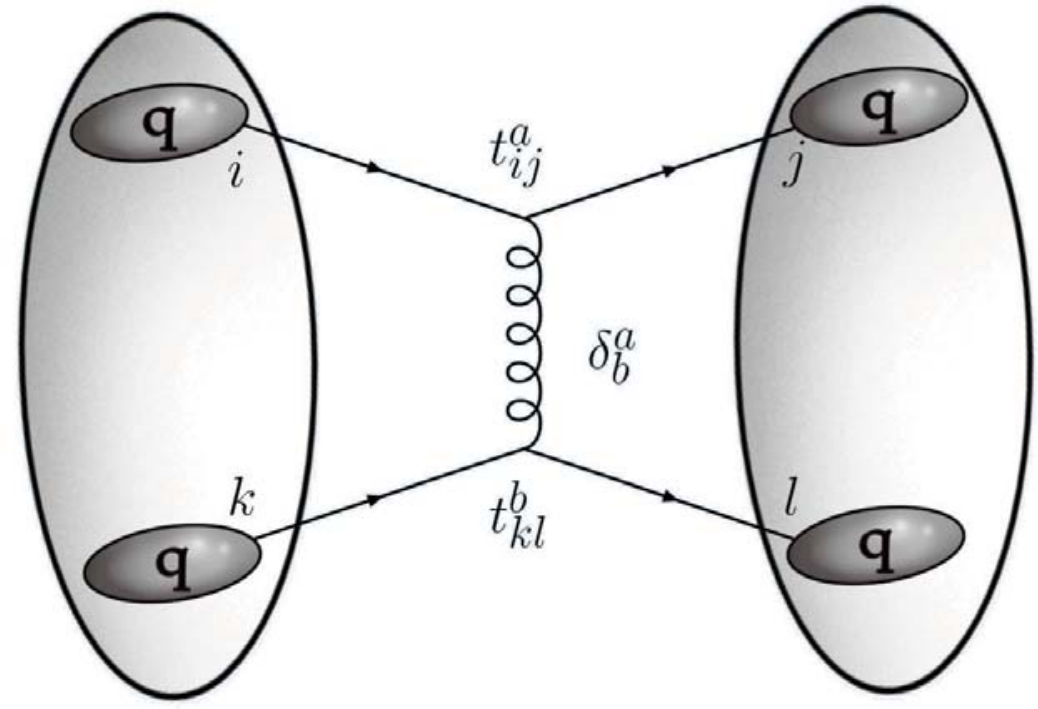}}
}
\caption{One-gluon exchange diagram for diquarks.
\label{ali:fig1}}
\end{figure}

The basic assumption of this  model is that diquarks are tightly bound colored objects and
they are the building blocks for forming tetraquark mesons and pentaquark baryons.
Diquarks, for which we use the notation $[qq]_c$, and interchangeably ${\mathcal Q}$,
have two possible SU(3)-color representations.
Since quarks transform as a triplet $\tt 3$ of color SU(3), the diquarks resulting from the
direct product $\tt 3 \otimes 3=\bar{3} \oplus 6$ are thus either a color anti-triplet $\tt \bar{3}$ or a
color sextet $\tt 6$. The leading diagram based on one-gluon exchange is
shown in Fig.~\ref{ali:fig1}.

The product of the SU(3)-matrices in Fig.~\ref{ali:fig1} can be decomposed as
\begin{equation}
\vspace{-0.3cm}
t^a_{ij}t^a_{kl}=
-\frac{2}{3}\underbrace{(\delta_{ij}\delta_{kl}-\delta_{il}\delta_{kj})/2}_{\rm{antisymmetric:\; projects}\enspace {  \bar{\mathbf 3} }}
+\frac{1}{3}\underbrace{(\delta_{ij}\delta_{kl}+\delta_{il}\delta_{kj})/2}_{\rm{symmetric:\; projects}\enspace { \mathbf 6}}~.
\end{equation}
\vspace*{3mm}

The coefficient of the antisymmetric $\tt \bar{3}$ representation is $-2/3$, reflecting that the two diquarks bind with a
strength half as strong as between a quark and an antiquark, in which case the corresponding coefficient is $-4/3$.
The symmetric $\tt 6$ on the other hand has a positive coefficient, +1/3,  reflecting a repulsion.
 Thus,  in working out the phenomenology,
 a diquark is assumed to be
an $SU(3)_c$-antitriplet, with the antidiquark a color-triplet. With this, we have two color-triplet fields, quark $q_3$ and anti-diquark $\overline{\mathcal Q}$ or $[\bar{q}\bar{q}]_{3}$, 
and two color-antitriplet fields, antiquark $\bar{q}_{\bar{3}}$ and diquark  ${\mathcal Q}$ or $[qq]_{\bar{3}}$,
from which the spectroscopy of the conventional and exotic  hadrons is built. However, the quarks and diquarks differ in an essential detail, namely the former are point-like objects but the latter are composite and have a hadronic size. 
This is of  crucial importance in determining the electromagnetic and strong couplings, and hence for the production cross-sections of multiquark states in leptonic and hadronic collisions. 

\begin{figure}[b]
\centerline{\includegraphics[width=8cm]{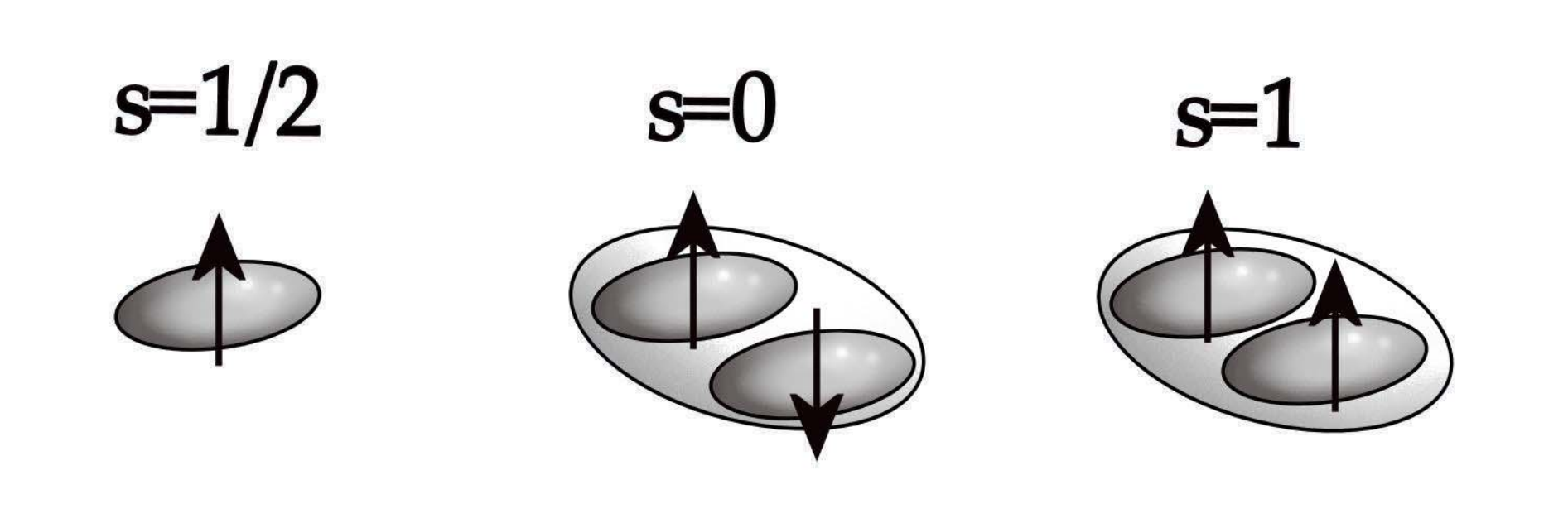}}
\caption{Quark and diquark spins.}
\label{ali:fig2}
\end{figure}

 Since quarks are spin-1/2 objects, a diquark has two possible spin-configurations, spin-0, with the
two quarks in a diquark having their spin-vectors anti-parallel, and spin-1, in which case the two quark spins
 are aligned, as shown in Fig.~\ref{ali:fig2}.
They were given the names ``good diquarks'' and ``bad diquarks'', respectively, by Jaffe~\cite{Jaffe:2004ph}, implying that in the former case, the two quarks bind, and in the latter, the binding is not as strong.
There is some support of this  pattern from lattice simulations for light diquarks~\cite{Alexandrou:2006cq}, 
in which correlations are studied in terms of the spatial distribution of the two quarks forming the diquark
in the background of the static quark. Phenomenological expectations that QCD dynamics favors the formation of good (spin-0) diquarks in color anti-triplet configuration is verified. It is exceedingly important to study on the lattice such correlations in
two-point functions, involving tetraquarks, which have been attempted but with inconclusive results so far.
 However, as the spin-degree of freedom decouples in the heavy quark systems, as can be shown
explicitly in the heavy quark effective theory context for heavy mesons and baryons~\cite{Manohar:2000dt},
 we expect that this decoupling
will also hold for heavy-light diquarks $[Q_i q_j]_{\bar{3}}$ with $Q_i=c, b; q_j=u,d,s$. So, for the heavy-light diquarks,
 both the spin-1 and spin-0 configurations are phenomenologically present. Also, the
diquarks in heavy baryons (such as $\Lambda_b$ and $\Omega_b$), consisting of a heavy quark and 
a light diquark,  both $J^p=0^+$ and $J^p=1^+$ quantum numbers of the diquark are needed to accommodate the observed baryon spectrum. 
In this review, we restrict ourselves to the heavy-light diquarks, though  heavy-heavy diquarks
 $[QQ]_{\bar{3}}$ $(Q=c,b)$, and the resulting tetraquark states $[QQ]_{\bar{3}}[\bar{Q}\bar{Q}]_3$ are
also anticipated and discussed in the literature~\cite{Karliner:2016zzc,Chen:2016ont,Wu:2016vtq}.

 Following the discussion above, we
construct the interpolating diquark operators for the two spin-states of such diquarks
 (here $Q=c,b$)~\cite{Maiani:2014aja}:
\vspace*{3mm}

%
\begin{tabular}{rlrcl}
$\textnormal{Scalar }$ &
$0^+$: &
$\mathcal Q_{i \alpha} $
&=& 
$\epsilon_{\alpha\beta\gamma}
(\bar{Q}_c^{\beta}\gamma_5
q_i^{\gamma}
-\bar{q}_{i_c}^\beta \gamma_5 Q^\gamma), \qquad ${{{\footnotesize {$\alpha, \beta, \gamma$: $SU(3)_C$  indices} }}}
\\
$\textnormal{Axial-Vector}$&
$1^+ $:
&
$\vec{\mathcal Q}_{i \alpha} $
&=&
$\epsilon_{\alpha\beta\gamma}(\bar{Q}_c^{\beta} \vec{\gamma} q_i^{\gamma}
+
\bar{q}_{i_c}^\beta \vec{\gamma} Q^\gamma).
$
\end{tabular}
%
%
\vspace*{2mm}

Here $ \bar{Q}_c $ and $\bar{q}_{i_c} $ are the charge conjugate fileds.
In the non-relativistic (NR) limit, these states are parametrized by Pauli matrices:
$ \Gamma^0 =
\frac{\sigma_2}{\sqrt{2}}$ for the scalar ($0^+$), 
and
$\;\;\vec{\Gamma} =
\frac{\sigma_2\vec{\sigma}}{\sqrt{2}}$  for the axial-vector  ($1^+$). 
A tetraquark state with total angular momentum $J$ may be described by the state vector
$\left\vert s_{\mathcal Q},s_{\bar{\mathcal Q}};~J\right\rangle$
showing the diquark spin $s_{\mathcal Q}$ and the antidiquark spin $s_{\bar{\mathcal Q}}$. 
There is no consensus on their names. We use the symbols $X_J$ for $J^{PC}=J^{++}$, $Y$ for $1^{--}$, and $Z$ for $1^{+-}$
states.
Thus, the tetraquarks with the following diquark-spin and angular momentum $J$ have the Pauli forms~\cite{Maiani:2014aja}: 
%
\begin{eqnarray}
\left\vert 0_{\mathcal Q},0_{\bar{\mathcal Q}};~0_{J}\right\rangle &=&\Gamma^0 \otimes \Gamma^0 ,  \notag \\
\left\vert 1_{\mathcal Q},1_{\bar{\mathcal Q}};~0_{J}\right\rangle &=&\frac{1}{\sqrt{3}}%
\Gamma^i \otimes \Gamma_i,  \notag 
\\
\left\vert 0_{\mathcal Q},1_{\bar{\mathcal Q}};~1_{J}\right\rangle &=&\Gamma^0 \otimes \Gamma^i ,  \notag \\
\left\vert 1_{\mathcal Q},0_{\bar{\mathcal Q}};~1_{J}\right\rangle &=&\Gamma^i \otimes \Gamma^0 ,  \notag \\
\left\vert 1_{\mathcal Q},1_{\bar{\mathcal Q}};~1_{J}\right\rangle &=&\frac{1}{\sqrt{2}}%
\varepsilon ^{ijk}\Gamma_j \otimes \Gamma_k. 
\end{eqnarray}
 
Whenever necessary, a subscript $c$ or $b$ is used to distinguish the $c\bar{c}$
and $b\bar{b}$ states.

\subsection{Non-relativistic Hamiltonian for tetraquarks with hidden charm}
\begin{figure}[b]
\centerline{\includegraphics[width=7cm]{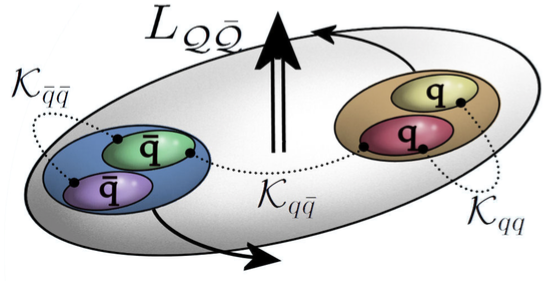}}
\vspace{-2mm}
\caption{Schematic diagram of a tetraquark in the diquark-antidiquark picture.}
\label{ali:fig3}
\end{figure}

For the heavy quarkonium-like exotic hadrons the non-relativistic limit is a good approximation.
This effective NR Hamiltonian has been proposed to calculate the tetraquark mass
 spectrum~\cite{Maiani:2004vq,Maiani:2014aja}
\begin{equation}
H_{\rm eff}=2m_{\mathcal Q}+H_{SS}^{(qq)}+H_{SS}^{(q\bar{q})}+H_{SL}+H_{LL}, 
\label{eq:H-eff-tetra}
\end{equation}
where $m_{\mathcal Q}$ is the constituent diquark mass, the second term above is the spin-spin interaction involving
the quarks (or antiquarks) in a diquark (or anti-diquark), the third term depicts spin-spin interactions
involving a quark and an antiquark in two different shells (i.e., in the two different diquark configurations), with the fourth and fifth terms being the
spin-orbit and the orbit-orbit interactions, involving the quantum numbers of the tetraquark, respectively.
For the $S$-states, these last two terms are absent. For  $Q=c$, the individual terms in $H_{\rm eff}$ are displayed below and illustrated in Fig.~\ref{ali:fig3}:
\begin{eqnarray}
H_{SS}^{(qq)}= 2(\mathcal{K}_{cq})_{\bar{3}}[(\mathbf{S}_{c}\cdot \mathbf{S}_{q})
+(\mathbf{S}_{\bar{c}}\cdot \mathbf{S}_{\bar{q}})],
\notag \\
&\hspace{-5.8cm} H_{SS}^{(q\bar{q})}=2(\mathcal{K}_{c\bar{q}})(\mathbf{S}_{c}
\cdot \mathbf{S}_{\bar{q}}+\mathbf{S}_{\bar{c}}\cdot \mathbf{S}_{q})
 +2 \mathcal{K}_{c\bar{c}} (\mathbf{S}_{c}\cdot \mathbf{S}_{\bar{c}})
+2 \mathcal{K}_{q\bar{q}} (\mathbf{S}_{q}\cdot \mathbf{S}_{\bar{q}}),
\notag \\
&\hspace{-11.3cm} H_{SL}  = 2 A_{\mathcal Q} (\mathbf{S}_{\mathcal{Q}}\cdot \mathbf{L}+\mathbf{S}_{\mathcal{\bar{Q}} }\cdot \mathbf{L}), 
\notag \\
&\hspace{-12cm} H_{LL} = B_{\mathcal Q} \frac{L_{\mathcal Q\bar{\mathcal Q}}(L_{\mathcal Q\bar{\mathcal Q}}+1)}{2}.  
\end{eqnarray}
Here $(\mathcal{K}_{cq})_{\bar{3}}$ parametrizes the spin-spin interaction between the charm $c$ and the light quark $q$ within a diquark
in a color anti-triplet configuration, and the various $(\mathcal{K}_{i\bar{j}})$ parametrize the spin-spin strengths between the quarks $i$ and the
antiquark $\bar{j}$, in the color-singlet configuration involving two different diquarks. The parameters $ A_{\mathcal Q}$ and
$B_{\mathcal Q}$ characterize the strength of the spin-orbit and the orbital angular force, respectively.  

The usual angular momentum algebra then yields the following form:
\begin{align}
H_{\rm eff}
&=2m_{\mathcal Q}
+ \frac{B_{\mathcal Q}}{2} \langle \bm{L}^2 \rangle
- 2a \langle \bm{L}\cdot \bm{S} \rangle
+ 2\kappa_{qc} \big[ \langle \bm{s_{q}}\cdot \bm{s_{c}} \rangle
 + \langle \bm{s_{\bar q}}\cdot \bm{s_{\bar c}} \rangle \big]
\nonumber\\
&=2m_{\mathcal Q}
- a J(J+1)
+ \bigg( \frac{B_{\mathcal Q}}{2} + a \bigg) L(L+1)
+ a S(S+1)- 3\kappa_{qc} \nonumber\\
&
+ \kappa_{qc}\big[ s_{qc}(s_{qc}+1) +
 s_{\bar{q}\bar{c}}(s_{\bar{q}\bar{c}}+1) \big].
\end{align}
The effective Hamiltonian given above can also be used for the tetraquark states involving a $b\bar{b}$ pair,
with appropriate rescaling of the parameters. 
\subsection{\boldmath{Low-lying $S$ and $P$-wave tetraquark states in the $c\bar{c}$ and $b\bar{b}$ sectors}}%

The states in the diquark-antidiquark basis $| s_{qQ}, s_{\bar{q}\bar{Q}}; S, L \rangle_J$
and in the $Q\bar{Q}$ and $q\bar{q}$ basis $| s_{q\bar{q}}, s_{Q\bar{Q}}; S', L' \rangle_J$ 
are related by Fierz transformation. The positive parity $S$-wave
tetraquarks are given in terms of the six
states listed in Table~\ref{ali:tbl1} (charge conjugation is defined for neutral states). These states
are characterized by the quantum number $L=0$, hence their masses depend on just two parameters the diquark mass,
$M_{00}$,  and $\kappa_{qQ}$, leading to  several testable predictions.
The $P$-wave states are listed in Table~\ref{ali:tbl2}. The first four of them have $L=1$, and the
fifth has $L=3$, and hence is expected to be significantly heavier.

\begin{table}
\begin{center}
\tbl{$S$-wave tetraquark states involving a $Q\bar{Q}$ pair in the two bases and their masses in the diquark model.}
\vspace*{2mm}
{\begin{tabular}{ccccc}
\hline\hline
Label & $J^{PC}$ & 
$| s_{qQ}, s_{\bar{q}\bar{Q}}; S, L \rangle_J$ &
$| s_{q\bar{q}}, s_{Q\bar{Q}}; S', L' \rangle_J$ &
Mass
\\
\hline
$X_0$ & $0^{++}$ & $| 0, 0; 0, 0 \rangle_0$ &
$\big( | 0, 0; 0, 0 \rangle_0 + \sqrt{3}| 1, 1; 0, 0 \rangle_0 \big)/2$ &
$M_{00} - 3\kappa_{qQ}$
\\
$X_0^\prime$ & $0^{++}$ & $| 1, 1; 0, 0 \rangle_0$ &
$\big( \sqrt{3}| 0, 0; 0, 0 \rangle_0 -| 1, 1; 0, 0 \rangle_0 \big)/2$ &
$M_{00} + \kappa_{qQ}$
\\
$X_1$ & $1^{++}$ &$\big( | 1, 0; 1, 0 \rangle_1 + | 0, 1; 1, 0 \rangle_1 \big)/\sqrt{2}$ &
$| 1, 1; 1, 0 \rangle_1$ &
$M_{00} - \kappa_{qQ}$
\\
$Z$ & $1^{+-}$ & 
$\big( | 1, 0; 1, 0 \rangle_1 - | 0, 1; 1, 0 \rangle_1 \big)/\sqrt{2}$ &
$\big( | 1, 0; 1, 0 \rangle_1 - | 0, 1; 1, 0 \rangle_1 \big)/\sqrt{2}$ &
$M_{00} - \kappa_{qQ}$
\\
$Z^\prime$ & $1^{+-}$ & $| 1, 1; 1, 0 \rangle_1$ &
$\big( | 1, 0; 1, 0 \rangle_1 + | 0, 1; 1, L' \rangle_1 \big)/\sqrt{2}$ &
$M_{00} + \kappa_{qQ}$
\\
$X_2$ &  $2^{++}$ & $| 1, 1; 2, 0 \rangle_2$ &
$| 1, 1; 2, L' \rangle_2$ &
$M_{00} + \kappa_{qQ}$
\\
\hline\hline
\end{tabular}
}
\label{ali:tbl1}
\end{center}
\end{table}
\begin{table}
\begin{center}
\tbl{$P$-wave ($J^{PC}=1^{--}$) tetraquark states  involving a $Q\bar{Q}$ pair in the two bases and their masses in the diquark model.}
\vspace*{2mm}
{\begin{tabular}{cccc}
\hline\hline
Label & 
$| s_{qQ}, s_{\bar{q}\bar{Q}}; S, L \rangle_J$ &
$| s_{q\bar{q}}, s_{Q\bar{Q}}; S', L' \rangle_J$ &
Mass
\\
\hline
$Y_1$ & $| 0, 0; 0, 1 \rangle_1$ &
$\big( | 0, 0; 0, 1 \rangle_1 + \sqrt{3}| 1, 1; 0, 1 \rangle_1 \big)/2$ &
$M_{00} - 3\kappa_{qQ} + B_Q$
\\
$Y_2$ & 
$\big( | 1, 0; 1, 1 \rangle_1 + | 0, 1; 1, 1 \rangle_1 \big)/\sqrt{2}$ &
$| 1, 1; 1, L' \rangle_1$ &
$M_{00} - \kappa_{qQ} + 2a + B_Q$
\\
$Y_3$ & 
$| 1, 1; 0, 1 \rangle_1$ &
$\big( \sqrt{3}| 0, 0; 0, 1 \rangle_1 -| 1, 1; 0, 1 \rangle_1 \big)/2$ &
$M_{00} + \kappa_{qQ} + B_Q$
\\
$Y_4$ & 
$| 1, 1; 2, 1 \rangle_1$ &
$| 1, 1; 2, 1 \rangle_1$ &
$M_{00} + \kappa_{qQ} + 6a + B_Q$
\\
$Y_5$ & 
$| 1, 1; 2, 3 \rangle_1$ &
$| 1, 1; 2, 1 \rangle_1$ &
$M_{00} + \kappa_{qQ} + 16a + 6B_Q$
\\
\hline\hline
\end{tabular}
}
\label{ali:tbl2}
\end{center}
\end{table}
%

The parameters appearing on the right-side columns of Tables~\ref{ali:tbl1} and~\ref{ali:tbl2} can be
determined using the masses of some of the observed $X,Y,Z$ states, and their
numerical values are given in Table~\ref{ali:tbl3}. Some parameters in
the $c\bar{c}$ and $b\bar{b}$ sectors can also be related using the heavy quark mass
scaling~\cite{Ali:2014dva}. 

\begin{table}
\begin{center}
\tbl{Numerical values of the parameters in $H_{\rm eff}$, obtained using some of the
$S$ and $P$-wave tetraquarks as input.}
\vspace*{2mm}
{\begin{tabular}{c|cc}
\hline\hline
& charmonium-like & bottomonium-like
\\
\hline
$M_{00}$ [MeV] & 3957 & 10630
\\
$\kappa_{qQ}$ [MeV] & 67 & 23
\\
$B_Q$ [MeV] & 268 & 329
\\
$a$ [MeV] & 52.5 & 26
\\
\hline\hline
\end{tabular}
}
\label{ali:tbl3}
\end{center}
\end{table}

\vspace*{-3mm}
\begin{table}
\begin{center}
\tbl{$J^{PC}$ quantum numbers of the $X,Y,Z$ exotic hadrons and their  masses from experiments and in the diquark-model.}
\vspace*{2mm}
{\begin{tabular}{c|c|cc|cc}
\hline\hline
&&
\multicolumn{2}{c|}{charmonium-like} & 
\multicolumn{2}{c}{bottomonium-like}
\\
Label & $J^{PC}$ & State & Mass [MeV] & State & Mass [MeV] 
\\
\hline
$X_0$ & $0^{++}$ & 
--- & 3756 &
--- & 10562
\\
$X_0'$ & $0^{++}$ &
--- & 4024 &
--- & 10652
\\
$X_1$ & $1^{++}$ & 
$X(3872)$ & 3890 &
--- & 10607
\\
$Z$ & $1^{+-}$ & 
$Z_c^+(3900)$ & 3890 &
$Z_b^{+,0}(10610)$ & 10607
\\
$Z'$ & $1^{+-}$ &
$Z_c^+(4020)$ & 4024 &
$Z_b^+(10650)$ & 10652
\\
$X_2$ & $2^{++}$ &
--- & 4024 &
--- & 10652
\\
\hline
$Y_1$ & $1^{--}$ &
$Y(4008)$ & 4024 &
$Y_b(10890)$ & 10891
\\
$Y_2$ & $1^{--}$ &
$Y(4260)$ & 4263 &
$\Upsilon(11020)$ &  10987
\\
$Y_3$ & $1^{--}$ &
$Y(4290)$ (or $Y(4220)$) & 4292 &
--- &  10981
\\
$Y_4$ & $1^{--}$ &
$Y(4630)$ & 4607 &
--- & 11135
\\
$Y_5$ & $1^{--}$ &
--- & 6472 &
--- & 13036
\\
\hline\hline
\end{tabular}
}
\label{ali:tbl4}
\end{center}
\end{table}
\vspace*{3mm}
 There are several predictions in
the charmonium-like sector, which, with the values of the parameters given in the tables above, are in the
right ball-park.
 In a modified scheme~\cite{Maiani:2014aja}, developed subsequently,
the parameters $(\kappa_{i\bar{j}})$ are set to zero, eliminating the $H_{SS}^{(q\bar{q})}$ term in
the effective Hamiltonian. With this,
 better agreement is reached with experiments. Despite this success, the continued experimental absence
of the two lowest-lying $0^{++}$ states, called $X_0$ and $X_0^\prime$, is puzzling. Perhaps,
they are below the threshold for strong decays and decay weakly, and thus have not been looked for.
 Alternative calculations of the tetraquark spectrum
based on diquark-antidiquark model have been carried out in other phenomenological schemes~\cite{Ebert:2005nc},
and in the QCD sum rule framework~\cite{Nielsen:2009uh,Albuquerque:2016znh}. All of them share the common feature
with the effective Hamiltonian approach discussed here, namely they all anticipate a very rich tetraquark spectroscopy.
So, if the diquark picture has come to stay, some dynamical selection rules are required to better understand the
observed spectrum.

The exotic bottomonium-like states are currently rather sparse.
The reason for this is that quite a few exotic candidate charmonium-like states were observed in the decays of 
$B$ hadrons. The  observed pentaquark states $P_c(4380)^+$ and $P_c(4450)^+ $ are
decay products of the $\Lambda_b$-baryon. However, as the top quark decays weakly before it gets a
chance to hadronize, exotic hadrons with  a hidden $b\bar{b}$ pair are not anticipated from the top
quark decays. Hence, they can only be produced in hadro- and electroweak  high energy processes,
which makes their detection a lot harder. Tetraquark states
with a single $b$ quark can, in principle, be searched for in the hadronic debris of a $b$-quark initiated jet,
or in the decays of the $B_c$ mesons~\cite{Ali:2016gdg}.
As the $c\bar{c}$ and $b\bar{b}$ cross-section at the LHC are very large, we 
anticipate that the exotic spectroscopy involving the open  and hidden heavy quarks is an area
where significant new results  will be reported by all the LHC experiments. Measurements of the
production and decay characteristics of exotica, such as the transverse-momentum distributions and polarization information, will go a long way in understanding the underlying dynamics. 
 
We now discuss the three candidate exotic states  observed so far in the bottomonium sector.
The hidden $b \bar{b}$ state
$Y_b(10890)$ with  $J^{\rm P}=1^{--}$  was discovered by Belle in 2007~\cite{Abe:2007tk} in the
 process $e^+e^- \to Y_b(10890) \to (\Upsilon(1S), \Upsilon(2S), \Upsilon(3S)) \pi^+ \pi^-$ just above the
$\Upsilon(10860)$. The branching ratios measured for these decays are about two orders of magnitude larger than anticipated from a similar dipionic transitions in the lower $\Upsilon(nS)$ states and in the $\psi^\prime$
(for a review and references to earlier work, see Brambilla {\it et al.}~\cite{Brambilla:2010cs}).
Also the dipion invariant mass distributions in the decays of $Y_b$ are  marked by the presence of
the resonances $f_0(980)$ and $f_2(1270)$. This state
was interpreted as a $J^{\rm PC}=1^{--}$ $P$-wave tetraquark~\cite{Ali:2009pi,Ali:2009es}.
Subsequent to this, a Van Royen-Weiskopf formalism was used~\cite{AHS:2010}
 in which direct electromagnetic couplings with
the diquark-antidiquark pair of the $Y_b$ was assumed. Due to the $P$-wave nature of the
$Y_b(10890)$,  and the hadronic size of the diquarks,
 the effective electromagnetic coupling is reduced. Hence,  the
production cross-section for $e^+e^- \to Y_b(10890) \to b\bar{b}$ is anticipated to be small, leading only to a small bump in the $R_b$-scan. However, due to the presence of a
$b\bar{b}$ and a light $q\bar{q}$ ($q=u,d $) pair in $Y_b(10890)$ in the valence approximation,
the decays  $Y_b(10890) \to \Upsilon(nS) \pi^+ \pi^-$ are Zweig-allowed. Since, there is practically no background to these final states
from the continuum $e^+e^-$ annihilation, as opposed to $e^+e^- \to b \bar{b}$, a bump in 
the  $e^+e^- \to  \Upsilon(nS) \pi^+ \pi^-$ due to $Y_b(10890)$ should be visible. This is, at least qualitatively, in
agreement with the Belle data~\cite{Santel:2015qga}. 

The model in which direct electromagnetic coupling with a  diquark-antidiquark pair is envisaged has experimentally
testable consequences. Among other implications, one expects large isospin-breaking effects, arising 
from the different electric charges of the $[bu]$ and $[bd]$ diquarks.  
In the tetraquark picture, the
$Y_b(10890)$ is the $b\bar{b}$ analogue of the $c\bar{c}$ state $Y(4260)$, also a $P$-wave, which  
is likewise found to have  a very small production cross-section in $e^+ e^- \to Y(4260) \to c \bar{c}$,
but which decays readily into $J/\psi \pi^+\pi^-$, reflecting the presence of a $c\bar{c}$ and a light $q\bar{q}$
pair in the Fock space of the $Y(4260)$.
Hence, the two $Y$-states have very similar production and decay characteristics, and both are $J^{PC}=1^{--}$
tetraquark candidates. There are other production mechanisms for tetraquarks in which the electromagnetic current couples
with the $c\bar{c}$ or $b\bar{b}$ quark pair, which before decaying picks up a light $q\bar{q}$ from the vacuum, resulting in a diquark-antidiquark pair. If the center-of-mass energy is close to a $J^{PC}=1^{--}$ tetraquark mass, one
expects a resonant production. However, the cross-section, due to the angular momentum barrier reflecting the $P$-wave
nature of the tetraquark, and  a small probability for the fragmentation $c\bar{c} \to [cq][\bar{c}\bar{q}]$ (or
$b\bar{b} \to [bq][\bar{b}\bar{q}]$), is expected to be small. In this case, as opposed to the direct electromagnetic coupling to the diquark-antidiquark pair, the production mechanism will respect isospin symmetry.

\begin{figure}[b!]
\centerline{\includegraphics[width=7cm]{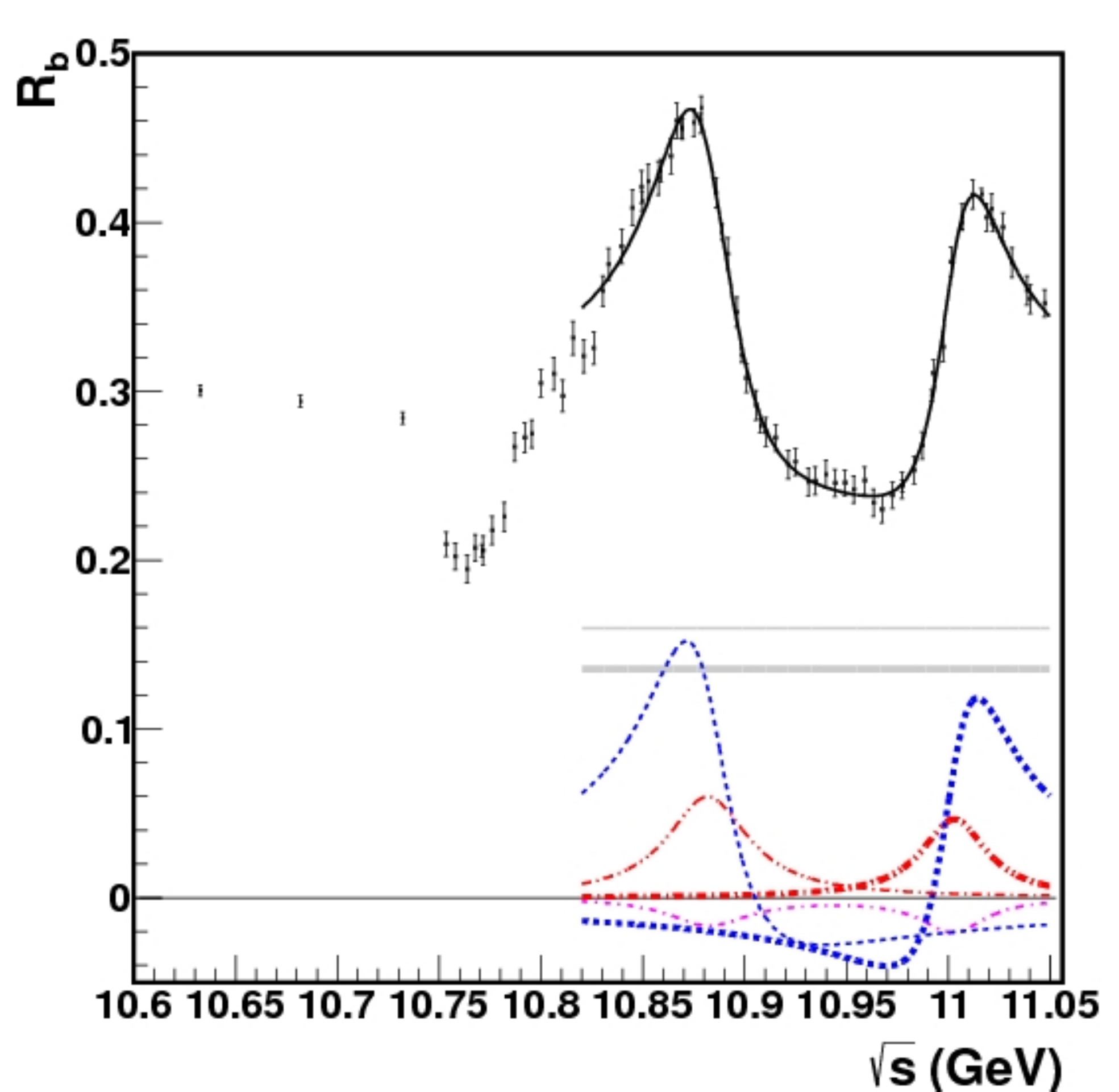}}
\caption{The ratio $R_b=\sigma(e^+e^- \to b\bar{b})/\sigma(e^+e^- \to \mu^+ \mu^-)$ in the $\Upsilon(10860)$ and $\Upsilon(11020)$ 
region. The components of the fit are depicted in the lower part of the figure:
total (solid curve), constant $\vert A_{ic}\vert^2$ (thin), $\vert A_c\vert^2$ (thick): for $\Upsilon(10860)$ (thin)
 and $\Upsilon(11020)$ (thick):
$\vert f \vert^2$ (dot-dot-dash), cross terms with $A_c$ (dashed), and two-resonance cross term (dot-dash).
 Here, $A_c$ and $A_{ic}$
are coherent and incoherent continuum terms, respectively (from Belle~\cite{Santel:2015qga}).
}
\label{ali:fig-rbplot}
\end{figure}

\begin{figure}[t!]
\centerline{\includegraphics[width=7cm]{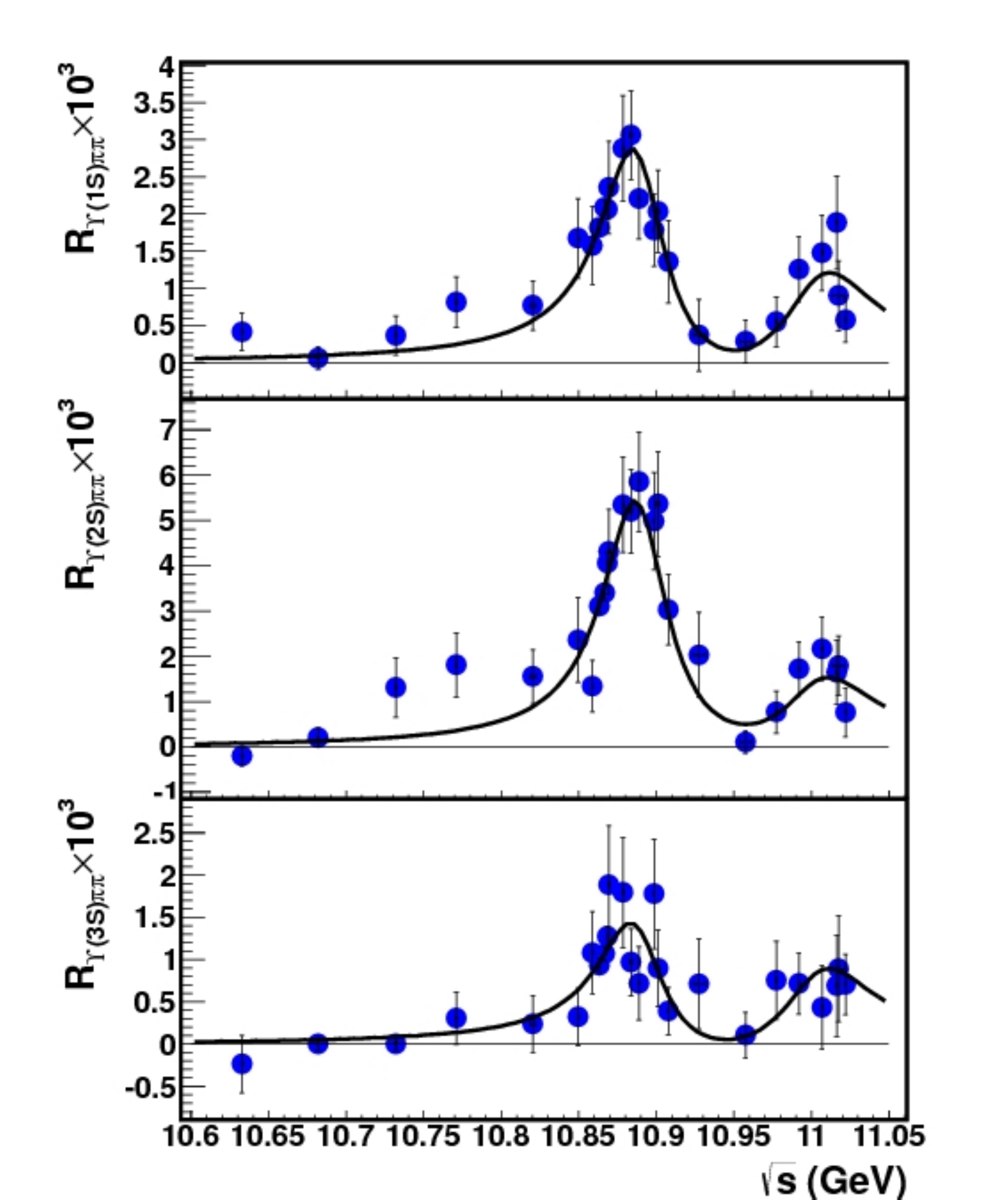}}
\caption{The ratio $R_{\Upsilon(nS) \pi^+\pi^-}= \sigma(e^+e^- \to \Upsilon(nS) \pi^+\pi^-)/\sigma (e^+ e^- \to \mu^+ \mu^-)$ 
in the $\Upsilon(10860)$ and $\Upsilon(11020)$ region (from Belle \cite{Santel:2015qga}).}
\label{ali:fig-fitC}
\end{figure}

The current status of $Y_b(10890)$ is, however, unclear. Subsequent to its discovery, Belle undertook
high-statistics scans to measure the ratio
$R_{b\bar{b}}=\sigma(e^+e^- \to b\bar{b})/\sigma (e^+ e^- \to \mu^+ \mu^-)$, and also more precisely the
ratios $R_{\Upsilon(nS) \pi^+\pi^-}= \sigma(e^+e^- \to \Upsilon(nS) \pi^+\pi^-)/\sigma (e^+ e^- \to \mu^+ \mu^-)$.
They are shown in Fig.~\ref{ali:fig-rbplot} and Fig.~\ref{ali:fig-fitC}, respectively. 
 The two masses, $M(\Upsilon(10860))_{b\bar{b}}$ measured through $R_{b\bar{b}}$, and $M(Y_b)$
 measured through $R_{\Upsilon(nS) \pi^+\pi^-}$, now differ by slightly more than 2$\sigma$, 
$M(\Upsilon(10860))_{b\bar{b}} -M(Y_b)= -9 \pm 4$ MeV.
 From the mass difference alone, these two could very well be just one and the same state,
namely the canonical $\Upsilon(5S)$ - an interpretation now adopted by the Belle
 collaboration~\cite{Santel:2015qga}. On the other hand, it is {\it the bookkeeping of the branching ratios
measured at or near the ``$\Upsilon(5S)$"}, which is enigmatic. The branching
ratios of the ``$\Upsilon(5S)$" measured by Belle are  saturated by the exotic states
 $(\Upsilon(nS) \pi^+\pi^-, h_b(mP) \pi^+\pi^-,
Z_b(10610)^\pm \pi^\mp, Z_b(10650)^\pm \pi^\mp$ and their isospin partners).
Another class of states consists of $ [B^* B^{(*)}]^\pm \pi^\mp $, which is found to originate exclusively from the
$ Z_b(10610)^\pm \pi^\mp$ and $ Z_b(10650)^\pm \pi^\mp$. Based on these measurements and assuming isospin symmetry, Belle has reported a cumulative
value ${\cal P}= 1.09 \pm 0.15$~\cite{Santel:2015qga}, where a value of ${\cal P}= 1$ corresponds to the saturation of the ``5S" amplitude by the contributing exotic channels (listed above). This leaves little room at ``5S" for other known final states, such as $B^{(*)}_{(s)}\bar{B}^{(*)}_{(s)}$, despite the fact these reactions have large cross sections measured in independent experiments, first reported by CLEO~\cite{pdg} and more recently by Belle itself. 

The reason for this mismatch is not clear;  Belle attributes it to the inadequate modeling of
$R_b$ due to several thresholds in this energy region. While this may eventually be the source of
the current "$ \Upsilon(5S)$" branching ratio puzzle,  an interpretation
of the Belle data based on two almost degenerate (in mass) resonances $\Upsilon(5S)$ and $Y_b(10890)$ is also a logical possibility, with
$\Upsilon(5S)$ having the decays expected for the bottomonium $S$-state above the $B^{(*)}\bar{B}^{(*)}$ threshold,
and the decays of $Y_b(10890)$, a tetraquark, being the source of the exotic states seen.  On the other hand,
no peaking structure at 10.9 GeV in the $R_b$ distribution is  seen in the Belle analysis, and an upper limit on
$ \Gamma_{ee} $ of 9 eV is set with a 90\% confidence level~\cite{Santel:2015qga}.   
 As  data taking starts in a
couple of years in the form of a new and expanded collaboration, Belle-II, cleaning up the current analysis in 
the $\Upsilon(10860)$ and $\Upsilon(11020)$ regions should be one of their top priorities.

\begin{figure}
\centerline{\includegraphics[width=12cm]{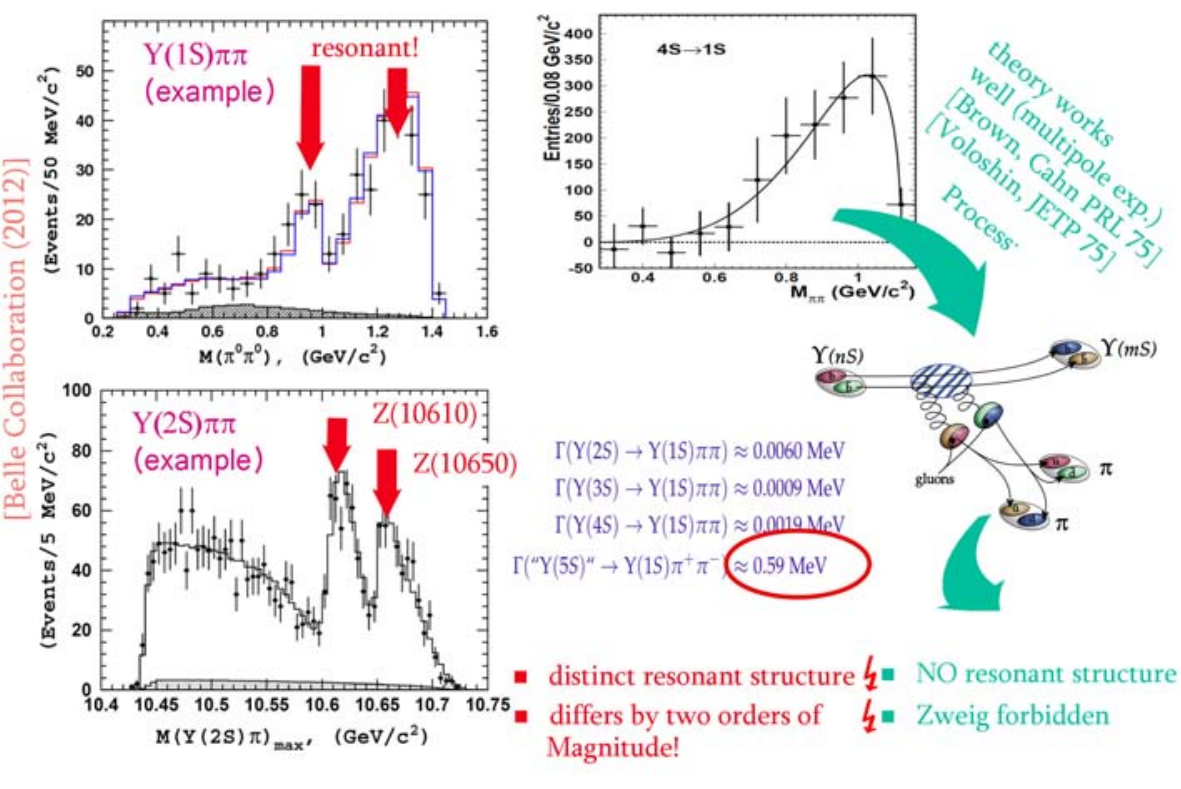}}
\vspace{-3mm}
\caption{Dipion invariant mass distribution in $\Upsilon(10860) \to \Upsilon(1S) \pi^0 \pi^0$ (upper left frame);
 the resonances indicated
 in the dipion spectrum correspond to the $f_0(980)$ and $f_2(1270)$;
the resonances $Z(10610)$ and $Z(10650)$ are indicated in
the $\Upsilon(2S) \pi^+$ invariant mass distribution from $\Upsilon(10860) \to \Upsilon(2S) \pi^+ \pi^-$ (lower left frame).
 The data are from the Belle collaboration~\cite{Belle:2011aa}.
The upper right hand frame shows the dipion invariant mass distribution in $\Upsilon(4S) \to \Upsilon(1S) \pi^+\pi^-$,
and the theoretical curve (with the references) is based on  the Zweig-forbidden process shown below. The measured decay widths 
from $\Upsilon(nS) \to \Upsilon(1S)\pi^+ \pi^-$ $nS=2S,3S,4S$ and $\Upsilon(10860) \to \Upsilon(1S) \pi^+\pi^-$
are also shown.}
\label{ali:fig4}
\end{figure}

 The hypothesis that 
 $\Upsilon(5S)$ and $Y_b(10890)$, while having the same $J^{\rm PC}=1^{--}$ quantum numbers and almost the same mass, are {\it different} states, is also hinted by the
drastically different decay characteristics of the dipionic transitions involving the
 lower quarkonia $S$-states, such as $\Upsilon(4S) \to \Upsilon(1S) \pi^+\pi^-$, on one hand, and similar decays of the
$Y_b$, on the other. These anomalies are seen both in the decay rates and in the dipion invariant mas spectra in the
$\Upsilon(nS)\pi^+\pi^-$ modes. The large branching ratios of $Y_b \to \Upsilon(nS) \pi^+\pi^-$,
 as well as of $Y(4260) \to J/\psi \pi^+\pi^-$, are due to the Zweig-allowed nature of these transitions,
as the initial and final states have the same valence quarks. The final state $\Upsilon(nS) \pi^+\pi^-$ in
$Y_b$ decays requires the excitation of a $q\bar{q}$ pair from the vacuum. Since, the light scalars
$\sigma_0$, $f_0(980)$ are themselves tetraquark candidates~\cite{Hooft:2008we,Fariborz:2008bd}, 
 they are expected to show up
 in the $\pi^+\pi^-$
invariant mass distributions, as opposed to the corresponding spectrum in the transition
$\Upsilon(4S) \to \Upsilon(1S) \pi^+\pi^-$ (see Fig.~\ref{ali:fig4}). 
 Subsequent discoveries~\cite{Belle:2011aa} of the charged states $Z_b^+(10610)$ and
$Z_b^+(10650)$, found in the decays $\Upsilon(10860)/Y_b \to Z_b^+(10610) \pi^-, Z_b^+(10650) \pi^-$, 
leading to the final states $\Upsilon(1S) \pi^+\pi^-$, $\Upsilon(2S) \pi^+\pi^-$, $\Upsilon(3S) \pi^+\pi^-$,
$h_b(1P)\pi^+\pi^-$ and $h_b(2P) \pi^+\pi^-$, also admit a tetraquark interpretation, as discussed below.

\subsection{\boldmath{Heavy-Quark-Spin Flip in $\Upsilon(10860) \to h_b(1P, 2P) \pi \pi$}}%
\begin{figure}
\centerline{\includegraphics[width=8cm]{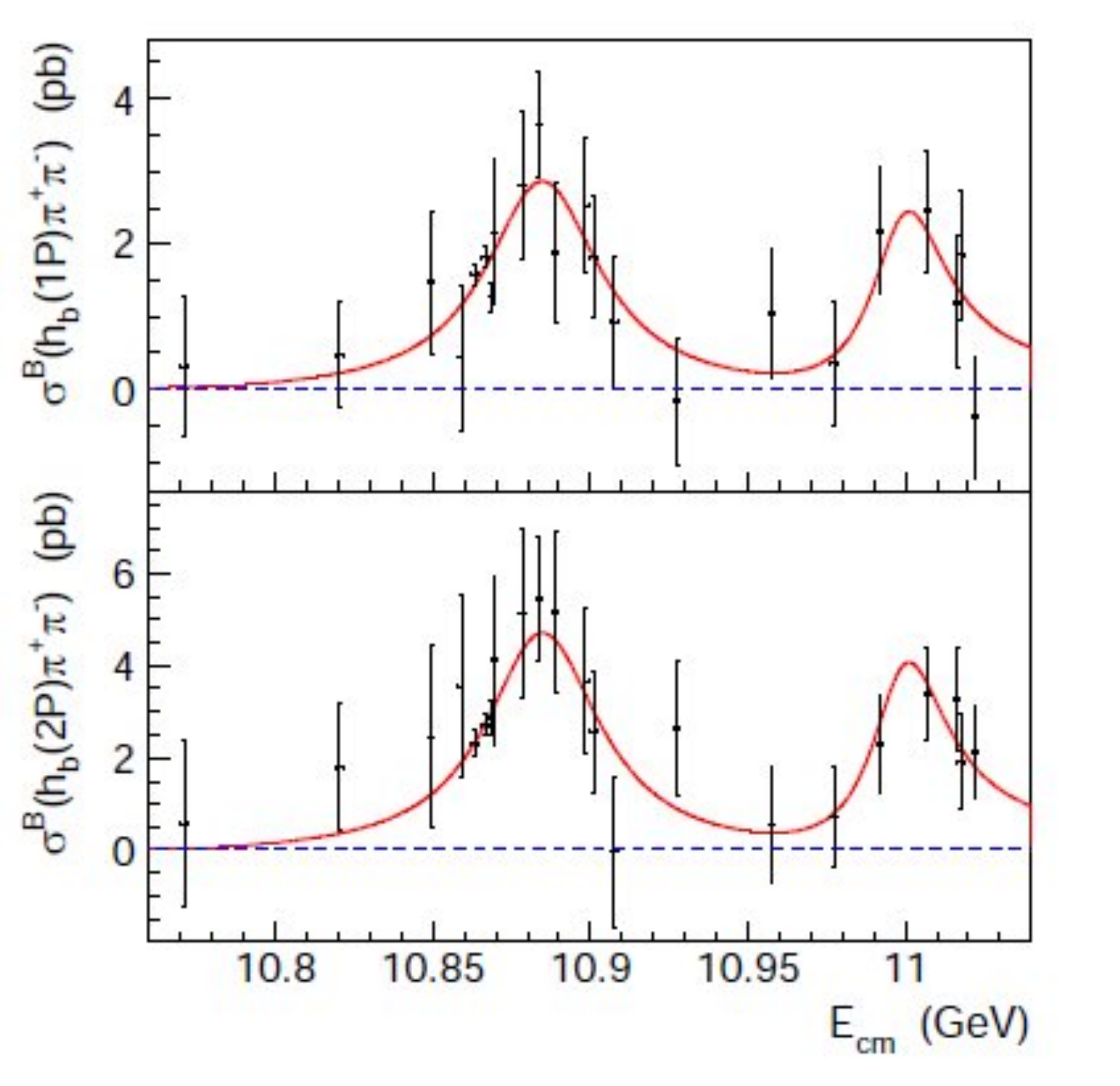}}
\vspace{-2mm}
\caption{$\sigma(e^+e^- \to h_b(1P) \pi^+\pi^-)$ and  $\sigma(e^+e^- \to h_b(2P) \pi^+\pi^)$-in the $\Upsilon(10860)$ and $\Upsilon(11020)$
 region(from Belle~\cite{Abdesselam:2015zza}).}
\label{ali:fig-hb}
\end{figure}

The cross-section $\sigma(e^+e^- \to  (h_b(1P),h_b(2P) \pi^+\pi^-)$  measured by Belle~\cite{Abdesselam:2015zza} is shown
in Fig.~\ref{ali:fig-hb}, providing clear evidence of production in the $\Upsilon(10860)$ and $\Upsilon(11020)$ region.
We summarize the relative rates and strong phases measured by Belle~\cite{Belle:2011aa} in the process $\Upsilon(10860) \to
 \Upsilon(nS) \pi^+\pi^-, h_b(mP) \pi^+\pi^-$, with $n=1,2,3$ and $m=1,2$ in Table~\ref{ali:tbl5}.
We use the notation $Z_b$ and $Z_b^\prime$ for the two charged $Z_b$ states. Here no assumption is
made about the nature of $\Upsilon(10860)$; it can be either $\Upsilon(5S)$ or $Y_b$. Of these, the decay $\Upsilon(10860)
\to \Upsilon(1S) \pi^+\pi^-$ involves both a resonant (i.e., via $Z/Z^\prime$) and a direct component, but the other four
are dominated by the resonant contribution. One notices that the relative normalizations are very similar and
the phases of the $(\Upsilon(2S), \Upsilon(3S)) \pi^+\pi^-$ differ by about $180^\circ$ compared to
the ones in $(h_b(1P, h_b(2P))\pi^+\pi^-$. At the first sight this seems to violate the heavy-quark-spin conservation,
as in the initial state $s_{b\bar{b}}=1$, which remains unchanged for the $\Upsilon(nS)$ in the final state, i.e., it
involves an $s_{b\bar{b}}=1 \to s_{b\bar{b}}=1$ transition, but as  $s_{b\bar{b}}=0$ for the $h_b(mP)$, this involves
an $s_{b\bar{b}}=1 \to s_{b\bar{b}}=0$ transition, which should have been suppressed, but is not
 supported by data. 
{\renewcommand{\arraystretch}{1.5}
\begin{table}
\begin{center}
\tbl{Relative normalizations and  phases for
 $s_{b\bar{b}}: 1 \to 1$~ and $1 \to 0$~transitions in 
$\Upsilon(10860)$ decays~\cite{Belle:2011aa}.}
\vspace*{2mm}
{\begin{tabular}{ l  c  c  c  c  c  }
\hline
   \hspace{-2mm} Final State \hspace{-2mm} & 
$\Upsilon(1S)\pi^+\pi^-$ & $\Upsilon(2S)\pi^+\pi^-$ & $\Upsilon(3S)\pi^+\pi^-$ &  
$h_b(1P)\pi^+\pi^-$ & \hspace{-2mm} $h_b(2P)\pi^+\pi^-$ \hspace{-2mm} \\
   \hline
 \hspace{-2mm} Rel. Norm.\hspace{-2mm} & 
\hspace{-2mm} $0.57\pm 0.21^{+0.19}_{-0.04}$ \hspace{-2mm} & 
\hspace{-2mm} $0.86\pm 0.11^{+0.04}_{-0.10}$ \hspace{-2mm} & 
\hspace{-2mm} $0.96\pm 0.14^{+0.08}_{-0.05}$ \hspace{-2mm} & 
\hspace{-2mm} $1.39\pm 0.37^{+0.05}_{-0.15}$ \hspace{-2mm} & 
\hspace{-4mm} $1.6^{+0.6+0.4}_{-0.4-0.6} $ \hspace{-4mm} \\
  \hspace{-2mm} Rel. Phase \hspace{-3mm} & 
$58\pm43^{+4}_{-9}$ & $-13\pm13^{+17}_{-8}$ & $-9\pm 19^{+11}_{-26}$ &  $187^{+44+3}_{-57-12}$ & 
\hspace{-4mm} $181^{+65+74}_{-105-109}$ \hspace{-4mm} \\[-1mm]
\hline
\end{tabular}
}
\label{ali:tbl5}
\end{center}
\end{table}
{\renewcommand{\arraystretch}{1}
\vspace{-1mm}
It has been shown that this contradiction is only apparent~\cite{Ali:2014dva}.

In the tetraquark picture, one has a triplet of $J^{PG}=1^{++}$ states, the  $Z_b$ and $Z_b^\prime$,
and another, yet to be discovered $X_b$, with $C=+1$. They  have the following form in the diquark-antidiquark spin representation:
\begin{eqnarray}
\ket{Z_b}  &=& \frac{\ket{1_{bq}, 0_{\bar{b}\bar{q}}} -  \ket{0_{bq}, 1_{\bar{b}\bar{q}}}}{\sqrt{2}},\nonumber\\
\ket{Z_b^\prime} &=& \ket{1_{bq}, 1_{\bar{b}\bar{q}}}_{J=1}, \nonumber\\
\ket{ X_b} &=&  \frac{\ket{1_{bq}, 0_{\bar{b}\bar{q}}} + \ket{0_{bq}, 1_{\bar{b}\bar{q}}}}{\sqrt{2}}.
\label{eq:tetraquark-Zb}
\end{eqnarray} 
These definitions correspond to $Z$, $Z^\prime$, and $X_1$ of Table \ref{ali:tbl1}, column 3, with $Q=b$.
Here $Z_b^\prime$ is the heavier one, with $M(Z_b^\prime) - M(Z_b)=2 \kappa_b \simeq 45$ MeV, consistent
with Table \ref{ali:tbl1}. This fixes
$\kappa_b$, which can also be estimated from the mass difference of the charged states in the charm sector $M(Z_c^\prime) - M(Z_c)=2 \kappa_c \simeq 120$ MeV, and the QCD expectations $\kappa_b:\kappa_c= m_c:m_b$.
 Expressing the states $Z_b$~and $Z_b^\prime$~in the
basis of definite $b\bar{b}$~and light quark $q\bar{q}$~spins, it becomes
evident that both the $Z_b$ and $Z_b^\prime$ have $s_{b\bar{b}}=1$ and $s_{b\bar{b}}=0$
components, 
\begin{eqnarray}
&&|Z_b\rangle=\frac{|1_{q\bar q},0_{b\bar b}\rangle-|0_{q\bar q},1_{b\bar b}\rangle}{\sqrt{2}},
~~|Z_b^\prime\rangle=\frac{|1_{q\bar q},0_{b\bar b}\rangle+|0_{q\bar q},1_{b\bar b}\rangle}{\sqrt{2}}.
\end{eqnarray}
Note, that this differs from the corresponding expressions given earlier in Eq. (\ref{eq:molecule-Zb}), which is based on the molecular interpretation of the states $Z_b$ and $Z_b^\prime$, but consistent with  the definitions of
$Z$ and $Z^\prime$ in Table \ref{ali:tbl1}.
It is conceivable that the subdominant spin-spin interactions may play a non negligible role in the
$b$-systems, as the spin-spin dominant interaction is suppressed by the large
$b$-quark mass. In this case the composition of the $Z_b$ and $Z_b^\prime$
indicated above  would be more general.

\begin{eqnarray}
&&|Z_b\rangle=\frac{\alpha |1_{q\bar q},0_{b\bar b}\rangle-\beta |0_{q\bar q},1_{b\bar b}\rangle}{\sqrt{2}},
~~|Z_b^\prime\rangle=\frac{\beta |1_{q\bar q},0_{b\bar b}\rangle+\alpha |0_{q\bar q},1_{b\bar b}\rangle}{\sqrt{2}}.
\label{eq:mixed-Zb-Zbp}
\end{eqnarray}
 Defining ($g$~is the effective couplings at the vertices $\Upsilon\, Z_b\, \pi$ and $Z_b\, h_b\, \pi$)
\begin{eqnarray}
&&g_{Z}\equiv g(\Upsilon \to Z_b\pi)g(Z_b\to h_b\pi)\propto -\alpha\beta\langle h_b|Z_b\rangle \langle Z_b|\Upsilon\rangle,\nonumber\\
&&g_{Z^\prime}\equiv g(\Upsilon \to Z_b^\prime\pi)g(Z_b^\prime \to h_b\pi)\propto \alpha\beta\langle h_b|Z^\prime_b\rangle
 \langle Z^\prime_b|\Upsilon\rangle,
\end{eqnarray}
we note that within errors, Belle data is consistent with the heavy quark spin conservation,
which requires~$g_Z=-g_{Z^\prime}$. The two-component nature of the $Z_b$ and $Z_b^\prime$ is also the feature
which was pointed out earlier for the $Y_b$ in the context of the direct transition $Y_b(10890) \to \Upsilon(1S) \pi^+\pi^-$.
 To determine the coefficients $\alpha$~and $\beta$, one has to resort to 
$s_{b\bar{b}}$: $1 \to 1 $~transitions
%
\begin{equation}
 \Upsilon(10860)\to  Z_b/Z_b^\prime+\pi\to  \Upsilon(nS)\pi\pi~(n=1,2,3).
 \end{equation}
The analogous effective couplings are
\begin{eqnarray}
&&f_{Z}=f(\Upsilon \to Z_b\pi)f(Z_b\to \Upsilon(nS)\pi)\propto |\beta|^2 \langle \Upsilon(nS)|0_{q\bar q},1_{b\bar b}\rangle \langle 0_{q\bar q},1_{b\bar b}|\Upsilon\rangle,\notag \\ 
&&f_{Z^\prime}=f(\Upsilon \to Z_b^\prime\pi)f(Z_b^\prime\to \Upsilon(nS)\pi)\propto |\alpha|^2 \langle \Upsilon(nS)|0_{q\bar q},1_{b\bar b}\rangle \langle 0_{q\bar q},1_{b\bar b} |\Upsilon\rangle.
\end{eqnarray}
 Dalitz analysis indicates that
 $\Upsilon(10860)\to  Z_b/Z_b^\prime+\pi\to  \Upsilon(nS)\pi\pi~(n=1,2,3)$~proceed
mainly through the resonances $Z_b$~and $Z_b^\prime$, though
 $\Upsilon(10860) \to  \Upsilon(1S)\pi\pi$~has a significant direct component,
expected in tetraquark interpretation of $\Upsilon(10860)$~\cite{AHS:2010}.
A comprehensive analysis of the Belle data including the direct and resonant
components is required to test the underlying dynamics, which is yet to be carried out.
However,  parametrizing the amplitudes in terms of two Breit-Wigners, one can determine the
ratio $\alpha/\beta$ from 
$\Upsilon(10860)\to  Z_b/Z_b^\prime+\pi\to  \Upsilon(nS)\pi\pi~(n=1,2,3)$. For the $s_{b\bar b}:1\to 1~{\rm transition}$, 
one obtains for the averaged quantities:
\begin{eqnarray}
\overline{{\rm Rel. Norm.}}= 0.85\pm 0.08=|\alpha|^2/|\beta|^2;
~~\overline{{\rm Rel. Phase}}= (-8\pm10)^\circ.
\end{eqnarray}
For the $s_{b\bar b}:1\to 0~{\rm transition}$, the corresponding quantities are
\begin{eqnarray}
\overline{{\rm Rel. Norm.}}= 1.4\pm 0.3;
~~\overline{{\rm Rel. Phase}}= (185 \pm 42)^\circ.
\end{eqnarray}
Within errors, the tetraquark assignment  with $\alpha=\beta=1$~is
 supported, i.e.,
\begin{eqnarray}
&&|Z_b\rangle=\frac{|1_{bq},0_{\bar b\bar q}\rangle-|0_{bq},1_{\bar b\bar q}\rangle}{\sqrt{2}},
~~|Z_b^\prime\rangle=|1_{b q},1_{\bar b\bar q}\rangle_{J=1},
\end{eqnarray}
and
\begin{eqnarray}
&&|Z_b\rangle=\frac{|1_{q\bar q},0_{b\bar b}\rangle-|0_{q\bar q},1_{b\bar b}\rangle}{\sqrt{2}},
~~|Z_b^\prime\rangle=\frac{|1_{q\bar q},0_{b\bar b}\rangle+|0_{q\bar q},1_{b\bar b}\rangle}{\sqrt{2}}.
\end{eqnarray}
It is interesting that a similar conclusion was drawn in the `molecular' interpretation~\cite{Bondar:2011ev}
 of the $Z_b$ and $Z_b^\prime$.

The Fierz rearrangement used in obtaining the econd of the above relations would put together the $b\bar{q}$ and $q\bar{b}$ fields,
yielding 

\begin{eqnarray}
&&|Z_b\rangle=|1_{b \bar{q}},1_{\bar {b} q}\rangle_{J=1},
~~|Z_b^\prime \rangle=\frac{|1_{b\bar{q}},0_{q \bar{b}}\rangle+|0_{b\bar{q}},1_{q \bar{b}}\rangle}{\sqrt{2}}.
\end{eqnarray}
Here, the labels $0_{b\bar{q}}$ and $1_{\bar{q}b}$ could be viewed as indicating $B$ and $B^*$ mesons, respectively,
leading to the prediction $Z_b \to B^* \bar{B}^*$ and $Z_b^\prime \to B \bar{B}^*$, which is not in agreement
 with the Belle data~\cite{Belle:2011aa}.
However, this argument rests on the conservation of the light quark spin, for which there is no theoretical
 foundation. Hence,
this last relation is not reliable. Since $Y_b(10890)$ and $\Upsilon(5S)$ are rather close in mass, and there
 is an issue with
the unaccounted {\it direct production} of the $B^* \bar{B}^*$ and $B \bar{B}^*$ states in the Belle data collected 
in their vicinity, we remark that the experimental situation is still in a state of flux and
look forward to its resolution with the upcoming  Belle-II data.

\subsection{Drell-Yan mechanism for vector exotica production at the LHC and Tevatron}
The exotic hadrons having $J^{\rm PC}=1^{--}$ can be produced at the Tevatron and LHC via the Drell-Yan
 process~\cite{Ali:2011qi}
$pp (\bar{p}) \to \gamma^* \to V +...$. The cases $V=\phi(2170), Y(4260), Y_b(10890)$ have been studied~\cite{Ali:2011qi}.
  With the other two hadrons already discussed earlier, we recall that the
 $\phi(2170)$ was first observed in the ISR process $e^+e^- \to \gamma_{\rm ISR} f_0(980) \phi(1020)$ by
 BaBar~\cite{Aubert:2006bu} and later confirmed by BESII~\cite{Ablikim:2007ab} and Belle~\cite{Shen:2009zze}.
 Drenska {\it et al.}~\cite{Drenska:2008gr} interpreted  $\phi(2170)$ as a $P$-wave tetraquark
 $[sq][\bar{s} \bar{q}]$ ($Y(2170)$). Thus,  all three vector exotica are assumed to be the first orbital 
excitation of diquark-antidiquark states with a hidden $s\bar{s}$, $c\bar{c}$ and $b\bar{b}$ quark content,
respectively. Since all three have very small branching ratios in a dilepton pair, they should be searched
for in the decay modes in which they have been discovered, all involving four charged particles,
which, in principle, can be detected in the experiments at hadron colliders.  The cross sections  for the processes
$p \bar{p}(p) \to \phi(2170) (\to  \phi(1020) f_0(980) \to K^+K^- \pi^+\pi^-)$,
  $p \bar{p}(p) \to Y(4260)(\to  J/\psi \pi^+\pi^- \to \mu^+\mu^- \pi^+\pi^-)$, and
 $ p \bar{p}(p) \to  Y_b(10890) (\to  \Upsilon (1S,2S,3S)\pi^+\pi^- \to \mu^+\mu^- \pi^+\pi^-)$,  
 at the Tevatron ($\sqrt s=$ 1.96 TeV) and the LHC are computed in ~\cite{Ali:2011qi}.
All these processes have measurable rates, and they should be searched for at the LHC. 

Summarizing the tetraquark discussion, we note that there are several puzzles in the $X,Y,Z$ sector.
 First, and foremost, a very rich spectrum of tetraquark states is predicted in the diquark scenario and the continued absence of many of the predicted states is enigmatic. The nature of the observed states
$J^{PC}=1^{--}$, $Y(4260)$ and  $Y_b(10890)$ is another open question, and whether they are related with each
other. Also, whether $Y_b(10890)$ and $\Upsilon(5S)$ are one and the same particle is still an open issue. 
 In principle, both  $Y(4260)$ and  $Y_b(10890)$ can be produced at the LHC and measured through
 the $J\psi \pi^+\pi^-$ and
$\Upsilon(nS) \pi^+\pi^-$ $(nS=1S,2S,3S)$ modes, respectively. Their hadroproduction cross-sections are unfortunately
 uncertain, but their (normalized) transverse momentum distributions will be quite revealing.
 As they are both $J^{PC}=1^{--}$ hadrons, they
can also be produced via the Drell-Yan mechanism and detected through their signature decay modes.
 The tetraquark interpretation of the charged
exotics $Z_b$ and $Z_b^\prime$ leads to a straight forward understanding of the relative rates
and strong phases of the heavy quark spin non-flip and
 spin-flip transitions in the decays $\Upsilon(10860) \to \Upsilon(nS) \pi^+\pi^-$ and 
 $\Upsilon(10860) \to h_b(mP) \pi^+\pi^-$, respectively.  However, these transitions can also be
 accommodated in the hadron molecule approach. 
 In the tetraquark picture, 
the corresponding hadrons in the charm sector $Z_c$ and $Z_c^\prime$ are related to their $b\bar{b}$ counterparts. 
A satisfactory dynamical formalism will have to deal with the unavoidable couplings of the physical states to the meson-meson
components whose threshold is close by. The closeness of the thresholds is bound to have an influence on the detailed
properties and the structure of the states. This is generally so, and the tetraquarks are no exception. Despite this, one
hopes that the charateristic features of strongly bound tetraquarks will remain discernible. 

A final comment is about the tetraquarks as candidates for the baryonium states. This is best illustrated in the case of the $L=1$ $Y$-states,
discussed earlier in the charmonium-like exotics. It was argued early on \cite{Cotugno:2009ys} that the state
$Y(4630)$, observed by Belle in 2008, in the decay mode $Y(4630) \to \Lambda_c \bar{\Lambda}_c$, with a width 
$\Gamma(Y(4639)= 92^{+41}_{-32} $ MeV, is probably the same as the state $Y(4660)$, seen in the decay
$Y(4660) \to \psi^\prime \pi^+ \pi^-$. The dominant decay mode is the $\Lambda_c \bar{\Lambda}_c$. This
data was interpreted as the first example of the charmed baryonium formed by four quarks. The general pattern that
the most natural decay of a tetraquark state, if allowed by phase space and other quantum numbers, is in a pair of
baryon-antibaryon, is anticipated also in the string-junction picture of the multiquark states \cite{ Rossi:2016szw},
and in the holography inspired stringy hadron (HISH) perspective \cite{Sonnenschein:2016ibx}. A corollary of this
picture is that the tetraquark states, very much like the $q\bar{q}$ mesons, are expected to lie on a Regge trajectory,
and predictions about a few excited states in the $s\bar{s}$, $c\bar{c}$, and the $b\bar{b}$ are available in the
literature \cite{Sonnenschein:2016ibx}. They should be searched for at the LHC. The Regge behavior of the excited tetraquark
states, if confirmed experimentally, would underscore the fundamental difference anticipated between the tetraquarks
and other competing scenarios, such as the kinematic cusps and hadron molecules, for which the Regge trajectories
are not foreseen.

%% file: Sec-Theoretical-models-for-pentaquarks-rev.tex
\section{Theoretical models for pentaquarks}
\label{sec:Theoretical-models-for-pentaquarks}
Pentaquarks remained elusive for almost a decade under the shadow of the botched discoveries
 of $\Theta(1540),\,\, \Phi(1860),\,\,\Theta_c(3100)$. This has definitely changed by the observation of $J/\psi p$ resonances consistent with pentaquark states in
$\Lambda_b^0 \to J/\psi K^- p$ decays by the LHCb collaboration~\cite{Aaij:2015tga}.
The measured distributions in the
invariant masses $m_{Kp}$ and  $m_{J/\psi p}$ are shown in Fig.~\ref{Pc2d} together with a model
comparison with two $P_c^+$ states.
A statistically good fit of the $m_{J/\psi p}$ distribution is consistent with the presence of two resonant
states, called $P_c(4450)^+$ and $P_c(4380)^+$, discussed earlier.
 Both of these states  carry a unit of baryonic number and have the valence quarks
$ P_c^+ = \bar c c u u d$. The preferred $J^{\rm P}$ 
assignments are $5/2^+$ for the $P_c(4450)^+$ and $3/2^-$ for the $P_c(4380)^+$. 

 The  Argand-diagram analysis in
the (${\rm Im}~A^{P_c}$ - ${\rm Re}~A^{P_c}$) plane found that the phase change in the amplitude is consistent
with a resonance for the $P_c(4450)^+$, but less so for the $P_c(4380)^+$, as shown in Fig.~\ref{DoubleArgand}.
The  phase diagram for the $P_c(4380)^+$ state  needs further study with more
data, but the resonant character of the $P_c(4450)^+$ state is very likely. This will be
contrasted with the corresponding phase diagram resulting from the assumption that
 $P_c(4450)^+$ is a kinematically-induced cusp state.

Following a pattern seen for the tetraquark candidates, namely their proximity to respective thresholds,
such as $D\bar{D}^*$ for the $X(3872)$, $B\bar{B}^*$ and $B^*\bar{B}^*$ for the $Z_b(10610)$ and
$Z_b(10650)$, respectively, also the two pentaquark candidates $P_c(4380)$ and $P_c(4450)$ lie close to several charm
meson-baryon thresholds~\cite{Burns:2015dwa}. The $\Sigma_c^{*+} \bar{D}^0$ has a threshold
 of $4382.3 \pm 2.4$ MeV, tantalizingly
close to the mass of $P_c(4380)^+$. In the case of $P_c(4450)^+$, there are several thresholds within
striking distance, $\chi_{c1}p (4448.93 \pm 0.07), \Lambda_c^{*+} \bar{D}^0 (4457.09 \pm 0.35),
\Sigma_c^+ \bar{D}^{*0} (4459.9 \pm 0.9)$, and $\Sigma_c^+ \bar{D}^0 \pi^0 (4452.7 \pm 0.5)$, where the masses are
in units of MeV. This has led to a number of hypotheses to explain the two $P_c$ states:
\begin{itemize}
\item $P_c(4380)$ and $P_c(4450)$  are baryocharmonia~\cite{Kubarovsky:2015aaa}.
\item Rescattering-induced kinematic effects are mimicking the resonances~\cite{Guo:2015umn,Liu:2015fea,Mikhasenko:2015vca}.
\item The state $P_c(4450)$ is a composite of  $\xi_{c1} p$ \cite{Meissner:2015mza}.
\item They are open charm-baryon and charm-meson bound 
states~\cite{Chen:2015moa,He:2015cea,Roca:2015dva,Chen:2015loa,Xiao:2015fia,Azizi:2016dhy}.
\item They are compact diquark-diquark-antiquark states~\cite{Maiani:2015vwa,Li:2015gta,Mironov:2015ica,Anisovich:2015cia,Ghosh:2015ksa,Wang:2015epa,Wang:2015ava,Ali:2016dkf},
with each component being a $\bar{3}$, yielding a color-singlet $\bar{c}[cq][qq]$ state. Another possibility is
via the sequential formation of compact color triplets, making up diquark-triquark
systems, yielding also color-singlet states~\cite{Lebed:2015tna,Zhu:2015bba}.
\item Finally, there are also studies of the LHCb pentaquarks as compact five quarks interacting through a
chromomagnetic hyperfine interaction, without diquark correlations \cite{Takeuchi:2016ejt,Park:2017jbn}.
\end{itemize}
In the baryocharmonium picture, the $P_c$ states are hadroquarkonium-type composites of $J/\psi$ and excited
nucleon states similar to the known resonances $N(1440)$ and $N(1520)$.  Photoproduction
of the $P_c$ states in $\gamma + p$ collisions is advocated as sensitive probe of this mechanism~\cite{Kubarovsky:2015aaa}.
 We shall  shortly discuss below the interpretation of pentaquarks as scattering-induced kinematic effects,
and as meson-baryon molecules, and review the compact diquark-based models in some detail.

\subsection{Pentaquarks as rescattering-induced kinematic effects}
Kinematic effects can result in a narrow
structure around the $\chi_{c1}p$ threshold. Two possible mechanisms shown in Fig.~\ref{ali:fig7} are:
 (a) 2-point loop with a 3-body
production~$\Lambda_b^0 \to K^- \,\chi_{c1}\, p $~followed by the rescattering
 process~$\chi_{c1}\, p \to J/\psi\, p$, and  
(b) in which $K^-\,p $~is produced from an intermediate~$\Lambda^*$~and the proton rescatters
 with the~$\chi_{c1}$~into a~{$J/\psi \, p$, as shown below.

\begin{figure}[b]
\centerline{\includegraphics[width=0.75\textwidth]{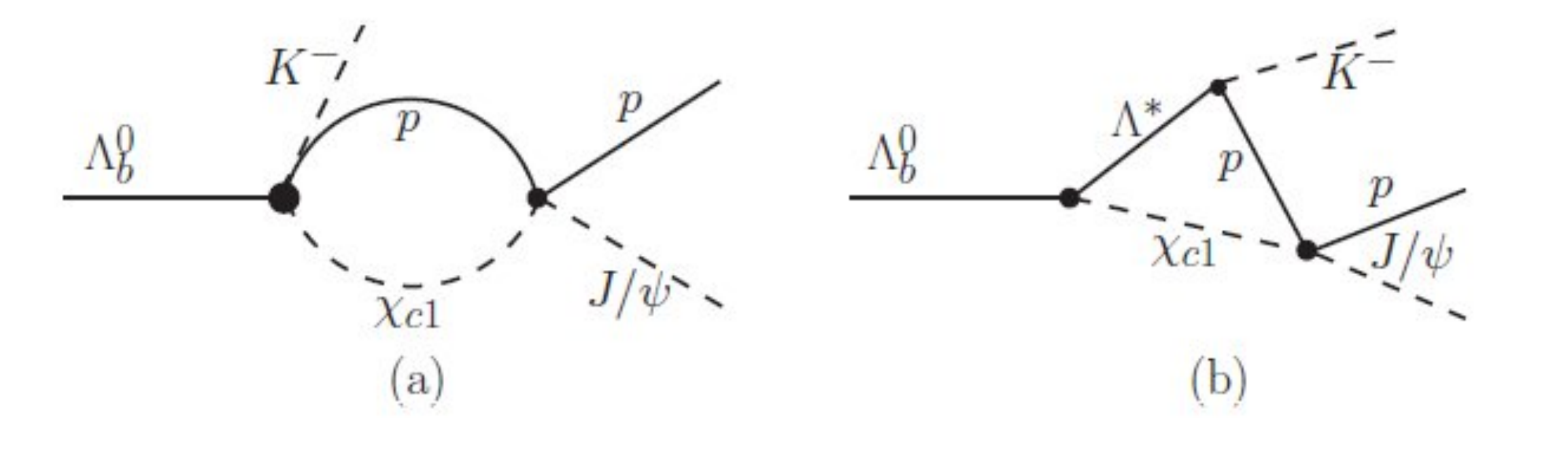}}
\vspace{-2mm}
\caption{The two scattering diagrams discussed in the text( from~\cite{Guo:2015umn}).}
\label{ali:fig7}
\end{figure}
 The amplitude for Fig.~\ref{ali:fig7}(a) can be expressed as
\vspace*{-2mm}
\begin{eqnarray}
&& G_{\Lambda}(E)=\int \frac{d^3q}{(2\pi)^3} \frac{\vec{q}^2 f_\Lambda(\vec{q}^2)}{E-m_p -m_{\chi_{c1}} -
 \vec{q}^2/(2 \mu)},
\end{eqnarray}
where $\mu$ is the reduced mass and $f_{\Lambda}(\vec{q}^2)=\exp(-2 \vec{q}^2/\Lambda^2$)
is a form factor to regularize the loop integral.
 Fitting the Argand diagram for the $P_c(4450)^+$ with ${\cal A}_{(a)}= N(b + G_{\Lambda}(E))$
 determines the
normalization $N$, the constant background $b$, and $\Lambda$. The integral can be solved
 analytically~\cite{Guo:2015umn}
\begin{eqnarray}
&& G_{\Lambda}(E)= \frac{\mu \Lambda }{(2\pi)^{3/2}} (k^2 + \Lambda^2/4)
+ \frac{\mu k^3}{2 \pi}\exp^{-2k^2/\Lambda^2} \left[{\rm erfc}(\frac{\sqrt{2} k}{\Lambda}) -i \right]
\end{eqnarray}
where $k=\sqrt{2 \mu(E-m_1 -m_2 +i\epsilon)}$. This function has a characteristic phase motion reflecting
the error function (erfc), as shown in Fig.~\ref{ali:fig-feng-kun}.
\begin{figure}
\centerline{\includegraphics[width=6cm]{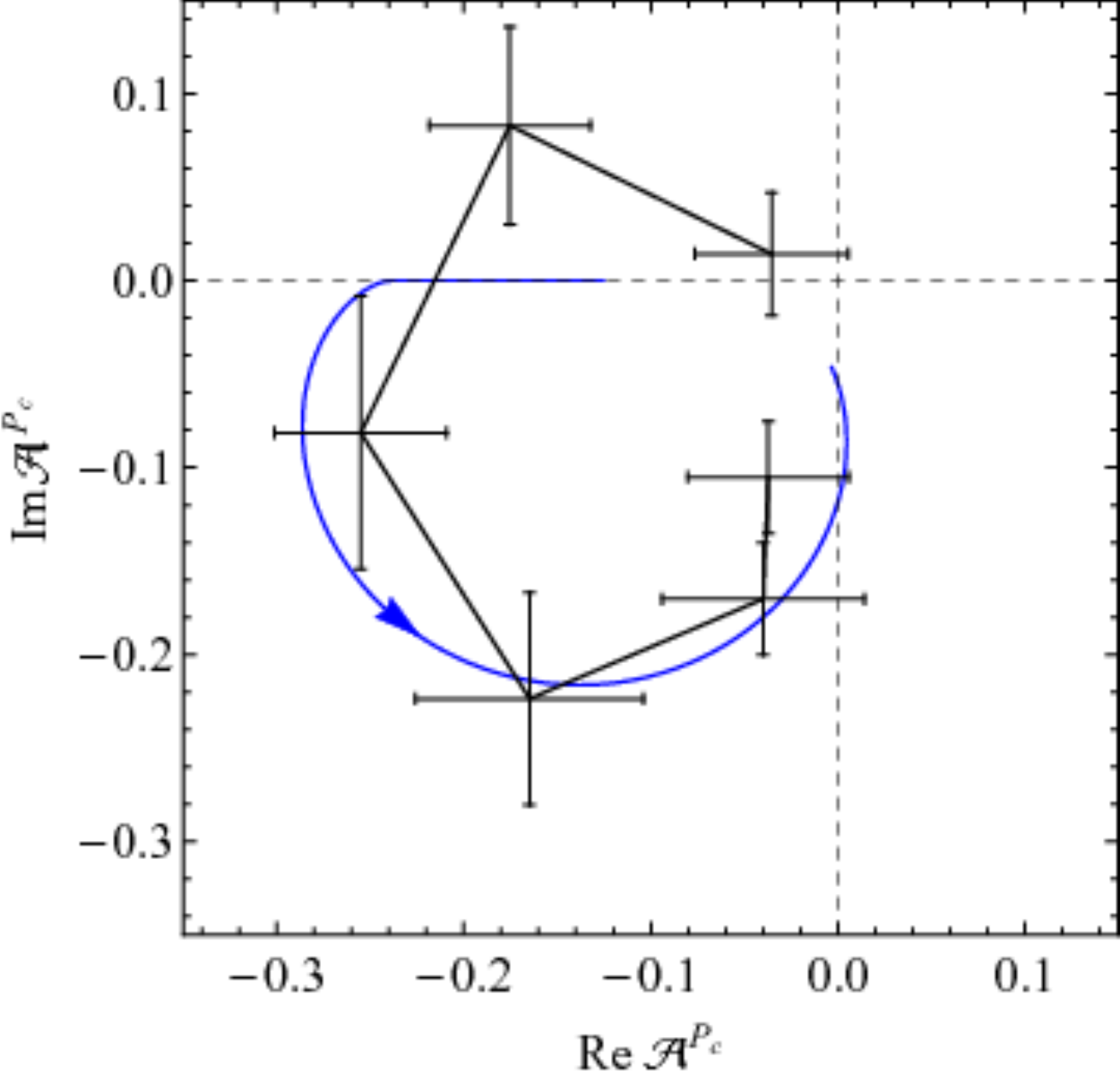}}
\caption{Fitted values of the real and imaginary parts of the amplitudes for the
$P_c(4450)^+$ using a Breit-Wigner formula with 
$M(\Gamma)$ of 4450(39) MeV~\cite{Aaij:2015tga}. The directed curve (blue) is the fit in the
cusp model.
(from~\cite{Guo:2015umn}).}
\label{ali:fig-feng-kun}
\end{figure}
 It differs from the Breit-Wigner fit, which is in excellent
agreement with the LHCb data~\cite{Aaij:2015tga}. The cusp-based fit also shows a counter-clockwise behavior in
the Argand diagram, but not for the two data points where the imaginary part of the cusp amplitude is zero.
 The absolute value of the amplitude in the cusp approach shows a resonant behavior, which
can be made to peak even more sharply at ${\rm Re}\sqrt{s}=4450$ ~MeV, if  the
 amplitude for Fig.~\ref{ali:fig7} (b) is included and assumed dominated by the $\Lambda^*(1890)$-exchange. However, it
is the phase motion, which is decisive in distinguishing a dynamical Breit-Wigner (or, for that matter a
Flatte~\cite{Flatte:1976xu} type) resonance and a kinematic-induced cusp behavior. More data is needed to
completely settle this difference in the case of $P_c(4450)^+$, but currently the Breit-Wigner fit is the preferred
description.

The singularities of the triangle loop integral from Fig.~\ref{ali:fig7} (b), describing the process $\Lambda_b \to J/\psi K^- p$ via
the $\Lambda^*$-charmonium-proton intermediate state, have been subsequently analyzed in detail \cite{Bayar:2016ftu}.
For the case of $\xi_{c1} p \to J/\psi p $, and the experimentally preferred quantum numbers $3/2^-$ or $5/2^+$, one
needs $P$- and $D$-waves, respectively, in the $\xi_{c1}p$, which, however, is at the threshold. This reduces the
strength of the contribution. In this case, it is concluded that the singularities cannot account for the observed narrow peak.

\subsection{Pentaquarks as meson-baryon molecules}
In the hadronic molecular interpretation, one identifies the $P_c(4380)^+$ 
with $\Sigma_c(2455)\bar{D}^*$~and the $P_c(4450)^+$ with $\Sigma_c(2520)\bar{D}^*$,
which are bound by meson exchanges. The underlying interaction for the case that the meson is a pion  can be expressed in terms of the 
effective Lagrangians~\cite{Chen:2015loa}:
\vspace*{-3mm}
\begin{align}
{\cal L}_{\cal P}  &=ig {\rm Tr} \left[\bar{H}_a^{(\bar{Q})}\, \gamma^\mu \, A^\mu_{ab} \, \gamma_5 H_b^{(\bar{Q})}\right],   \nonumber \\
{\cal L}_{\cal S}  &= -\frac{3}{2} g_1 \epsilon^{\mu \lambda \nu \kappa} v_\kappa {\rm Tr }
\left[\bar{\cal S}_\mu \, A_\nu \, {\cal S}_\lambda \right], 
\end{align}
\vspace*{2mm}
which are built using the heavy quark and chiral symmetries.
Here $H_a^{(\bar{Q})} = [P_a^{*(\bar{Q})\,\mu}\gamma_\mu - P_a^{(\bar{Q})} \gamma_5](1-\slashed{v})/2$
is a pseudoscalar and  vector charmed meson multiplet~$(D,D^*)$, $v$ being the four-velocity vector
 $v=(0,\vec{1})$, 
 ${\cal S}_\mu= 1/\sqrt{3}(\gamma_\mu + v_\mu) \gamma^5 {\cal B}_6 + {\cal B}^*_{6\mu}$ stands
for the charmed baryon multiplet, with~${\cal B}_6 $ and ${\cal B}^*_{6\mu} $ corresponding to
 the $J^P=1/2^+ $ and $J^P=3/2^+$ in~$6_F$ flavor representation, respectively.
 $A_\mu$~is an axial-vector current, containing a pion chiral multiplet, defined as
$A_\mu=1/2(\xi^\dag \partial_\mu \xi -\xi \partial_\mu \xi^\dag)$, with $\xi=\exp(i \mathbb {P}/f_\pi)$,
with $\mathbb {P}$ an $SU(2)$ matrix containing the pion field, and $f_\pi=132$ MeV.
This interaction Lagrangian is
used to work out effective potentials, energy levels and wave-functions
of the $ \Sigma_c^{(*)} \bar{D}^*$ systems, shown in Fig.~\ref{ali:chen-1}. In this picture,
$P_c(4380)^+$ is a $\Sigma_c \bar{D}^*$~$(I=1/2, \, J=3/2)$~molecule, and
$P_c(4450)^+$ is a $\Sigma^*_c \bar{D}^*$~$(I=1/2, \, J=5/2)$ molecule (top left and right frames, respectively).

 Apart from accommodating the two
observed pentaquarks, this framework
predicts two additional hidden-charm molecular pentaquark states,
$\Sigma_c \bar{D}^*$~$(I=3/2,\, J=1/2)$ and $\Sigma^*_c \bar{D}^*$~$(I=3/2,\, J=1/2)$ (bottom left and right frames), 
 which are isospin partners of $P_c(4380)^+$~and~$P_c(4450)^+$, respectively, 
decaying into $\Delta(1232) J/\psi$ and $\Delta(1232) \eta_c$.
\begin{figure}[b!]
\centerline{\includegraphics[width=0.6\textwidth]{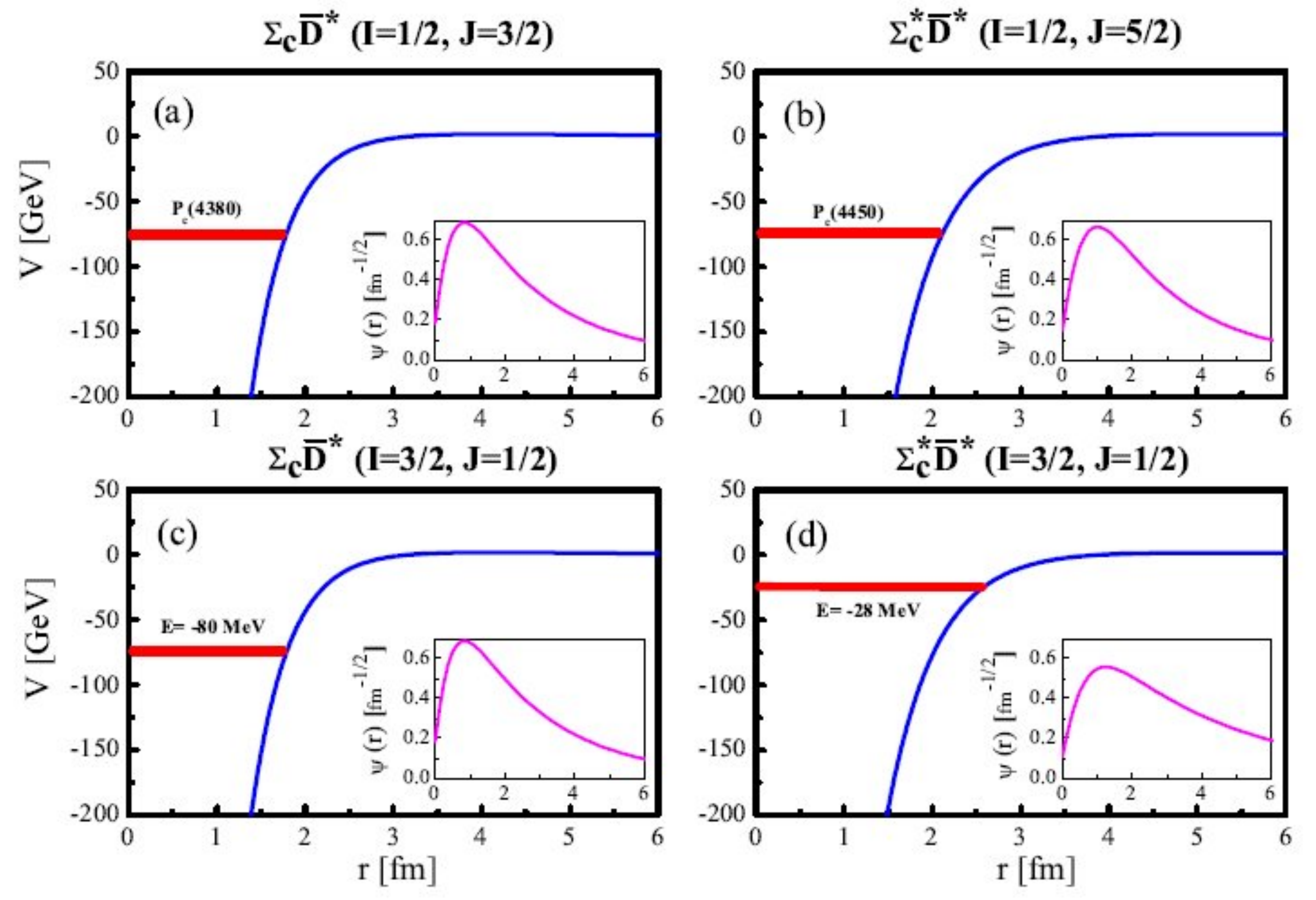}}
\vspace{-2mm}
\caption{Effective potentials, $V$ (GeV), energy levels, thick (red) lines, 
and wave-functions, $\psi(r)$, of the
$\Sigma_c^{(*)} \bar{D}^*$ system (from~\cite{Chen:2015loa}).}
\label{ali:chen-1}
\end{figure}
 In addition, 
a rich pentaquark spectrum of states for the hidden-bottom~$ (\Sigma_b B^*, \Sigma^*_b B^* )$,
$B_c$-like~$ (\Sigma_c B^*, \Sigma^*_c B^*)$ and  $(\Sigma_b \bar{D}^*, \Sigma_b^* \bar{D}^*)$
with well-defined~{$(I,J)$ is predicted.

\subsection{Pentaquarks in the compact diquark models}
  In the paper by Maiani~{\it et al.}~\cite{Maiani:2015vwa} on the pentaquark interpretation
  of the  LHCb data on $\Lambda_b^0 \to J/\psi\; p\; K^-$ decay, which is mainly discussed here,
 the assigned internal quantum numbers are:
  $P_c^+(4450)= \{\bar{c} [cu]_{s=1} [ud]_{s=0}; L_{\mathcal{P}}=1,J^{\rm P}=\frac{5}{2}^+ \}$ 
  and  $P_c^+(4380)= \{\bar{c} [cu]_{s=1} [ud]_{s=1}; L_{\mathcal{P}}=0, J^{\rm P}=\frac{3}{2}^- \}$. 
  Taking into account the mass differences due to the orbital angular momentum and the light diquark spins,
  the observed mass difference between the two $P_c^+$ states of about 70 MeV is approximately reproduced.
  The crucial assumption is that the two diagrams for the decay
 $\Lambda_b^0 \to J/\psi\; p \;K^-$  in which the $ud$-spin in
 $\Lambda_b^0$ goes over to the $[ud]$-diquark spin in the pentaquark, Fig.~\ref{ali:fig-maiani}(A),
 and the one in which the 
 $ud$-spin is shared among the final state pentaquark and a meson, generating a light diquark $[ud]$ having 
  spin-0 and spin-1, Fig.~\ref{ali:fig-maiani}(B), are treated at par. This is a dynamical assumption, and remains to be
tested. 
\begin{figure}[b]
\centerline{\includegraphics[width=12cm]{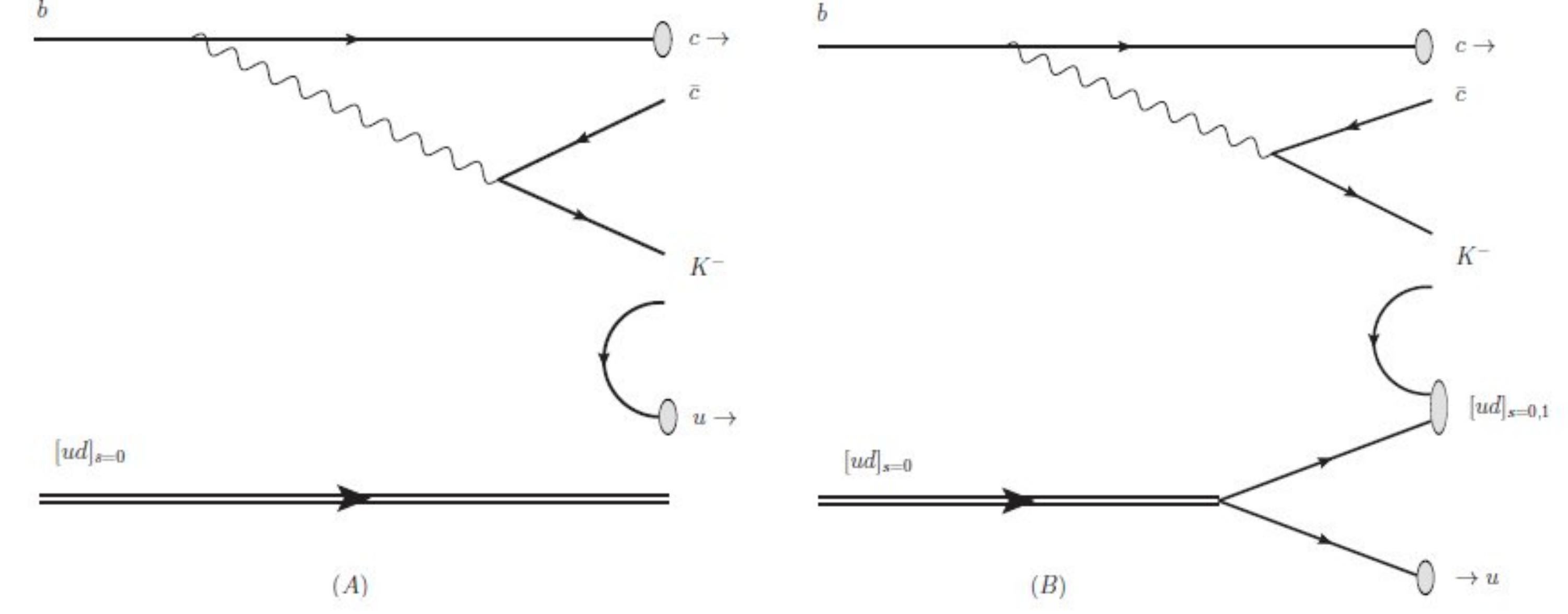}}
\caption{Two mechanisms for the decays $\Lambda_b^0 \to J/\psi  K^- p$ in the diquark picture (from~\cite{Maiani:2015vwa}).}  
\label{ali:fig-maiani}
\end{figure}

\subsection{$SU(3)_F$ structure of pentaquarks}

Concentrating on the quark flavor of the pentaquarks $\mathbb {P}_c^+=\bar{c}cuud$, they are of two
 different types~\cite{Maiani:2015vwa}:
\begin{align}
\mathbb {P}_u &= \epsilon^{\alpha \beta \gamma}\bar c_\alpha\, [cu]_{\beta, s=0,1}\, [u d]_{\gamma,s=0,1},
\\
\mathbb {P}_d &= \epsilon^{\alpha \beta \gamma}\bar c_\alpha\, [cd]_{\beta, s=0,1}\, [u u]_{\gamma,s=1}, 
\end{align}
the difference being that the $\mathbb {P}_d$ involves  a $[uu]$ diquark, and the Pauli exclusion principle
implies that this diquark has to be in an $SU(3)_F$-symmetric representation. 
This leads to two distinct $SU(3)_F$~series of pentaquarks
\begin{align}
\mathbb {P}_A &= \epsilon^{\alpha \beta \gamma}\left\{\bar c_{\alpha}\, [cq]_{\beta, s=0, 1}\,
 [q^\prime q^{\prime \prime}]_{\gamma, s=0}, L\right\}\nonumber 
= \mathbf { 3 \otimes \bar{3}= 1 \oplus 8},\nonumber\\
\mathbb {P}_S &= \epsilon^{\alpha \beta \gamma}\left\{\bar c_{\alpha}\, [cq]_{\beta, s=0, 1}\,
 [q^\prime q^{\prime \prime}]_{\gamma, s=1}, L\right\}
= \mathbf { 3 \otimes 6= 8 \oplus 10}.
\end{align}
 For $S$~waves, the first and the second series have the angular momenta 
\begin{align}
\mathbb {P}_A(L=0) &:~~~J=1/2(2),~3/2(1),\\
\mathbb {P}_S(L=0) &:~~~J=1/2(3),~3/2(3),~5/2(1),
\end{align}
where the multiplicities are given in parentheses. One assigns $\mathbb {P}(3/2^-)$ to 
the $\mathbb {P}_A$ and $\mathbb{P}(5/2^+)$ to the $\mathbb {P}_S$~series of pentaquarks~\cite{Maiani:2015vwa}.

The decay amplitudes of interest of a $b$-baryon ${\cal B}$ to an octet of pseudoscalar meson ${\cal M}$ and
a pentaquark with a hidden $c\bar{c}$, ${\cal P}$, can be generically written as%
\begin{equation}
\mathcal{A} = \left\langle \mathcal{PM}\left\vert H^W_{\text{eff}}\right\vert 
\mathcal{B}\right\rangle, \label{production-amplitude}
\end{equation}
where, $ H^W_{\text{eff}}$ is the effective weak Hamiltonian inducing the Cabibbo-allowed
$\Delta I=0, \Delta S =-1$ transition $b \to c \bar{c} s$, and the Cabibbo-suppressed $\Delta S =0$ transition $b \to c \bar{c} d$.
The $SU(3)_F$~based~analysis of the decays $\Lambda_b \to \mathbb {P}^+ K^- \to (J/\psi\, p) K^-$ goes as follows.
With respect to $SU(3)_F$, $\Lambda_b(bud)~\sim~\bar{3}$ and it is an isosinglet $I=0$. 
Thus, the weak non-leptonic Hamiltonian for $b\to c\bar{c}q$ $(q=s,d)$ decays is:
\begin{equation}
 H^W_{\text{eff}} = \frac{4 G_F}{\sqrt{2}} \left[ V_{cb} V_{cq}^* ( c_1 O_1^{(q)} +   c_2  O_2^{(q)} ) \right].
\label{weak-hamiltonian}
\end{equation}
Here, $G_F$ is the Fermi coupling constant, $V_{ij}$ are the CKM matrix elements, and 
$c_i$ are the Wilson coefficients of  the operators $O_1^{(q)}$ ($q=d,\; s  $), defined as
\begin{equation}
  O_1^{(q)}=  (\bar{q}_\alpha c_\beta)_{V-A} (\bar{c}_\alpha b_\beta)_{V-A};\;\;\;
  O_2^{(q)}=  (\bar{q}_\alpha c_\alpha)_{V-A} (\bar{c}_\beta b_\beta)_{V-A},
\label{tree-operators}
\end{equation}
where $\alpha$ and $\beta$ are $SU(3)$ color indices, and $V-A= 1 -\gamma_5$ reflects that the
charged currents are left-handed, and the penguin amplitudes are ignored.
With $M$ a nonet of $SU(3)$ light mesons $(\pi, K, \eta, \eta^\prime)$, the weak transitions  
$ \langle \mathbb {P}, M | H_{\rm W} |\Lambda_b\rangle$ requires $\mathbb{ P}+ M$  to be
in $8 \oplus 1$ representation.
Recalling the $SU(3)$ group multiplication rules
\begin{align}
8 \otimes 8~\, &= 1 \oplus 8 \oplus 8 \oplus 10 \oplus \overline{10} \oplus 27,\nonumber\\
8 \otimes 10 &= 8 \oplus 10 \oplus 27 \oplus 35,  
\end{align}
the decay $ \langle \mathbb { P}, M| H_{\rm W} |\Lambda_b\rangle$ can be realized
 with $\mathbb {P}$ in
either an octet ${\tt 8}$ or a decuplet ${\tt 10} $.
The discovery channel $\Lambda_b \to \mathbb {P}^+ K^- \to J/\psi p K^-$ corresponds 
to $\mathbb {P}$ in an  octet ${\tt 8}$.

\subsection{An effective Hamiltonian for the hidden charm pentaquarks}
Keeping the basic building blocks of the pentaquarks 
to be quarks and diquarks, we  follow  here the template in which the
 two $P_c$ states are assumed to be made from five quarks, consisting of two highly
correlated diquark pairs, and an antiquark. For the  present discussion, it is an anti-charm quark $\bar{c}$ which
is correlated with the two
diquarks  $[cq]$ and $[q^{\prime}q^{\prime \prime}]$, where $q,q^{\prime},q^{\prime \prime}$ can be $u$ or $d$.
The tetraquark formed by the diquark-diquark $([cq]_{\tt \bar{3}}[q^{\prime}q^{\prime \prime}]_{\tt \bar{3}})$ is a color-triplet
object, following from ${\tt \bar{3} \times \bar{3} = \bar{6} + 3}$, with orbital and spin quantum numbers,
 denoted by $L_{\Q\Q}$ and $S_{\Q\Q}$, 
which combines with the color-anti-triplet ${\tt \bar{3}}$ of the
 $\bar{c}$  to form an overall color-singlet pentaquark,
with the corresponding quantum numbers $L_{\P}$ and $S_{\P}$. This is 
shown schematically in Fig.~\ref{ali:fig-pentaquark-model}.

\begin{figure}
\vspace*{-5mm}
\centerline{\includegraphics[width=9.2cm]{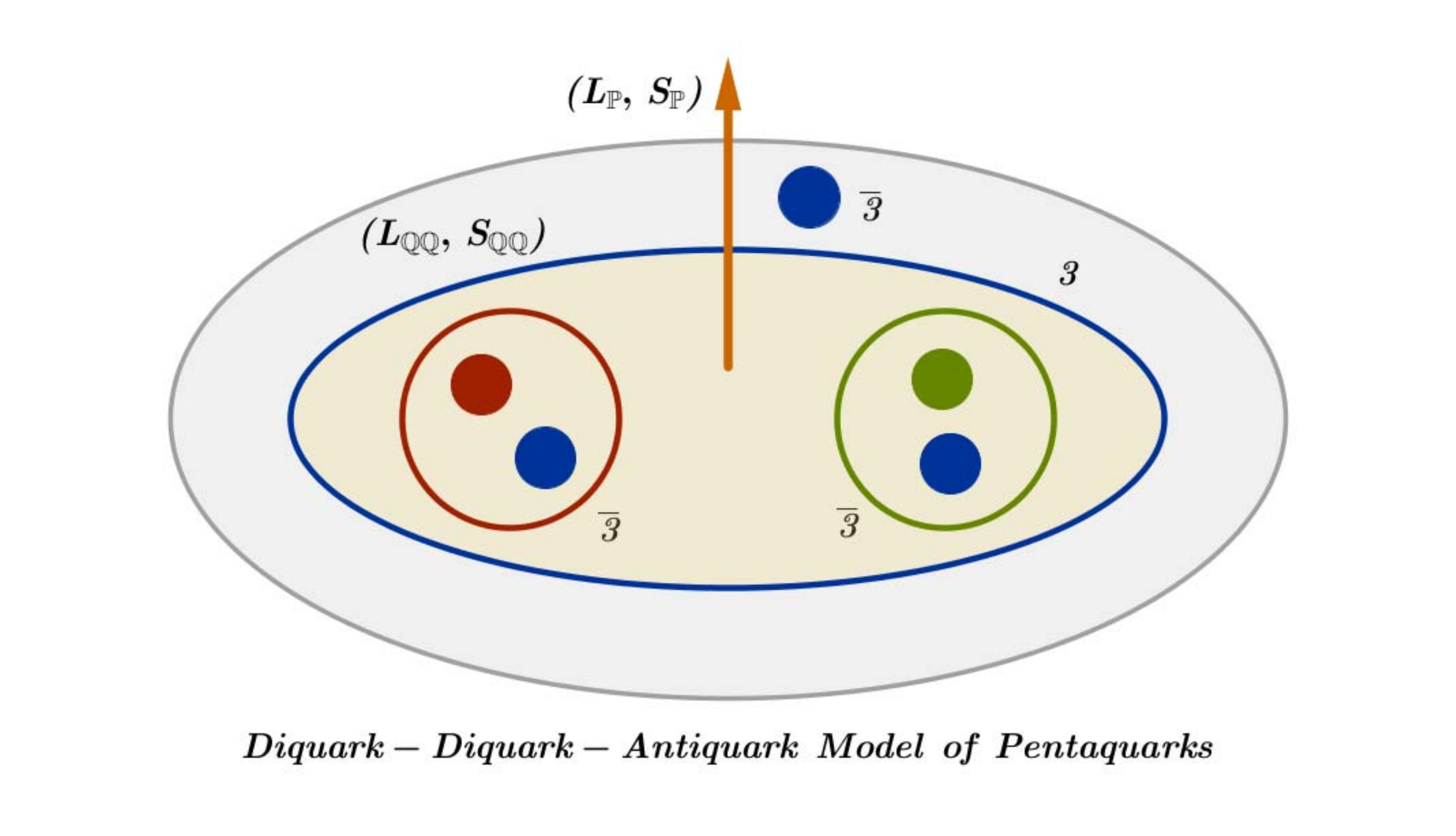}}
\caption{$SU(3)$-color quantum numbers of the diquarks, tetraquark and antiquark are indicated, together
with the orbital and spin quantum numbers of the tetraquark and pentaquark (from~\cite{Ali:2016dkf}).}
\label{ali:fig-pentaquark-model}
\end{figure}
\vspace*{3mm}

An effective Hamiltonian based on this picture is constructed~\cite{Ali:2016dkf}, extending the underlying tetraquark 
Hamiltonian developed for the $X,Y, Z$ states~\cite{Maiani:2004vq}. It involves the
constituent diquarks masses, $m_{[cq]}$, $m_{[q^{\prime}q^{\prime
    \prime}]}$, the spin-spin interactions between the quarks in each diquark shell, and the spin-orbit
and orbital angular momentum of the tetraquarks. To this are added 
 the charm quark mass $m_c$,  the spin-orbit and the orbital terms of the pentaquarks.
%
\begin{equation}
H=H_{[\mathcal{Q}\mathcal{Q}^{\prime}]} + H_{\bar{c}[\mathcal{Q}\mathcal{Q}^\prime]} + H_{S_{\P} L_{\P}} + H_{L_{\P} L_{\P}}\label{main-Hamiltion} ,
\end{equation}
where the diquarks $[cq]$ and $[q^{\prime}q^{\prime \prime}]$ are denoted by  $\mathcal{Q}$ and $\mathcal{Q}
^{\prime}$ having masses $m_{\mathcal{Q}}$ and $m_{\mathcal{Q}^{\prime}}$,
respectively.  $L_{\P}$ and $S_{\P}$ are the orbital
angular momentum and the spin of the pentaquark state, and the quantities 
$A_{\mathcal{P}}$ and  $B_{\mathcal{P}}$  parametrize the strength of their
spin-orbit  and orbital angular momentum couplings,  respectively. The individual terms in the
 Hamiltonian \eqref{main-Hamiltion} are given in~\cite{Ali:2016dkf}.

The mass formula for the pentaquark state with the ground state tetraquark ($%
L_{{\mathcal{Q}}{\mathcal{Q}}^{\prime }}=0)$ can be written as 
\begin{equation}
M=M_{0}+\frac{B_{\mathcal{P}}}{2}L_{\mathcal{P}}(L_{\mathcal{P}}+1)+2A_{%
\mathcal{P}}\frac{J_{\mathcal{P}}(J_{\mathcal{P}}+1)-L_{\mathcal{P}}(L_{%
\mathcal{P}}+1)-S_{\mathcal{P}}(S_{\mathcal{P}}+1)}{2}+\Delta M
\label{mass-formula}
\end{equation}%
where $M_{0}=m_{\mathcal{Q}}+m_{\mathcal{Q}^{\prime }}+m_{c}$ and $\Delta M$ is the mass term
that arises from different spin-spin interactions. With the tetraquark in 
$L_{{\mathcal{Q}}{\mathcal{Q}}^{\prime }}=1$, one has to add the two terms
given above with their coefficients $A_{{\mathcal{Q}}{\mathcal{Q}}^{\prime }}
$ and $B_{{\mathcal{Q}}{\mathcal{Q}}^{\prime }}$. In this work, we restrict ourselves to the $S$-wave tetraquarks.

 For $L_{\mathcal{P}}=0$, the pentaquark states are classified in terms
of the diquarks spins, $S_{\mathcal{Q}}$ and $S_{\mathcal{Q}^{\prime }}$; the spin of anti-charm
quark is $S_{\bar{c}}=1/2$. There are
four $S$-wave pentaquark states for $J^{P}=\frac{3}{2}^{-}$ and a single
state with $J^{P}=\frac{5}{2}^{-}$. 
For $J^{P}=\frac{3}{2}^{-}$, we have the following states\footnote{For a similar classification in the diquark-triquark
picture, see\cite{Zhu:2015bba}.}:
\begin{eqnarray}
|0_{\mathcal{Q}},1_{\mathcal{Q}^{\prime }},\frac{1}{2}_{\bar{c}};\frac{3}{2}\rangle _{1} &=&%
\frac{1}{\sqrt{2}}[\left( \uparrow \right) _{c}\left( \downarrow \right)
_{q}-\left( \downarrow \right) _{c}\left( \uparrow \right) _{q}]\left(
\uparrow \right) _{q^{\prime }}\left( \uparrow \right) _{q^{\prime \prime
}}\left( \uparrow \right) _{\bar{c}}  \notag \\
|1_{\mathcal{Q}},0_{\mathcal{Q}^{\prime }},\frac{1}{2}_{\bar{c}};\frac{3}{2}\rangle _{2} &=&%
\frac{1}{\sqrt{2}}[\left( \uparrow \right) _{q^{\prime }}\left( \downarrow
\right) _{q^{\prime \prime }}-\left( \downarrow \right) _{q^{\prime }}\left(
\uparrow \right) _{q^{\prime \prime }}]\left( \uparrow \right) _{c}\left(
\uparrow \right) _{q}\left( \uparrow \right) _{\bar{c}}  \notag \\
|1_{\mathcal{Q}},1_{\mathcal{Q}^{\prime }},\frac{1}{2}_{\bar{c}};\frac{3}{2}\rangle _{3} &=&%
\frac{1}{\sqrt{6}}\left( \uparrow \right) _{c}\left( \uparrow \right)
_{q}\{2\left( \uparrow \right) _{q^{\prime }}\left( \uparrow \right)
_{q^{\prime \prime }}\left( \downarrow \right) _{\bar{c}}-[\left( \uparrow
\right) _{q^{\prime }}\left( \downarrow \right) _{q^{\prime \prime }}+\left(
\downarrow \right) _{q^{\prime }}\left( \uparrow \right) _{q^{\prime \prime
}}]\left( \uparrow \right) _{\bar{c}}\}  \notag \\
|1_{\mathcal{Q}},1_{\mathcal{Q}^{\prime }},\frac{1}{2}_{\bar{c}};\frac{3}{2}\rangle _{4} &=&%
\sqrt{\frac{3}{10}}[\left( \uparrow \right) _{c}\left( \downarrow \right)
_{q}+\left( \downarrow \right) _{c}\left( \uparrow \right) _{q}]\left(
\uparrow \right) _{q^{\prime }}\left( \uparrow \right) _{q^{\prime \prime
}}\left( \uparrow \right) _{\bar{c}}-\sqrt{\frac{2}{15}}\left( \uparrow
\right) _{c}\left( \uparrow \right) _{q}\{\left( \uparrow \right)
_{q^{\prime }}\left( \uparrow \right) _{q^{\prime \prime }}\left( \downarrow
\right) _{\bar{c}}  \notag \\
&&+[\left( \uparrow \right) _{q^{\prime }}\left( \downarrow \right)
_{q^{\prime \prime }}+\left( \downarrow \right) _{q^{\prime }}\left(
\uparrow \right) _{q^{\prime \prime }}]\left( \uparrow \right) _{\bar{c}}\},
\label{Spinj3by2}
\end{eqnarray}%
and the spin representation  corresponding  to $J^{P}=\frac{5}{2}^{-}$ state is:
\begin{equation}
|1_{\mathcal{Q}},1_{\mathcal{Q}^{\prime }},\frac{1}{2}_{\bar{c}};\frac{5}{2}\rangle =\left(
\uparrow \right) _{c}\left( \uparrow \right) _{q}\left( \uparrow \right)
_{q^{\prime }}\left( \uparrow \right) _{q^{\prime \prime }}\left( \uparrow
\right) _{\bar{c}}.  \label{spinj5by2}
\end{equation}
The masses for the four $S$-wave pentaquark states with
$J^{P}=\frac{3}{2}^{-}$ and a single state with $J^{P}=\frac{5}{2}^{-}$ 
in terms of the parameters of the effective Hamiltonian 
are given in Table \ref{Tab-Spectrum-1},
 where we
 label the states as $\mathcal {P}_{X_i}$. The corresponding 
five $P$-wave pentaquark states with $L_{\mathcal{P}}=1$ and $J^{P}=\frac{5}{2}^{+}$  are labeled as $\mathcal {P}_{Y_i}$ in
Table  \ref{Tab-Spectrum-1}.
\footnotesize
\begin{table}[tb]
\caption{$S$ ($P$)- wave pentaquark states $\mathcal{P}_{X_{i}}$ ($\mathcal{P}_{Y_{i}}$)
and their spin- and orbital angular momentum quantum numbers. The subscripts  $\mathcal{Q}$ and 
$\mathcal{Q}^{\prime}$ represent the heavy $[cq]$ and light $[q^{\prime}
q^{\prime \prime}]$ diquarks, respectively. In the expressions for the masses of the
$\mathcal{P}_{Y_{i}}$ states, the terms $M_{\mathcal{P}_{X_{i}}}= M_0 + \Delta M_i$ with $i=1,...,5$.}
\vspace*{2mm}
\begin{tabular}{llllllllll}
\hline
Label  &
 $|S_{\mathcal{Q}},S_{\mathcal{Q}^{\prime }};L_{\mathcal{P}},J^{P}\rangle _{i}
$ & Mass & Label & $|S_{\mathcal{Q}},S_{\mathcal{Q}^{\prime }};L_{\mathcal{P}%
},J^{P}\rangle _{i}$ & Mass \\ \hline
$\mathcal{P}_{X_{1}}$ & $|0_{\mathcal{Q}},1_{\mathcal{Q}^{\prime }},0;%
\frac{3}{2}^{-}\rangle _{1}$ & $M_{0}+\Delta M_{1}$ & $\mathcal{P}_{Y_{1}}$
& $|0_{\mathcal{Q}},1_{\mathcal{Q}^{\prime }},1;\frac{5}{2}^{+}\rangle _{1}
$ & $M_{\mathcal{P}_{X_{1}}}+3A_{\mathcal{P}}+B_{\mathcal{P}}$ \\ \hline
$\mathcal{P}_{X_{2}}$ & $|1_{\mathcal{Q}},0_{\mathcal{Q}^{\prime }},0;%
\frac{3}{2}^{-}\rangle _{2}$ & $M_{0}+\Delta M_{2}$ & $\mathcal{P}_{Y_{2}}$
& $|1_{\mathcal{Q}},0_{\mathcal{Q}^{\prime }},1;\frac{5}{2}^{+}\rangle _{2}
$ & $M_{\mathcal{P}_{X_{2}}}+3A_{\mathcal{P}}+B_{\mathcal{P}}$ \\ \hline
$\mathcal{P}_{X_{3}}$ & $|1_{\mathcal{Q}},1_{\mathcal{Q}^{\prime }},0;%
\frac{3}{2}^{-}\rangle _{3}$ & $M_{0}+\Delta M_{3}$ & $\mathcal{P}_{Y_{3}}$
& $|1_{\mathcal{Q}},1_{\mathcal{Q}^{\prime }},1;\frac{5}{2}^{+}\rangle _{3}
$ & $M_{\mathcal{P}_{X_{3}}}+3A_{\mathcal{P}}+B_{\mathcal{P}}$ \\ \hline
$\mathcal{P}_{X_{4}}$ & $|1_{\mathcal{Q}},1_{\mathcal{Q}^{\prime }},0;%
\frac{3}{2}^{-}\rangle _{4}$ & $M_{0}+\Delta M_{4}$ & $\mathcal{P}_{Y_{4}}$
& $|1_{\mathcal{Q}},1_{\mathcal{Q}^{\prime }},1;\frac{5}{2}^{+}\rangle _{4}
$ & $M_{\mathcal{P}_{X_{4}}}+3A_{\mathcal{P}}+B_{\mathcal{P}}$ \\ \hline
$\mathcal{P}_{X_{5}}$ & $|1_{\mathcal{Q}},1_{\mathcal{Q}^{\prime }},0;%
\frac{5}{2}^{-}\rangle _{5}$ & $M_{0}+\Delta M_{5}$ & $\mathcal{P}_{Y_{5}}$ & $|1_{\mathcal{Q}},1_{\mathcal{Q}^{\prime }},%
\frac{1}{2}_{\bar{c}},1;\frac{5}{2}^{+}\rangle _{5}$ & $M_{\mathcal{P}%
_{X_{5}}}-2A_{\mathcal{P}}+B_{\mathcal{P}}$ \\ \hline
\end{tabular}%
\label{Tab-Spectrum-1}
\end{table}
\normalsize
 $\Delta M_{i}$ are defined in~\cite{Ali:2016dkf}, where also the various input parameters are given. The resulting mass spectrum
 of the $S$- and $P$-wave pentaquarks, with $ J^P=(3/2^-, 5/2^-) $ and $ J^P=5/2^+ $, respectively, and having the
quark flavor content  $\bar{c}[cq][qq]$, $\bar{c}[cq][sq]$, 
$\bar{c}[cs][qq]$,  $\bar{c}[cs][sq]$, and $\bar{c}[cq][ss]$, is given in Table \ref{Tab-Spectrum-2}. Later, for
ease of writing, the labels $c_1$,...,$c_5$ will be used for these quark flavor combinations. 
Thus, for each of the $c_i$, the masses of the $S(P)$-wave pentaquark states $X_j(Y_j)$,
$j=1,...,5$  can be read off from this table.
In working out the masses, isospin-symmetry is used in that the small $m_d-m_u $ mass difference is ignored. Thus,
the pentaquark states with the quark content $\bar{c}[cu][ud]$ and $\bar{c}[cd][ud]$ are mass degenerate. These
states will be denoted subsequently by a subscript $P_p$ and $P_n$, respectively, and the notation is such that the
 light-quark content of the pentaquark is represented by the corresponding light baryon.
 
  To make the notation  clear,
 let us consider the decay $\Lambda_b^0 \to P_c(4450)^+ (\to J/\psi\; p)\; K^-$. In our notation, this decay is
 expressed as $\Lambda_b^0 \to P_p^{\{Y_2\}_{c_1}}  (\to J/\psi\; p)\;  K^-$, with the diquark-spin and angular momentum
 quantum numbers given by the entry ${\cal P}_{Y_2}$ in Table \ref {Tab-Spectrum-1}, and its mass is given by the
 entry for $c_1 = \bar{c}[cq][qq]$ ($4450 \pm 57$ MeV ) in Table \ref {Tab-Spectrum-2}.  The isospin-related decay (which is not yet seen)
 in our notation is $\Lambda_b^0 \to P_n^{\{Y_2\}_{c_1}}  (\to J/\psi\; n)\;  \bar{K}^0$, where the pentaquark
 $ P_n^{\{Y_2\}_{c_1}} $ is the neutral partner of $ P_c(4450)^+ $, having $J^P=5/2^+$, and degenerate in mass.
 By isospin, their decay rates are also the same. The masses of the 
 entire $SU(3)_F$ multiplets of pentaquarks with the given $J^P$ quantum numbers and their decays are worked
 out using this notation. 
 \begin{table*}[tbp]
\caption{Masses of the $S$- and $P$-wave pentaquarks, $\mathcal{P}_{X_i}$ and $\mathcal{P}_{Y_i}$ and
having the $J^P$ quantum numbers given in Table \ref{Tab-Spectrum-1}, (in MeV) formed through different diquark-diquark-anti-charm quark combinations in type-I diquark model. The quoted errors are obtained from the uncertainties in the input parameters in the effective Hamiltonian. The light-quark content is given explicitly (with $q=u$ or $d$)
 (From~\cite{Ali:2016dkf}).}
\vspace*{2mm}
\begin{center}
\begin{tabular}{llllll}
\hline
$\mathcal{P}_{X_{i}} \quad$& $\mathcal{P}_{X_{1}}$ & $\mathcal{P}_{X_{2}}$
& $\mathcal{P}_{X_{3}}$ &$\mathcal{P}_{X_{4}} $  &$\mathcal{P}_{X_{5}}
$ \\ 
\hline
 $\bar{c}[cq][qq]$ & $4133\pm55$ & $4133\pm55$ &
$4197\pm55$ & $4385\pm55$ & $4534\pm55$   \\ 
 $\bar{c}[cq][sq]$ & $4115\pm58$ & $4138\pm 47$ & $ 4191\pm 53$ & $4324\pm 47$ & $4478\pm 47$  \\ 
 $\bar{c}[cs][qq]$ & $4365\pm 55$  & $ 4390\pm 42$ &$ 4443\pm 49$  & $4578\pm 43$ & $4727\pm 42$ \\ 
 $\bar{c}[cs][sq]$ & $4313\pm 47$ & $ 4382\pm 45$ & $ 4434\pm 51$ & $4568\pm 46$ & $4721\pm 45$\\ 
 $\bar{c}[cq][ss]$ & $4596\pm 47$ & $ 4664\pm 46$ & $4721\pm 51$ & $4853\pm 46$ & $5006\pm 45$\\ \hline 
$\mathcal{P}_{Y_{i}} \quad$& $\mathcal{P}_{Y_{1}}$ & $\mathcal{P}_{Y_{2}}$
& $\mathcal{P}_{Y_{3}}$ &$\mathcal{P}_{Y_{4}} $  &$\mathcal{P}_{Y_{5}}$ \\ 
\hline
 $\bar{c}[cq][qq]$ & $4450\pm 57 $ & $4450\pm 57$ & $4515\pm 57 $ & $4702\pm 58$ & $4589\pm 56$ \\ 
 $\bar{c}[cq][sq]$ & $ 4432\pm61$  & $4456\pm 50$ & $4508\pm 56$ & $4642\pm 50$ & $4532\pm 48$  \\ 
 $\bar{c}[cs][qq]$ & $ 4682\pm 57$ & $4708\pm 46$ & $4760\pm 52$ & $4895\pm 47$ & $4782\pm 44$  \\ 
$\bar{c}[cs][sq]$ & $4603\pm 51$ & $ 4699\pm 49$ & $ 4752\pm 54$ &$4885\pm 49$ & $4776\pm 47$ \\ 
 $\bar{c}[cq][ss]$ & $4913\pm 51$ & $ 4981\pm 49$ & $5038\pm 54$  & $5170\pm 49$ & $5061\pm 47$ \\ 
\hline
\end{tabular}
\end{center}
\label{Tab-Spectrum-2}
\end{table*}%
In addition to these, also the spectroscopy of the pentaquarks having $ J^P=1/2^\pm $ and the quark flavor content as shown in
Table  \ref{Tab-Spectrum-2} have been worked out in the compact diquark picture. However, as none of these states have so
far been discovered, we restrict the discussion to the  $ J^P=(3/2^-, 5/2^-) $ and $ J^P=5/2^+ $ pentaquarks, since two such
candidates have been observed by the LHCb.
These tables illustrate that the spectrum of pentaquark states in the compact diquark model is very rich. Apart from the other predicted
states, there is a state, $\mathcal{P}_{{X}_4}$, which is predicted to have a mass around 4385 MeV, having the  quantum numbers 
 $|1_{\mathcal{Q}},1_{\mathcal{Q}^{\prime }},0;\frac{3}{2}^{-}\rangle $.  This agrees with the mass of the observed state $P^+_c(4380)$.
  Likewise, the state $P^+_c(4550)$, having $J^P=\frac{5}{2}^+$ can be identified with the state $\mathcal{P}_{{Y}_2}$ in the
second row of Table \ref{Tab-Spectrum-1},
having the quantum numbers $|1_{\mathcal{Q}},0_{\mathcal{Q}^{\prime }},1;\frac{5}{2}^{+}\rangle $.
 We recall that these two states have the same internal quantum numbers as in Ref.~
 Maiani \textit{et al.}~\cite{Maiani:2015vwa}: 
\begin{align}
P_c(4380)^+= \mathbb {P}^+(3/2^-) &= \left\{\bar c\, [cq]_{s=1} [q^\prime q^{\prime \prime}]_{s=1}, L=0\right\},\nonumber 
\\
P_c(4450)^+= \mathbb {P}^+(5/2^+)&=\left\{\bar c\, [cq]_{s=1} [q^\prime q^{\prime \prime}]_{s=0}, L=1\right\}.
\end{align}
\subsection{\boldmath{$b$-baryon decays to pentaquarks and  heavy quark symmetry}}
The pentaquark states reported by the LHCb are produced in  $\Lambda_b^0$ decays, $\Lambda_b^0 \to \P^{+}\;K^{-}$, where $\P$ denotes a generic pentaquark state. QCD has a symmetry in the heavy quark limit, i.e., for
 $m_b \gg \Lambda_{\rm QCD}$, $b$-quark becomes a static color source \cite{Manohar:2000dt}.
 In this limit, the angular momentum of the light degrees of freedom, i.e., of the $[ud]$ diquark, is conserved. Hence,
 the light diquark spin becomes a
 good quantum number, constraining the states which can otherwise be produced in $\Lambda_b$ decays.  
 The $b$-baryon decays to pentaquarks
 having a  $c\bar{c}$ component are also presumably subject to the selection rules following from heavy quark
 symmetry. Thus, the state  $\mathcal{P}_{{X}_4}$ (identified with $P_c(4380)^+$ in
~\cite{Maiani:2015vwa}) is unlikely to be produced in
$\Lambda_b$ decays, as it has the ``wrong'' light-diquark spin number. On the other hand, there is a 
  lower mass state $\mathcal{P}_{{X}_2}$ present in the spectrum, having the correct flavor and spin quantum numbers
  $|1_{Q},0_{Q^{\prime }},0;\frac{3}{2}^{-}\rangle$,  with  a mass of about
  4130 MeV, which we expect to be produced in $\Lambda_b$ decays. One could argue that the mass estimates
  following from the assumed effective Hamiltonian are in error by a larger amount than quoted in~\cite{Ali:2016dkf}.
  However, as already stated, the mass difference
  between the $J^P=\frac{5}{2}^+$ and $J^P=\frac{3}{2}^-$ pentaquarks, having the right quantum numbers 
  $|1_{\mathcal{Q}},0_{\mathcal{Q}^{\prime }},1;\frac{5}{2}^{+}\rangle $ and $|1_{\mathcal{Q}},0_{\mathcal{Q}^{\prime }},0;\frac{3}{2}^{-}\rangle$
   is expected to be around 340 MeV, yielding a mass for
  the  lower-mass $J^P=\frac{3}{2}^-$ pentaquark state of about 4110 MeV. The two estimates are compatible with
  each other, and we advocate to search for this state in the LHCb data. Among the ten states listed in 
  Table \ref{Tab-Spectrum-1}, only the ones called  $\mathcal{P}_{{X}_2}$ and 
 $\mathcal{P}_{{Y}_2}$ are allowed as the $\Lambda_b$ decay products. 
 
%

%
\subsection{\boldmath{Weak decays with $ \mathbb {P}$ in decuplet representation}}
 Decays involving the decuplet ${\tt 10} $ pentaquarks may also occur, if the 
light diquark pair having spin-0 $[ud]_{s=0}$ in $\Lambda_b$ gets broken to produce a spin-1~light
 diquark $[ud]_{s=1}$. In this case, one would also observe the decays of $\Lambda_b$, such as
\begin{align}
\Lambda_b \to \pi \mathbb {P}_{10}^{(S=-1)} & \to \pi(J/\psi \Sigma(1385)),\nonumber\\
\Lambda_b \to K^+ \mathbb {P}_{10}^{(S=-2)} & \to K^+ (J/\psi \Xi^-(1530)). 
\label{eq:decuplets}
\end{align}
These decays are, however, disfavored by the heavy-quark-spin-conservation selection rules.
The extent to which this rule is compatible with the existing data on $B$-meson and $\Lambda_b$ decays can
be seen in the PDG entries. Whether the decays of the pentaquarks are also subject to the same selection rules
is yet to be checked, but on symmetry grounds, we do expect it to hold. Hence, the observation (or not) of these
decays will be quite instructive.

 Apart from the $\Lambda_b(bud)$}, several other $b$-baryons,
such as $\Xi_b^0(usb)$, $\Xi_b^-(dsb)$ and $\Omega_b^-(ssb)$ undergo weak decays. These $b$-baryons are characterized
by the spin of the light diquark, as shown below, making their isospin ($I$) and
strangeness ($S$) quantum numbers explicit as well as their light diquark $J^{\rm P}$ quantum numbers. 
\begin{figure}[b]
\centerline{\includegraphics[width=0.5\textwidth]{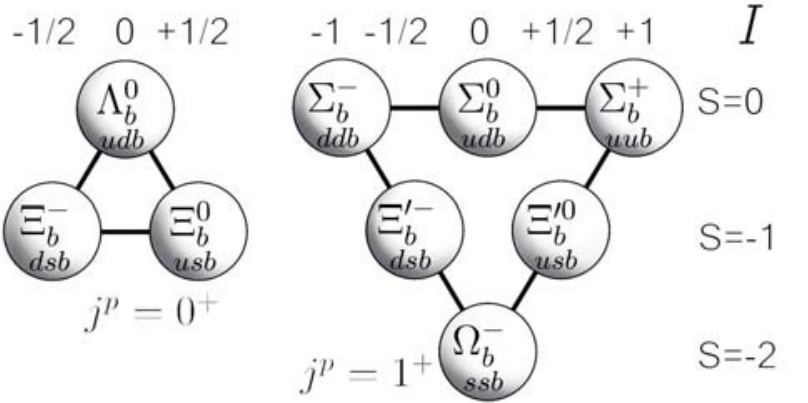}}
\caption{$b$-baryons with the light diquark  spins $J^p=0^+$ (left) and $J^p=1^+$ (right).}
\label{ali:fig10}
\end{figure}
The $c$-baryons are likewise characterized similarly.
Examples of bottom-strange $b$-baryon in various charge combinations,
 respecting $\Delta I=0,\, \Delta S=-1$ are:
 \vspace*{-2mm}
\begin{align}
\Xi_b^0(5794) & \to K (J/\psi \Sigma(1385)),
\label{eq:b-s-baryon}
\end{align}
\vspace*{2mm}
which corresponds to the formation of the pentaquarks with the spin configuration
$ \mathbb {P}_{10} (\bar{c}\, [cq]_{s=0,1}\, [q^\prime s]_{s=0,1})$ 
with $(q,q^\prime=u,d)$.

The above stated considerations have been extended involving the entire $SU(3)_F$ multiplets
entering the generic decay amplitude
$ \langle  {\cal P}  {\cal M}|H_{\rm eff} | {\cal B}\rangle,$
 where ${\cal B}$ is the 
$SU(3)_F$ antitriplet $b$-baryon, shown in the left frame of Fig.~\ref{ali:fig10},
${\cal M}$ is the $3\times 3$ pseudoscalar meson matrix\\
\begin{center}
$\qquad \qquad \mathcal{M}_{i}^{j}=\left( 
\begin{array}{ccc}
\frac{\pi ^{0}}{\sqrt{2}}+\frac{\eta _{8}}{\sqrt{6}} & \pi ^{+} & K^{+} \\ 
\pi ^{-} & -\frac{\pi ^{0}}{\sqrt{2}}+\frac{\eta _{8}}{\sqrt{6}} & K^{0} \\ 
K^{-} & \bar{K}^{0} & -\frac{2\eta _{8}}{\sqrt{6}}%
\end{array}%
\right)$,
\label{eq:meson-matrix}
\end{center}
and ${\cal P}$ is a pentaquark state belonging to an octet with definite
$J^P$, denoted as a $3\times 3$ matrix $J^P$, ${\cal P}^i_j(J^{\rm P})$,\\
\begin{center}
$\qquad \mathcal{P}_{i}^{j}\left( J^{P}\right) =\left( 
\begin{array}{ccc}
\frac{P_{\Sigma ^{0}}}{\sqrt{2}}+\frac{P_{\Lambda }}{\sqrt{6}} & P_{\Sigma
^{+}} & P_{p} \\ 
P_{\Sigma ^{-}} & -\frac{P_{\Sigma ^{0}}}{\sqrt{2}}+\frac{P_{\Lambda }}{%
\sqrt{6}} & P_{n} \\ 
P_{\Xi ^{-}} & P_{\Xi ^{0}} & -\frac{P_{\Lambda }}{\sqrt{6}}%
\end{array}%
\right) ,
\label{eq:penta-matrix}$\\
\end{center}
\vspace*{2mm}
 or a decuplet ${\cal P}_{ijk}$ (symmetric in the indices), with
${\cal P}_{111}= \Delta^{++}_{10},...,{\cal P}_{333}=\Omega_{10}^-$.
(see Guan-Nan Li {\it et al.}~\cite{Li:2015gta} for a detailed list of the component fields
and $SU(3)_F$-based relations among decay widths).
The two observed pentaquarks are denoted as $P_p(3/2^-)$ and $P_p(5/2^+)$.

Estimates of the $SU(3)$ amplitudes require a dynamical model, which will be
lot more complex to develop than the factorization-based models for the two-body $B$-meson
decays, but, as argued in the literature, SU(3) symmetry can be used to relate
different decay modes. Using heavy quark symmetry, which reduces the number of Feynman
diagrams to be calculated, they are worked out in~\cite{Ali:2016dkf}. Thus, the decay
$\Lambda_b^0 \to J/\psi p K^-$ and $\Lambda_b^0 \to J/\psi p \pi^-$ have just one dominant Feynman diagram each,
the one in which the $[ud]$ diquark in $\Lambda_b^0$ retains its spin. The ratio of the branching fraction 
${\cal B}(\Lambda_b^0 \to J/\psi p \pi^-)/{\cal B}(\Lambda_b^0 \to J/\psi p  K^-)=0.0824 \pm 0.0024 \pm 0.0042$~\cite{Aaij:2014zoa}
is consistent with the expectations from Cabibbo suppression. This ratio should also hold for the resonating part of
the amplitudes, namely if one replaces the $J/\psi p$ by $P_c(4450)^+$, and likewise for the $J^P=3/2^-$
$P_c^+$ state. This is hinted by the current LHCb measurements~\cite{Aaij:2016ymb}.

%
\begin{table}[h]
\begin{center}
\caption{{Estimate of the ratios of the decay widths for $b$-baryon decays into hidden-charm pentaquarks $\Gamma(\mathcal{B}(\mathcal{C})\to \mathcal{P}^{5/2}\mathcal{M})/\Gamma(\Lambda^{0}_b \to P_{p}^{5/2}K^{-})$ for $\Delta S=1$ transitions (upper part
of the table) and the Cabibbo-suppressed $\Delta S=0$ transitions (lower part of the table)
(from~\cite{Ali:2016dkf}). }}
\vspace{2mm}
\label{Relative-Rates1}
\begin{tabular}{lclcc}
\hline \\[-4mm] $\Delta S = 1$ & & & \\[1.5mm]
\hline
~Decay Process & $\Gamma/\Gamma(\Lambda^{0}_b \to P_{p}^{5/2}K^{-})$ & Decay Process & $\Gamma/\Gamma(\Lambda^{0}_b \to P_{p}^{5/2}K^{-})$ \\[1.5mm] 
$\Lambda _{b}\rightarrow P_{p}^{\{Y_{2}\}_{c_{1}}}K^{-}$ & $%
1$  &  $\Xi _{b}^{-}\rightarrow P_{\Sigma ^{-}}^{\{Y_{2}\}_{c_{2}}}  \bar{K}^{0}$
& $2.07$  \\[1.5mm] 
$\Lambda _{b}\rightarrow P_{n}^{\{(Y_{2}\}_{c_{1}}}\bar{K}^{0}$ & $%
1$ & $\Xi _{b}^{0}\rightarrow P_{\Sigma ^{+}}^{\{Y_{2}\}_{c_{2}}} K^{-}$ & $%
2.07$ \\[1.5mm] 
$\Lambda _{b}\rightarrow P_{\Lambda^{0}}^{\{Y_{2}\}_{c_{3}}} \eta^{\prime}$ 
& $0.03 $  & $\Lambda _{b}\rightarrow P_{\Lambda^{0}}^{\{Y_{2}\}_{c_{3}}} \eta$ 
& $0.19 $ \\[1.5mm] 
 $\Xi _{b}^{-}\rightarrow P_{\Sigma^{0}}^{\{Y_{2}\}_{c_{2}}}  K^{-}$ & $%
1.04$ & $\Xi _{b}^{-}\rightarrow P_{\Lambda ^{0}}^{\{Y_{2}\}_{c_{2}}}  K^{-}$ & $%
0.34$  \\[1.5mm] 
$\Omega _{b}^{-}\rightarrow P_{\Xi_{10} ^{-}}^{\{Y_{3}\}_{c_{5}}} \bar{K}^{0}$
& $0.14$  & $\Omega _{b}^{-}\rightarrow P_{\Xi_{10} ^{0}}^{\{Y_{3}\}_{c_{5}}} K^{-}$ & $%
0.14$  \\[1.5mm]  
\hline 
$\Delta S = 0$ & & & \\[1.5mm]
\hline
~Decay Process & $\Gamma/\Gamma(\Lambda^{0}_b \to P_{p}^{5/2}K^{-})$ & Decay Process & $\Gamma/\Gamma(\Lambda^{0}_b \to P_{p}^{5/2}K^{-})$ \\[1.5mm] 
\hline
$\Lambda _{b}\rightarrow P_{p}^{\{Y_{2}\}_{c_{1}}} \pi ^{-}$  & $%
0.08$ &$\Lambda _{b}\rightarrow P_{n}^{\{Y_{2}\}_{c_{1}}} \pi ^{0}$ &$0.04  $ \\[1.5mm] 
$\Lambda _{b}\rightarrow P_{n}^{\{Y_{2}\}_{c_{1}}} \eta$  & $%
0.01$ &$\Lambda _{b}\rightarrow P_{n}^{\{Y_{2}\}_{c_{1}}} \eta^{\prime}$ &$0 $ \\[1.5mm] 
 $\Xi _{b}^{-}\rightarrow P_{\Xi ^{-}}^{\{Y_{2}\}_{c_{4}}} K^{0}$ & $0.02$ & $\Xi _{b}^{-}\rightarrow P_{\Sigma ^{0}}^{\{Y_{2}\}_{c_{2}}}  \pi ^{-}$ & $0.08$ \\[1.5mm] 
 $\Xi _{b}^{-}\rightarrow P_{\Sigma ^{-}}^{\{Y_{2}\}_{c_{2}}} \eta$ &$0.02$ &$\Xi _{b}^{-}\rightarrow P_{\Sigma ^{-}}^{\{Y_{2}\}_{c_{2}}} \eta^{\prime}$ &$0.01 $ \\[1.5mm] 
$\Xi _{b}^{-}\rightarrow P_{\Sigma ^{-}}^{\{Y_{2}\}_{c_{2}}} \pi ^{0}$ & $%
0.08$  & $\Xi _{b}^{0}\rightarrow P_{\Sigma ^{0}}^{\{Y_{2}\}_{c_{2}}} \pi ^{0}$ & $%
0.04$ \\[1.5mm] 
$\Xi _{b}^{0}\rightarrow P_{\Lambda ^{0}}^{\{X_{2}\ (Y_{2})\}_{c_{2}}} \eta$  & $0.01$ & $\Xi _{b}^{0}\rightarrow P_{\Lambda ^{0}}^{\{Y_{2}\}_{c_{2}}} \eta^{\prime}$ & $%
0.01$ \\[1.5mm] 
$\Xi _{b}^{0}\rightarrow P_{\Lambda ^{0}}^{\{Y_{2}\}_{c_{2}}} \pi ^{0}$  & $0.01$ & $\Omega _{b}^{-}\rightarrow P_{\Xi_{10} ^{-}}^{\{Y_{3}\}_{c_{5}}} \pi ^{0}$ & $%
0.01$  \\[1.5mm] 
$\Omega _{b}^{-}\rightarrow P_{\Xi_{10}^{0}}^{\{Y_{3}\}_{c_{5}}} \pi ^{-}$ & $%
0.02$ & &\\ \hline 
\end{tabular}%
\end{center}
\end{table}

Examples of the weak decays in which the initial $b$-baryon has a spin-1 light diquark, i.e. $J^{\rm P}=1^+$,
which is retained in the transition, are provided by the $\Omega_b$ decays.
The $s\bar{s}$ pair in $\Omega_b$ is in the symmetric ${\tt 6}$ representation
of $SU(3)_F$ with spin 1 and is expected to produce decuplet pentaquarks in association
with a $\phi$ or a kaon~\cite{Maiani:2015vwa} 
\begin{align}
\Omega_b(6049) & \to \phi (J/\psi\, \Omega^-(1672)), K (J/\psi\, \Xi(1387)).
\label{eq:omegab}
\end{align}
These correspond, respectively, to the formation of the following pentaquarks ($q=u,\, d $):
\begin{align}
&\mathbb {P}^-_{10} (\bar{c}\, [cs]_{s=0,1}\, [ss]_{s=1}),
\mathbb {P}_{10} (\bar{c}\, [cq]_{s=0,1}\, [ss]_{s=1}).
\label{eq:omegab-diquark}
\end{align}
These transitions are expected on firmer theoretical footings, as the initial $[ss]$ diquark
in $\Omega_b$~is left unbroken. Again, a lot  more transitions can be found relaxing this condition, which
would involve a $J^{\rm P}=1^+ \to 0^+$ light diquark, but they are anticipated to be suppressed. 

The ratios of $\Gamma(\mathcal{B}(\mathcal{C})\to \mathcal{P}^{5/2}\mathcal{M})/\Gamma(\Lambda^{0}_b \to P_{p}^{5/2}K^{-})$ for $\Delta S=1$  and the Cabibbo-suppressed $\Delta S=0$ transitions are given in
Table \ref{Relative-Rates1}. The suppression factor is $(V_{cd}/V_{cs})^2 $. Note that the pentaquark state $P_p^{5/2}$
denotes the state $P_c(4450)^+$ with $J^P=5/2^+$.  The corresponding ratios involving the $J^P=3/2^-$ pentaquark
states are given in Table \ref{Relative-Rates2}.

\begin{table}[h]
\begin{center}
\caption{ {Estimate of the ratios of the decay widths  
$\Gamma(\mathcal{B}(\mathcal{C}) \to \mathcal{P}^{3/2}\mathcal{M})/\Gamma(\Lambda^{0}_b \to P_{p}^{{\{X_2\}}_{c_1}}K^{-})$ for $\Delta S=1$ transitions (upper part
of the table) and the Cabibbo-suppressed $\Delta S=0$ transitions (lower part of the table)
 }(from~\cite{Ali:2016dkf}). }
\label{Relative-Rates2}
\vspace{3mm}
\begin{tabular}{ccccc}
\hline 
$\Delta S = 1$ & & & \\[1.5mm] 
\hline
Decay Process & $\Gamma/\Gamma(\Lambda^{0}_b \to P_{p}^{{\{X_2\}}_{c_1}K^{-}})$ & Decay Process & $\Gamma/\Gamma(\Lambda^{0}_b \to P_{p}^{{\{X_2\}}_{c_1}}K^{-})$ \\[1.5mm] 
$\Lambda _{b}\rightarrow P_{p}^{\{X_{2}\}_{c_{1}}}K^{-}$ & $%
1$  &  $\Xi _{b}^{-}\rightarrow P_{\Sigma ^{-}}^{\{X_{2}\}_{c_{2}}}  \bar{K}^{0}$
& $1.38 $ \\[1.5mm]
$\Lambda _{b}\rightarrow P_{n}^{\{X_{2}\}_{c_{1}}}\bar{K}^{0}$ & $%
1 $ & $\Xi _{b}^{0}\rightarrow P_{\Sigma ^{+}}^{\{X_{2}\}_{c_{2}}} K^{-}$ & $%
1.38$ \\[1.5mm]
$\Lambda _{b}\rightarrow P_{\Lambda^{0}}^{\{X_{2}\}_{c_{3}}} \eta^{\prime}$ 
& $0.17 $  & $\Lambda _{b}\rightarrow P_{\Lambda^{0}}^{\{X_{2}\}_{c_{3}}} \eta$ 
& $0.22 $ \\[1.5mm]
 $\Xi _{b}^{-}\rightarrow P_{\Sigma^{0}}^{\{X_{2}\}_{c_{2}}}  K^{-}$ & $%
0.69$ & $\Xi _{b}^{-}\rightarrow P_{\Lambda ^{0}}^{\{X_{2}\}_{c_{2}}}  K^{-}$ & $%
0.23$ \\[1.5mm]
$\Omega _{b}^{-}\rightarrow P_{\Xi_{10} ^{-}}^{\{X_{3}\}_{c_{5}}} \bar{K}^{0}$
& $0.24$  & $\Omega _{b}^{-}\rightarrow P_{\Xi_{10} ^{0}}^{\{X_{3}\}_{c_{5}}} K^{-}$ & $%
0.24$  \\[1.5mm]
\hline 
$\Delta S = 0$ & & & \\[1.5mm]
\hline
Decay Process & $\Gamma/\Gamma(\Lambda^{0}_b \to P_{p}^{{\{X_2\}}_{c_1}K^{-}})$ & Decay Process & $\Gamma/\Gamma(\Lambda^{0}_b \to P_{p}^{{\{X_2\}}_{c_1}}K^{-})$ \\[1.5mm]
$\Lambda _{b}\rightarrow P_{p}^{\{X_{2}\}_{c_{1}}} \pi ^{-}$  & $%
0.06$ &$\Lambda _{b}\rightarrow P_{n}^{\{X_{2}\}_{c_{1}}} \pi ^{0}$ &$0.03  $ \\[1.5mm]
$\Lambda _{b}\rightarrow P_{n}^{\{X_{2}\}_{c_{1}}} \eta$  & $%
0.01 $ &$\Lambda _{b}\rightarrow P_{n}^{\{X_{2}\}_{c_{1}}} \eta^{\prime}$ &$0.01 $ \\[1.5mm]
 $\Xi _{b}^{-}\rightarrow P_{\Xi ^{-}}^{\{X_{2}\}_{c_{4}}} K^{0}$ & $0.02$ & $\Xi _{b}^{-}\rightarrow P_{\Sigma ^{0}}^{\{X_{2}\}_{c_{2}}}  \pi ^{-}$ & $0.03$ \\[1.5mm]
 $\Xi _{b}^{-}\rightarrow P_{\Sigma ^{-}}^{\{X_{2}\}_{c_{2}}} \eta$ &$0.02$ &$\Xi _{b}^{-}\rightarrow P_{\Sigma ^{-}}^{\{X_{2}\}_{c_{2}}} \eta^{\prime}$ &$0.01 $ \\[1.5mm]
$\Xi _{b}^{-}\rightarrow P_{\Sigma ^{-}}^{\{X_{2}\}_{c_{2}}} \pi ^{0}$ & $%
0.04$  & $\Xi _{b}^{0}\rightarrow P_{\Sigma ^{0}}^{\{X_{2}\}_{c_{2}}} \pi ^{0}$ & $%
0.02$ \\[1.5mm]
$\Xi _{b}^{0}\rightarrow P_{\Lambda ^{0}}^{\{X_{2}\}_{c_{2}}} \eta$  & $0$ & $\Xi _{b}^{0}\rightarrow P_{\Lambda ^{0}}^{\{X_{2}\}_{c_{2}}} \eta^{\prime}$ & $%
0$ \\[1.5mm]
$\Xi _{b}^{0}\rightarrow P_{\Lambda ^{0}}^{\{X_{2}\}_{c_{2}}} \pi ^{0}$  & $0.01$ & $\Omega _{b}^{-}\rightarrow P_{\Xi_{10} ^{-}}^{\{X_{3}\}_{c_{5}}} \pi ^{0}$ & $%
0.01$ \\[1.5mm]
$\Omega _{b}^{-}\rightarrow P_{\Xi_{10}^{0}}^{\{X_{3}\}_{c_{5}}} \pi ^{-}$ & $%
0.02 $ & &\\ \hline 
\end{tabular}%
\end{center}
\end{table}

\section{Summary}
In summary, with the discoveries of the $X,Y,Z$ and $P_c$ states a new era of hadron spectroscopy 
is upon us. In addition to the well-known  $q\bar{q}$ mesons and $qqq$ baryons,
there is increasing evidence that the hadronic world is multi-layered, in the form of tetraquark mesons,
pentaquark baryons, and likely also the hexaquarks (or $H$ dibaryons)~\cite{Maiani:2015iaa}.
 However, the underlying dynamics is far from being understood, and the real issue is how the
various constituents of an exotic multiquark state rearrange themselves. The two competing pictures
are hadron molecules and compact diquak models, with $Q\bar{Q}g$ hybrids and glueballs also anticipated.
 Thresholds near the resonances do play
a role in the phenomenology, and in some cases kinematic-induced cusp effects may also be a viable template.
It is plausible, perhaps rather likely, that no single mechanism fits all the observable states,
and the exotic hadrons  may find their abode in competing theoretical frameworks.
The case of diquark models in this context was reviewed here in more detail.
 Existence proof on the lattice of diquark correlations in some of the tetra- and pentaquark
states discussed  here would be a breakthrough and keenly awaited.   
In the meanwhile, phenomenological models built within constrained theoretical frameworks are 
unavoidable. They and experiments will guide us how to navigate through this uncharted territory.

%% file: Sec-Acknowledgments.tex
We thank  Luciano Maiani,  Christoph Hanhart, Antonello Polosa and Gerrit Schierholz for helpful discussions. Our colleagues in the Belle and LHCb experiments provided a great deal of input and useful discussions. S. Stone thanks the U. S. National Science Foundation for support.